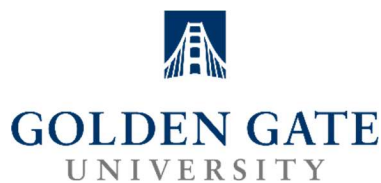

**Mitigate or Fail: How Risk Management Shapes Cybersecurity Competency**

*https://doi.org/10.31237/osf.io/rf8xj_v1*

Jeffrey T. Gardiner, MBA, Doctoral Candidate

Doctor of Business Administration (DBA)
Dissertation

Submitted to:
***Edward S. Ageno School of Business***
***Golden Gate University***
*San Francisco, CA*

March 20, 2026

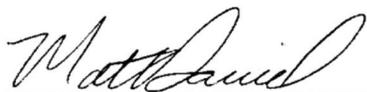

[Chair: Dr. Matt Daniel]

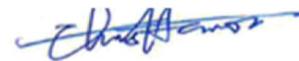

[Co-Chair**:** Dr. Christian Harrison]







**Contents**













**List of Tables**









## List of Figures





**Dedication**

*For Angie, who never stopped believing, and for my daughters, who are always in my heart.*



## Acknowledgements

I must acknowledge that this dissertation was completed in the shadow of a cancer diagnosis received midway through the writing process. Treatment delayed data collection by approximately 6 months.

I am deeply grateful to my dissertation Chair, Dr. Matt Daniel, whose patience, thoughtful feedback, and personal encouragement — particularly during my illness — made the completion of this work possible. To my Co-Chair, Dr. Christian Harrison, thank you for your consistent support and scholarly guidance throughout this journey.

Special recognition is owed to my friend and colleague Austin Page, who contributed generously to the technical development of this study. His generous investment of time in reviewing and improving my NLP code was indispensable to the competency analysis at the heart of this study. The NLP analysis of the NICE cybersecurity framework would not have been achievable without his involvement.

To my fiancée, Angie Lafontaine: You were never entirely convinced this undertaking was sensible, and perhaps you were right. Yet your advocacy never wavered. You stood with me through every chapter, every revision, and every setback — including the most significant one. Your curiosity about this work and your willingness to listen, even when the subject matter was far removed from everyday life, consistently reminded me of why the work mattered. I am profoundly fortunate to have you beside me.

 Finally, that I was able to resume and ultimately finish this work at all is a testament not only to the support of those named above, but to a restoration I attribute to God alone. It is to Him that I owe the strength that carried me through, and it is to His glory that I dedicate the completion of this effort.

*Soli Deo Gloria.*



# Abstract

Contemporary cybersecurity governance assumes that professionals apply formal risk-exposure reasoning. Yet organizational failures persist despite substantial technical investment in tools, staff and credentialing. This study investigates the structural origin of that paradox. The findings suggest that cybersecurity speaks the language of risk, but its structural training has shaped it to think in terms of threats. The two are not the same. A sequential mixed-methods design integrated four independent analyses: semantic similarity-based Natural Language Processing (NLP) applied to the NIST NICE Framework v2.0.0 (2,111 TKS statements); Structural Equation Modelling (SEM; n = 126 cybersecurity professionals); a control group comparison (n = 133 general professionals); and thematic coding of seven senior cybersecurity leadership interviews. Four convergent findings emerged. First, NLP analysis found that "likelihood" and "probability" (necessary ingredients for gauging risk) each appear zero times across 2,111 TKS statements; risk management content accounts for only 4.5% of high-confidence semantic classifications, ranking 18th of 29 competency domains. NICE codifies threat-management operations while primarily invoking risk vocabulary at the category level, indicating a framework oriented toward threat management rather than formal risk analysis. Second, SEM confirmed that training exposure significantly predicts risk management competence both directly ($\beta$ = .406, p < .001) and indirectly through conceptual salience ($\beta$ = .223, p < .001), yielding a total effect of $\beta$ = .629. However, the theoretically four-dimensional risk competency construct collapsed into a single undifferentiated factor (a phenomenon this study terms epistemic compression), demonstrating that practitioners internalize the framework's cognitive architecture. Third, cybersecurity professionals demonstrated no measurable advantage over the general professional population in foundational risk reasoning (Cohen's d = 0.16, p = .205); only 11.9% achieved high differentiation. Fourth, all seven senior leaders expect their teams to apply Likelihood × Impact risk calculus, yet five did not articulate the formula they require of others. These findings converge on a single structural conclusion: cybersecurity has taken on a professional form as a threat management discipline, adopting borrowed risk vocabulary. The study advances a three-level structural explanation (Training Architecture → Cognitive Internalization → Organizational Consequence) and concludes that effective remediation requires fundamental redesign of professional formation, not curriculum reform at the margins.

**Chapter 1 – Background & Problem**

In the modern digital world, organizations and individuals face an ever-expanding cybersecurity threat landscape marked by rapidly advancing technologies and increasing interconnectivity (Deloitte, 2019). Cyberattacks have become a persistent and evolving threat, no longer confined to sporadic incidents but escalating into systematic and sophisticated attacks (Choo, 2011) that target a wide range of sectors, including small businesses, large multinational corporations, and government agencies (Hess, 2021; IBM, 2024a). These attacks are frequent and devastating, with large-scale data breaches becoming an almost daily occurrence.  High-profile cyber incidents, such as the 2021 Colonial Pipeline attack, demonstrate the catastrophic consequences of cybersecurity failures (Easterly, 2023). Organizations across sectors face a growing need for cybersecurity professionals to defend against evolving threats (Crumpler & Lewis, 2019; Deloitte, 2019).

According to IBM's 2024 Cost of a Data Breach Report, the global average data breach cost surged by 10% over the previous year, reaching an alarming USD 4.88 million (IBM, 2024a). Dreyer et al. (2018) found that global cost estimations of the direct cost of cybercrime, though highly sensitive to input parameters, ranged from $275 billion to $6.6 trillion, and total GDP costs ranged from $799 billion to $22.5 trillion. The costs of these breaches are staggering.

## 1.1    Introduction

These figures reflect a broader pattern of financial and operational damage: breaching just 1 million records can cost companies millions in remediation efforts, legal fees, and regulatory fines. (IBM, 2024a) reported that the average cost of a Canadian data breach in 2024 was CAD 7 million.  When small and medium-sized businesses fail, 82% of the time, it is due to cash-flow issues (Hagen & Sutter, 2019; Vance, 2023).  Cybercrime Magazine reports that 60% of small and medium-sized businesses (SMBs)



that experience a cyberattack close within 6 months, primarily due to financial strain and operational disruption (Johnson III, 2018). Cybersecurity failures stemming from data breaches pose a non-negligible threat to businesses, as their financial impacts can cause cash flow problems. Cybersecurity failures, at their core, stem from technical risk-management failures within the field. For instance, the Equifax breach exemplified a failure to implement effective risk management strategies despite employing advanced technical measures (Fruhlinger, 2020). The intricate dance of digital defence is not merely about identifying threats and repelling intrusions—it is about foresight, anticipation, and the meticulous orchestration of risk mitigation strategies before calamity strikes.

This framing is not merely an academic proposition. The United States federal government's authoritative guidance on information security explicitly establishes this relationship as a matter of policy. The National Institute of Standards and Technology (NIST) Special Publication 800-12 Revision 1, An Introduction to Information Security (2017), states that security controls "must be commensurate with the risk and magnitude of harm resulting from unauthorized access, use, disclosure, disruption, modification, or destruction of information" (NIST, 2017, Section 2.5). This principle — that every security control decision is a calibrated response to a specific level of risk — treats cybersecurity not as a technical discipline that occasionally interfaces with risk, but as an applied risk management activity in which risk-commensurate judgment is the primary professional obligation. The leading federal cybersecurity authority established this normative standard; it serves as the benchmark for measuring the contemporary cybersecurity workforce's competency profile.

While many might argue that smaller businesses are particularly vulnerable due to limited resources, the reality is that even the most well-funded companies are not immune. High-profile breaches involving large, multinational organizations—such as Yahoo, Equifax, and Facebook—demonstrate that even substantial cybersecurity budgets and personnel do not guarantee immunity



from attack. These examples underscore a critical issue: despite significant investments in technical security measures, many companies fail to manage cyber risks effectively, leading to costly breaches that damage their reputations and financial stability. Why? An illustrative analogy helps clarify this paradox. These companies are like owners of advanced surgical robots who lack medical training. Although they possess highly sophisticated tools, without foundational knowledge of anatomy, diagnosis, and systemic function, they are ill-equipped to recognize underlying conditions or respond effectively when complications arise. Similarly, businesses may deploy advanced cybersecurity technologies without understanding how risk propagates across the enterprise or how to diagnose and prioritize cyber risk in business terms.

### 1.1.1   *Setting the Stage: The Growing Cybersecurity Threat Landscape*

This persistent technical failure highlights a deeper question: What is the relationship between a successful cybersecurity program and the essential risk management competency? Effective cybersecurity is not merely about implementing technical controls (Antonucci, 2017; Hess, 2021; Parekh et al., 2018; Wilkinson, 2020); it is fundamentally about risk-based decision-making (Braumann, 2018; Stine et al., 2020; Wangen, 2016; Webb et al., 2014; Wilkinson, 2020), which includes identifying, assessing, quantifying, and mitigating risks in alignment with an organization's broader business objectives. Yet, many cybersecurity training programs emphasize technical skills—such as cryptography, penetration testing, and network security (Chowdhury & Gkioulos, 2021; Shreeve et al., 2021).  Risk management is a linchpin among the multifaceted skill sets essential to cybersecurity, often overlooked yet fundamentally indispensable (Braumann, 2018). An effective cybersecurity posture demands more than reactive measures; it necessitates a proactive, calculated approach that continuously assesses, fortifies, and pre-empts technical risks before they metastasize into full-blown crises. Contrary to the



established risk doctrine across leading enterprise frameworks, which defines risk as the interaction of likelihood and impact (Li et al., 2020), the profession has not converged on this formulation in practice— a gap with direct implications for training design.

Moreover, as the digital landscape becomes more complex, companies must move beyond technical defences and embrace strategic, risk-based cybersecurity approaches. Just as a child first learns to crawl, stand, walk, and eventually run, they learn to manage risk ("Do not chase the ball on the road!"). Risk management development follows initial skill development as the more complex aspects of cybersecurity become consequential. Information-age companies have access to advanced technology but often lack the foundational risk-management knowledge needed to navigate the complexities of today's cyber threats (Goode, 2018). This deficiency is evident in the news (Schall & Oni, 2019). More shockingly, Oltramari and Kott (2018) found that cybersecurity practitioners do not merely define risk differently; they operationalize an alternative construct. In their study, practitioners structured 78% of risk descriptions as an adversarial tuple: a description of the system state, the vulnerability present, the exploit available, the available follow-on exploitation options, and potential counteractions — with no reference to the probability of occurrence or the magnitude of loss. This formulation is not a variant of expected-loss reasoning; it contains no mechanism for cross-risk prioritization, no basis for communicating residual exposure to organizational leadership, and no logic for situating cybersecurity risk within an enterprise risk register. The profession has not merely adopted imprecise terminology — it has replaced a governance-compatible risk ontology with an adversarial-state frame that, while operationally rational at the tactical level, is structurally incommensurable with the enterprise risk functions cybersecurity professionals are increasingly expected to perform.

More alarmingly, this definitional ambiguity is not limited to the concept of risk. Craigen et al. (2014) identified a fundamental lack of uniformity in how cybersecurity is defined, documenting nine



distinct, often subjective definitions in active use. As they conclude, "there is a fundamental lack of uniformity when it comes to defining cybersecurity. A cybersecurity definition is genuinely in the eye of the beholder." The absence of a shared definitional foundation reinforces disciplinary fragmentation and may impede the development of a coherent risk-management knowledge base.

The leading risk frameworks, including NIST risk management Framework (2011), Factor Analysis of Information Risk (FAIR)(Jones, 2006), Committee of Sponsoring Organizations of the Treadway Commission's Enterprise risk management (COSO ERM)(2004), and International Standards Organization (ISO-31000) (2018), all define 'risk' as the expected value of loss — specifically, the combination (typically the product) of the potential impact (severity) and the likelihood (probability) of a loss event occurring. Notably, this framing of risk as the interaction of impact and likelihood is not merely a contemporary managerial convention but a historically grounded definition. As Li et al. (2020) explain, in the risk management and actuarial literature, "risk describes a loss of a certain magnitude weighted by the probability of its occurrence," raising questions about whether cybersecurity education meaningfully operationalizes this well-established conception or merely invokes its terminology. As a result of this misconception, cybersecurity professionals are often ill-equipped to recognize, prioritize, and address cyber risks effectively, leaving their organizations vulnerable to attack.

Importantly, this **L x I** definition of risk transcends academic consensus; it serves as the operational foundation for the federal risk management architecture that legally situates cybersecurity practice. NIST Special Publication 800-39, Managing Information Security Risk (2011), states that "risk management is a comprehensive process that requires organizations to: (i) frame risk, (ii) assess risk, (iii) respond to risk, and (iv) monitor risk on an ongoing basis" and explicitly frames this as "a comprehensive, organization-wide activity" in which risk-based decision-making is "integrated into every aspect of the organization" (NIST, 2011, p. 7). Complementarily, NIST Special Publication 800-37 Revision



2, Risk Management Framework for Information Systems and Organizations (2018b), operationalizes these principles through a structured process that "integrates security, privacy, and cyber supply chain risk management activities" across the organizational life cycle (NIST, 2018b, p. 1). Taken together, these documents establish that the normative architecture already treats cybersecurity as an applied specialization of risk management, not a parallel technical discipline but a subordinate, functionally integrated component of enterprise-wide risk governance. This dissertation does not ask whether cybersecurity constitutes risk management; instead, it investigates why the workforce fails to operationalize a standard already defined by authoritative policy.

Executives and board members who lack deep immersion in the day-to-day realities and technical nuances of emerging risks sharpen the edge of this crisis when they stumble through their roles without a basic grasp of risk management. Khansa and Liginlal (2007) have argued that regulatory compliance pressures have moved information security into corporate boardrooms, thus providing economic justification for information security firms to innovate. Even so, a study by the Harvard Law School Forum on Corporate Governance found that most corporate boards fail to comprehend the technical risks associated with cybersecurity, often underestimating the critical role technology plays in creating value (Sumner et al., 2024). This lack of awareness at the board level contributes to the overall failure to implement effective risk management strategies, leaving organizations vulnerable to catastrophic breaches (Ormazabal et al., 2010; Sumner et al., 2024). The question remains whether training programs adequately prepare cybersecurity professionals to apply Risk management principles in practice.

This dissertation addresses this gap by examining how extensively cybersecurity training programs integrate Risk management principles and whether those integrations foster workplace competency. Specifically, it explored whether these programs equip professionals with the



competencies to manage cyber risks in real-world settings. By systematically analyzing the presence and depth of risk management content in cybersecurity curricula, this research sought to provide actionable insights into how training programs can be improved to better prepare professionals for the challenges posed by today's dynamic cybersecurity landscape.

### 1.1.2   Considerations Driving this Investigation

Four principal considerations drive this investigation:

**I.     Is There Sufficient Empirical Evidence about the Inclusion of Risk Management in Cybersecurity Training Programs?**

Although tens of thousands of cybersecurity training programs exist—from academic courses to industry certifications—there is little empirical evidence demonstrating that these programs effectively incorporate and convey risk management concepts (Švábenský et al., 2020).  Some researchers in the field emphasize the need for more empirical evidence. (Baiden, 2024; Goode, 2018; Hess, 2021)Most research focuses on overarching technical cybersecurity competencies rather than evaluating the comprehensiveness of risk management instruction in training materials.

**II.     Is there an Excessive Focus on Technical Proficiencies to the Detriment of Strategic Cognition?**

A cursory review of thousands of cybersecurity training courses and research by Parekh et al. (2018)  suggests that such programs emphasize technical proficiency rather than strategic, risk-based decision-making (Goode, 2016). Oltramari and Kott (2018) find that professionals focus on system configurations and vulnerabilities rather than on risk. Webb et al. (2014) suggest that organizations perform information security risk identification perfunctorily, making it one of the top six reasons cybersecurity efforts fail. This disparity may lead to the emergence of professionals with advanced



expertise in implementing security controls, but who cannot perform fundamental risk assessments or formulate proactive mitigation strategies. In such circumstances, security may hinder the implementation of business information technology.

### III.    Is There a Sufficient Systematic Evaluation of Training Content?

Systematic research has not evaluated the extent to which cybersecurity training programs incorporate risk management or whether cybersecurity practitioners use those skills in practice.  Cabaj et al. (2018) analyze cybersecurity master's programs to identify the topics covered and their distribution across courses, but only to map that distribution. Wangen's (2016) findings strongly imply that a systematic investigation into the relevance and representativeness of existing training content and practices in the field is long overdue. This research employed Natural Language Processing (NLP) techniques to quantitatively assess the prevalence and adequacy of risk management concepts in cybersecurity training curricula, building on the work by Schatz et al. (2017).  This approach is not new (Petersen et al., 2020; Piesarskas et al., 2019). Pandey et al. (2017a) provide an example of NLP use in an educational context.

### IV.    There Appears to be Ambiguity in Training Development Between Training Programs and Workforce Requirements.

Workforce competency frameworks, including the NIST NICE, the European Union's ECSF, and the SPARTA Cybersecurity Skills Framework, outline the essential skills required for various cybersecurity roles (NICCS, 2024; Petersen et al., 2020; Piesarskas et al., 2019). Nevertheless, given the evident lack of skills in the workforce, it remains unclear whether training programs adequately equip professionals with the risk management competencies inherent to these positions. This research investigated competency deficiencies by aligning training content with workforce expectations.



### *1.1.3    Purpose and Objectives of the Study*

This dissertation examined how effectively cybersecurity training programs integrate risk management principles and how risk management appears in practice. Specifically, it sought to determine whether these programs equip professionals with the necessary functional skills to identify, assess, mitigate, and respond to cyber threats.

This research addressed the following key objectives.  It:

1. Assessed the extent to which cybersecurity training programs incorporate risk management principles.

2. Determined the impact of risk management training on cybersecurity professionals' ability to manage cyber risks.

3. Compared the risk management competencies of cybersecurity-trained professionals with those of untrained individuals.

4. Identified potential gaps in risk management instruction across various cybersecurity education frameworks.

5. Provided recommendations for improving the integration of risk management concepts in cybersecurity curricula.

### *1.1.4    Research Approach and Methodology*

This study adopted a mixed-methods approach to evaluate the effectiveness of cybersecurity training in risk management education (Creswell, 2014; Teddlie & Tashakkori, 2009). I present a more detailed rationale for selecting a mixed-methods approach in Chapter 3 (Methodology).  Even so, to outline the general approach, this approach employs:



1. **Natural Language Processing (NLP) Analysis.** This analysis applied NLP techniques to cybersecurity training frameworks (specifically NIST NICE) to quantify the extent to which risk management concepts appear.  This approach is common in management research (Kang et al., 2020; Mavrogiorgos et al., 2022a). This analysis analyzed knowledge statements, tasks, and competency areas (KSAs) outlined in the NIST NICE, ECSF, and SPARTA frameworks.

2. **Qualitative Insights from Cybersecurity Professionals.** I surveyed cybersecurity professionals and interviewed executive leaders (and other managers of cybersecurity teams) to evaluate how working professionals apply risk management principles in practice and to understand leadership expectations of cybersecurity professionals' capabilities. These data sources provided insights into the extent to which cybersecurity training translates into practical, risk-based decision-making (Creswell, 2014) and gauge leadership expectations.

The selection of NLP as an analytical method is grounded in its demonstrated utility for educational content analysis, as discussed in Pandey et al. (2017b), since its use was similar to that described in Mavrogiorgos et al. (2022a). These studies illustrate how NLP can identify patterns and themes related to professional competencies, making it a suitable technique for assessing the presence of risk management concepts in training programs.  This approach equips us to systematically evaluate whether existing training programs adequately prepare professionals for risk-based decision-making in cybersecurity operations by combining quantitative content analysis with qualitative professional insights (Mavrogiorgos et al., 2022b; Pandey et al., 2017b).

### 1.1.5   *Expected Contributions*



Research, such as that by Baiden (2024) and Wilkinson (2020), has identified the critical role of risk management in enhancing organizational resilience and improving the performance of cybersecurity programs. Hess (2021) notes that different organizations often lack a consistent understanding of cybersecurity goals. Similarly, Dawson and Thomson (2018) advocate a more holistic approach to cybersecurity workforce development, extending beyond technical skills to encompass broader aspects of the field. They suggest that defining the necessary knowledge, skills, and attributes for a successful cyber workforce is more complex than simply listing technical skills. Shreeve et al. (2021) point out that employers increasingly need judgment, decision-making, and contextual understanding. Yet, according to Oltramari and Kott (2018), cybersecurity professionals focus on threats rather than risks (see also Libicki et al. (2015)). Does this stem from training?

Therefore, this research contributed to the fields of cybersecurity education, workforce development, and organizational resilience by:

a) Providing empirical evidence on the effectiveness of cybersecurity in risk management education.

b) Identifying curriculum gaps that may hinder professionals' ability to conduct cyber-risk-related responsibilities.

c) Informing policymakers and educators on improving cybersecurity workforce competencies and outcomes using risk frameworks.

d) Enhancing workforce readiness by aligning training programs with industry risk management expectations in cybersecurity.

Understanding whether cybersecurity training effectively incorporates risk management is crucial for enhancing cyber resilience at both the individual and organizational levels. This research helped bridge the gap between theoretical cybersecurity education and the practical skills necessary to



manage cyber threats effectively in real-world scenarios. Enhancing risk management training not only equips cybersecurity professionals with the skills to identify and mitigate threats but also empowers organizations to develop proactive, strategic security measures that reduce vulnerabilities and improve overall cyber resilience (Baiden, 2024; Dawson & Thomson, 2018; Goode, 2018; Hess, 2021; Shreeve et al., 2021; Wilkinson, 2020).

Recognizing cybersecurity as both a technical and leadership challenge, this research bridges the gap between technical training and strategic decision-making. This study sought to contribute to cybersecurity workforce development and executive corporate risk governance strategies by examining risk-based decision-making as a core competency in cybersecurity. The dissertation consists of five chapters. Chapter 1 introduces the research problem, the study's purpose, and its significance. Chapter 2 reviews and synthesizes the relevant literature on cybersecurity training, workforce competency frameworks, and risk management. Chapter 3 outlines the research methodology, including the study design, data collection, and analysis techniques. Chapter 4 presents the results of the empirical analyses, and Chapter 5 discusses the findings, implications, limitations, and recommendations for future research.

## 1.2    Background

This Section provides background for this research by examining the evolution of cybersecurity training, the role of risk management in cybersecurity, and the challenges of aligning cybersecurity education with industry needs.

### 1.2.1    The Evolution of Cybersecurity Training

Cybersecurity education and training have undergone a dramatic transformation in lockstep with the relentless escalation of cyber threats (Crumpler & Lewis, 2019). Initially, cybersecurity training



focused heavily on technical skills like encryption, firewalls, intrusion detection systems, and penetration testing. These skills were vital as organizations faced relatively straightforward cyber threats. However, as the digital ecosystem has become increasingly complex, so too have the threats that organizations face (Chowdhury & Gkioulos, 2021; Deloitte, 2022). Today's cyberattacks are multifaceted, targeting systems, networks, and the underlying business processes and human behaviours that drive them (Hess, 2021; IBM, 2024a; Piesarskas et al., 2019).

While cybersecurity training has evolved to address more complex threats, many programs prioritize technical proficiency over a strategic, risk-based approach. Shreeve et al. (2021) observe that cybersecurity education disproportionately emphasizes technical capability and that existing curricula do not align with real-world organizational needs. The need for technical skills remains undeniable; however, it is becoming increasingly clear that future cybersecurity professionals must also think critically about risk—identifying, assessing, quantifying, justifying control choice, and developing strategies to mitigate it. The frequent reports of high-profile breaches suggest that companies are failing at risk management. Accordingly, this study assumes that a critical gap exists in many cybersecurity curricula, particularly in industry certifications and academic programs, where risk management concepts are sometimes included but not adequately integrated into training.

While professionals are often prepared to implement security controls, questions remain about whether they receive systematic training in risk assessment, risk-based prioritization, and strategic decision-making—critical components of an effective cybersecurity strategy. Risk management controls are only adequate if they match the risks they aim to address. Should training programs fail to integrate risk management principles fully, many cybersecurity professionals will remain alarmingly ill-equipped to address the full spectrum of modern threats, thereby leaving organizations perilously vulnerable to high-profile breaches.



### 1.2.2    The Importance of Risk Management in Cybersecurity

Risk management is central to effective cybersecurity. Baiden (2024) found that the use of risk management frameworks was the dominant mediator—far surpassing all other factors—in enabling robust implementation of cybersecurity strategy. This final point underscores the paramount importance of structured risk management for achieving superior cybersecurity outcomes. Baiden (2024) demonstrated that integrating risk management practices into cybersecurity strategies is the chief approach to strengthening organizational defences against cyber threats.  Similarly, Hess (2021) demonstrated that applying enterprise risk management across all sectors and business functions is essential for a holistic approach to defence; securing systems as an enterprise, rather than limiting security to a single Section or system, is critical for measuring and achieving cybersecurity effectiveness. Companies can understand their environment and ensure proper risk management alignment by using an RMF. The dynamic allocation of resources based on risk rather than speculation is crucial for success. Finally, Webb et al. (2014) showed that failing to base cybersecurity on risk management accounted for four of the top six reasons cybersecurity programs fail. These findings suggest that cybersecurity is about technical defences and strategic decision-making informed by risk analysis.

Risk management systematically anticipates, assesses, quantifies, and mitigates risks before they escalate into crises. As organizations increasingly rely on interconnected systems and data-driven operations, a structured approach to risk management has become more critical than ever. Frameworks such as the NIST Risk Management Framework (RMF), Enterprise Risk Management (ERM), and ISO 27001 help organizations integrate cybersecurity into their broader risk management strategies, providing a comprehensive approach to managing the full spectrum of cybersecurity risks (Petersen et al., 2020).



While organizations must adopt risk management frameworks, it remains unclear whether formal education systematically trains cybersecurity professionals in risk management methodologies. If cybersecurity professionals lack exposure to structured risk management frameworks, they may struggle to prioritize and effectively respond to cyber threats, resulting in weaker cybersecurity postures within organizations.

Nevertheless, despite the growing recognition of the importance of risk management in cybersecurity, its integration into cybersecurity training programs remains unclear. While some training programs address these concepts, many programs treat risk management as a secondary consideration rather than a core competency. Treating risk management as a secondary consideration raises concerns about whether cybersecurity professionals are sufficiently equipped to identify, prioritize, control and respond to risks effectively in real-world scenarios (Shreeve et al., 2021). In many cases, training programs teach cybersecurity professionals to implement technical controls and tools but do not equip them to assess the risks that necessitate these controls in the first place. As a result, organizations may deploy cutting-edge security solutions only to face breaches because professionals did not adequately identify, prioritize, quantify, or mitigate the risks (with appropriate controls) in the first place.

This lack of clarity raises some critical concerns:

- If research has established that risk management frameworks improve cybersecurity outcomes, why is there no systematic research assessing whether cybersecurity training programs teach these frameworks?

- Could companies' widespread struggles with cybersecurity be partially explained by inadequate training in risk-based approaches?

These issues are problematic for small businesses and large multinational corporations, many of which have robust technical capabilities. High-profile data breaches involving large companies, such as



Yahoo, Equifax, and Facebook, underscore that even organizations with substantial cybersecurity budgets often struggle to manage risk effectively. These breaches, which frequently involve millions of exposed records, demonstrate that technical security measures alone are insufficient to protect against the increasingly sophisticated tactics of cybercriminals. Instead, a comprehensive, risk-based approach is essential to prevent such breaches.

Despite the widespread adoption of these frameworks, it remains unclear whether formal education systematically trains cybersecurity professionals in risk management methodologies. Professionals without exposure to structured risk management frameworks may struggle to prioritize and respond effectively to cyber threats, resulting in weaker cybersecurity postures within organizations.

### 1.2.3    The Tension Between Technical Training and Risk Management Principles

Cybersecurity training programs often prioritize technical skills over risk management principles (Choo, 2011; Goode, 2018; Shreeve et al., 2021). While professionals must understand how to defend networks and protect systems, assessing the potential risks these systems face and making strategic decisions based on those risks is equally vital. As the volume of personal data and sensitive information stored online increases, assessing the risks to data privacy and the likelihood of breaches becomes paramount. While some programs address these concepts, they often treat cybersecurity risk management as a secondary consideration rather than a core competency.

This potential training gap is problematic, particularly given the constantly changing risk landscape. New technologies, such as cloud computing and the Internet of Things (IoT), introduce new attack vectors, and cybercriminals continually innovate their tactics to exploit these vulnerabilities. Organizations need cybersecurity professionals to defend against known threats and emerging risks in



this rapidly evolving environment. Without proper risk management training, cybersecurity professionals may lack the foresight to adapt their strategies to evolving threats.

Moreover, the absence of risk management training carries profound and far-reaching organizational consequences. Webb et al. (2014) linked four of the top six reasons for cybersecurity program failures to poor risk management practices, including perfunctory risk identification and risk assessments that ignored business context. These findings underscore the importance of training cybersecurity professionals in robust risk management methodologies that extend beyond technical skills.

### 1.2.4    A Call to Action: Addressing the Training Questions

The research by Lawrence Baiden (2024) further underscores the importance of risk management in cybersecurity, showing that the most successful cybersecurity programs incorporate comprehensive risk management frameworks. These programs integrate risk management principles at every stage, from risk identification and assessment to the development and implementation of mitigation strategies. Baiden's study underscores the key insight that organizations that prioritize risk management in their cybersecurity strategies tend to achieve better outcomes in resilience and response to cyberattacks.

This study examines whether cybersecurity professionals receive adequate training in these critical Risk management principles. If the answer is no, then the failure of many organizations to prevent and respond to cyberattacks may not be due to a lack of technical expertise but rather the absence of risk management training in cybersecurity curricula. Given the staggering financial and operational toll of data breaches, closing this training gap is no longer an academic luxury—it is an urgent business imperative.



### *1.2.5    Identifying Questions about Cybersecurity Training Competencies*

Although numerous cybersecurity training programs exist—including academic courses, professional certifications, and industry-led workshops—there is limited empirical evidence assessing how well these programs teach Risk management principles. Several key challenges hinder the development of risk-aware cybersecurity professionals.

1.    **Lack of Specific Standardized Risk Management Training Competencies**

The National Initiative for Cybersecurity Education (NICE) Cybersecurity Workforce Framework (U.S.) (Petersen et al., 2020), European Cybersecurity Skills Framework (ECSF)(European Union Agency for Cybersecurity, 2022), and the Strategic Programs for Advanced Research and Technology in Europe (SPARTA) Cybersecurity Skills Framework (Europe)(Piesarskas et al., 2019) outline cybersecurity competencies but do not classify risk management as a standalone competency. This lack of emphasis leads to inconsistent risk management coverage in training programs.

2.    **Overemphasis on Technical Skills**

Many cybersecurity programs prioritize technical proficiencies (e.g., penetration testing, intrusion detection, and vulnerability exploitation) over strategic risk management, incident prevention, and response. As a result, professionals may lack the ability to identify, assess, and holistically mitigate risks (Goode, 2018; Shreeve et al., 2021).

3.    **Unclear Link Between Training and Industry Needs**

Cybersecurity professionals often acquire risk management skills through on-the-job experience rather than formal training. This gap suggests a disconnect between expectations for workforce competency and the outcomes of educational programs.

4.    **Limited Empirical Research on Training Effectiveness**



Existing studies primarily evaluate cybersecurity training based on technical proficiency metrics rather than the effectiveness of risk management instruction. This lack of systematic research makes it difficult to assess whether training adequately prepares professionals for risk-based cybersecurity operations.

### 1.2.6   Justification for This Research

The importance of risk management in cybersecurity is well-documented, with studies demonstrating that risk-based frameworks enhance security outcomes (Baiden, 2024). However, a glaring and critical gap persists in the literature on whether cybersecurity training programs truly integrate these risk management frameworks into their curricula. This disconnect between cybersecurity education and real-world needs may be a critical factor in why many organizations struggle with cybersecurity challenges.

As cyber threats continue to grow in sophistication, organizations must ensure that cybersecurity professionals are technically skilled and equipped with robust risk management competencies. If formal education does not systematically train cybersecurity professionals in risk management approaches, they may lack the competencies necessary to:

a) Estimate risk as a structured function of likelihood and impact, applying expected loss reasoning to quantify exposure and support prioritization decisions consistent with enterprise risk doctrine (Courtney, 1977; Department of Commerce, 1979; NIST, 2012).

b) Identify, compare, and rank cyber risks using differentiated exposure analysis, distinguishing between threats of varying probability and severity of consequences rather than treating risks as undifferentiated technical problems requiring equivalent responses.



c) Construct and communicate defensible risk rationales to stakeholders — including governing bodies and senior leadership — articulating why specific risks warrant priority treatment based on their estimated likelihood, potential impact, and organizational consequence.

d) Select, justify, and sequence controls according to their expected risk-reduction effect, evaluating whether proposed measures reduce exposure through impact attenuation, likelihood reduction, or both, and calibrating investment proportionally to residual risk.

e) Align cybersecurity risk assessments with broader organizational risk appetite, tolerance thresholds, and strategic objectives, ensuring that security decisions are legible within enterprise risk governance frameworks rather than siloed within technical operations.

f) Apply structured, iterative risk management methodologies — including NIST SP 800-39, the NIST RMF, ISO 27005, and the FAIR model — across the full risk management lifecycle: framing, assessment, response, and ongoing monitoring.

The empirical necessity of this study is grounded in four converging imperatives. First, despite the foundational role of risk management in enterprise security governance, no prior research has systematically examined whether cybersecurity training curricula substantively incorporate structured risk management principles—specifically the probabilistic reasoning, likelihood–impact estimation, and comparative exposure logic that enterprise risk doctrine requires. Second, while competency frameworks such as NIST NICE, ECSF, and SPARTA are widely used as proxies for curriculum content and workforce development by educators, the degree to which these frameworks operationalize risk management as a discrete and assessable professional competency ( rather than an assumed byproduct of technical instruction ) has not been empirically evaluated by scholars or industry - against standardized workforce competency requirements. Third, the precise location and depth of risk management instruction gaps within academic and professional training programs remains



unquantified; without systematic, content-level analysis, remediation efforts risk targeting symptoms rather than structural curriculum deficiencies. Taken together, these three imperatives establish the conditions for a fourth: the generation of evidence-based recommendations capable of meaningfully reforming cybersecurity risk management education. Rather than offering prescriptive advocacy, this study provides an empirical foundation for training designers, certification bodies, and institutional leaders to make defensible, theory-grounded decisions about curriculum investment, workforce development priorities, and governance-level accountability for risk competency outcomes.

This study systematically evaluated cybersecurity training programs through Natural Language Processing (NLP) analysis and qualitative insights from professionals. It provided valuable empirical data on the effectiveness of integrating risk management concepts into cybersecurity education.

This research helped bridge the gap between cybersecurity education and industry demands, ultimately enhancing workforce preparedness and strengthening organizational resilience. However, this research is not just about filling an academic gap—it has the potential to reveal a missing link in cybersecurity workforce development. If risk management frameworks improve cybersecurity outcomes, yet training programs do not teach professionals these frameworks, this study could provide key insights into a fundamental flaw in how education prepares cybersecurity professionals for the workforce.

This research empirically assessed whether cybersecurity training programs incorporate Risk management principles and provide evidence-based recommendations to improve cybersecurity education. By doing so, it sought to contribute to the development of a more resilient cybersecurity workforce, ensuring that professionals possess both technical skills and strategic, risk-based decision-making capabilities. These requirements follow from broader discussions on the need for risk



management training in cybersecurity, as emphasized by Baiden (2024) and Hess (2021), who argue for integrating risk management into cybersecurity strategies.

Building on the background that outlines the evolution of cybersecurity training and its heavy emphasis on technical skills, a critical gap emerges. While organizations have advanced their technical defences, there is mounting concern that these training programs do not adequately prepare professionals for structured risk-based decision-making. This observation raises the question: Are current training programs equipping professionals with the essential risk management competencies to mitigate cyber threats effectively? This question serves as the foundation for the following problem statement.

### 1.3    Problem Statement

Despite the increasing sophistication and frequency of cyber threats, cybersecurity training programs designed to equip professionals with the skills to safeguard organizations are not sufficiently equipping professionals with the risk management competencies needed to protect organizations. While these programs typically focus on technical expertise—such as proficiency in network security, cryptography, and penetration testing—they often fail to integrate Risk management principles (Baiden, 2024; Dawson & Thomson, 2018; Goode, 2018; Hess, 2021; Wangen, 2016; Webb et al., 2014; Wilkinson, 2020). This gap is concerning, as risk management is critical to effective cybersecurity. Oltramari and Kott (2018) document that this divergence is not due to imprecise terminology. In their study, practitioners structured 78% of risk descriptions as adversarial tuples (centred on system state, vulnerability, and exploit path) with no reference to the probability of occurrence or the magnitude of loss. This definition is a structurally different construct from $L \times I$ reasoning: it cannot produce the cross-risk prioritization, residual exposure communication, or cost-justified control selection that enterprise



governance requires. The implication is not simply that practitioners define risk loosely; it is that the dominant practitioner construct is organizationally inoperable for the strategic risk management and governance functions cybersecurity professionals are increasingly expected to perform. Marotta and Madnick (2020) further emphasize this by suggesting that many cybersecurity breaches stem from poor risk assessment and prioritization, and by pointing to a lack of strategic decision-making in existing training programs. Webb et al. (2014) found that poor risk management accounted for four of the top six endemic deficiencies in information security practice.

Problem Statement: ***Insufficient integration of strategic risk-management decision-making into cybersecurity training programs leaves professionals unprepared to address modern cyber threats, thereby increasing organizational vulnerability.***

### 1.3.1    Identification of the Problem

The lack of structured risk management in cybersecurity education has immediate and severe implications. Increasingly common and severe data breaches result in substantial financial losses, operational downtime, and costly remediation efforts. The costs incurred from these breaches include lost business, operational downtime, and the financial burden of post-breach remediation, including legal and regulatory fines. Cybersecurity training programs often fail to equip professionals with the expertise necessary to counter the increasing sophistication and frequency of cyber threats, leaving organizations dangerously vulnerable.

### 1.3.2    Who is Affected

Daily, we see evidence of breaches and cybersecurity failures in the news. Such incidents underscore the catastrophic financial and reputational damage organizations suffer when they fail to manage cyber risks appropriately. The fact that some of the costliest cyberattacks in history have



targeted large, multinational companies with substantial cybersecurity budgets compounds the financial ramifications of these breaches. However, even these organizations have faced breaches despite their investments. High-profile cases such as the Yahoo and Equifax breaches underscore a potential flaw in the cybersecurity industry. Despite substantial investments in cybersecurity infrastructure, these organizations struggled to manage their risks effectively. A contributing factor to these failures is the unquestioned assumption that cybersecurity professionals are adequately trained to manage cyber risks (ISC[2], 2019, 2024). This result raises the question of whether these organizations—and others—could have been better prepared if the cybersecurity practices had incorporated a stronger foundation in Risk management principles.

**Table 1**

*Examples of Costly Multinational Data Breaches*

| Company | Estimated Cost / Impact | Key Takeaway |
|---|---|---|
| UnitedHealth Group (Change Healthcare) | **$2.3 Billion+** (2024) | Perhaps the best modern example. Despite being a Fortune 5 company, a single breach paralyzed the U.S. healthcare payment system for weeks. |
| Jaguar Land Rover (JLR) | **£1.5 Billion** (~$1.9B) (2025/26) | A massive ransomware attack on its supply chain caused significant cash outflows and disrupted production across its global footprint. |
| Equifax | **$1.38 Billion** (2017) | A massive financial services firm that had to pay nearly $700 million in settlements and was mandated to spend another **$1 billion** on security upgrades. |
| MGM Resorts | **$100 Million+** (2023) | Shows operational cost: the company lost $100M in just a few weeks after attackers took hotel systems, slot machines, and booking sites offline. |
| Yahoo | **$350 Million** (2013–2016) | While the legal settlement was $117M, the breach famously knocked $350 million off the company's sale price to Verizon. |



This issue directly impacts cybersecurity professionals, who may lack the essential risk-management skills to anticipate, assess, mitigate, and respond to cyber threats. As these professionals are responsible for safeguarding organizations' digital infrastructure, their inability to manage cyber risks appropriately exposes organizations to material risk. Regardless of an organization's size or resources, failing to integrate risk management into cybersecurity training leaves it vulnerable to increasingly complex and costly cyberattacks.

### 1.3.3    *What is Known and Unknown about Cybersecurity Training's Relationship to Risk Management*

Situating this study within the existing literature requires a structured examination of what prior research has established and where meaningful gaps persist. The following subsections distinguish between empirically supported findings and areas where evidence remains insufficient, inconclusive, or absent.

#### 1.3.3.1  What is Known:

Four findings characterize the current state of knowledge regarding the relationship between cybersecurity training and risk management practice.

I.    Risk Management Frameworks Improve Cybersecurity Outcomes

Empirical research indicates that structured risk management frameworks have a positive impact on cybersecurity outcomes. Baiden (2024) demonstrates that organizations that implement risk-based approaches exhibit greater resilience against cyber threats. Antonucci (2017) supports this, highlighting that mature cyber risk management strategies improve threat detection, response efficiency, and organizational resilience.

II.    Cybersecurity Training Appears to Prioritize Technical Skills Over Risk Management



Research indicates that cybersecurity training programs primarily focus on technical skills, such as network security, cryptography, intrusion detection, and penetration testing (Caulkins et al., 2018). Švábenský et al. (2020) conducted a systematic review of cybersecurity education research. Most training programs emphasize attack and defence strategies but provide little structured instruction in risk assessment, threat prioritization, or mitigation planning. Goode (2016) compared training methodologies and found that risk management is often an afterthought in technical cybersecurity courses.

III.   Enterprise Risk Management (ERM) and Compliance Drive Cybersecurity Governance

Many organizations adopt cybersecurity risk management frameworks primarily to meet regulatory requirements rather than for strategic foresight (Baiden, 2024; Marotta & Madnick, 2020; Wilkinson, 2020). Marotta and Madnick (2020) found that cybersecurity programs often align with compliance-driven frameworks (e.g., ISO 27001, NIST RMF, and GDPR). Still, these do not always translate into practical risk-based decision-making in cybersecurity operations.

IV.   Industry Faces a Shortage of Cybersecurity Professionals with Risk-Based Decision-Making Skills

Organizations struggle to hire cybersecurity professionals who are technically proficient and trained in Risk management principles (Hess, 2021; ISC[2], 2019; NICCS, 2024; Oltramari & Kott, 2018; Parekh et al., 2018; Petersen et al., 2020). Dreyer et al. (2018) estimated that cyber risk is among the least understood financial risks for enterprises, suggesting that professionals may lack adequate training in risk mitigation strategies.



**1.3.3.2  What is Unknown (Research Gaps):**

Against this body of established findings, three substantive gaps in the literature remain unresolved, each of which directly motivates the research questions and hypotheses developed in this study.

I.  Lack of certainty about the inclusion of Risk management principles within cybersecurity training programs

There is uncertainty about the sufficiency of incorporating structured Risk management principles into cybersecurity training programs and whether it leads to professional competence. To what extent do cybersecurity training programs incorporate structured Risk management principles found in industry-accepted Risk Management Frameworks, as reflected in training NICE, ECSF, and SPARTA Knowledge Statements (KSAs)?

No systematic research has quantified the presence of risk management instruction in cybersecurity training programs (Caulkins et al., 2018). Chowdhury and Gkioulos (2021) note that key risk management principles are often missing from cybersecurity training frameworks, suggesting a potential gap in training programs. Rather than assuming that risk management competence emerges implicitly from technical instruction, this study measures the structural presence of formal risk management principles in cybersecurity training frameworks. It then evaluates whether exposure to those principles is associated with differentiated professional competence.

II.  Lack of clarity about whether professional training in cybersecurity adequately emphasizes risk-based decision-making and differentiated risk management competence.



It remains unclear whether professional cybersecurity training adequately emphasizes risk principles, particularly risk-based decision-making. Are professionals being trained in structured risk-based decision-making, or does training content lack sufficient emphasis on risk principles?

While Baiden (2024), Hess (2021), and Antonucci (2017) show that risk management improves cybersecurity resilience, no research examines whether real-world cybersecurity incidents result from a lack of risk management training (though Oltramari and Kott (2018) imply this). Marotta and Madnick (2020) suggest that many cybersecurity breaches result from poor risk assessment and prioritization, yet it remains unclear whether training deficiencies contribute to these failures.

III.   Alignment between Training Programs and Workforce Needs

The degree to which cybersecurity training programs align with workforce needs is unknown. Do the NICE, ECSF, and SPARTA frameworks differ in how they integrate risk management, potentially leading to inconsistent workforce competencies?

Although Petersen et al. (2020) and the SPARTA Cybersecurity Skills Framework (Piesarskas et al., 2019) provide competency models, no large-scale study has mapped cybersecurity training outcomes to industry needs. The National Initiative for Cybersecurity Careers and Studies (NICCS, 2024) tracks cybersecurity certification courses based on NIST's NICE framework. However, it does not evaluate their effectiveness in teaching risk-based decision-making skills.

IV.   The impact enhanced risk management training has on workforce readiness and organizational resilience.

The potential impact of enhanced risk management training on workforce readiness and organizational resilience remains uncertain. Does the absence (or presence) of risk management content in training programs correlate with professionals' ability to apply risk-based security measures in real-world cybersecurity roles?



Dreyer et al. (2018) argue that organizations struggle with quantifying cyber risk, but whether better training could help close this gap remains unknown. No studies have tested whether modifying cybersecurity curricula to emphasize risk management improves workforce preparedness.

V. The structural measurement of risk management content in cybersecurity training frameworks

No prior research has systematically quantified the structural presence of risk management content in cybersecurity training frameworks using Natural Language Processing. By applying NLP to the 2,111 TKS statements of the NIST NICE Framework v2.0.0, this study provides a novel empirical approach to measuring the presence at the architectural scale.

While existing literature consistently demonstrates that robust risk management frameworks enhance cybersecurity resilience, research has not systematically investigated the extent to which cybersecurity training programs incorporate these principles. This research gap is critical, as inadequate or inconsistent risk management training may contribute to the ongoing challenges organizations face in mitigating cyber threats effectively. If training programs do not systematically teach cybersecurity professionals structured risk-based decision-making, their ability to assess, prioritize, and rationalize their assessments, and ultimately respond to threats, may suffer. This study sought to address this gap by empirically assessing the integration of risk management principles within cybersecurity curricula and evaluating how well these curricula align with the evolving needs of the cybersecurity workforce.

### 1.3.4 Consequences of Continuing the Problem

If this gap in risk management education persists, the financial, operational, and reputational consequences will escalate. The IBM 2024 Cost of a Data Breach report  (2024a) reveals a 10% increase in the global average data breach cost, now reaching USD 4.88 million. The rise in breach costs highlights



a fundamental issue: Organizations suffer catastrophic financial and reputational damage when they fail to manage cyber risks appropriately. This deficit is particularly alarming given that high-profile breaches, such as those involving Yahoo and Equifax, have occurred despite these companies' substantial investments in cybersecurity infrastructure. Oltramari and Kott (2018) and Marotta and Madnick (2020) suggest that these failures stem from a lack of risk management training, leading to poor decision-making, inadequate threat prioritization, and ineffective mitigation strategies. These failures highlight a fundamental flaw in current cybersecurity training, which prioritizes technical skills over risk-management competencies (Dawson & Thomson, 2018; Goode, 2016, 2018). As the threat landscape continues to grow more complex, organizations will remain vulnerable to devastating breaches if training programs do not adequately prepare cybersecurity professionals to manage cyber risks strategically.

These consequences establish the practical stakes of the research problem and motivate the study's focus on the relationship between training architecture and professional competence. The following Section examines the significance of this research and its potential contributions to cybersecurity workforce development.

Without a structured risk management foundation, cybersecurity professionals may struggle to prioritize risks effectively, resulting in weaker organizational security postures. This dissertation sought to bridge this knowledge gap by empirically assessing the extent to which cybersecurity training frameworks embed formal risk management principles—and by testing whether exposure to those principles predicts differentiated competence in risk assessment, prioritization, control selection, and risk communication.

## 1.4    Significance of this Research



The significance of this study lies in its potential to address critical gaps in integrating formal risk management principles into cybersecurity training and to demonstrate how those gaps manifest as measurable differences in professional risk management competence. As cyber threats become increasingly sophisticated, the ability to identify, assess, mitigate, and manage risks is no longer optional—it is an essential competency for cybersecurity professionals. Despite growing recognition of the importance of risk management in cybersecurity, empirical research on whether current training programs adequately prepare professionals to make risk-based decisions in real-world cybersecurity challenges remains limited.

This study is significant in three interrelated respects. Conceptually, it challenges the prevailing view of cybersecurity as a predominantly technical discipline by empirically examining whether cybersecurity training frameworks operationalize it as an applied form of risk management rather than a parallel technical function. In practice, it addresses a persistent paradox: organizations continue to experience material cybersecurity failures despite sustained and increasing investment in cybersecurity capabilities. Methodologically, it provides a systematic, scalable approach to analyzing the presence, structure, and emphasis of risk-related competencies in cybersecurity curricula, employing mixed-methods empirical research. This dissertation is especially timely and consequential for several critical reasons.

### 1.4.1 Enhance Organizational Cyber-resilience

One of the most pressing challenges organizations face today is the increasing frequency and severity of cyberattacks. According to IBM's 2024 Cost of a Data Breach Report (2024a), the global average cost of a data breach has reached USD 4.88 million, a 10% increase from the previous year. This financial burden, which includes the cost of lost business, operational disruption, legal fees, and



regulatory fines, can threaten the survival of small and medium-sized enterprises, many of which lack the resources to recover from such incidents. These consequences underscore the critical need for cybersecurity professionals who can deploy technical defences and make informed, risk-based decisions to prevent and mitigate these costly incidents.

Furthermore, the failure of large organizations with substantive cybersecurity budgets—such as Yahoo, Jaguar Land Rover, MGM Resort, and Equifax—to prevent catastrophic data breaches underscores the need to integrate risk management frameworks into cybersecurity programs. This research sought to determine whether training adequately prepared professionals to manage cyber risks; if not, it proposed solutions to address this deficiency. Enhancing cybersecurity professionals' training in risk management could prevent future breaches, safeguard sensitive data, and mitigate financial impacts on businesses.

The research revealed that despite current training efforts, risk management is not yet a core component of the curriculum. Consequently, cybersecurity professionals may understand the vocabulary of risk without possessing the risk cognition (risk thinking) necessary to apply those concepts in high-stakes environments. It highlighted a key area for improvement and offered actionable recommendations for educators, policymakers, and organizations to better equip cybersecurity professionals with the competencies needed to protect their organizations effectively.

### 1.4.2    Contributions to Cybersecurity Education

The lack of systematic risk management training in cybersecurity programs remains a gap that researchers have not yet thoroughly examined. This research filled that gap by examining the extent to which existing curricula incorporate Risk management principles. By investigating how cybersecurity education teaches risk management - and whether programs teach it at all - these findings offer a clear



roadmap for modernizing cybersecurity education to address the escalating complexity of global security threats and identify critical pathways for realigning professional training with the sophisticated risk profiles of modern enterprises.

The study also contributed to the academic literature by examining the relationship between risk management education and real-world cybersecurity outcomes. It explored how integrating structured risk management frameworks, such as NIST RMF or ISO 27001, into training programs can enhance cybersecurity professionals' ability to identify, assess, and mitigate risks effectively. This contribution leads to further academic research on best practices for integrating risk frameworks into educational programs.

### 1.4.3    Workforce Development and Professional Competence

Cybersecurity professionals face a rapidly growing skills gap, with an estimated 4 million jobs left unfilled globally (Crumpler & Lewis, 2019; Piesarskas et al., 2019).  While technical skills remain in high demand, organizations are also recognizing the need for professionals proficient in risk management. This study's findings may provide an essential step toward defining the specific competencies cybersecurity professionals need to meet industry standards and address the dynamic nature of cyber threats.

This study found that cybersecurity training risk management is inadequate; it informs the development of industry standards and guidelines for cybersecurity education, ultimately leading to more effective training programs and certifications. By adopting these findings, organizations can close the skills gap and empower their workforce to navigate a complex cybersecurity environment.

### 1.4.4    Policy and Education Implications



The research carries strategic weight for policymakers and educators tasked with workforce development. Given the global focus on cybersecurity, the industry will increasingly require professionals capable of managing complex risks. By identifying gaps in current cybersecurity training programs, this research provides valuable insights for developing policies and standards to ensure that cybersecurity education aligns with the evolving needs of the workforce.

Furthermore, the findings from this study inform curriculum developers as they integrate Risk management principles into cybersecurity education frameworks. This outcome resonates powerfully against the backdrop of the growing centrality of risk management in global cybersecurity frameworks, notably the NIST RMF and ISO 27001 (Baiden, 2024; S. Gates et al., 2012). This study can improve the alignment of training programs with industry standards, contributing to a more effective and cohesive cybersecurity education ecosystem.

### 1.4.5 Broader Societal and Economic Impact

The research has significant societal and economic implications at a broader level. Cyberattacks pose a significant threat to individual organizations, national security, public safety, and the global economy. The study aimed to contribute to the development of a more resilient cybersecurity workforce capable of effectively protecting critical infrastructure, including that in sectors such as finance, healthcare, and government, where cyber threats pose a direct risk to public safety.

By enhancing the integration of risk management into cybersecurity training programs, this study sought to strengthen the overall cybersecurity posture of both public and private organizations, thereby reducing the economic impact of cybercrime and improving the security of the digital systems on which society increasingly relies. Ultimately, this research aimed to ensure that cybersecurity



professionals are adequately prepared to address the challenges posed by the rapidly evolving cyber threat landscape, thereby enhancing organizational resilience and national security.

By bridging the gap between theoretical cybersecurity education and practical, risk-based decision-making, this study sought to develop a more resilient cybersecurity workforce capable of addressing the evolving challenges posed by cyber threats. As organizations and governments continue to rely on interconnected digital systems, the need for professionals who strategically manage cyber risks will grow, making this research critical to future cybersecurity success. From a scholarly perspective, this study contributes to the literature by demonstrating that the structural integration of formal risk management principles within cybersecurity training frameworks is associated with measurable variation in professional risk management competence. This area has mainly been assumed rather than measured. From a practical perspective, the findings may inform curriculum development, workforce training strategies, and policy decisions to strengthen risk-based decision-making in cybersecurity practice.

### 1.5    Originality of this Research

This study makes an original contribution to knowledge across the dimensions identified by Phillips and Pugh (2015) and Guetzkow, Lamont, and Mallard (2004): it conducts empirical work not previously performed, applies existing methods to a new domain, and produces a new theoretical synthesis.

No prior study has applied NLP content analysis to cybersecurity training frameworks at the task, knowledge, and skill statement levels to quantify risk management integration, nor has it used SEM to test the cognitive transmission pathway from training architecture to professional competency within a cybersecurity-specific population. The convergence of these quantitative strands with cross-group



benchmarking and structured leadership interviews to produce a single multi-method causal explanation is itself without precedent in this literature. Theoretically, the study extends Competency-Based Learning Theory through the concept of epistemic compression — the condition in which training produces positive but undifferentiated professional orientation without generating the domain-differentiated cognitive schemas required for practice. This mechanism is generalizable beyond cybersecurity to any professional domain in which the training architecture fails to reflect the cognitive structure of its subject matter.

The study's novelty does not lie in observing that cybersecurity faces a risk management problem, which practitioner literature has long noted, but in the empirical precision that demonstrates the structural mechanism tracing the causal chain.

### 1.6    Research Questions

Given these gaps in the cybersecurity workforce's ability to manage risks effectively, this dissertation examines the integration of risk management training into cybersecurity programs and the practical application of these principles by professionals in the field. Specifically, it addresses the following research questions. The primary objective of this study was to explore the integration of Risk management principles into cybersecurity training programs and to assess the extent to which these programs equip professionals with the competencies needed to manage cyber risks effectively. Based on the identified gap in training—where cybersecurity education often prioritizes technical skills over strategic, risk-based decision-making—this research sought to answer the following primary and secondary research questions:

### *1.6.1    Primary Research Questions*



**RQ1**: *How effective are cybersecurity training programs in incorporating risk management principles?*

This question assessed whether existing cybersecurity training programs effectively integrate risk management concepts and equip professionals with the skills to assess, mitigate, and respond to cyber risks. Given the increasing frequency and financial impact of data breaches, this question was crucial for assessing whether training programs adequately prepared professionals to address the challenges they face in real-world practice. Answering this question requires two complementary analytical approaches: structural analysis of whether risk management content is present within training frameworks, and outcome measurement of whether training exposure produces demonstrable professional competence.

**RQ2**: *How well do these programs equip professionals with the competencies required for risk management in real-world settings?*

This research question evaluates the bridge between classroom theory and operational practice. It focuses the research on the efficacy of training in fostering the risk cognition needed for real-world risk mitigation. Because robust risk management directly prevents catastrophic breaches, verifying that professionals can execute these strategies is a critical requirement for the industry. It probes whether training programs adequately prepare professionals to address the dynamic nature of cyber threats.

**RQ3**: *What gaps exist in cybersecurity curricula, particularly in relation to risk management?*

If an analysis of the NIST NICE, ECSF, and SPARTA frameworks reveals significant gaps in their integration of risk management, closing these gaps would ensure that cybersecurity curricula evolve alongside the practical demands of industry and modern cybersecurity operations, ultimately enhancing the professional competency of the workforce.



### 1.6.2    Secondary Research Questions

**RQ4***: To what extent do organizational mandates implicitly require cybersecurity professionals to apply structured likelihood–impact risk reasoning and differentiated stages of risk assessment, and does this demand mirror their actual skill sets?*

This secondary research question examined the risk management responsibilities implicitly assigned to cybersecurity professionals within organizations and the extent to which these expectations align with their demonstrated risk management competencies. Given persistent organizational failures to manage cybersecurity risk effectively, this question focused on whether cybersecurity roles are structured and governed in ways that support strategic, risk-based decision-making or primarily emphasize operational control execution. In doing so, the question highlights the implications of role design and governance expectations for integrating cybersecurity into enterprise risk management and organizational decision-making.

**RQ5**: *How do the foundational risk-reasoning competencies of cybersecurity-trained professionals compare to those of non-cybersecurity professionals?*

This question provides an external benchmark for interpreting the competence levels observed within the cybersecurity sample by comparing them to those of a non-cybersecurity professional population. In doing so, it explores differences in risk management competencies between cybersecurity and non-cybersecurity professionals. It sought to determine whether specialized cybersecurity training leads to higher risk management competency. The question investigates if those with cybersecurity training possess more advanced skills in identifying, assessing, quantifying and mitigating risks than those outside the cybersecurity field, who may lack such training.

By addressing these questions, this study assessed the presence, depth, and effectiveness of risk management content in cybersecurity education and measured its impact on professional competency.



The findings provided valuable insights into how cybersecurity training aligns with real-world risk management needs and offered actionable recommendations to improve workforce development strategies.

### 1.7    Research Hypotheses

The research questions articulated above examine whether cybersecurity training frameworks structurally embed risk management and whether this integration produces differentiated professional competence that mirrors enterprise risk doctrine.

To empirically evaluate these questions, this study operationalizes each research question as a testable hypothesis. The hypotheses below specify measurable relationships among training content (IV1/IV1x), perceived relevance and conceptual salience (MeV1), demonstrated risk-based behaviours (DV1–DV4), and organizational expectations.

Primary hypotheses address the structural representation of risk management within training frameworks and its relationship to professional competence. Secondary hypotheses examine organizational expectations and benchmarking differences between cybersecurity and non-cybersecurity professionals. Together, these hypotheses translate the research questions into empirically testable propositions across the NLP analysis, Structural Equation Modelling (SEM), and qualitative leadership interviews. The study pairs each hypothesis with a corresponding null hypothesis, organizing them by research question and analytic method.

Chapter 3 details the construct definitions (IV1, IV1x, MeV1, DV1–DV4) within the Section title: Taxonomy of Cybersecurity Competencies and Variable Operationalization.

### *1.7.1    Primary Hypotheses*



To provide a note about the following notation: $H_{1n}$ denotes the alternative hypothesis for Research Question n.

#### 1.7.1.1 RQ1: Structural Integration of Risk Management in Training Frameworks

The first primary hypothesis set addresses whether established cybersecurity workforce frameworks structurally embed risk management at the competency level, as formalized below.

*Objective:*

To analyze the representation of risk management competencies within cybersecurity training frameworks (NICE, ECSF, SPARTA) using NLP-based content analysis.

*Alternative Hypothesis ($H_{11}$)*

Cybersecurity training frameworks that integrate risk management principles will exhibit a statistically significant representation of risk-related Knowledge, Skill, and Ability (KSA) statements (IV1) relative to other competency domains.

*Null Hypothesis ($H_{01}$)*

Cybersecurity training frameworks allocate fewer Knowledge, Skills, and Abilities (KSAs) statements to risk-related competency domains than to other domains.

*Rationale:*

If workforce frameworks structurally embed risk management, they should proportionately represent it across the competency architecture. Underrepresentation would indicate a curricular gap in the formal integration of risk.



### 1.7.1.2 RQ2: Translation of Training into Professional Risk Management Competence

The second primary hypothesis set shifts the analytic focus from framework architecture to professional outcome, examining whether structured training exposure translates into measurable risk management competence in practice.

*Objective:*

To assess whether exposure to risk-integrated training predicts differentiated real-world risk management competence among cybersecurity professionals.

*Alternative Hypothesis (H$_{12}$)*

If cybersecurity training integrates risk management competencies (IV1x), then professionals will demonstrate higher risk-based behaviours (DV1–DV4), because training increases the conceptual salience and perceived relevance of risk management (MeV1), thereby strengthening real-world risk decision-making competence.

*Null Hypothesis (H$_{02}$)*

The integration of risk management competencies in cybersecurity training programs (IV1x) does not significantly influence professionals' real-world risk management competence (DV1–DV4), either directly or indirectly through perceived relevance (MeV1).

*Rationale:*

Grounded in competency-based learning theory, structured exposure to likelihood–impact reasoning should produce differentiated behavioural competence. The absence of an effect would suggest that training does not meaningfully translate into applied risk capability.



### 1.7.1.3 RQ3: Identification of Curricular Gaps

The third primary hypothesis set addresses proportionality, specifically, whether risk management occupies a structurally appropriate share of the competency architecture relative to other domains within cybersecurity curricula.

***Objective:***

To evaluate whether cybersecurity curricula embed risk management proportionately relative to other competencies.

***Alternative Hypothesis (H$_{13}$)***

If cybersecurity curricula inadequately address risk management, then the RISK competency (IV1) will be significantly underrepresented relative to other competency domains, indicating a curricular gap.

***Null Hypothesis (H$_{03}$)***

There are no significant differences in the representation of risk management competencies compared with other competency domains in cybersecurity curricula.

***Rationale:***

Risk management, as defined by formal likelihood–impact reasoning, should function as a cross-cutting analytic framework. Structural underrepresentation would indicate an architectural imbalance in the training design.

### 1.7.2 Secondary Hypotheses

Whereas the primary hypotheses address deficiencies internal to the training architecture, the secondary hypotheses examine how those deficiencies manifest at the organizational and professional



levels, specifically whether employers implicitly assume competencies that training has not reliably produced and whether cybersecurity professionals are empirically distinguishable from non-specialists on risk management measures.

### 1.7.2.1 RQ4: Organizational Expectations and Alignment with Structured Risk Management Competence

This hypothesis set evaluates the alignment between organizational expectations and the specific risk management competencies that current training and curricula equip cybersecurity professionals with.

***Objective***

To evaluate whether organizations implicitly expect cybersecurity professionals to perform functions consistent with formal strategic risk management frameworks, and whether those expectations align with demonstrated competencies.

***Alternative Hypothesis ($H_{14}$)***

Organizations implicitly assign cybersecurity professionals responsibilities consistent with formal strategic risk management frameworks, assuming these professionals demonstrate skilled and differentiated risk management competencies aligned with likelihood–impact reasoning.

***Null Hypothesis ($H_{04}$)***

Organizations do not assign cybersecurity professionals strategic risk management responsibilities or expect them to demonstrate differentiated risk management competencies; instead, they limit these roles to the execution of technical or operational controls.

***Rationale:***



This hypothesis evaluates whether organizational expectations for cybersecurity professionals align with formal risk management doctrine. Specifically, it tests whether leaders presume that cybersecurity professionals apply structured likelihood–impact reasoning when managing cyber risk. The hypothesis does not examine how such expectations emerge, but rather whether they exist and align with demonstrated competencies. This hypothesis also tested the expectation–capability gap identified by Schall (2019).

### 1.7.2.2 RQ5: Benchmarking Cybersecurity Professionals Against Non-Cybersecurity Professionals

To address **RQ5**, the study benchmarks cybersecurity practitioners' risk management competencies against those of non-cybersecurity professionals using the following hypotheses.

*Objective:*

To compare the risk management competence of cybersecurity professionals with that of non-cybersecurity professionals.

*Alternative Hypothesis ($H_{15}$)*

Cybersecurity professionals will demonstrate significantly higher risk management competence than non-cybersecurity professionals.

*Null Hypothesis ($H_{05}$)*

There is no significant difference in risk management competence between cybersecurity professionals and non-cybersecurity professionals.

*Rationale:*

If specialized cybersecurity training meaningfully enhances structured risk reasoning, cybersecurity professionals should outperform their non-cyber counterparts on measures of differentiated risk-management competence.



### *1.7.3    Variables*

This research examines the integration of risk management competencies into cybersecurity training frameworks and assesses their impact on professional competencies in real-world decision-making. Using Natural Language Processing (NLP) to analyze the content of industry-standard frameworks (NIST NICE, ECSF, and SPARTA) and survey/interview responses from cybersecurity professionals, this research assesses the extent to which risk management concepts are embedded in training programs and translated into applied competencies.

The research model variables fell into four categories: Independent Variable (IV), Mediating Variable (MeV), Dependent Variable (DV), and Control Variable (CV). These constructs reflected the structural equation model (SEM) tested in this dissertation. The following Section outlines the variables used in this model, explains how each was measured, and discusses its significance for addressing the research questions and hypotheses. These variables are central to understanding how cybersecurity risk management training competencies shape professionals' competencies.

#### 1.7.3.1  Independent Variables (IVs)

Independent variables are factors manipulated or categorized in this research model to assess their effect on the risk management Competency (the dependent variable). In this research model, the following IVs are central:

1. **IV1: Competencies Reflected in KSA Statements**

    - Definition: The degree to which cybersecurity training frameworks incorporate risk management concepts within their Knowledge, Skill, and Ability (KSA) statements.

    - Rationale: Training frameworks establish learning requirements for professionals. When these frameworks incorporate risk management principles, they facilitate practitioners' acquisition of risk-based competencies.



- Measurement Approach: NLP analysis quantified the frequency and depth of risk management terms and concepts in KSA statements across the NIST NICE, ECSF, and SPARTA frameworks.

### 1.7.3.2 Mediating Variables (MeV)

Mediating variables explain how independent variables influence dependent variables. The research model incorporates the following mediating variable:

2. **MeV1:** Conceptual Salience and Relevance of Risk Management

- Definition: The degree to which professionals perceive risk management as central to their cybersecurity role.

- Rationale: Perceived relevance explains the mechanism by which training content (IV1) influences competence. If professionals see risk management as relevant, they are more likely to apply it in practice.

- Measurement Approach: Likert-scale survey items (e.g., "risk management competencies are central to effective cybersecurity decisions").

### 1.7.3.3 Dependent Variables (DVs)

Dependent variables are the outcomes that this research model sought to explain or predict as a function of the independent and control variables. The following DVs were measured:

3. **DV1: Risk-based assessment behaviour**

- Definition: Ability to identify and analyze threats using likelihood × impact.

- Measurement: Survey items assessing the use of risk analysis in professional tasks.

4. **DV2: Risk-based prioritization behaviour**



- Definition: Ability to prioritize cybersecurity actions based on risk severity rather than counts or alerts.

- Measurement: Survey items capturing prioritization practices.

5. **DV3: Control selection justified by risk reduction**

- Definition: Choosing security controls because they reduce the likelihood and/or impact of risk.

- Measurement: Survey questions linking controls to their risk-reduction function.

6. **DV4: Risk communication and decision rationale**

- Definition: Ability to communicate, the process by which risk was assessed (calculated), residual risk, trade-offs, and control selection rationale to stakeholders.

- Measurement: Survey items capturing communication of risk-based reasoning.

## 1.7.3.4 Control Variables (CVs)

The research model incorporates control variables to account for factors that may influence the relationship between KSA competencies and risk management Competency, ensuring that confounding variables do not skew the observed effects. The research model accounts for the following control variables:

7. **CV1: Years of Experience in Cybersecurity**

- Definition: The number of years the professional has worked in cybersecurity.

- Measurement: Self-reported categorical or continuous scale.

8. **CV2: Prior risk management Learning**

- Definition: Previous exposure to risk management concepts through self-study, certifications, formal training, or on-the-job training.

- Measurement: Self-report (Yes/No + optional elaboration).



### 1.7.3.5 Conclusion

This Section outlines the key variables used in the research model, clarifying how each was measured and its role in examining the relationship between KSA competencies and risk management Competencies. The independent, dependent, and control variables measure the extent to which cybersecurity training programs integrate risk management competencies and assess how professionals apply these principles in real-world settings. The moderating and mediating variables further explained the processes underlying these relationships.

**Figure 1**
*Theoretical Causal Logic: Structural Equation Model (SEM)*

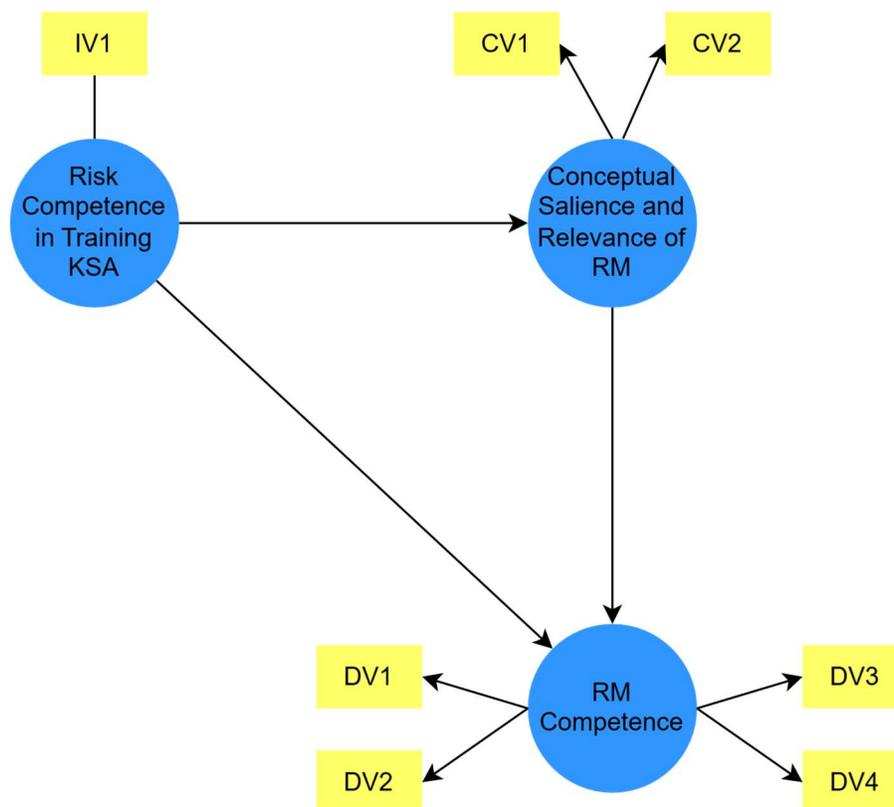



### *1.7.4    Conceptual Logic Model of Variable Relationships (Path Model)*

This study's Conceptual Logic Model follows a causal chain (Hair et al., 2011, 2019) that illustrates how different types of variables (independent, dependent, control, and mediating) interact to explain the development of risk management Competency. Figure 1 illustrates the structural relationships among these variables:

#### 1.7.4.1  Summary Description

This dissertation's conceptual logic model posits that KSA competencies embedded in training content are key predictors of risk management Competency. Mediating variables, such as the perceived relevance of risk management (RM), influence these relationships. Control variables, such as years of experience, professional and prior learning, are included to ensure that external factors do not confound the relationships between training content and competency outcomes. By understanding how these variables interact, the study illuminates how cybersecurity training programs can more effectively integrate risk management competencies and prepare professionals to apply them effectively.

### 1.8    Assumptions

Several assumptions underpin this study regarding the research design and the topic's context. These assumptions are necessary to establish the foundation of the investigation and ensure the validity and reliability of the research process. The following assumptions are central to the study:

### *1.8.1    Inconsistent Integration of Risk Management Principles in Cybersecurity Training*

The research framework presumes that cybersecurity training includes risk management competencies as a core component. However, the study also recognizes that significant variations exist in how deeply and consistently curricula incorporate these principles. This assumption enables an



investigation into how effectively programs incorporate risk management and whether such training empowers professionals to manage cyber risks effectively.

If this assumption does not hold — meaning training programs uniformly and adequately embed risk management principles — then the research problem would be overstated, and the NLP analysis would likely reveal a proportional distribution of risk management content across competency domains. Treating this as an empirical question rather than a settled one directly motivates the study's design.

### 1.8.2    *Professional Self-Reports Provide a Reliable Reflection of Training Experience*

This study assumes that participants provided transparent and accurate reflections of their training and professional experiences during surveys and interviews, and that those reflections reliably mirror the competencies their training produced. This assumption is necessary because the study's methodology depends on self-reported data to assess whether cybersecurity training adequately prepares professionals to exhibit risk-based competencies. If respondents systematically misrepresent their training experiences or competency levels, the findings may not accurately reflect the effectiveness of current training programs.

Although self-reporting involves inherent subjective elements, several design features mitigate this risk. Langhan et al. (2022) demonstrate that structured surveys using competency-based questioning reduce response bias by anchoring responses to specific, observable behaviours rather than general self-evaluations. Consistent with this, Goode (2016) found that cybersecurity professionals' self-assessments of technical competency tend to align with objective evaluations when survey instruments are appropriately structured. Additionally, the use of anonymized responses reduces social desirability bias, increasing the likelihood that participants reported their actual levels of preparedness rather than idealized accounts of their training experience. The study, therefore, assumes that professional insights



collected under these conditions accurately reflect practical training experiences and their subsequent influence on risk management competencies.

### 1.8.3    Relevance of Risk Management to Real-World Cybersecurity Practices

This study further assumes that risk management training aligns with the practical requirements and real-world settings of cybersecurity professionals. This assumption underpins the rationale for assessing the effectiveness of current training programs in preparing professionals for the risk-based decision-making required to navigate the complex threat landscape. This assumption is central to the research because it underpins the study's hypothesis that risk management training influences cybersecurity decision-making. Prior research supports this assumption by demonstrating that structured risk management training improves cybersecurity risk assessment and incident response. Baiden (2024) found that professionals who receive formal risk management training exhibit strategic decision-making in risk assessment and threat mitigation.

If this assumption does not hold—meaning, there is no correlation between risk management training content and professional competency—this may indicate a fundamental disconnect between cybersecurity education and real-world workforce preparedness. This disconnect suggests that risk management training is either insufficient in depth or not effectively translated into practical application, with broader implications for cybersecurity workforce development.

### 1.8.4    Theoretical Frameworks Are an Accurate Representation of Cybersecurity Competencies

The theoretical frameworks and competency models used in this study, such as NIST NICE, ECSF, and SPARTA, are assumed to accurately represent the skills and competencies required for effective cybersecurity practice. This study employs NLP techniques to investigate the presence and depth of risk management content in training materials, aligning with a theoretical framework that integrates risk



management theories with educational content analysis. These frameworks provide a structured approach to identifying and categorizing the competencies required to manage cyber risks effectively. This research adopts these models on the assumption that they effectively measure the presence of risk management competencies in training curricula.

### 1.8.5   Training Content Reflects Industry Frameworks and Workplace Standards

This study assumes that certification-based training programs (as mapped by NICCS) and academic cybersecurity programs accurately reflect how cybersecurity professionals acquire risk management skills. The NICCS framework maps certification courses that conform to the NICE framework and, by extension, to NICE Knowledge Statements (KSAs), reinforcing that these programs are structured to align with workforce expectations.  These frameworks serve as industry-recognized standards for defining the knowledge, skills, and abilities (KSAs) required for various cybersecurity roles. The training provided by the 14,696 cybersecurity training programs (listed in NICCS), which build their curricula from these frameworks, should accurately reflect the structure and content of the KSAs, as they develop their curricula from them. By analyzing these frameworks, the study assumes that it can obtain a representative understanding of how cybersecurity professionals acquire their competencies (Caulkins et al., 2018).  These competencies include alignment with recognized cybersecurity frameworks and compliance with professional standards, ensuring the training is up to date and relevant to industry needs.

### 1.8.6   Cybersecurity Training Programs Will Continue to Evolve

This study interprets its findings through the assumption that the examined training landscape remains a sufficiently stable institutional structure for cross-Sectional analysis. Rapid, fundamental changes to the architecture of cybersecurity training between data collection and interpretation would



limit the applicability of the findings; however, the institutional inertia characteristic of large workforce development frameworks — including NICE, which underwent its most recent major revision in 2023 — supports the stability assumption for the study period.

By clearly articulating and justifying these assumptions, the study ensures that its methodology remains valid, its findings are interpretable, and its conclusions are grounded in established research and industry best practices. These assumptions establish the interpretive scope for the study's findings and anchor the methodology introduced in Chapter 3.

### 1.9    Limitations of the Study

While this study sought to provide valuable insights into integrating risk management competencies in cybersecurity training programs, every research study has inherent limitations that may affect the interpretation of its findings. This study is no exception. These limitations reflect constraints that could impact the generalizability and scope of the findings. By acknowledging these limitations, this Section contextualizes the study's results and establishes clear boundaries for the research. Although these limitations exist, the research addresses them through data triangulation and rigorous coding, thereby minimizing their influence where possible. The key limitations of the study are as follows:

#### 1.9.1    *Scope of Data Collection*

The study primarily analyzed cybersecurity training content and feedback from professionals who have completed these training programs. While this data provided essential insights, its limitations included the availability and accessibility of relevant training materials, as well as the willingness of professionals to participate in surveys and structured interviews (Creswell, 2014). The diversity of training programs and professionals included in the study may not fully represent all forms of cybersecurity education or every region, potentially limiting the breadth of the findings.



### 1.9.2    Potential Biases in Self-Reported Survey and Interview Data

The study used self-report measures to assess professionals' perceptions. Although this approach yields unique first-hand insights, it also carries the risk of subjective bias or inaccuracies. This limitation is standard in survey- and interview-based studies, as demonstrated by Podsakoff et al. (2003), who discussed common biases in self-reported data. Professionals may overestimate or underestimate the effectiveness of their training, particularly regarding risk management knowledge and its application in real-world settings. This limitation was mitigated by cross-referencing self-reported data with objective analyses of training content to validate the findings. This study addressed this limitation using structured, competency-based questions rather than subjective self-evaluations. Additionally, NLP-based analysis of survey responses compared self-reported competency against risk management terminology usage, providing a secondary validation measure.

### 1.9.3    Dependence on the NICE & ECSF Frameworks as Proxies for Cybersecurity Training

A primary limitation of this study is that it relies on the NIST NICE, ECSF, and SPARTA frameworks as proxies for cybersecurity training content. While educational institutions use these frameworks to define curricula (competencies) for tens of thousands of certification courses, the models may not fully capture risk management training occurring elsewhere. There is significant variation in the structure, content, and quality of cybersecurity training programs, ranging from academic degrees to industry certifications and vendor-led workshops. The study may not fully capture the nuances of these diverse programs, particularly those operating in non-traditional or less-regulated environments.

This variability may affect the consistency of the findings across different training types and industry sectors. This limitation means that the study's findings may not generalize to all forms of cybersecurity training, particularly in organizations that adopt customized risk management frameworks



not covered by NICE or SPARTA. The study cross-referenced findings with NICCS mappings and analyzed survey responses to determine whether professionals acquire risk management concepts through alternative means, thereby mitigating this limitation.

### 1.9.4    Focus on Existing Training Models

The study evaluates existing cybersecurity training models and their integration of risk management competencies. As such, it may not account for emerging trends in cybersecurity education or for new methodologies that are not yet widely adopted. The rapidly evolving nature of the cybersecurity field means that training programs may undergo significant changes over time, and the results of this study may not fully reflect future developments or innovations in cybersecurity education.

### 1.9.5    Limited Generalizability to Non-Western Contexts

While the study focused on training programs in Western contexts (primarily North America and Europe), it could not fully account for the diversity of cybersecurity training systems in other regions. The needs, resources, and approaches to cybersecurity training in non-Western countries or developing regions may differ, limiting the direct applicability of the study's findings to these contexts. However, the core findings on the need for cybersecurity risk management education are expected to have global relevance, despite contextual differences. The study acknowledges that different regulatory environments prioritize regional variations in cybersecurity training and risk management competencies.

### 1.9.6    Impact of External Factors

External factors, such as organizational culture, leadership support, and the specific cybersecurity threats organizations face, may influence the effectiveness of risk management training. While this study explored the role of training programs, it cannot fully isolate these external factors or



measure their impact on the success of cybersecurity practices. Other studies have looked at this

(Aldasoro et al., 2022; Braumann, 2018; Khansa & Liginlal, 2007; Webb et al., 2014). While the study

examined leadership perspectives and organizational dynamics, the research design did not allow for full

control over these external factors.

### 1.9.7    *Time and Resource Constraints*

The study is constrained by time and resource limitations, particularly in data collection and

analysis. Analyzing a large volume of training materials and conducting in-depth interviews with

professionals across various sectors requires significant resources. As a result, the sample size and scope

may be limited, potentially affecting the generalizability of the findings. Future research could expand

the sample size or investigate specific industry sectors in greater detail to provide a more

comprehensive understanding of the findings.

### 1.9.8    *Complexity of Measuring Risk Management Knowledge*

Risk management is a broad, multidimensional concept that encompasses strategies, tools, and

decision-making processes. Measuring the depth of risk management knowledge is challenging because

training programs vary significantly in how they teach these principles. This study sought to quantify risk

management education using Natural Language Processing (NLP) techniques and professional feedback;

however, the subjective nature of assessing competence in complex areas, such as risk assessment and

mitigation, presents challenges. This study focuses exclusively on cybersecurity training frameworks and

risk management competencies, excluding broader organizational policies. It does not evaluate

individual organizational security programs, measure the effectiveness of technical controls, or assess

the operational maturity of specific cybersecurity tools. By establishing these delimitations, this study

prioritizes the analysis of training materials and professional skills over broad organizational outcomes.



### *1.9.9 Limitations*

While this study has inherent limitations, each identified constraint is addressed with appropriate mitigation strategies to minimize its impact. By clearly defining the research boundaries, the study provides a transparent evaluation of how cybersecurity training programs integrate risk management and how those competencies appear in the workforce. The study addresses these limitations during the interpretation of results to ensure that all conclusions align with the established research scope.

### 1.10 Organization of the Dissertation

This dissertation comprises five chapters, moving from the initial theoretical framework to the final synthesis of Quantitative and Qualitative findings. Chapter 1 has established the problem context, research questions, hypotheses, and the study's significance. Chapter 2 presents the theoretical and empirical literature underpinning the study, including Competency-Based Learning Theory, risk management frameworks, and research on the adequacy of cybersecurity training. Chapter 3 describes the mixed-methods research design, including the NLP content analysis methodology, survey instrument, SEM specification, and interview protocol, together with sampling strategy, data collection procedures, and ethical considerations. Chapter 4 presents the empirical results of all four analytical strands — NLP content analysis, SEM, cross-group benchmarking, and leadership interview analysis — in a descriptive, technically precise manner, without interpretive commentary. Chapter 5 interprets the findings in relation to the research questions and hypotheses, develops their theoretical implications, discusses practical significance for training design and organizational governance, and identifies directions for future research.



### 1.11   Definition of Terms

The following key terms related to cybersecurity, risk management, and cybersecurity training are defined to ensure clarity and consistency throughout this dissertation. These definitions align with recognized cybersecurity frameworks, industry standards, and academic literature, ensuring that all terms are used precisely and consistently in the research.

#### *1.11.1   Cybersecurity*

**Cybersecurity:** Protecting organizations by managing technical risks associated with using, storing, processing, transmitting, and destroying information assets. (NIST Glossary, 2025)

**Cyberattack:** A malicious act intended to compromise the security of digital infrastructure, including computer networks, systems, and data. Cyberattacks may include hacking, malware infections, ransomware, denial-of-service (DoS) attacks, and phishing campaigns. (NIST Glossary, 2025)

**Vulnerability:** A system weakness that facilitates unauthorized access, malicious activity, or harm to an organization. Vulnerabilities can result from software flaws, misconfigurations, or human errors. (NIST Glossary, 2025)

**Threat:** A potential danger to a computer system or network, such as a malicious actor, cyberattack, system failure, or natural disaster. (NIST Glossary, 2025)

**Data Breach:** An incident in which unauthorized individuals or systems gain access to sensitive or confidential data, typically resulting in its exposure, theft, or misuse. Data breaches can have severe financial, operational, and reputational consequences for organizations. They often result from insufficient risk management practices, technical defences, or human error. (NIST Glossary, 2025)



**Incident Response:** The structured process of identifying, containing, managing, and recovering from a cybersecurity breach or other cyber incident. Incident response involves threat detection, mitigation, forensic analysis, and reporting. (NIST Glossary, 2025)

### 1.11.2  Risk Management

**Risk:** The probability and potential impact of an uncertain event causing harm or loss to an organization, its assets, systems, or individuals. (NIST Glossary, 2025)

**Cybersecurity Risk:** The likelihood of a threat exploiting a vulnerability to cause harm or disrupt an organization's security posture. (NIST Glossary, 2025)

**Risk Assessment:** The systematic process of identifying, quantifying, analyzing, and evaluating the threats, determining their likelihood and impact, using this information to estimate risk, and prioritizing them for mitigation. (NIST Glossary, 2025)

**Risk Management:** In cybersecurity, risk management involves identifying, assessing, prioritizing, and mitigating risks that could harm an organization's digital infrastructure or operations (COSO, 2004; International Organization for Standardization, 2018; NIST, 2011). It involves structured decision-making to minimize the likelihood and impact of potential cyber threats, ensuring that the organization can function effectively in the face of uncertainty. Risk management in cybersecurity involves:

- **Identifying and categorizing threats** – Recognizing potential risks, including cyber threats, human errors, and system failures.

- **Assessing threats leading to risk** – Analyzing weaknesses that cyber threats could exploit.

- **Quantifying or calculating risk** (likelihood and impact) – Measuring the probability and severity of a risk occurring. Quantifying risk based on these two measures (Li et al., 2020).



- **Prioritizing response based on risk** – Determining the most critical risks that require immediate mitigation.

- **Controlling risk** – Implementing security measures and controls to reduce a threat's likelihood or impact, thereby reducing risk.

- **Developing mitigation strategies** – Designing proactive approaches to reduce, transfer, or eliminate risks, considering the likelihood and impact of cyber threats.

**Risk Management Framework (RMF)**: Risk Management Frameworks (RMFs) are structured methodologies that enable organizations to identify, assess, quantify, and mitigate risks systematically  (COSO, 2004; International Organization for Standardization, 2018; NIST, 2011). They provide a foundation for consistent risk management practices, emphasizing continuous improvement and adaptation to changing threats. Examples include NIST and ISO frameworks. NIST SP 800-12r1 (2017) provides the foundational normative statement underlying the RMF architecture, stipulating that security controls must be "commensurate with the risk and magnitude of harm" — a principle that positions every technical security decision as an exercise in calibrated risk judgment rather than standardized technical compliance (NIST, 2017, Section 2.5).

**Risk-Based Decision-Making**: Risk-based decision-making involves evaluating potential risks, considering both their likelihood and potential impact (COSO, 2004; International Organization for Standardization, 2018; Jones, 2006; NIST, 2011). Cybersecurity involves prioritizing threats and vulnerabilities based on their severity and the resources required to mitigate them. Effective risk-based decision-making enables cybersecurity professionals to allocate resources efficiently and respond to the most critical threats.

**Cybersecurity risk management Competencies:** Cybersecurity risk management competencies refer to the knowledge, skills, and abilities (KSAs) necessary for individuals to identify,



assess, quantify, and mitigate cyber risks effectively  (COSO, 2004; International Organization for

Standardization, 2018; NIST, 2011). These competencies include understanding risk assessment

methodologies, threat prioritization, impact analysis, and developing strategies to minimize risk

exposure.

      **Risk Tolerance:** The level of risk an organization is willing to accept to pursue its objectives.

(COSO, 2004; International Organization for Standardization, 2018; NIST, 2011)

      **Risk Appetite***:* The amount and type of risk an organization must take to achieve its strategic

goals. (COSO, 2004; International Organization for Standardization, 2018; NIST, 2011)

      **Risk Mitigation***:* Actions taken to reduce the probability or impact of a risk to an acceptable

level. These may include technical security controls, policy changes, and employee training. (COSO,

2004; International Organization for Standardization, 2018; NIST, 2011)

      **Risk Transfer***:* The process of shifting the responsibility for a risk to another party, such as

through cybersecurity insurance, outsourcing, or third-party agreements. (COSO, 2004; International

Organization for Standardization, 2018; NIST, 2011)

      **Risk Avoidance***:* A risk management strategy that eliminates risk by not engaging in

activities that could lead to its occurrence. (COSO, 2004; International Organization for

Standardization, 2018; NIST, 2011)

      **Risk Acceptance***:* A risk management strategy in which an organization acknowledges a risk

but chooses not to take action to mitigate or transfer it. (COSO, 2004; International Organization for

Standardization, 2018; NIST, 2011)

      **Risk Hedging***: A* strategy of taking protective measures to offset exposure to potential losses

due to risk by compensating for that loss in other ways. In cybersecurity, risk hedging involves



diversifying security controls to mitigate potential risks. (COSO, 2004; International Organization for Standardization, 2018; NIST, 2011)

**Risk Management Competency**: The integrated ability to identify and categorize threats, assess vulnerabilities by determining the likelihood and impact of adverse events, calculate and prioritize risks, control risk exposures, and develop effective mitigation strategies to protect organizational assets. (COSO, 2004; International Organization for Standardization, 2018; NIST, 2011)

**Conceptual Salience (of risk management)**: The degree to which risk management competencies are cognitively accessible, subjectively important, and perceived as relevant by an individual when interpreting cybersecurity problems and making professional judgments. Conceptual salience captures how strongly decision-makers mentally foreground risk-based reasoning (e.g., likelihood-impact trade-offs, prioritization, and control justification). In this study, conceptual salience functions as a mediating construct linking exposure to risk-integrated cybersecurity training with observable risk management competence in practice.

### 1.11.3  Cybersecurity Training

**Cybersecurity Training**: Structured programs, courses, or certifications designed to equip individuals with the skills and knowledge to protect digital infrastructure.

**Professional Certification:** A credential signifying that an individual has demonstrated cybersecurity proficiency by meeting established knowledge and skills standards. Examples include Certified Information Systems Security Professional (CISSP), Certified Ethical Hacker (CEH), and CompTIA Security+.



**Continuing Education:** Ongoing professional development to maintain and enhance cybersecurity and risk management skills. This development may include workshops, industry conferences, certification renewals, and advanced training courses.

**Competency Gap:** A competency gap refers to the discrepancy between the cybersecurity skills required for a job role and the skills currently possessed by professionals in the field, as well as those taught in training programs (Goode, 2018; Parekh et al., 2018; Wilkinson, 2020). This gap can result from inadequate training or failure to incorporate essential risk management competencies into cybersecurity curricula. Competency gaps indicate a need for better alignment between training content and real-world cybersecurity needs.

**Risk-Based Cybersecurity Leadership:** The strategic decision-making process where executive leaders integrate cybersecurity risk management into enterprise risk management (ERM). This approach involves prioritizing cybersecurity investments, fostering a culture of risk awareness, and aligning cybersecurity initiatives with business objectives to ensure adequate protection. Effective cybersecurity leadership requires technical risk assessment skills and strategic business acumen (NIST, 2024).

### 1.11.4  Cybersecurity Workforce Frameworks

**Cybersecurity Competency Framework:** A structured model that defines the knowledge, skills, and abilities required for various roles in the cybersecurity field. NIST's NICE, ECSF, and SPARTA frameworks guide educational institutions, industry certifications, and organizations seeking to develop and measure cybersecurity competencies. They also play a key role in identifying the risk management competencies that educators should integrate into training programs. (European Union Agency for Cybersecurity, 2022; Petersen et al., 2020; Piesarskas et al., 2019)



**NIST NICE Framework:** A US cybersecurity workforce framework developed by the National Institute of Standards and Technology (NIST) to standardize the Knowledge, Skills, and Abilities (KSAs) required for cybersecurity roles. (Petersen et al., 2020)

**The European Cybersecurity Skills Framework (ECSF)**: ENISA developed the ECSF, a cybersecurity workforce framework, to provide a common understanding of the skills, competencies, and knowledge required for various cybersecurity roles across Europe. (European Union Agency for Cybersecurity, 2022)

**SPARTA Cybersecurity Framework:** This European cybersecurity workforce framework categorizes cybersecurity training across critical infrastructure sectors, including energy, finance, healthcare, and digital infrastructure. It is based on the JRC Cybersecurity Skills Framework and focuses on sector-specific cybersecurity training needs. (European Commission, 2019; Piesarskas et al., 2019)

**KSAs (Knowledge, Skills, and Abilities):** Core components of cybersecurity workforce frameworks that define what professionals need to know (Knowledge), what they should be able to do (Skills), and how they apply those skills in practice (Abilities). (European Union Agency for Cybersecurity, 2022; Petersen et al., 2020)

**Professional Development:** Professional development refers to ongoing education, training, and skill-building that enhances an individual's expertise and career progression in a specific field. In cybersecurity, professional development includes certification programs, workshops, seminars, and courses designed to keep professionals up to date on the latest threats, technologies, and best practices.

**Cybersecurity Workforce Development:** The strategic process of training and preparing cybersecurity professionals to meet industry and government security needs. Workforce development initiatives often involve public-private partnerships, competency-based learning, and standardization



efforts, such as NICE, ECSF, and SPARTA. (European Commission, 2019; European Union Agency for Cybersecurity, 2022; Petersen et al., 2020)

### *1.11.5  Analytical Methods*

**Natural Language Processing (NLP):** A branch of artificial intelligence (AI) that enables computers to analyze, understand, and derive meaning from human language  (Jurafsky & Martin, 2021). This study employs NLP techniques, including topic modelling, sentiment analysis, and text classification, to assess the presence of risk management concepts in cybersecurity training frameworks and professional discourse.

**Professional Discourse Analysis:** A method for analyzing written and spoken communication among professionals to identify trends, terminology usage, and conceptual understanding within a field (Gee, 2014). This study employs discourse analysis to examine survey and interview responses and assess whether cybersecurity professionals incorporate risk management concepts into their discussions and decision-making.



## Chapter 2 – Literature Review

This literature review synthesizes existing research on integrating risk management within cybersecurity training and education, focusing on the gaps this dissertation sought to address.

Despite the growing importance of risk management in mitigating cyber threats, much of the current educational content remains technical, with limited focus on the broader strategic and risk-based decision-making necessary to confront evolving cybersecurity challenges. Therefore, this review ensures that the new research presented in this effort builds on prior work while acknowledging gaps, contradictions, and unanswered questions in the existing literature. This review establishes the case for a new investigation and demonstrates how it advances the field's understanding.

### 2.1    Background and Significance

The role of risk management in cybersecurity training has gained increasing attention in the literature due to the evolving complexity of cyber threats (Švábenský et al., 2020). While cybersecurity training programs have historically focused on technical competencies such as network defence, encryption, and vulnerability exploitation, there is growing recognition that risk-based decision-making is essential to cybersecurity effectiveness (Chowdhury & Gkioulos, 2021). Despite this, a gap remains in the literature regarding how formal cybersecurity training incorporates structured risk management competencies. This gap aligns with RQ1, which sought to determine how cybersecurity training programs incorporate risk management competencies and how this integration affects professionals' ability to make risk-based decisions in practice.

The normative basis for this integration is not a matter of scholarly debate. The United States federal government's authoritative guidance on information security has explicitly defined cybersecurity practice as a risk management activity for more than a decade. NIST Special Publication 800-12 Revision



1, An Introduction to Information Security (2017), establishes the foundational policy principle that security controls must be "commensurate with the risk and magnitude of harm resulting from unauthorized access, use, disclosure, disruption, modification, or destruction of information" (NIST, 2017, Section 2.5). This principle does not treat risk management as an optional governance overlay; it positions calibrated risk judgment as the primary criterion for every security control decision. By positioning risk management as an 'organization-wide activity,' NIST SP 800-39 (2011) establishes the requirement to integrate risk-based logic across all organizational functions (p. 7). NIST SP 800-37 Revision 2, Risk Management Framework for Information Systems and Organizations (2018), further operationalizes this requirement through a structured lifecycle process that integrates security decisions with organizational risk governance at every stage. Taken together, these documents establish that the federal normative architecture has already resolved the definitional question this literature review examines. Cybersecurity is not a discipline that should engage with risk management — it is, by authoritative policy definition, an applied instantiation of it. The research problem is not whether this standard exists, but why the workforce has not fully operationalized it.

Several studies have used or referenced workforce competency frameworks, such as the NIST NICE, ECSF, and SPARTA frameworks, but have not systematically analyzed their emphasis on risk management competencies (Cabaj et al., 2018; Dawson & Thomson, 2018; European Union Agency for Cybersecurity, 2022; Goode, 2018; Parekh et al., 2018; Petersen et al., 2020; Piesarskas et al., 2019; Stine et al., 2020).

Because there are thousands of cybersecurity courses—and it is impractical to review every syllabus—this study uses the NIST NICE Framework and the European Cybersecurity Framework as a standardized representation of what these courses claim to teach. By analyzing how the NICE Framework embeds risk management across its task, skill, and knowledge statements, we can effectively



measure the curriculum content of a broad range of training programs aligned with NICE and compare it to professional experience. Similarly, Europe's SPARTA Framework provides a comparable mapping of work roles and competencies, serving as a parallel reference. While not every European certification relies on SPARTA, it nonetheless provides a structured, recognized model for linking risk management concepts to workforce competencies. Examining these frameworks serves as a practical proxy for gauging whether the 'officially prescribed' content for cybersecurity training included robust risk management coverage, which served as a baseline for professional practice.

### 2.1.1    Purpose and Objectives

This literature review evaluates existing research on integrating risk management competencies in cybersecurity training programs and identifies gaps in current educational methodologies. Specifically, this review examines how workforce competency frameworks, such as NICE, ECSF, and SPARTA, address risk management and whether cybersecurity training programs adequately prepare professionals for risk-based decision-making. Furthermore, this review lays the groundwork for the study's empirical analysis by establishing the theoretical context for evaluating risk management content in cybersecurity curricula. By systematically reviewing the literature, this chapter sought to determine how cybersecurity education aligns with workforce expectations regarding risk management competencies.

### 2.1.2    Scope and Limitations

This literature review covers research related to cybersecurity training programs, workforce competency frameworks (NICE, SPARTA), risk management education, and professional competencies in cyber risk assessment and mitigation. The reviewed literature primarily spans the last five to ten years, with foundational works included for historical context. The geographical scope primarily focuses on North America and Europe, using the NICE, ECSF, and SPARTA frameworks. Methodologically, this



review incorporates quantitative studies, qualitative analyses, and policy reviews related to the effectiveness of cybersecurity training.

Despite its broad scope, this review does not assess individual training programs; instead, it relies on structured competency frameworks, such as NICE, ECSF, and SPARTA (Marotta & Madnick, 2020). Additionally, given the proprietary nature of some training curricula, the literature primarily reflects publicly available data. While this study incorporates a broad range of perspectives, the emphasis remains on US and EU-based cybersecurity frameworks.

Although this review examines NICE, ECSF, and SPARTA as the three principal international cybersecurity workforce frameworks, the full NLP pipeline described in Chapter 3 — which classifies statements into 29 competency categories using a structured large language model — was applied exclusively to NICE v2.0.0. This decision reflects a structural difference among the frameworks: NICE organizes its content as 2,111 discrete, individually classifiable Task, Knowledge, and Skill (TKS) statements, enabling statement-level semantic classification. ECSF organizes its content as narrative role profiles across 12 professional roles, and SPARTA as sector-level competency descriptions; neither framework presents a volume of discrete, consistently structured statements amenable to the same pipeline. However, a direct lexical analysis of all three frameworks — searching for the probabilistic reasoning vocabulary that characterizes formal risk management, specifically the terms "likelihood," "probability," "probabilistic," and "expected loss" — reveals the same structural absence across all three: zero occurrences in NICE (2,111 TKS statements), zero occurrences in the ECSF Role Profiles (12 role profiles), and zero occurrences in SPARTA D9.1. This cross-framework convergence suggests that the absence of a probabilistic reasoning vocabulary identified through NLP classification in NICE is not an artifact of that framework's architecture but rather a structural feature of the field's dominant



workforce frameworks across jurisdictions. Chapter 4 reports the full NLP findings for NICE, alongside lexical confirmation from the ECSF and SPARTA.

## 2.2    Theoretical Framework

This study's theoretical foundation draws from enterprise risk management (ERM) frameworks (COSO, NIST RMF, ISO 31000:2018) and educational theories (constructivist learning, competency-based training) to evaluate the incorporation of risk management competencies within cybersecurity curricula. Specifically, this study's theoretical foundation is grounded in risk management theory, competency-based learning frameworks, and workforce development models. These frameworks provide both a methodological and a conceptual lens for assessing the effectiveness of cybersecurity education.

### 2.2.1    Risk Management Theory

As defined by ISO 31000:2018 and the NIST risk management Framework (RMF), risk management theory emphasizes a structured, systematic approach to identifying, quantifying, assessing, and mitigating risks (Antonucci, 2017; Baiden, 2024; International Organization for Standardization, 2018; Perera, 2019). These frameworks base their definitions of risk (RISK = LIKELIHOOD x IMPACT) on earlier financial expected value calculations (Department of Commerce, 1979, p. 10). As detailed in the Section title 'The Historical Convergence of Risk as Expected Loss', this likelihood × impact formulation did not originate in contemporary cybersecurity practice but was codified in federal information systems standards following Courtney's (1977) expected-loss model and its formalization in FIPS PUB 65 (Department of Commerce, 1979). The mathematical structure of risk as probability-weighted loss, therefore, reflects a longstanding analytical tradition rather than a recent managerial convention. This historical grounding is significant because it establishes that structured likelihood estimation is not optional within risk doctrine; it is foundational to prioritization and control selection.



This risk framework provides a lens for evaluating the inclusion of risk-based decision-making principles in cybersecurity training programs. Braumann's (2018) work on the importance of risk awareness within ERM supports the idea that a strong understanding of risk is crucial across domains, including cybersecurity. Organizations that incorporate risk management frameworks demonstrate improved resilience against cyber threats (Baiden, 2024), yet it remains unclear whether cybersecurity education sufficiently embeds these principles (Marotta & Madnick, 2020).

Critically, this theoretical framework is not only an academic construction; it is consistent with the analytical consensus across independent governance frameworks. NIST SP 800-39 (2011) defines risk management for information systems as a four-component process — framing risk, assessing risk, responding to risk, and monitoring risk on an ongoing basis — and explicitly requires practitioners to execute this process simultaneously at the organizational, mission, and system levels (NIST, 2011, pp. 7–9). This tiered governance structure implies that cybersecurity practitioners at any level must be able to reason about risk in terms consistent with the organizational frame within which their work sits. NIST SP 800-37 Revision 2 (2018) operationalises this requirement through a risk management Framework that "integrates security, privacy, and cyber supply chain risk management activities into the system development life cycle" (NIST, 2018, p. 1), treating risk-based judgment not as a senior leadership function but as a technical practitioner obligation embedded in every phase of system design, implementation, and operation. NIST SP 800-12 Revision 1 (2017) provides the foundational policy statement that links this framework to individual control decisions: security measures must be "commensurate with the risk and magnitude of harm," a requirement that presupposes the practitioner's capacity to estimate both dimensions before selecting any countermeasure (NIST, 2017, Section 2.5). This normative architecture replaces academic ideals with published, operational policy requirements; these mandates dictate how we evaluate contemporary cybersecurity training. The



question this study addresses — whether training produces the differentiated risk competence that governance demands — is therefore a question of compliance with a standard the federal policy architecture has already defined.

### 2.2.2    The Historical Convergence of Risk as Expected Loss

Risk, as it applies to cybersecurity, has never been a term borrowed from another discipline. The operational definition that underlies every major enterprise and information systems governance framework—risk as the interaction of likelihood and impact—was codified in federal information systems doctrine before the modern cybersecurity profession existed. It did not migrate into cybersecurity from finance or actuarial science. It was present at the origin.

Understanding this matters not as historical context, but because it establishes what risk-based cybersecurity practice has always required analytically. When training programs treat risk as a governance label or a control checklist rather than a structured estimation problem, they are not offering a simplified version of risk management. They are omitting its central function.

The foundational moment in federal information systems risk doctrine came in 1977, when Robert H. Courtney Jr., writing in support of the US Department of Commerce Federal Information Processing Standards Task Group, articulated the structural requirement for security decision-making. Courtney (1977) argued that two estimable parameters must underpin any security decision: the probability of an undesirable event occurring and the magnitude of its impact on organizational objectives, expressed in comparable units. Their interaction yields annualized exposure—expected loss per unit time. In this formulation, risk was not a label applied to danger; it was a computable expectation designed to support cost-justified control selection. The purpose of risk management, as



defined in this first codification, was to answer a specific question: given what we stand to lose and how likely we are to lose it, which controls are worth implementing?

This definition represented a deliberate conceptual advance over earlier federal guidance. FIPS PUB 31 (Department of Commerce, 1974) employed the language of "risk management" in the context of physical security safeguards for automatic data processing facilities, yet did not define risk in probabilistic terms. Risk was discussed descriptively—in relation to threats and protective measures—without being operationalized as an expected loss estimate. The language of risk management existed before its analytical structure. Courtney's contribution was the structure.

FIPS PUB 65 (Department of Commerce, 1979) subsequently formalized that structure. Information system risk was framed explicitly as the probability of loss multiplied by its magnitude, embedding expected-value reasoning directly into federal data processing policy. Crucially, this formalization appeared in the same policy tradition that later produced the frameworks that now govern professional cybersecurity practice. Cybersecurity did not have the expected loss logic grafted in from elsewhere; rather, the information systems governance tradition from which cybersecurity emerged embedded it from the start.

The subsequent propagation of this structure across enterprise risk frameworks confirms continuity rather than introducing competing definitions. ISO 31000 (2018) defines risk as the effect of uncertainty on objectives—linguistically broader, but analytically unchanged. ISO's risk assessment process explicitly requires evaluation of both the likelihood of events and the magnitude of their consequences. NIST SP 800-39 (2011) defines risk as a function of the likelihood of a threat event and the resulting impact on organizational operations, assets, and individuals—language that is native to information systems governance. NIST SP 800-30 (2012) operationalizes this further, providing a structured methodology for estimating threat event frequency, vulnerability likelihood, and



consequence severity as prerequisites for any control selection decision. COSO's Enterprise Risk Management Framework (2004) requires organizations to estimate both the likelihood and severity of potential events before prioritizing responses. The FAIR model (Jones, 2006) returns most explicitly to the expected value tradition, decomposing risk into Loss Event Frequency and Loss Magnitude—a direct restatement of the probabilistic logic first embedded in federal information systems standards.

The structural consistency across these frameworks is analytically significant. Vocabulary varies, and impact categories broaden beyond financial loss to include operational, legal, reputational, and societal harm. But the core requirement has remained unchanged across more than four decades of evolving policy and practice: risk assessment requires estimating likelihood and impact, and their interaction enables prioritization, trade-off analysis, and cost-justified control selection. No major enterprise risk framework has abandoned this requirement.

This consistency reflects a deeper analytical tradition rooted in early probability theory that extended into Knight's (1921) distinction between measurable risk—in which researchers can estimate probability distributions—and irreducible uncertainty, where they cannot. This distinction is important for practice: where estimation is possible, decision-makers are obligated to estimate rather than narrate. Hubbard (2009) emphasizes that risk must be measurable and that uncertainty becomes actionable only when expressed probabilistically. Hubbard and Seiersen (2017), writing specifically about cybersecurity, go further; they demonstrate that the field possesses sufficient historical and actuarial data to support structured estimation, arguing that the failure to measure cybersecurity risk stems from poor analytical practice rather than an inherent property of the domain.

In actuarial and financial risk theory, risk is defined as probability-weighted loss (Li et al., 2020). The federal codification of risk in information systems governance, from Courtney (1977) through FIPS PUB 65, drew on this same tradition. The expected loss structure is mathematically orthodox and cross-



disciplinary. Its presence in cybersecurity governance is not a recent managerial imposition; it is the original architecture.

This historical record has a direct implication for how cybersecurity training should be understood and evaluated. If information systems risk management has always required structured estimation of likelihood and impact, then a training program that develops technical competence without developing that estimation capability is not offering a simplified form of risk-based practice. It is omitting the analytical function that distinguishes risk management from threat response. The concern this study addresses is not that cybersecurity professionals have failed to adopt lessons from an adjacent discipline. It is that training programs have not adequately developed the competencies that cybersecurity's own governance tradition has required from the beginning.

For this study, risk is operationalized consistently with this cross-framework consensus: risk reflects the structured interaction of likelihood and impact, and its management requires the capacity to estimate, compare, and act on that interaction. This definition does not introduce a novel interpretation. It applies the definition embedded in information systems governance doctrine since 1977, one that every major enterprise risk framework has continued to require.

### 2.2.3    *Competency-Based Learning Frameworks*

Competency-based learning (CBL) theory posits that professional education should align with clearly defined skillsets and knowledge areas relevant to real-world tasks (Petersen et al., 2020). The NICE Framework, developed by the National Institute of Standards and Technology, categorizes cybersecurity roles into specific knowledge, skills, and abilities (KSAs). Similarly, the SPARTA Framework in Europe maps sector-specific cybersecurity competencies. By leveraging these competency-based frameworks, this study evaluates how cybersecurity curricula structure risk management content.



### *2.2.4    Workforce Development and Training Models*

Workforce development theories emphasize aligning training programs with industry needs to ensure that professionals are adequately prepared to meet workforce demands (Caulkins et al., 2018). Any gap between cybersecurity education and practical risk management competencies would suggest that training models may not fully align with employer expectations. This study applies workforce development models to assess whether cybersecurity training effectively prepares professionals to perform cyber risk assessments and implement mitigation strategies.

## 2.3    Summary of Literary Source Analysis

This study draws on a diverse range of literature sources to ensure a comprehensive and balanced review of existing research. This study categorizes these sources by document type, publication setting, knowledge type, methodology, and thematic classification. The distribution of sources provides insights into the breadth and depth of existing research on cybersecurity training, risk management, and workforce development.

### *2.3.1    Categorization of Literature*

This study categorizes the reviewed literature into the following thematic areas:

1. **Cybersecurity Training & Education (CSTE)**: This category focuses on studies related to cybersecurity skills training, workforce development, and education curricula.

2. **Cyber Risk & Threats (CRTH)**—This category includes studies on cyber risk, threat perception, cybercrime, and security incidents.



3. **Enterprise & Organizational Risk Management** (**EORM**)—This category includes sources that examine enterprise risk management (ERM), corporate governance, and organizational risk oversight.

4. **Cybersecurity Governance & Compliance** (**CGOV**) – This category covers regulatory frameworks, compliance, and security governance strategies.

5. **Risk Management Frameworks & Methodologies** (**RMFM**) – This category discusses theoretical and practical risk management models and frameworks.

6. **Quantitative & Economic Risk Analysis** (**QERA**) – This category focuses on financial risk modelling, cost analysis, and quantitative approaches to cybersecurity risk.

7. **Information Security & Technical Controls** (**ISTC**) – This category addresses technical security measures, system vulnerabilities, and security mechanisms.

8. **Behavioural and Psychological Aspects of Security (BAPS)**—This category explores human factors, security awareness, and behavioural influences on risk management (Haag et al., 2021; Sommestad et al., 2015).

| Literature Category | | |
|---|---|---|
| Behavioural & Psychological Aspects of Security | 4 | 3% |
| Cybersecurity Governance & Compliance | 7 | 5% |
| Cyber Risk or Threats | 10 | 8% |
| Cybersecurity Training & Education | 22 | 17% |
| Enterprise & Organizational Risk Management | 16 | 12% |
| Information Security & Technical Controls | 23 | 18% |
| Literature Review & Research | 4 | 3% |
| Quantitative & Economic Risk Analysis | 7 | 5% |
| Risk Management Frameworks & Methodologies | 35 | 27% |
| Sector-Specific Cybersecurity & Risk | 2 | 2% |
| **Totals:** | **130** | **100%** |



9. **Sector-Specific Cybersecurity & Risk (SSCR)** – This category examines cybersecurity and risk management within specific industries (e.g., healthcare, finance, critical infrastructure).

10. **Literature Review & Research (LITR) –** This category provides insight into research, research methodologies and the current state of knowledge.

The statistics of the thematic areas for the literature cited in this study were as follows:

This body of literature served as the foundation for this dissertation, which analyzed how cybersecurity training programs incorporated risk management competencies. The presence of substantial research in Risk Management Frameworks & Methodologies (27%), Information Security & Technical Controls (18%), Cybersecurity Training & Education (17%), and Enterprise Risk Management (12%) underscores the academic and professional recognition of risk-based cybersecurity approaches. However, the comparatively lower representation of Behavioural & Psychological Aspects of Security (3%) and Cyber Risks and Threats (8%) suggests that research on cybersecurity skill development and human-centred risk management remains underdeveloped.

| Document Types | Count | Percentage |
|---|---|---|
| Book | 16 | 12% |
| Book Section | 2 | 2% |
| Conference Paper | 15 | 12% |
| Forum Post | 2 | 2% |
| Journal Article | 61 | 47% |
| Report | 24 | 18% |
| Thesis | 5 | 4% |
| Other | 2 | 2% |
| Blogpost | 2 | 2% |
| Webpage | 1 | 1% |
| **Totals:** | **130** | **100%** |

| Publication setting | Count | Percentage |
|---|---|---|
| Academic | 66 | 51% |
| Engineering | 1 | 1% |
| Industry | 13 | 10% |
| Mixed | 18 | 14% |
| Practitioner | 26 | 20% |
| Public | 3 | 2% |
| Other | 3 | 2% |
| **Totals:** | **130** | **100%** |

**2.3.2    *Literature Distribution and Characteristics***



The literature reviewed for this dissertation spans diverse document types, publication sources, and methodological approaches.  The distribution and Characteristics of the literature that has informed this study are detailed here.

**Document Types**: Journal articles (47%) and reports (18%) constitute a significant portion of the sources, while books (12%) and conference papers (12%) also contribute to the academic discourse.

**Publication Settings**: Academic sources constitute the largest share (51%), followed by industry, mixed settings, and practitioner-focused literature.

**Knowledge Type**: The majority of sources are applied studies (24%) and theoretical studies (22%), with a smaller subset comprising empirical studies (19%), reports (16%), and mixed studies (11%).

**Methodology:** A balanced mix of qualitative (25%), quantitative (23%), and not applicable (N/A – 33%) ensures methodological diversity.

**Time Span:** The literature spans 1970-2025, with an average publication year of 2013, reflecting the field's continuous evolution.

### 2.3.3    *Relevance of Literature to this Study*

This review categorizes and analyzes the literature used here, ensuring a structured approach to assessing the role of cybersecurity training in preparing professionals for risk-based decision-making.

| Methodology | | |
| --- | --- | --- |
| Qualitative | 33 | 25% |
| Quantitative | 30 | 23% |
| Mixed | 24 | 18% |
| N/A | 43 | 33% |
| **Totals:** | **130** | **100%** |

| Year | |
| --- | --- |
| Oldest | 1970 |
| Newest | 2025 |
| Average | 2013 |

| Knowledge Type | Count | Percentage |
| --- | --- | --- |
| Applied | 31 | 24% |
| Theoretical | 29 | 22% |
| Empirical | 25 | 19% |
| LitReview | 6 | 5% |
| Report | 16 | 12% |
| Technical | 9 | 7% |
| Mixed | 14 | 11% |
| **Totals:** | **130** | **100%** |



This classification also highlights gaps in current research, thereby supporting the rationale for further investigation into integrating risk management competencies into cybersecurity curricula.

### 2.4   Review of Existing Literature

Now, understanding the nature of the literature that informs this dissertation, what does that literature tell us, and what themes emerge?

#### 2.4.1   *Cybersecurity as an Enterprise Risk Management Problem*

The literature consistently frames cybersecurity as an applied form of enterprise risk management, emphasizing the identification, assessment, prioritization, and treatment of risk rather than the isolated implementation of technical controls. The industry widely recognizes frameworks such as ISO 31000, NIST RMF, and enterprise risk management (ERM) as foundational to effective cybersecurity practice. Baiden (2024) and Antonucci (2017) demonstrate that organizations that adopt formal risk management approaches experience improved cybersecurity resilience; however, it remains unclear how effectively training programs embed these frameworks. Significantly, Baiden (2024) goes beyond descriptive association and empirically identifies risk management as the primary mediating factor through which cybersecurity capabilities translate into improved organizational resilience and program performance. Perera (2019) similarly argues that organizations adopting ERM strategies tend to demonstrate improved risk awareness and resilience against cyber threats.  Importantly, enterprise risk management frameworks are not solely conceptual governance models; they are structured analytical systems that estimate the likelihood of adverse events and the magnitude of their consequences. This probabilistic structure has deep roots in federal information systems doctrine and financial risk theory. Therefore, framing cybersecurity as an ERM function implicitly requires applying structured likelihood–impact reasoning rather than solely implementing technical safeguards.



This framing transcends scholarly consensus; the authoritative policy documents governing federal information security codify this definition as a mandatory standard across the United States. NIST SP 800-39 (2011) defines the risk management process as an organization-wide function and explicitly requires leaders to integrate risk-based decision-making across strategic, operational, and tactical levels, rather than confining it to a technical security function (NIST, 2011, pp. 7–9). This organizational integration requirement directly implies that cybersecurity practitioners are not technical implementers who occasionally consult risk documentation; they are contributors to an enterprise-level risk governance process. NIST SP 800-37 Revision 2 (2018b) reinforces this expectation through a risk management Framework structured around six steps — Prepare, Categorize, Select, Implement, Assess, and Authorize — each of which requires risk-calibrated judgment rather than rule-following compliance. NIST SP 800-12 Revision 1 (Nieles et al., 2017), the foundational introductory document for the entire NIST information security guidance family, establishes the governing principle: all security measures must be "commensurate with the risk and magnitude of harm" posed by each specific threat scenario (Nieles et al., 2017, Section 2.5). The implication is unambiguous — security control selection is not a technical process of template application; it is an analytical process of risk-calibrated judgment. While the literature describes cybersecurity as an "applied form" of enterprise risk management, these policy documents go further: They establish normative expectations that diverse jurisdictions and sectors reflect in their own standards. This study anchors the research problem in an explicit, existing normative standard, investigating whether the workforce meets established policy rather than a mere theoretical ideal.

Empirical studies further highlight the gap between the conceptual emphasis and the practical application. Scala et al. (2019) identify risk management as fundamental to addressing cybersecurity threats, yet note its inconsistent application in professional training. Similarly, Stine et al. (2020)



emphasize the importance of aligning cybersecurity activities with enterprise risk management competencies to ensure coherence with organizational objectives. Despite the need for this emphasis, Zwilling (2022) observes that academic and professional literature continues to focus predominantly on tools, controls, and frameworks, with risk management frequently referenced but rarely developed as an operational competency. As Zwilling notes, "There is still a gap between the companies' investment in cyber security controls and their willingness to invest in CISOs' skills" (Zwilling, 2022, pp. 11, 27).

This gap has implications for leadership and workforce development. Libicki et al. (2015) observe that CISOs, vendors, and the broader security community often prioritize threats over risk-based decision-making, suggesting an emphasis on identifying adversarial activity rather than estimating probabilistic exposure and comparative impact. This observation raises concerns about the depth of risk-management expertise among cybersecurity leaders. At the same time, workforce competency frameworks such as NICE, ECSF, and SPARTA do not explicitly classify risk management as a core competency (European Union Agency for Cybersecurity, 2022; Petersen et al., 2020), which may reinforce deficiencies in formal training. While ERM frameworks such as COSO ERM and ISO 31000 provide structured methodologies for integrating cybersecurity into organizational risk governance (Hess, 2021; Wilkinson, 2020), evidence suggests that risk awareness alone is insufficient if risk is not embedded into routine professional judgment (Cains et al., 2022). These findings directly inform RQ2, which examines the extent to which cybersecurity training materials align with established risk management frameworks and prepare professionals for risk-based decision-making in practice.

### 2.4.2    Conceptual Ambiguity of Risk in Cybersecurity

The literature widely describes cyber risk and threat assessment as foundational components of cybersecurity practice, intended to enable organizations to identify, analyze, and mitigate potential



threats before they materialize into incidents. Wilkinson (2020) and Hess (2021) note that structured

methodologies such as NIST RMF, ISO 27005, and Factor Analysis of Information Risk (FAIR) provide

systematic approaches for assessing cyber risk. Despite the availability of these frameworks, the

literature reveals persistent ambiguity in how risk is conceptualized and operationalized in cybersecurity

contexts.

However, established risk doctrine within information systems governance defines risk as the

structured interaction of likelihood and impact. Against this historical baseline, ambiguity in

cybersecurity discourse does not merely reflect variation in terminology; it reflects inconsistency in

treating risk as a probabilistic construct or as a descriptive synonym for threat or hazard. Oltramari and

Kott (2018) empirically illustrate this conceptual ambiguity. They find that cybersecurity practitioners

routinely describe and reason about cyber risk using idiosyncratic, practice-specific language that

conflicts with the formal risk definitions within established risk management frameworks. In many cases,

this reasoning emphasizes system states, vulnerabilities, or adversarial capabilities without explicitly

estimating their likelihood or impact magnitude, thereby weakening the probabilistic structure

embedded in enterprise risk doctrine.

Several studies emphasize the growing importance of threat intelligence and proactive

assessment strategies in cybersecurity operations and training (Scala et al., 2019). Nevertheless, this

emphasis often conflates risk assessment with threat identification and technical analysis, rather than

explicit evaluation of likelihood, impact, and risk-based prioritization. Dawson and Thomson (2018)

observe that many cybersecurity professionals develop risk assessment capabilities informally through

on-the-job experience, suggesting that formal education frequently under-specifies how risk should be

defined, measured, and used to guide decision-making. This inconsistency contributes to variation in



how risk management competencies are emphasized across training models, supporting RQ4's focus on differences in professional discourse and instructional depth.

The increasing adoption of technology-driven cybersecurity solutions further reinforces this ambiguity. Advances in artificial intelligence, machine learning, and automation have significantly improved threat detection, anomaly identification, and incident response capabilities (Sun et al., 2019). While these technologies enhance operational effectiveness and support broader goals of cyber resilience, defined as the ability to prepare for, withstand, and recover from cyber incidents, they also risk shifting attention away from human risk judgment. As attackers adapt to evade automated controls, over-reliance on AI-driven defences may blur the distinction between threat management and risk management (Szterenfeld, 2022). Collectively, the literature suggests that although risk terminology is pervasive, its conceptual boundaries remain blurred, with risk frequently subsumed under technical or threat-centric perspectives rather than treated as a distinct probabilistic decision-making construct grounded in likelihood–impact estimation. This divergence raises important questions about whether cybersecurity training explicitly develops the analytical competencies required by established risk management frameworks. This observation aligns with RQ5, which sought to identify patterns in professional discourse regarding the practical application of risk management competencies and the influence of emerging technologies on this understanding.

### 2.4.3    Workforce Competency Frameworks as Proxies for Training Content

Cybersecurity frameworks play a central role in shaping cybersecurity education, professional training programs, and workforce expectations. Widely adopted frameworks, such as the National Institute of Standards and Technology (NIST) Cybersecurity Framework (CSF) and ISO/IEC 27001, provide structured approaches to improving cybersecurity risk management and governance within



organizations. As a result, these frameworks increasingly influence how cybersecurity competencies are defined, taught, and assessed across academic, certification, and industry-led training programs.

The literature, however, suggests that translating framework guidance into cybersecurity education and training remains inconsistent. Marotta and Madnick (2020) highlight the benefits of aligning cybersecurity curricula with standards such as NIST and ISO to improve workforce readiness, while noting variation in how comprehensively educational programs reflect these standards. This observation is directly relevant to RQ3, which examines how cybersecurity professionals' perceptions of their training align with the actual content embedded within recognized competency frameworks. In contrast, AlDaajeh et al. (2022) find that national cybersecurity strategies often lack alignment with international standards, leading to fragmentation and inconsistent professional competencies across jurisdictions.

Workforce competency frameworks further formalize expectations for cybersecurity roles and skills. Frameworks such as the NIST NICE Cybersecurity Workforce Framework and the SPARTA Cybersecurity Skills Framework articulate the knowledge, skills, and abilities expected of cybersecurity professionals across a range of functions and sectors. Industry research by ISC[2] (2019, 2024) and SANS (2025) underscores the growing demand for cybersecurity professionals capable of risk-based decision-making; however, the literature indicates that training programs frequently overemphasize technical proficiency and underemphasize structured risk management instruction (Petersen et al., 2020; Shreeve et al., 2021). For example, Shreeve et al. (2021) demonstrate that many programs prioritize tools, controls, and hands-on defensive or offensive techniques, while providing limited attention to strategic, contextual, and risk-based reasoning.

Consequently, many cybersecurity professionals report acquiring risk assessment and decision-making skills primarily through on-the-job experience rather than formal education, contributing to



misalignment between training programs and industry needs (Dreyer et al., 2018). Despite the

proliferation of cybersecurity certifications and training initiatives, the literature continues to identify a

shortage of professionals with well-developed risk management competencies, reinforcing calls for

reforming cybersecurity education curricula (Chowdhury & Gkioulos, 2021).

### 2.4.4    *Misalignment Between Cybersecurity Training and Risk Management Competencies*

Information security governance is widely recognized as a critical component of effective

cybersecurity management, ensuring that security activities align with organizational objectives,

regulatory requirements, and enterprise risk priorities. Governance-oriented frameworks such as

ISO/IEC 27001 and the NIST Cybersecurity Framework (CSF) provide structured approaches for managing

information security at an organizational level; however, the literature suggests that cybersecurity

education and training do not consistently reflect the principles within these frameworks.

Schinagl and Shahim (2020) describe the evolution of information security governance from a

predominantly technical function toward a strategic, board-level concern, emphasizing the need to

integrate governance models with enterprise risk management (ERM) to support effective risk

mitigation. Webb et al. (2014) further evidence this misalignment by identifying poor or perfunctory risk

management as a dominant factor in cybersecurity program failure, accounting for most of the top

reasons initiatives fail despite substantial investment in technical controls. Within the established

doctrine of risk management outlined earlier in this chapter, such deficiencies would imply inadequate

estimation of likelihood, impact, and comparative exposure rather than merely insufficient

documentation or policy alignment. In other words, cybersecurity program failure may reflect

underdeveloped probabilistic reasoning capacity in addition to governance shortcomings. Importantly,

within established risk doctrine, "poor risk management" does not simply refer to insufficient



governance oversight or documentation; it reflects inadequate estimation of likelihood, impact, and comparative exposure. Historically, risk management required structured probability-weighted analysis. If cybersecurity training emphasizes compliance, frameworks, and technical safeguards without developing these analytical competencies, misalignment may stem not only from organizational structure but from underdeveloped risk estimation capability.

This perspective aligns with RQ4, which examines how cybersecurity professionals conceptualize and communicate risk management competencies within leadership and training contexts. Similarly, Sumner et al. (2024) characterize cybersecurity as an emerging governance challenge, noting that deficiencies in executive understanding or governance structures can exacerbate organizational vulnerability to cyber risk.

The literature further suggests that cybersecurity training can increase risk when programs do not sufficiently develop governance and risk management competencies. Moore et al. (2020) caution that cybersecurity education focused primarily on technical skill acquisition may increase institutional risk if not accompanied by adequate emphasis on governance, ethics, and controls, observing that "cyber risk for an institution that teaches cybersecurity goes beyond baseline cybersecurity and includes behavioural risks that develop as student populations acquire and practice newfound cybersecurity skills" (Moore et al., 2020, p. 2). Collectively, these findings indicate a persistent misalignment between the technical focus of cybersecurity training programs and the risk management and governance competencies required for effective professional practice. Given that enterprise risk frameworks rely on structured likelihood–impact estimation to support prioritization and resource allocation, insufficient emphasis on these analytical competencies may contribute to organizational vulnerability even when technical controls are robust. This misalignment may include insufficient emphasis on structured



likelihood–impact estimation, which underpins prioritization and resource allocation within enterprise risk management frameworks.

### 2.4.5    *Summary of Review of Existing Literature*

Collectively, the literature reviewed in this Section reveals four interrelated themes that underpin this study. First, cybersecurity is consistently positioned in theory as an application of enterprise risk management, emphasizing structured approaches to identifying, assessing, prioritizing, and treating risk. Second, despite the widespread use of risk terminology, the literature demonstrates persistent conceptual ambiguity, with risk frequently conflated with threats, vulnerabilities, or technical controls rather than treated as a distinct probabilistic decision-making construct grounded in structured likelihood–impact estimation. Third, cybersecurity workforce competency frameworks and standards, including NICE, ECSF, and SPARTA, emerge as influential proxies for training content, shaping how cybersecurity knowledge and skills are defined and taught across educational and professional contexts. Finally, the literature highlights a misalignment between cybersecurity training programs and the risk management and governance competencies required for effective professional practice, suggesting that training programs often develop technical capability without emphasizing risk-based judgment, leadership, and governance. Together, these themes underscore the need to empirically examine how cybersecurity training integrates risk management and how this integration influences professional competency in practice.

### 2.5    Discussion & Key Findings

This Section synthesizes and interprets the findings of the literature review in relation to the study's research questions and theoretical framework. As established earlier in this chapter, enterprise risk doctrine—rooted in federal information systems standards and subsequent enterprise



frameworks—defines risk as the structured interaction of likelihood and impact. The NIST policy

documents examined in this chapter — SP 800-12r1 (2017), SP 800-39 (2011), and SP 800-37 Revision 2

(2018) — collectively establish that this doctrine is not merely theoretical: it constitutes the operative

standard against which professional cybersecurity practice in information systems governance contexts

is formally measured. The analysis that follows interprets the literature against that standard. This

probabilistic foundation serves as the analytical baseline for the following interpretation of the theme.

Rather than reiterating individual sources, the discussion is organized around four cross-cutting themes

that emerged from the review: the positioning of cybersecurity as an enterprise risk management

function, persistent ambiguity in the way the industry conceptualizes cyber risk, the role of workforce

frameworks as proxies for training content, and the resulting misalignment between cybersecurity

training and applied risk management competencies. Together, these themes clarify how current

training practices may fail to develop the risk-based decision-making capabilities required in professional

cybersecurity practice and establish a clear rationale for the empirical investigation presented in

Chapter 3.

### 2.5.1    Risk Management as the Missing Integrative Capability

The literature reviewed indicates that cybersecurity training continues to emphasize technical

skills, while risk management competencies remain underrepresented. As established earlier in this

chapter, risk management doctrine within information systems governance is grounded in structured

estimation of likelihood and impact. Therefore, underrepresentation of risk competencies may not

simply reflect limited exposure to governance frameworks, but limited development of probabilistic

reasoning and exposure analysis skills required for prioritization and resource allocation. Risk-based

reasoning is frequently overshadowed by technical instruction, despite evidence that effective



cybersecurity practice requires analytical thinking, decision-making under uncertainty, and an understanding of organizational context (Shreeve et al., 2021). While workforce competency frameworks such as NICE, ECSF, and SPARTA categorize cybersecurity roles across technical and governance domains, the extent to which these frameworks explicitly integrate structured risk management education remains unclear. Cybersecurity training often assumes risk competence rather than validating it, reinforcing a disconnect between technical capability and risk-based decision-making (Zwilling, 2022). This misalignment suggests that cybersecurity education may insufficiently prepare professionals to apply risk management competencies—particularly structured likelihood–impact estimation and comparative exposure analysis—in real-world contexts.

### 2.5.2    *Conceptual and Linguistic Ambiguity of Cyber Risk*

The literature further reveals persistent ambiguity in how cyber risk is conceptualized and communicated within the profession. Oltramari and Kott (2018) document that the dominant practitioner risk construct (used by 78% of professionals in their study) is an adversarial tuple organized around system state, vulnerability, exploit path, and available counteractions, with no probabilistic or magnitude components. This risk construct is not definitional imprecision; it is a structurally different construct that cannot produce the risk-prioritization outputs enterprise governance requires. Similarly, Cains et al. (2022) find that cybersecurity discourse frequently references risk without a consistent definition or operationalization. Although these studies highlight deficiencies in how cyber risk is understood, they do not examine whether such shortcomings originate in cybersecurity education and training. When considered against the historical and cross-framework convergence of risk as a probabilistic interaction between likelihood and impact (See Section: The Historical Convergence of Risk as Expected Loss), this ambiguity becomes analytically significant. It suggests not merely variation in



terminology, but potential inconsistency in whether practitioners treat risk as a measurable construct capable of supporting prioritization and trade-off decisions. This gap underscores the need to investigate how formal training programs introduce, frame, and reinforce risk management competencies, particularly whether they explicitly develop structured likelihood–impact reasoning consistent with established enterprise risk frameworks.

### 2.5.3    *Workforce Frameworks and the Assumption of Risk Competence*

Despite widespread adoption of workforce competency frameworks, empirical research on the explicit inclusion of risk management competencies in cybersecurity training programs remains limited. While educators treat frameworks such as NICE, ECSF, and SPARTA as proxies for curriculum content, few studies have systematically assessed whether these frameworks operationalize risk management skills as a core professional competency. Moreover, the mere inclusion of the term "risk" within a competency framework does not guarantee alignment with the probabilistic foundations of enterprise risk doctrine. This literature review has shown that effective risk management historically required explicit estimation of likelihood and impact. If frameworks reference risk descriptively—without specifying structured exposure analysis or prioritization logic—they may implicitly assume risk competence rather than define and develop it. Zwilling (2022) notes that prior research has largely avoided hypothesis-driven evaluation of skill gaps, calling for empirical examination of deficiencies in executive and professional risk management capabilities. Similarly, Shreeve et al. (2021) identify misalignment between cybersecurity education and organizational needs but stop short of evaluating whether training explicitly develops risk management competencies grounded in established risk theory, including structured likelihood–impact estimation and comparative exposure analysis.

### 2.5.4    *Implications of Training–Competency Misalignment*



The consequences of these gaps extend beyond individual skill development to organizational resilience and risk exposure. Research suggests that insufficient risk-based decision-making contributes significantly to cybersecurity failures and organizational vulnerability (Perera, 2019). Moore et al. (2020) caution that cybersecurity education may inadvertently increase institutional risk when training programs develop technical capabilities without emphasizing governance, ethics, and controls. Case studies further indicate that organizations lacking structured risk management training experience greater financial losses, operational disruption, and reputational damage following cyber incidents (Marotta & Madnick, 2020). However, empirical research directly linking these outcomes to deficiencies in cybersecurity education remains limited. Addressing this gap, the present study applies Natural Language Processing (NLP) techniques to systematically evaluate how cybersecurity training frameworks incorporate risk management competencies, building on prior methodological work by Schatz et al. (2017) and Mavrogiorgos et al. (2022b). Consistent with prior findings (Baiden, 2024; Goode, 2018), this study hypothesizes that cybersecurity training programs insufficiently emphasize risk management and fail to adequately prepare professionals for risk-based decision-making in practice ($H_{13}$). By addressing these gaps, this research not only evaluates the extent to which cybersecurity training incorporates structured risk management education but also explores how these gaps may contribute to broader organizational risks. These findings could provide valuable insights for curriculum development, workforce training strategies, and policy recommendations to strengthen individual and institutional cybersecurity resilience.

## 2.6    Summary of Literature Review Findings

This literature review highlights key themes related to cybersecurity training, workforce competency frameworks, and the integration of risk management competencies. Despite growing



recognition of the importance of risk management in cybersecurity, current research often overlooks integrating risk management competencies into training curricula. Few studies have employed systematic methods, such as NLP, to evaluate the prevalence of these principles within existing training programs (Baiden, 2024; Švábenský et al., 2020). This study sought to address these gaps by employing a mixed-methods approach to examine the incorporation of risk management competencies across several key cybersecurity frameworks, including NIST, NICE, and SPARTA. Existing research emphasizes the predominance of technical skill development, with significant attention devoted to network security, encryption, incident response, and ethical hacking. However, cybersecurity training programs inconsistently integrate structured risk management education. This potential gap suggests the need for further empirical research to determine the extent of risk management instruction within the cybersecurity curricula.

The findings from this review reinforce the importance of assessing whether current cybersecurity training programs adequately prepare professionals for risk-based decision-making (Caulkins et al., 2018). Workforce competency models, such as NICE, ECSF, and SPARTA, provide structured approaches to cybersecurity education; however, it remains unclear whether they sufficiently emphasize risk management competencies. Additionally, while cybersecurity education often aligns with compliance-based frameworks, it may not necessarily equip professionals with the strategic risk-assessment skills required in dynamic threat environments.

This study employs Natural Language Processing (NLP) techniques to analyze cybersecurity training frameworks and systematically address these gaps. This research sought to provide data-driven insights into how cybersecurity education aligns with workforce expectations by examining the extent to which risk management is represented and emphasized in established competency models. This study is significant because it sought to bridge the gap between cybersecurity training and real-world, risk-based



decision-making, ultimately contributing to the development of effective curricula and workforce readiness strategies. In doing so, the study evaluates whether contemporary cybersecurity training maintains continuity with the probabilistic foundations of risk embedded in federal standards and enterprise risk management doctrine.

The next chapter will outline the research methodology and detail the approach to analyze risk management integration in cybersecurity training programs.



**Chapter 3 – Research Methods**

This chapter describes the mixed-methods methodology used to evaluate the integration of risk management within cybersecurity training and its relationship to professional practice. Using the 29 competency areas of the NIST NICE framework as an analytical structure, Tasks, Knowledge, and Skills (TKS) statements were categorized via an NLP pipeline to assess the presence and distribution of risk management content across training frameworks. This content analysis operationalizes the measurement of training emphasis (RQ1–RQ3).

To examine whether exposure to risk-embedded training translates into real-world professional competence, survey data from cybersecurity professionals were analyzed using a theory-driven structural equation model grounded in competency-based learning theory. The model tested whether training exposure influences risk management competence directly and indirectly through perceived relevance, thereby linking training content to observed work practices.

The study uses structural equation modelling because this approach allows simultaneous estimation of direct and mediated pathways. The research question was not limited to whether risk content exists in training, but whether such exposure translates into competence through cognitive mechanisms of perceived relevance and conceptual salience. A regression-based approach would be insufficient to test this mediated structure, whereas SEM enables evaluation of the training → salience → competence pathway within a unified theoretical model.

The chapter further details participant selection, ethical considerations, methodological limitations, and mitigation strategies employed to ensure reliability and validity.

Given the mixed nature of this study, combining both qualitative and quantitative methods enables a comprehensive examination of the research questions, specifically the presence of risk management concepts within training materials and their practical application among cybersecurity



professionals. The study employed a mixed-methods approach, combining Natural Language Processing

(NLP) techniques with survey and interview data to comprehensively analyze training content and its

alignment with workforce competency frameworks, particularly the NIST NICE framework.

Given the breadth and variability of cybersecurity training materials across academic,

certification-based, and industry-led programs, traditional manual content analysis becomes impractical.

The decision to use NLP stems from the need to efficiently analyze large volumes of text and objectively

assess the presence of risk management concepts. NLP provided a scalable, data-driven solution for

identifying patterns, themes, and gaps, with a primary focus on integrating risk management into

training materials. This approach works well with the rationale for a mixed-methods approach provided

by Creswell (2014) and Teddlie and Tashakkori (2009).

Beyond issues of scale, cybersecurity training frameworks differ in structure, terminology, and

granularity, making direct manual comparison across frameworks conceptually inconsistent and difficult

to replicate. A purely qualitative review risks interpretive drift and lacks cross-framework

standardization. The use of NLP enabled the normalization of terminology, the systematic categorization

of TKS statements, and the quantifiable comparison across heterogeneous training sources. This

approach applied consistent analytical criteria to evaluate differences in risk management emphasis,

thereby eliminating subjective interpretation.

This approach followed a systematic structure to ensure rigour and replicability. It begins with

defining the research design and then provides a detailed description of the data sources, including how

NLP categorized and quantified the presence of risk management among the 29 competency areas of

the NIST NICE framework. The process involved categorizing Tasks, Knowledge, and Skills (TKS)

statements into one of the 29 competency areas, with an emphasis on evaluating risk management

content across these categories. The chapter details the NLP pipeline, participant selection, and ethical



considerations. It concludes with a discussion of the methodology's limitations and the strategies employed to mitigate potential biases and inaccuracies, ensuring the reliability of the results. This study assessed risk management competency by surveying cybersecurity professionals and developing a structural equation model linking training competency to work practices.

### 3.1    Research Philosophy & Approach

This study is grounded in a pragmatist philosophical worldview. The research problem is simultaneously structural, whether risk management cognition is architecturally absent from training frameworks; cognitive, whether that absence reproduces itself in professional reasoning; and organizational, whether the gap between what professionals can do and what leaders assume they can do produces governance consequences. No single methodological tradition is adequate to span that span. Quantitative measurement of training content, latent variable modelling of professional competence, and thematic analysis of leadership cognition are not alternatives to one another; they address different levels of the same causal chain. Pragmatism, therefore, provides the philosophical warrant for a mixed-methods design that integrates NLP content analysis, survey-based SEM, cross-group benchmarking, and qualitative interview analysis, treating each strand as a complementary source of evidence for a common explanatory claim (Creswell & Clark, 2018; Morgan, 2014).

The study adopts an abductive reasoning approach (Peirce, 1878; Timmermans & Tavory, 2012). Competency-Based Learning Theory and risk management theory provided the a priori basis for the hypotheses, specified a priori (from Competency-Based Learning Theory and risk management theory), which this study tested through confirmatory SEM. However, the research problem carries genuine structural uncertainty. If training frameworks are architecturally deficient in risk management content, the downstream effects on professional cognition may not follow the patterns that existing theory



predicts with clean precision. A purely deductive design assumes the theory is sufficient to anticipate all meaningful patterns in the data; a purely inductive design foregoes the theoretical grounding that makes the findings interpretable. Abduction accommodates both; it begins with theoretically motivated hypotheses and remains open to extending the explanatory framework where the data warrant it. This iterative logic reflects the exploratory-confirmatory hybrid nature of the design. This approach is appropriate because prior research has not empirically tested the causal chain from training architecture to professional competence in this domain.

### 3.2    Research Design

The research design was a mixed-methods design, combining Natural Language Processing (NLP) techniques with qualitative insights from cybersecurity professional surveys and semi-structured interviews (Creswell, 2014). The design sought to evaluate the extent to which cybersecurity training programs integrate risk management competencies and align with workforce competency frameworks, specifically the NIST NICE, ECSF, and SPARTA frameworks.  By employing a mixed-methods research design, this study captures both the structural integration of risk management concepts in training and the practical competencies cybersecurity professionals demonstrate in the workplace. This approach is appropriate because it enables quantitative analysis of training content and professional responses while also supporting qualitative interpretation of decision-making and risk conceptualization.

The research adopts an explanatory sequential design. The NLP analysis is conducted first to identify patterns and potential gaps in risk management training content, followed by qualitative interviews to explore and contextualize these findings. The quantitative analysis focused on identifying the extent of coverage of risk management competencies in training materials. In contrast, the qualitative analysis gathered professional insights into how these principles are applied and perceived in



practice. Both quantitative and qualitative analyses used the same 29 competency areas to ensure

consistency across the study, enabling a comprehensive comparison between training content and real-

world professional practice.

With this overall design in mind, the following Sections detail the participants, data collection

methods, and analytical strategies that ensured our study directly addresses the research questions.

### 3.2.1 Quantitative Component: NLP Content Analysis

The quantitative phase applied Natural Language Processing (NLP) techniques to analyze textual

data from cybersecurity training frameworks and educational materials. Specifically, the analysis

focused on knowledge, skills, and abilities (KSAs) statements that underpin course curricula. This phase

of the study sought to identify and measure the presence of risk management competencies by

examining patterns of key terminology, thematic clusters, and the alignment between training content

and established competency frameworks, including the NIST NICE, ECSF, and SPARTA frameworks. The

NLP analysis categorized each TKS statement into one or more of the 29 competency areas, using

semantic similarity measures to quantify the extent to which training materials incorporate risk

management competencies.

Since the 29 competency areas are primarily independent, the analysis focused solely on risk

management content within the risk management category, without assessing its integration with other

categories, such as Threats and Vulnerabilities, Incident Management and Response, or Cybersecurity

Strategy and Governance.

### 3.2.2 Qualitative Component: Professional Insights

The qualitative phase involved collecting survey and interview data from cybersecurity

professionals across various sectors, including management and other relevant areas. These insights



provide context regarding how risk management competencies are perceived and applied in practice. Responses are analyzed using NLP-based discourse analysis to measure the frequency and context of risk-related terminology in professional communication. In addition, this research mapped these qualitative insights onto the same 29 competency areas used in the quantitative and NLP analyses. This shared competency architecture functioned not merely as a classification scheme but also as a cross-method analytical framework, enabling structural comparison among training content, professional cognition, and leadership expectations. By anchoring all data sources to a unified competency ontology, the study evaluated whether workforce frameworks architecturally embed risk management and whether that structure aligns with observed professional behaviour and executive assumptions.

### 3.2.3    Justification for the Mixed-Methods Approach

This study employs a mixed-methods approach, leveraging objective NLP analysis of training content alongside professional insights to evaluate how training translates into real-world risk management competencies. The integration of methods addresses a fundamental translation challenge: the movement from formal curriculum content to professional cognition and from cognition to observable workplace practice. Document analysis alone cannot establish whether the emphasis on training meaningfully shapes professional decision-making. Similarly, survey data alone cannot determine whether perceived competence is grounded in formal training structures. The mixed-methods design was therefore necessary to evaluate alignment across structural, cognitive, and behavioural levels of analysis.

The quantitative component (NLP analysis) evaluated the extent to which cybersecurity curricula incorporate risk management concepts. In contrast, the qualitative component (surveys and interviews) examined professionals' perceptions of the extent to which these concepts are applied



effectively in practice.  Teddlie and Tashakkori (2009) and Creswell (2014) outline the advantages of combining quantitative and qualitative approaches in educational studies.

The analytical approach to framework analysis evolved during the course of the study. Initial framing described the work as "NLP analysis," which accurately characterizes the methodological category: a systematic, computational analysis of text corpora to classify and quantify semantic content. As implementation progressed, the specific instrument used to perform that classification was an LLM-assisted pipeline — the GPT-4 model accessed via the OpenAI Python SDK, with structured outputs enforced through Pydantic schema validation. These terms are hierarchically related rather than synonymous: LLM-assisted classification is the instrument; NLP analysis is the broader methodological framework within which that instrument operates. Throughout this dissertation, "NLP analysis" refers to this computationally implemented approach. Where the specific instrument is relevant — particularly in the description of the classification pipeline in this chapter — the LLM-assisted implementation is identified explicitly. This distinction is maintained consistently across Chapters 3, 4, and 5

Natural Language Processing (NLP) techniques offer a systematic, objective method for measuring the coverage of risk management competencies across cybersecurity training frameworks, explicitly evaluating how thoroughly these frameworks incorporate such principles. Both Kang et al. (2020) and Pandey et al. (2017b) highlight the advantages of using a technique such as these to generate metrics.  Mavrogiorgos et al. (2022b) provide a strong example of how researchers employ NLP to mine large datasets.  While quantitative analysis using NLP provided insights into the extent and distribution of risk management content, the qualitative component—comprising surveys and semi-structured interviews—helps explain the real-world impact of these training programs on professionals' competencies (Creswell, 2014). By focusing on the 29 competency areas, the qualitative data



contextualized the extent to which the risk management training content aligns with the competencies and experiences professionals need in practice.

### 3.2.4    Variable–Hypothesis–Method Alignment

To ensure conceptual and methodological traceability, Table 2 provides a consolidated alignment of research questions, hypotheses, variables, and analytic methods. This matrix clarifies how each hypothesis is operationalized and tested across the NLP analysis, Structural Equation Modelling, qualitative interview analysis, and comparative benchmarking procedures.

**Table 2**

*Variable–Hypothesis–Method Alignment*

| Q | Hypo-thesis | Independent Variable(s) | Dependent Variable(s) | Method |
|---|---|---|---|---|
| Q1 | $H_{11}$ | IV1 (RISK KSA frequency via NLP) | Comparative frequency across domains | NLP content analysis |
| Q2 | $H_{12}$ | IV1x (Training exposure) + MeV1 (Mediator) | DV1–DV4 (Risk-based behaviours) | SEM (WLSMV) |
| Q3 | $H_{13}$ | IV1 (RISK frequency) | Proportional representation vs other competencies | NLP comparative analysis |
| Q4 | $H_{14}$ | Organizational expectations (coded interviews) | Demonstrated competence (DV1–DV4 structure) | Qualitative thematic analysis + cross-case comparison |
| Q5 | $H_{15}$ | Professional group (Cyber vs Non-cyber) | Risk Management Competence Composite | Group comparison (t-test / SEM comparison) |

### 3.2.5    Methodological Framework

The research framework is grounded in the principles of risk management theory (Hess, 2021; Wilkinson, 2020), competency-based learning (Wangen, 2016), and workforce development models (Dawson & Thomson, 2018; Goode, 2018). This alignment ensures that the study's findings are relevant to both academic and practical applications. Risk management theory provides the theoretical foundation for understanding how organizations and professionals should identify, assess, and manage



cybersecurity risks. The competency-based learning framework aligns with the 29 competency areas used in the study, ensuring that the research directly links training content with workforce competencies, specifically in risk management.

### 3.2.5.1 Taxonomy of Cybersecurity Competencies and Variable Operationalization

The study uses the NICE Framework Competency Areas (NIST, 2023) as the primary analytical framework. NICE's status as the most comprehensive North American cybersecurity framework justified its selection; it provides the detailed, operationalized competency definitions across Knowledge, Skills, and Tasks necessary for this analysis. While the ECSF and SPARTA frameworks provided a baseline for cross-framework consistency, the analysis utilized NICE as the central taxonomy. These (European) regional frameworks mirror NICE's structure and do not add unique competency domains to the study's scope. NICE is sufficiently comprehensive to serve as a representative analytical baseline for examining the integration of risk management competencies within cybersecurity training.

The resulting 29 competency areas provide a standardized taxonomy for categorizing training content and professional responses throughout the study. These competencies provide a taxonomy for analyzing training content and professional insights gathered throughout the study. The 29 competency areas cover a broad spectrum of cybersecurity competencies and serve as the foundation for evaluating the integration of risk management competencies into cybersecurity training and practice.

The 29 competency areas are as follows:

**Table 3**
*Taxonomy of Cybersecurity Competencies*

| Category ID | NIST Competency Area ID | Competency Area (Domain) | Competency Area Description |
|---|---|---|---|
| ACCO | NF-COM-001 | Access Controls | This Competency Area describes a learner's capabilities to define, manage, and monitor the roles and secure access privileges of |
| | | | authorized individuals who access protected data and resources, and understand the impact of different types of access controls. |
| AISC | NF-COM-002 | Artificial Intelligence (AI) Security | This Competency Area describes a learner's capabilities to secure Artificial Intelligence (AI) against cyberattacks, to contain AI effectively during deployment, and to mitigate the threats posed by malicious AI use. |
| ASST | NF-COM-003 | Asset Management | This Competency Area describes a learner's capabilities to conduct and maintain an accurate inventory of all digital assets, including identifying, developing, operating, maintaining, upgrading, and disposing of assets. |
| CLOU | NF-COM-004 | Cloud Security | This Competency Area describes a learner's capabilities for securing shared-responsibility technology stacks, including cloud services across IaaS, PaaS, and SaaS models. It emphasizes shared responsibility, cloud-native controls, and multi-tenant risk considerations. |
| COMM | NF-COM-005 | Communication Security | This Competency Area describes a learner's capabilities to secure the transmission, broadcasting, switching, control, and operation of communications and related network infrastructures. |
| CRYP | NF-COM-006 | Cryptography | By anchoring all data sources to a unified competency ontology, the study evaluated whether workforce frameworks architecturally embed risk management and whether that structure aligns with observed professional behaviour and executive assumptions. |
| RESI | NF-COM-007 | Cyber Resilience | This Competency Area describes a learner's capability related to architecting, designing, developing, implementing, and maintaining the trustworthiness of systems that use or leverage cyber resources to anticipate, withstand, recover from, and adapt to adverse conditions, stresses, attacks, or compromises. |
| DEVS | NF-COM-008 | DevSecOps | This Competency Area describes a learner's capabilities related to integrating security into continuous integration/continuous delivery (CI/CD) pipelines and development workflows. It emphasizes automation, configuration management, and the integration of security into delivery processes. |
| OPSC | NF-COM-009 | Operating System (OS) Security | This Competency Area describes a learner's capabilities to install, administer, troubleshoot, back up, and recover Operating Systems (OS), including in simulated environments. |
| OTSC | NF-COM-010 | Operational Technology (OT) Security | This Competency Area describes a learner's capabilities to improve and maintain the security of Operational Technology (OT) systems while addressing their unique performance, reliability, and safety requirements. |
| SCSY | NF-COM-011 | Supply Chain Security | This Competency Area describes a learner's capabilities to analyze and control digital and physical risks presented by technology products or services purchased from parties outside your organization. |
| RISK | NF-COM-012 | Cyber risk management | This Competency Area describes a learner's capabilities related to assessing, prioritizing, and treating cybersecurity risks using likelihood-and-impact models. It emphasizes risk-based decision-making across assets, systems, and enterprises, informed by technical vulnerability data. |
| MASO | NF-COM-013 | Malware & Software Analysis | This Competency Area describes a learner's capabilities for analyzing malicious code and artifacts using static/dynamic techniques, and for reverse-engineering software/protocols to understand behaviour, capabilities, and countermeasures. |
| SEHA | NF-COM-014 | Secure Hardware & Embedded Systems | This Competency Area describes a learner's capabilities for designing, configuring, and evaluating security for hardware, firmware, and embedded/edge systems (e.g., secure boot, TPM, trusted execution) throughout their lifecycles. |
| DEOP | NF-COM-015 | Detection Operations | This Competency Area describes a learner's capabilities in engineering and optimizing detection. It includes designing and tuning SIEM rules, IDS/IPS signatures, correlation logic, analytics pipelines, and adversary detection use cases; validating detection efficacy; and applying advanced analytics to identify abnormal behaviours and adversary techniques. |
| DFOR | NF-COM-016 | Digital Forensics | This Competency Area describes a learner's capabilities for collecting, preserving, and analyzing digital evidence to support investigations and incident response. It emphasizes chain of custody, artifact recovery, and evidentiary integrity. |
| SESO | NF-COM-017 | Secure Software Development | This Competency Area describes a learner's capabilities in applying secure coding practices and design principles throughout the software lifecycle, including threat modelling, code review, and dependency management. |
| SOPS | NF-COM-018 | Security Operations | This Competency Area describes a learner's capabilities related to monitoring, triaging, escalating, and responding to security events and incidents. It emphasizes playbook execution, containment and recovery coordination, and reporting in operational environments. |
| THMA | NF-COM-019 | Threat Management | This Competency Area describes a learner's capabilities related to identifying, analyzing, and remediating system vulnerabilities and configuration weaknesses. It emphasizes scanning, patching, and prioritization of flaws, often using technical severity and exploitability data. |
| CGOV | NF-COM-020 | Cybersecurity Governance | This Competency Area describes a learner's capabilities for establishing strategies, policies, standards, and oversight mechanisms aligned with business objectives and risk appetite. |
| THRT | NF-COM-021 | Cyber Threat Intelligence | This Competency Area describes a learner's capabilities related to collecting, analyzing, and disseminating intelligence on adversary campaigns, TTPs, and emerging threats to inform defence strategies. |
| DATA | NF-COM-022 | Data Security & Protection | This Competency Area describes a learner's capabilities for classifying, handling, and protecting data throughout its lifecycle (e.g., encryption, DLP, masking, retention) to preserve CIA and privacy expectations. |
| SYSA | NF-COM-023 | Systems Security Architecture | This Competency Area describes a learner's capabilities related to designing and evaluating secure systems, architectures, and control patterns. It emphasizes applying principles such as zero trust, segmentation, and layered defences. |
| PENN | NF-COM-024 | Offensive Security | This Competency Area describes a learner's capabilities related to planning and conducting authorized adversary emulation, penetration |
| | | | testing, and red team operations. It emphasizes simulating real-world attacker behaviours to assess defences. |
| AWAR | NF-COM-025 | Security Awareness | This Competency Area describes a learner's capabilities to recognize and mitigate human-centric risks by understanding threats, social engineering, and organizational security practices. It emphasizes user behaviour and culture across the workforce. |
| PLVI | NF-COM-026 | Platform & Virtualization Security | This Competency Area describes a learner's capabilities related to securing compute platforms, hypervisors, and containerized environments. It emphasizes hardening, isolation, and lifecycle management of virtualized infrastructure. |
| EMTE | NF-COM-027 | Emerging Technology Risk Management | This Competency Area describes a learner's capabilities for evaluating and addressing the security implications of rapidly evolving technologies (e.g., quantum impacts, IoT/edge; beyond model-specific AI tasks addressed in AI Security) to anticipate/mitigate novel risks. |
| CYBR | NF-COM-028 | Cybersecurity Foundations | This Competency Area describes a learner's capabilities in applying foundational cybersecurity concepts, including basic management of confidentiality, integrity, availability, defence-in-depth, and basic threat modelling. It provides the baseline knowledge needed for specialized domains. |
| TEST | NF-COM-029 | Security Testing | This Competency Area describes a learner's capabilities related to conducting verification, validation, and assurance of systems and applications. It emphasizes test design, fuzzing, and the evaluation of controls for effectiveness, separate from adversary-emulation activities. |

The workforce development models inform the study's practical implications by highlighting the skills and knowledge needed for professionals to perform effectively in cybersecurity.

By focusing on these frameworks, the study ensures its findings are theoretically grounded and applicable to cybersecurity workforce development. It primarily emphasizes risk management competencies as a crucial area of professional training.

**3.2.5.2 Structural Equation Model Specification**

The study's conceptual model is specified using structural equation modelling (SEM) to test whether risk management content in cybersecurity training programs contributes to professionals' real-world competencies in applying risk management principles. The model is built directly on the hypotheses outlined in Chapter 1 and the theoretical framework established in Chapter 2, with



Competency-Based Learning Theory and risk management Theory providing the foundation. I refined the SEM specification during analysis.

### 3.2.5.3 Constructs and Relationships

1. **Independent Variable (IV1)**: Competencies in KSA Statements.

This variable captures the presence of risk management competency in training frameworks (e.g., NIST NICE, ECSF, SPARTA). The study operationalizes this concept through natural language processing (NLP) to analyze knowledge, skills, and abilities (KSAs) statements. While IV1 represents the framework-level embedding of risk management competencies identified through NLP analysis, IV1x serves as an exogenous predictor in the structural equation model, acting as an individual-level proxy for participants' exposure to those competencies through their training experiences.

2. **Mediator Variable (MeV1)**: Conceptual Salience and Relevance of Risk Management.

This construct reflects participants' perceptions of risk management as an integral part of cybersecurity practice. It mediates the relationship between IV1 and the latent outcome construct, aligning with prior findings that professionals often misinterpret or underweight the role of risk management in decision-making (Oltramari & Kott, 2018; Webb et al., 2014).

3. **Dependent Variable** (DV latent): Risk Management Competence in Real-World Scenarios.

Defined as a latent factor measured by four observable behaviours, each corresponding to a key stage in the risk management cycle:

1) **DV1**: Risk-based assessment behaviour using impact-likelihood calculation

2) **DV2**: Risk-based prioritization behaviour

3) **DV3**: Control selection justified by risk reduction

4) **DV4**: Risk communication and decision rationale



### 3.1.5.4 Theoretical Specification of Risk Management Competence.

Enterprise risk theory provides the basis for modelling risk management competence as four distinct abilities. Consistent with COSO ERM, ISO 31000, NIST RMF, FAIR, and the historical definition of risk as likelihood × impact (Department of Commerce, 1979; Li et al., 2020), competence was conceptualized as the ability to (1) assess threats and estimate likelihood and impact, (2) prioritize risks based on calculated severity, (3) select controls that reduce likelihood and/or impact, and (4) communicate residual risk and trade-offs. This specification reflects a normative benchmark derived from enterprise risk theory, which assumes that risk reasoning involves cognitively distinct analytical stages. These indicators operationalize the integration of likelihood × impact into practice, consistent with risk management frameworks above.

4. **Control Variables**.

Two control variables account for background variance in professional development:

1) **CV1**: Years of experience in cybersecurity.

2) **CV2**: Prior informal risk management learning (self-study, certifications, or experiential learning).

Both controlling variables predict MeV1 in the model, with a covariance between them.

### 3.2.5.4 Model Representation

The Lavaan syntax below represents the basic hypothesized model.

```
#######################
# Measurement model
#######################

# IV: self-reported exposure to risk-embedded training content
IV1x =~ IV1x_1 + IV1x_2 + IV1x_3 + IV1x_4
```



```
# Mediator
MeV1 =~ MeV1_1 + MeV1_2 + MeV1_3 + MeV1_4 + MeV1_5 + MeV1_6

# DVs as 1st-order factors
DV1 =~ DV1_1 + DV1_2 + DV1_3 + DV1_4 + DV1_5
DV2 =~ DV2_1 + DV2_2 + DV2_3 + DV2_4 + DV2_5
DV3 =~ DV3_1 + DV3_2 + DV3_3 + DV3_4 + DV3_5
DV4 =~ DV4_1 + DV4_2 + DV4_3 + DV4_4 + DV4_5

# 2nd-order competence factor
RM_Competence =~ DV1 + DV2 + DV3 + DV4

#########################
# Variance constraints (Heywood case handling)
#########################

MeV1 ~~ 0.001*MeV1
DV1  ~~ 0.001*DV1

#########################
# Structural model
#########################

MeV1 ~ a*IV1x + CV1_YEARS + CV1_SENIORITY + CV2_RM_LRN1 + CV2_RM_LRN2 +
CV2_RM_LRN_CHK1
RM_Competence ~ b*MeV1 + cprime*IV1x

#########################
# Control covariances (optional)
#########################

CV1_YEARS ~~ CV1_SENIORITY
CV2_RM_LRN1 ~~ CV2_RM_LRN2 + CV2_RM_LRN_CHK1
CV2_RM_LRN2 ~~ CV2_RM_LRN_CHK1

#########################
# Effects
```



```
########################

indirect := a*b
direct   := cprime
total    := (a*b) + cprime
```

This specification enabled testing of both the direct effect of training content on competence and the indirect effect mediated by perceived relevance.

### 3.2.5.5  Variable–Theory Alignment

Table 4 summarizes how each variable aligns with the hypotheses and elements of risk management theory:

**Table 4**
*Theory Alignment Table (SEM Model)*

| Construct | Theoretical Role | Conceptual Definition | Operationalization | Hypothesis Link |
|---|---|---|---|---|
| IV1x – Risk-Embedded Training Exposure | Independent Variable | The degree to which formal cybersecurity training explicitly integrates risk management competencies (likelihood × impact reasoning) | Latent construct measured by IV1x_1–IV1x_4 (self-reported exposure indicators) | $H_{11}$, $H_{12}$ |
| MeV1 – Conceptual Salience of Risk | Mediator | The extent to which participants perceive risk management as relevant to cybersecurity practice | Latent construct measured by MeV1_1–MeV1_6 | $H_{11}$ (mediation pathway) |
| DV1 – Risk-Based Assessment | First-order latent DV | Likelihood estimation behaviour | DV1_1–DV1_5 | $H_{12}$ |
| DV2 – Risk Prioritization | First-order latent DV | Prioritization using risk calculus | DV2_1–DV2_5 | $H_{12}$ |
| DV3 – Control Selection | First-order latent DV | Control selection justified by risk reduction | DV3_1–DV3_5 | $H_{12}$ |
| DV4 – Risk Communication | First-order latent DV | Risk-based explanation of decisions | DV4_1–DV4_5 | $H_{12}$ |
| *RM_Competence* | Second-order latent DV | Integrated real-world risk management competence | Measured by DV1–DV4 | $H_{12}$ |
| CV1_YEARS | Control | Years of cybersecurity experience | Observed variable | Control |
| CV1_SENIORITY | Control | Organizational seniority level | Observed variable | Control |
| CV2_RM_LRN1–CHK1 | Control | Prior formal/informal risk learning | Observed variables | Control |

### 3.3    Participants and Sample

This Section describes the target population and sampling strategy, detailing the sample size, inclusion/exclusion criteria, and the survey instrument development and validation process. The target population for this study consists of cybersecurity professionals working across a range of industries and organizational roles. A purposive sampling strategy was employed to recruit participants with relevant cybersecurity experience, ensuring representation across technical, governance, and leadership functions. Participants were required to have at least one year of professional experience in a cybersecurity-related role. The analysis excluded individuals without professional cybersecurity experience or whose work did not involve cybersecurity decision-making.

The training materials analyzed in this study include syllabi, course content, certification programs, and associated competencies from frameworks like NIST NICE, ECSF, and SPARTA. These materials were selected based on their relevance to the study's core competencies and their prevalence



in cybersecurity education programs, to assess their alignment with risk management competencies across the 29 competency areas.

The qualitative data sample included cybersecurity professionals in risk management roles and industry leaders. Participants were selected using a purposive sampling strategy to ensure they had firsthand experience applying risk management competencies in cybersecurity contexts. This sample represented a cross-Section of individuals and organizations engaged in cybersecurity activities. The target sample size was 90 participants for interviews and 90 respondents for surveys, as detailed below.

### 3.3.1 Study Population

The study targets three primary populations to gain a comprehensive understanding of risk management integration across training content and professional practice:

**Cybersecurity Professionals**. Individuals with professional certifications, training, or job responsibilities in cybersecurity. This group provided insights into how **professionals integrate** risk management competencies into practice. The study surveyed and interviewed professionals from various roles (e.g., Risk Analysts, Security Consultants) to assess how they apply and perceive risk management and to identify gaps between training and real-world application.

**Non-Cybersecurity Professionals (Control Group)**. This study recruited working professionals outside the cybersecurity domain as a benchmark comparison group. These participants hold structured decision-making responsibilities and encounter conditions of uncertainty in their professional roles, but possess no formal cybersecurity training or certification. Comparing this group against the cybersecurity professional sample allows the study to evaluate whether cybersecurity training produces a measurable and substantive advantage in foundational risk reasoning — specifically, the capacity to assess likelihood and consequence, prioritize by impact, and differentiate actions based



on their risk-reduction logic — or whether such reasoning reflects general professional cognition that training does not reliably differentiate. The Section titled '**Control Group Survey Instrument**' describes the adapted instrument used with this group and the rationale for the eight-item foundational scale retained for cross-group analysis.

**Executive Leadership (Management)**. Senior leaders and executives within organizations provided insights into how they perceive the role of their cybersecurity professionals in risk management. Specifically, this group helped gauge whether management views cybersecurity professionals primarily as technical specialists or individuals responsible for managing and mitigating cybersecurity risks within the organization.

This analysis examined training content for competencies from the following sources:

- **NIST NICE Cybersecurity Workforce Framework (US)**: Knowledge Statements (KSAs), Competency Categories, and mapped training content.

- **ECSF (European Cybersecurity Skills Framework)**: Published by ENISA, it provides a common understanding of cybersecurity roles, competencies, skills, and knowledge needed in Europe.

- **SPARTA Cybersecurity Skills Framework (Europe)**: Sector-specific competency frameworks for critical infrastructure.

- **NICCS Cybersecurity Education & Training Catalogue**: Publicly accessible records of NICE-aligned training programs.

- **Supplementary Materials**: Course syllabi, certification outlines, and curricula from industry-led training programs (e.g., SANS, ISC2).

This population and these sources provide the foundation for understanding how educators teach, professionals apply, and executives perceive risk management concepts across diverse contexts.



The study provided a comprehensive overview of risk management integration and its practical implications, involving cybersecurity professionals, non-professionals, and management.

### 3.3.2    Sampling Method

The study employs a stratified purposive sampling method to ensure representation across professional experience levels and organizational roles within the cybersecurity domain. Patton (2015) justifies the use of purposive sampling by explaining its rationale in qualitative studies, such as the semi-structured interviews and surveys employed in this study. This approach

Following Creswell (2014), the study designed this approach to capture various perspectives on risk management competencies and their integration into training programs and professional practice. The study stratified participants into the following categories:

**Senior Leadership**: C-level executives and board members involved in cybersecurity governance and strategic decision-making. This group provided insights into how risk management is viewed at the organizational level, particularly whether cybersecurity professionals are considered technical specialists or key contributors to overall enterprise risk management.

**Mid-Level Management**: Team managers and department heads are responsible for overseeing cybersecurity operations within organizations. This group offered a perspective on how organizations integrate risk management into daily operations and how management expects that training programs equip cybersecurity professionals to handle risk effectively.

**Practitioners**: Front-line cybersecurity professionals with NICE-aligned certifications are responsible for directly implementing cybersecurity measures and risk management strategies. This group helped assess the practical application of risk management competencies in day-to-day tasks and the alignment between training content and real-world needs.



**Non-Professionals**: Individuals without formal cybersecurity training or experience, serving as a baseline comparison group. This category enables measurement of the general public's awareness of risk management competencies, highlighting the gap between professional training and general understanding.

This stratification enabled comparisons across organizational levels and between professional and non-professional groups, allowing the study to comprehensively examine how risk management competencies are perceived, taught, and applied across contexts. By examining the viewpoints of leaders and practitioners and comparing professional expertise with that of non-professionals, the study assessed the alignment of training content with real-world risk management practices.

**Figure 2**

*G\*Power Supplementary Reference: Conservative Lower Bound*

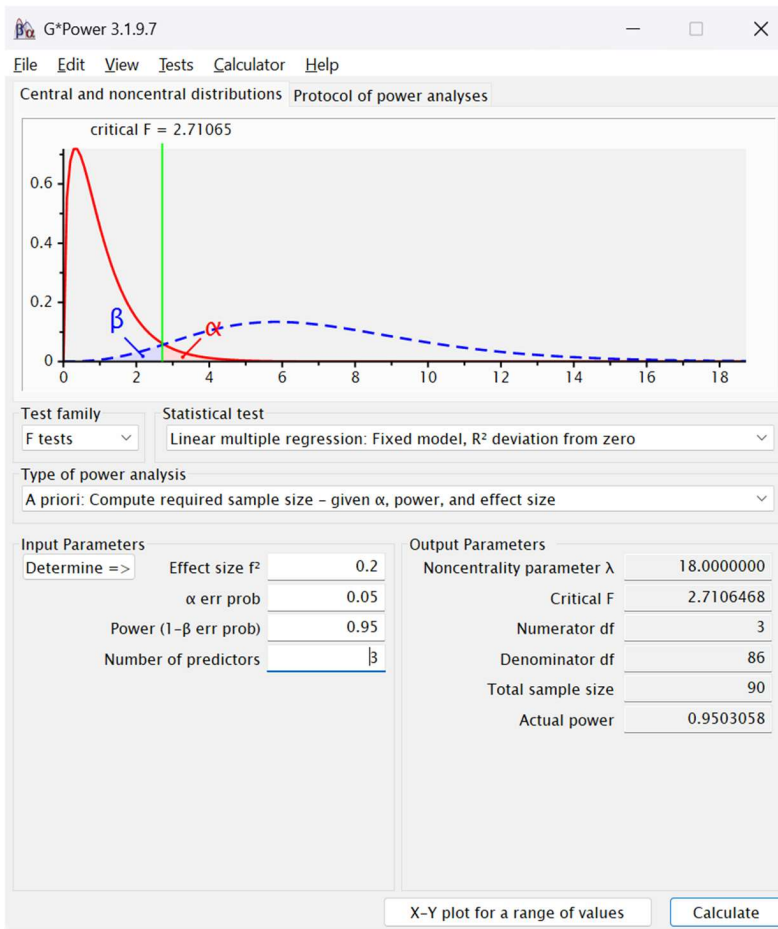

### 3.3.3  *Sample Size Calculation*

The primary quantitative analysis employs Structural Equation Modelling (SEM) using the weighted least squares mean- and variance-adjusted (WLSMV) estimator. Regression-based power analysis tools fail to establish sample size adequacy for SEM because estimation depends on specific model parameters. This



section, therefore, establishes sample size adequacy using criteria appropriate to the method employed. Accordingly, this study rejected regression-based power estimation in favour of SEM-specific criteria, which provide the sole basis for determining sample size adequacy.

This study adopts the primary criterion established by Wolf et al. (2013) and uses their simulation-derived minimum sample size recommendations for covariance-based SEM. For models comprising three to five latent constructs with moderate-to-strong indicator loadings — calibrated to achieve a power of 0.80 — Wolf et al. (2013) recommend a minimum of 70 to 130 cases. The standardized factor loadings observed in this study range from 0.722 to 0.902 for the primary constructs, placing the model within the moderate-to-strong category. The obtained sample of n = 126 satisfies the Wolf et al. (2013) criterion for this model structure.

For supplementary reference, Figure 2 presents a G*Power output based on a linear multiple regression specification (F-test, $f^2 = 0.02$, $\alpha = .05$, power = 0.95, three predictors), yielding a minimum sample size of n = 90. This estimate is not the primary adequacy justification for the SEM but serves as a conservative lower bound for detecting the bivariate relationships embedded within the structural paths. The SEM-specific criterion described above serves as the authoritative basis for assessing sample size adequacy in this study.

### 3.3.4    *Survey Instrument Development and Pre-Test Pilot*

The study used a custom survey instrument to assess participants' understanding, perceptions, and applications of cybersecurity risk management competencies. The survey included questions measuring variables related to integrating risk management competencies into training programs, and the interviews explored the real-world applications of risk management by professionals (Baiden, 2024; Webb et al., 2014; Wilkinson, 2020).



**3.3.4.1 Survey Development**

This research used a Qualtrics survey instrument developed to measure the constructs defined in the structural equation model shown in Figure 1 on page 59. Survey items were original rather than adapted from existing instruments. However, they were grounded in established theoretical frameworks — specifically Competency-Based Learning Theory and risk management Theory — and informed by prior empirical work on risk perception, decision-making, and competency frameworks (e.g., NIST NICE, ECSF, SPARTA). The design of this survey instrument aligns with best practices and incorporates three core components:

1. **Risk Management Knowledge**: Questions designed to assess participants' familiarity with fundamental risk management concepts, such as risk identification, risk assessment or calculation (using likelihood and impact), risk analysis rationale (why, or on what basis, is a high risk 'high'), control selection (and risk mitigation strategies). This Section was aligned with the 29 competency areas, ensuring the evaluation captures a broad spectrum of risk management knowledge.

2. **Risk Management in Training**: Questions to evaluate participants' perceptions of how the curriculum integrated risk management competencies. This Section helped assess whether participants believe their training (whether academic or certification-based) adequately covered risk management concepts and whether these principles align with the 29 competency areas.

3. **Demographic Information**: Basic details about participants, such as job role, years of experience, educational background, and familiarity with specific cybersecurity training frameworks. This analysis allowed for subgroup analyses and provided context for interpreting responses to the other survey Sections.



Each survey item was mapped to a SEM variable (see Appendix B). For example:

1. DV1 (Risk-based assessment) items measure the ability to estimate likelihood and impact when analyzing cybersecurity events to quantify risk from a risk matrix.

2. DV2 (Risk-based prioritization) items capture prioritization of actions based on risk severity rather than volume or tool-generated scores.

3. DV3 (Control selection justified by risk reduction) items capture behaviours where professionals choose controls because they reduce likelihood and/or impact.

4. DV4 (Risk communication & rationale) items measure the ability to communicate risk trade-offs and residual risk to stakeholders.

5. MeV1 (Conceptual Salience and Relevance of RM) items assessed whether participants perceive risk management as central to cybersecurity work.

6. Control variables (CV1, CV2) were self-reported items that assessed years of experience and prior informal learning.

The survey used a 5-point Likert scale for all attitudinal and behavioural items (Strongly Disagree to Strongly Agree, or Rarely to Always), consistent with best practices in educational and organizational research.

### 3.3.4.2 Control Group Survey Instrument

To test Hypothesis $H_{15}$ — whether cybersecurity professionals demonstrate a measurable and substantive advantage in foundational risk reasoning relative to professionals without cybersecurity-specific training — this study administered a parallel survey instrument to a non-cybersecurity professional control group. The control group instrument serves a single, bounded purpose: to supply an external reference point for the risk-reasoning levels observed within the primary cybersecurity sample.



***Instrument Design Rationale***

The control group instrument draws its construct logic from the same theoretical foundation as the primary survey — Competency-Based Learning Theory and the likelihood × impact definition of risk — but adapts item language to remove cybersecurity-domain specificity. The primary survey measures risk reasoning as enacted in cybersecurity professional practice; the control group instrument measures the same underlying reasoning processes as enacted in general professional decision-making contexts. This distinction, rather than the construct, governs all item-level adaptation decisions.

This study selected this design approach over a full-instrument equivalence approach for two reasons. First, administering the full 30-item SEM instrument to non-cybersecurity professionals would introduce systematic construct contamination: items referencing safeguards, threat actors, controls, or security-specific risk events would measure domain familiarity alongside reasoning capacity, confounding group comparisons. Second, the study's evaluative question requires a domain-neutral measurement surface. Adapting items to remove field-specific references — while preserving the logical structure of the reasoning process each item targets — meets this requirement without abandoning construct alignment between the two instruments.

***Item Administration and Block Structure***

The control group instrument utilizes an eight-item foundational scale that targets domain-neutral reasoning processes across three sub-dimensions: risk-based assessment (ASMT_1–3), risk-based prioritization (PRIO_1–3), and risk-based control selection (ACTN_1–2). These eight items constitute the cross-group comparison instrument. Two additional items — CG_DV4_1 and CG_DV4_2, targeting residual outcome evaluation and risk communication — were administered for supplementary



descriptive purposes but were not part of the foundational scale by design. Table 5 presents all ten

administered items.

**Table 5**

*Control Group Survey Items Correspondence to Primary Survey Constructs*

| CG Code | Primary Survey Code | Sub-Dimension | Control Group Item | Primary Survey Item (Paraphrased) |
|---|---|---|---|---|
| CG_DV1_1 | ASMT_1 | Risk-Based Assessment | I deprioritize work items with lower expected impact, even when others give them attention. | Changes in how often an issue occurs or how disruptive it appears influence how professionals prioritize that issue in practice. |
| CG_DV1_2 | ASMT_2 | Risk-Based Assessment | When new information changes the likelihood of a problem, it affects how I assess its priority. | When new information changes the estimated likelihood of a problem, it affects how professionals assess its priority. |
| CG_DV1_3 | ASMT_3 | Risk-Based Assessment | I document and explain why I chose one course of action over another, detailing the specific risk factors that drove the decision. | Changes in operating conditions influence how professionals evaluate the severity of potential consequences. |
| CG_DV2_1 | PRIO_1 | Risk-Based Prioritization | When something appears urgent but would have a limited effect on outcomes, I prioritize based on expected impact rather than urgency alone. | In cases of low-impact but urgent issues, business impact dictates the final priority, overriding urgency. |
| CG_DV2_2 | PRIO_2 | Risk-Based Prioritization | If the expected effects of an issue change, I reconsider earlier prioritization decisions. | Shifts in expected operational effects trigger a review of previous prioritization decisions. |
| CG_DV2_3 | PRIO_3 | Risk-Based Prioritization | Differences in expected outcomes justify prioritizing one item over another. | Differences in expected outcomes justify prioritizing one item over another in practice. |
| CG_DV3_1 | ACTN_1 | Risk-Based Control Selection | Differences in how possible actions reduce the frequency of a problem or its disruption influence which option I recommend. | Differences in how safeguards affect likelihood and impact influence the controls that practitioners recommend. |



| CG_DV3_2 | ACTN_2 | Risk-Based Control Selection | My recommendations differ depending on whether an action mainly reduces frequency, reduces consequences, or both. | Control recommendations differ depending on whether the selected control reduces likelihood, reduces impact, or both. |
|---|---|---|---|---|
| *CG_DV4_1* | *RSID_1* | *Residual Outcomes & Communication, (supplementary†)* | *After choosing an approach, I consider the potential outcomes should it fail to work as expected.* | *After safeguards are selected, I consider what outcomes remain possible if they do not work as intended.* |
| *CG_DV4_2* | *RSID_2* | *Residual Outcomes & Communication (supplementary†)* | *When explaining a decision to others, I describe trade-offs in terms of consequences and outcomes rather than the action taken.* | *When discussing security decisions with non-technical colleagues, I explain trade-offs in terms of consequences and outcomes.* |

**Note.** This Section presents paraphrased item text for clarity, while the Qualtrics instruments provide the complete versions used during data collection.

† Items $CG\_DV4\_1$ and $CG\_DV4\_2$ provide supplementary descriptive data and remain separate from the eight-item foundational scale used in the cross-group analysis. The preceding Section details the construct-comparability rationale.

The design decision to exclude the residual outcomes and communication items from the foundational scale rests on a specific, falsifiable methodological criterion: cross-group measurement invariance requires that respondents in both groups interpret the items as measuring the same underlying construct. Two features of the RSID block place it outside that criterion.

First, RSID_1 reference safeguards and frames residual risk evaluation within the context of control deployment decisions—a cognitive task that presupposes cybersecurity operational knowledge. CG_DV4_1 adapts the language to a general decision context, but the shift moves the item from evaluating residual risk after control selection to evaluating general consequences after any chosen action. These are structurally analogous but not psychologically equivalent, and the cognitive work each demands varies systematically across domains.

Second, RSID_2 explicitly specifies security decisions and non-technical colleagues as the communication context, making it a direct measure of cybersecurity-specific practice. CG_DV4_2 removes the security framing, but the resulting item targets a general professional communication behaviour that activates different response schemata across the two populations. Treating responses



from both groups as equivalent would risk measuring different cognitive processes under the same label.

The eight foundational items contain no security-specific artifacts, professional roles, or domain terminology. Each targets a reasoning process — assessment of probability and consequence, prioritization by impact over urgency, and differentiation of actions by their effect on frequency or severity — that operates across professional decision contexts independently of cybersecurity knowledge. Section 4.7.1 reports the confirmatory factor analysis and cross-group metric invariance tests that empirically verify this assumption.

### *Sampling and Administration*

The study recruited control group participants from professional networks outside the cybersecurity domain, targeting individuals in structured decision-making roles who routinely assess and prioritize work items under conditions of uncertainty. The control group survey administered the same five-point Likert response format and identical scale anchors as the primary instrument. Section 4.7.1 reports the control group comparison analysis and documents the control group's distributional properties alongside those of the cybersecurity professional sample.

### 3.3.4.3  Pilot Testing and Validation

Before deployment, the instrument underwent pilot testing with a small group of cybersecurity professionals (n≈10). Feedback focused on clarity, terminology, and usability. This process confirmed that items were understandable across diverse roles and levels of experience. These participants provided feedback on question wording, structure, and survey length, with particular attention to the clarity and relevance of the assessed risk management concepts.

Analysis of the pilot test data accomplished the following:

- Identify any problematic or ambiguous questions.



- Calculate preliminary summary statistics to ensure initial responses are valid and representative.

- Validate the sample size assumptions to determine the appropriate number of participants for the complete survey.

Following revisions based on pilot feedback, I submitted the survey instrument to the Golden Gate University Institutional Review Board (IRB). Following IRB approval, I then distributed the survey to the broader participant pool.

### 3.3.4.4 Interview Protocol Development

A semi-structured interview protocol complemented the survey, yielding qualitative insights into how professionals and leaders conceptualize cybersecurity risk management. The protocol is original, grounded in the risk management literature, and designed to complement the survey by probing themes of relevance, decision-making, and leadership framing.

The interview questions mirror the constructs of the Concept Logic Model (Figure 1, p. 59) to ensure theoretical consistency. The protocol targets leaders' understanding of risk as likelihood × impact, risk-prioritization logic, and control trade-offs or as a technical exercise. For example, the protocol prompted leaders to discuss:

A. *"In your organization, how is cyber risk typically defined? When you think of risk, do you think in terms of likelihood and impact, or in other terms?"*

B. *"When evaluating a potential cybersecurity threat, how do teams estimate how likely it is to occur and what its impact will be on the business?"*

C. *"How are cybersecurity risks communicated to executives or boards? Are likelihood and impact explicitly articulated?"*



D. *"When selecting between alternative security controls or mitigation strategies, how are trade-offs evaluated? Is expected risk reduction considered?"*

Appendix A contains the interview guide.

To ensure clarity, reliability, and construct alignment, a pre-test pilot was conducted with a subset of 10 cybersecurity professionals. These participants provided feedback on question wording, structure, and survey length, with particular attention to the clarity and relevance of the assessed risk management concepts.

### 3.3.5  Recruitment Strategy

Participant recruitment was conducted through multiple channels to ensure a representative sample of cybersecurity professionals across diverse sectors. Recruitment efforts focused on gathering insights into how Risk management principles are perceived and applied across training content and professional practice. The recruitment strategy included:

**Professional Associations**: This research leveraged collaborations with organizations like CUCCIO, NICE, ISC2, and ISACA to reach professionals who have received cybersecurity training aligned with the NIST NICE framework or comparable standards.

**Cybersecurity Conferences**: Recruitment occurred at virtual and in-person cybersecurity training and education events. These events provide opportunities to engage with a diverse range of cybersecurity professionals and ensure representation in the sample.

**Direct Outreach**: The recruitment process involved sending targeted invitations to professionals listed in publicly accessible directories (e.g., LinkedIn, NICCS) who have relevant experience in cybersecurity and risk management. This outreach helped ensure



that professionals from various sectors, with diverse expertise and experience in risk management, are included.

These recruitment efforts emphasized the importance of research participation to improve cybersecurity training outcomes and ensure that professionals effectively integrate risk management principles into their training and professional practice.

### 3.3.6    Data Collection Process

Participants completed the survey through the secure, web-based platform Qualtrics. Survey responses were anonymized and stored in an encrypted database to ensure confidentiality and data security. The study also used Natural Language Processing (NLP) to extract and analyze textual content from training frameworks, following the methodology described in the 'Data Collection Methods' Section. Specifically, the NLP analysis focused on identifying and categorizing Risk management principles within training materials, using the 29 cybersecurity competency categories. This analysis enabled the study to assess how risk management is integrated into cybersecurity training content, utilizing frameworks such as NICE, ECSF, and SPARTA.

### 3.3.7    Inclusion Criteria

The research protocol required the following inclusion criteria to focus the research question on the studied population and ensure internal and external validity. The study established internal validity by selecting participants most likely to provide reliable and relevant data. The study established external validity by ensuring that its results apply to a well-defined, relevant group: cybersecurity professionals.

The inclusion criteria are as follows:



a) **Criteria 1**: <u>Cybersecurity professionals with at least one year of experience in risk-based cybersecurity operations</u>. This criterion ensured participants were familiar with Risk management principles and their application in real-world cybersecurity roles.

b) **Criteria 2**: <u>Cybersecurity training materials explicitly or implicitly linked to NICE, ECSF, or SPARTA frameworks or referenced in the NICCS framework</u>. This criterion allowed the analysis to focus on training content aligned with established frameworks, ensuring consistency and relevance.

c) **Criteria 3**: <u>Non-professionals with no formal cybersecurity training (for benchmark comparison)</u>. This criterion established a baseline group for comparing the public's understanding of Risk management principles with that of professionals.

These criteria facilitated data analysis by selecting a relatively homogeneous group of participants, thereby enabling the study to more effectively measure perceptions and applications of risk management concepts across different professional contexts. By clearly defining inclusion and exclusion criteria, the study's findings can be replicated, thereby enhancing the research's credibility and reproducibility.

### 3.3.8 Exclusion Criteria

The research protocol required the following exclusion criteria to control for potential confounding factors that could affect the results, thereby reducing bias and enhancing the study's internal validity. The exclusion criteria are as follows:

a) **Criteria 1**: <u>Cybersecurity Professionals</u>.  This criterion excludes professionals who have not participated in formal training programs or do not apply cybersecurity management principles in their daily work.



b) **Criteria 2**: <u>Training programs not aligned with NICE, ECSF, or SPARTA curricula</u>. This criterion

ensures that the analysis focuses solely on training content relevant to the NIST NICE, ECSF,

and SPARTA frameworks, which are central to the study's competency categories and risk

management competencies.

c) **Criteria 3**: <u>Survey respondents who provide incomplete or inconsistent answers</u>: This

criterion ensured that inconsistent or incomplete responses do not skew the data. Excluding

these respondents helped ensure that the data collected accurately reflects participants'

perceptions and experiences.

These exclusion criteria strengthened the study's internal validity, ensuring that the

independent variables were likely to have caused the observed effects. (e.g., risk management training

competency), as outlined in Chapter 1.

### 3.3.9 *Justification of the Sampling Approach*

The study's sampling approach is carefully aligned with the research objectives to ensure that

the data collected accurately reflects the experiences of professionals actively engaged in cybersecurity

operations and individuals unfamiliar with the field. This distinction is crucial for identifying gaps

between curriculum content, professional understanding and judgement, and the practical application

of risk management.

The research design exhibits a high degree of crystallization, with well-defined research

questions and a transparent methodological approach. The structured nature of the study ensures that

both quantitative and qualitative components contribute to a coherent, replicable analysis, providing

clarity and precision in the findings. The research identified measurable gaps in the integration of formal



Risk management principles within cybersecurity training curricula, as well as corresponding gaps in professionals' demonstrated risk management competencies in practice.

The primary purpose of this research is to improve cybersecurity outcomes by identifying training gaps and influencing the development of cybersecurity curricula. Petersen et al. (2020) on the NICE framework, the European Union Agency for Cybersecurity (2022), and the SPARTA framework (Piesarskas et al., 2019) all include risk management to some degree.  By strengthening the profession's understanding of risk management principles, this study sought to equip cybersecurity professionals to manage risk-based challenges better. Training programs can more effectively equip cybersecurity professionals by explicitly operationalizing risk as the interaction of likelihood and impact, and by embedding structured risk assessment, prioritization, and trade-off analysis into applied practice.

The timing of this study is particularly relevant, given the increasing frequency and financial impact of cybersecurity breaches. Szterenfeld (2022) reported that the number of cyberattacks has consistently risen in recent years, placing greater pressure on organizations to enhance their cybersecurity defences. IBM (2024a) similarly documented a notable increase in the cost of data breaches, highlighting the growing financial risks associated with insufficient cybersecurity preparedness. These trends underscore the urgency of researching cybersecurity professionals' training, particularly their understanding of Risk management principles (Dawson & Thomson, 2018; Wilkinson, 2020).

The study is theoretically grounded in the premise that insufficient risk management training may adversely affect cybersecurity practices (Webb et al., 2014). If training frameworks reference risk without operationalizing likelihood × impact reasoning, professionals predictably exhibit undifferentiated or heuristic risk cognition rather than structured assessment, prioritization, and trade-off analysis. This research examined the potential gap between training content and professional



practice, aiming to provide insights to enhance cybersecurity education. By demonstrating a correlation between risk management education and cybersecurity outcomes, the study sought to contribute to the development of more effective cybersecurity curricula.

The availability of structured cybersecurity training frameworks such as NICE, ECSF, and SPARTA supports the feasibility and accessibility of the research. These frameworks provide publicly accessible text data for NLP analysis, facilitating the systematic evaluation of training content. The independent, dependent, mediating, and moderating variables have been carefully defined and can be quantitatively measured. The NICCS catalogue includes 14,696 certification courses mapped to the NIST NICE Workforce Framework.  The NICCS site provided the certification courses for this analysis. Alignment with NICE competencies allowed for an analysis of NICE Knowledge, Skills, and Abilities (KSAs) as a structural proxy for the risk management content and competency expectations embedded across these broader programs. Web scraping techniques, as described by Mitchell (2015), were used to gather this data efficiently. Additionally, the study population is accessible through partnerships with professional associations and cybersecurity communities, ensuring a sufficient survey response rate.

The methodological structure adopted in this study is both rigorous and practical, enabling the collection of reliable data on the integration of Risk management principles within cybersecurity training programs. By triangulating NLP analysis with survey responses from both practitioners and non-practitioners, the study provides a comprehensive perspective on the current state of risk management training and its impact on workforce competency.

### 3.4    Data Collection Methods

The data collection process gathers, preprocesses, and analyzes textual content from publicly available cybersecurity training materials, as well as survey responses from cybersecurity professionals



and non-professionals. This multi-source approach ensures a comprehensive assessment of the integration of Risk management principles within cybersecurity curricula and the resulting influence on professional competency.

Two primary sources provide the study's Data:

1. Training materials from established cybersecurity frameworks represent the curriculum of tens of thousands of training courses and industry educational programs.

2. Cybersecurity professionals and non-professionals completed a structured survey and participated in semi-structured interviews (Creswell, 2014).

### 3.4.1    Primary Data Sources

The study drew on cybersecurity training frameworks and educational repositories aligned with industry standards and best practices. The following sources provided the primary datasets for NLP-based content analysis:

1. **NIST NICE Cybersecurity Workforce Framework (U.S.)**: The NICE framework outlines the knowledge, skills, and abilities (KSAs) necessary for various cybersecurity roles. This analysis examined Knowledge Statements from the NICE framework to measure the presence of risk management principles (taught competencies) and to compare their relative emphasis with that of technical topics. This analysis assessed how NICE-aligned training content integrates risk management.

2. **European Cybersecurity Skills Framework (ECSF)**: Published by ENISA and inspired by the SPARTA Cybersecurity Skills Framework, the ECSF provides a common understanding of cybersecurity roles, competencies, skills, and knowledge needed in Europe. ECSF was analyzed



to assess the integration of Risk management principles within European training frameworks and compared with NIST NICE where relevant.

3. **SPARTA Cybersecurity Skills Framework (Europe)**: The SPARTA framework categorizes cybersecurity training needs across critical infrastructure sectors. The study analyzed sectoral ratings to identify potential differences in training priorities between Europe and North America, focusing on how sectors such as banking, energy, and healthcare address risk management.

4. **NICCS Cybersecurity Education & Training Catalogue (U.S.)**: The National Initiative for Cybersecurity Careers and Studies (NICCS) website provides publicly available information on 14,696 industry-recognized certification courses. Web-scraping techniques similar to those described by Mitchell (2015) yielded 14,696 records of certification courses, confirming the extent to which industry-recognized programs declare alignment with the NICE framework. A crosswalk analysis connects these certification course curricula to NIST NICE Knowledge, Skills, and Abilities (KSAs). This mapping established that NICE serves as the structural competency architecture underpinning the industry-recognized certification ecosystem. Accordingly, this analysis treated NICE KSAs as a proxy for the competency expectations within these aligned programs.

5. **Certification Syllabi and Training Materials**: A detailed review of industry-recognized certifications (CISSP, CRISC, CISM, Security+, and CEH) confirmed their alignment with the NIST NICE framework or the ECSF. Although the study initially considered performing NLP analysis on course syllabi, the heterogeneity and limited availability of content rendered systematic analysis untenable. Consequently, this study used the NICE framework (and related workforce frameworks) as the primary analytic unit because they provide the structural competency



architecture for these certifications. This analysis evaluated risk management's coverage in certifications widely recognized in the cybersecurity field.

6. **Higher Education and Industry-Led Training Materials**: This dissertation reviewed industry-sponsored cybersecurity training programs (e.g., SANS Institute, ISC2) and higher education courses to compare risk management coverage across academic and industry contexts.  This research also notes that Dawson and Thomson (2018) conducted a review that highlighted the NSA has identified over 200 colleges and universities in the United States whose cybersecurity curricula align with the NICE framework (Also see: (Bicak et al., 2014; Cabaj et al., 2018; Chandrinos, 2023; S. Gates et al., 2012; Moore et al., 2020; Švábenský et al., 2020; Woodward et al., 2013)). These sources helped determine the variations in risk management instruction across industry-led and educational sectors.

Together, these sources provide a broad and diverse dataset for evaluating the coverage and depth of Risk management principles in cybersecurity training. By combining publicly available frameworks (e.g., NICE, SPARTA, ECSF) with industry-led materials, the study provided a comprehensive view of how training programs integrate risk management and align with professional practices.

This study employed Natural Language Processing (NLP) techniques, specifically topic modelling and word frequency analysis, to identify key themes and concepts related to risk management within cybersecurity curricula and analyze the integration of risk management concepts in the above training materials. Python was used with the spaCy library for NLP analysis, specifically for tokenization and text processing. Meanwhile, the study utilized scikit-learn for topic modelling and clustering. The NLP analysis output was statistically analyzed using regression models to identify correlations between the presence of risk management content and the perceived effectiveness of training in real-world



applications. These techniques enabled an objective, data-driven assessment of the prevalence of Risk management principles across multiple training frameworks.

### 3.4.1.1 Validity and Reliability

A pre-distribution review ensured the survey instrument's clarity and alignment with the study constructs, supporting overall validity and reliability. The use of Natural Language Processing (NLP) provides a systematic and replicable method for analyzing training content, reducing subjective interpretation and enhancing analytical consistency across frameworks. During the main data collection, the analysis included psychometric validation and reliability assessment using Cronbach's alpha for each multi-item construct (DV latent, MeV1). Confirmatory factor analysis (CFA) established the construct validity of the measurement model before SEM testing.

To ensure validity and reliability in the NLP analysis, I employed data triangulation by using multiple sources of training materials and ensured consistent text preprocessing (tokenization and stop-word removal) to minimize biases. Additionally, I applied inter-rater reliability checks by having multiple coders assess the same corpus of training materials.

### 3.4.2    Survey Data Collection

Survey data provided complementary insights into participants' perceptions and understanding of risk management concepts. Survey participants answered Likert-scale questions to measure their perception of how effectively their training programs integrated risk management principles. In-depth interviews explored participants' real-world application of these principles, using open-ended questions to elicit detailed accounts of their experiences and insights.

Data collection commenced once the survey instrument had been developed and pre-tested. The survey reached a broad participant pool, including both cybersecurity professionals and non-



professionals. The Qualtrics platform distributed the survey, providing a secure, web-based environment that ensures ease of access, participant confidentiality, and data integrity.

Key elements of the data collection process include:

a) **Online Administration**: The survey was accessible via Qualtrics, allowing participants to complete it conveniently. This platform ensures data security by anonymizing responses and encrypting datasets, thereby maintaining the privacy and confidentiality of all participants.

b) **Participant Groupings**: The survey targeted two primary groups:

   a. Cybersecurity professionals with experience in risk-based cybersecurity operations.

   b. Non-professionals without formal cybersecurity training serve as a benchmark comparison group. The Section entitled "Control Group Survey Instrument describes the control group instrument design and the rationale for retaining the items.

c) **Survey Distribution**: Invitations to participate were sent via email or provided through cybersecurity associations (e.g., NICE, ISACA, ISC2) and cybersecurity communities (e.g., LinkedIn, NICCS).

d) **Data Integrity and Monitoring**: The survey was closely monitored during the collection process to ensure the completeness and reliability of the responses. The study enforced mandatory responses for all questions in Qualtrics, eliminating inconsistent records. During analysis, I discarded any incomplete surveys to maintain data integrity.

The data collection process targeted risk management principles and workplace competencies to generate a dataset of quantitative and qualitative insights into their professional integration. Qualitative data from interviews and survey responses were analyzed using a combination of Natural Language Processing (NLP) and qualitative coding. Qualitative data from interviews and survey responses were analyzed using the directed thematic coding approach described in the Leadership



Interview Coding Section. The Leadership Risk Cognition Codebook (LRCC) guided the coding process. R's statistical functions facilitated reliability checks and ensured coding consistency across the dataset. These codes were then refined and validated through manual review, potentially using the RQDA package in R to organize and manage the coding process. The study utilized R's statistical functions to conduct inter-rater reliability checks, ensuring consistency. This combined approach allowed for automated theme discovery and in-depth qualitative interpretation, providing a comprehensive understanding of how cybersecurity professionals apply risk management competencies in their work.

### 3.4.2.1 Validity and Reliability

For qualitative data, validity was ensured through member checking, where participants reviewed their interview transcripts to confirm accuracy. I achieved reliability by harmonizing transcripts from two independent services, Fireflies and Otter.ai, to ensure an accurate representation of the interview data. This approach ensured data integrity. I compared and reconciled the automated transcripts from Fireflies and Otter.ai, creating a single, verified version for thematic analysis.

### 3.4.3    Survey Pre-Test and Validation

A pre-test pilot validated the survey instrument before the full-scale distribution. During this phase, ten cybersecurity professionals tested the instrument to evaluate the clarity, reliability, and relevance of the survey questions, particularly those focused on risk management competencies.

Key steps in this process included:

a) **Pilot Testing**: 10 professionals completed the survey and provided feedback on the question wording, structure, and length. Their feedback identified any areas of confusion, ambiguity, or misalignment regarding the risk management focus.



b) **Data Review**: The pilot test analysis identified patterns and anomalies within the initial responses. This analysis also enabled the calculation of preliminary summary statistics to assess the initial distribution of risk management knowledge and perceptions.

c) **Validation of Reliability and Validity**: Pilot test feedback helped refine the survey questions, ensuring they accurately capture participants' understanding of risk management principles. These adjustments improved clarity, minimized ambiguity, and ensured the survey adequately addressed all aspects of risk management.

d) **Sample Size Assumptions**: The pilot data also helped validate assumptions about sample size for the more extensive study, ensuring sufficient statistical power to detect significant relationships between risk management training and professional competency.

After making necessary revisions based on pre-test feedback, the survey instrument was finalized and submitted to the IRB.

### 3.4.4   IRB Approval

After finalizing the survey instrument through pre-testing and validation, I submitted it to the Golden Gate University Institutional Review Board (IRB) for ethical review and approval. The IRB submission ensures that the study adheres to ethical standards and protects participant confidentiality, privacy, and rights.

Key steps in the IRB approval process included:

a) **Ethical Review**: The IRB reviewed the survey instrument and the overall study protocol to ensure the research adheres to the highest ethical standards. This review examines participant consent processes, data storage protocols, and confidentiality measures.



b) **Data Privacy and Confidentiality**: The Qualtrics platform anonymizes and encrypts survey responses, protecting participants' data and ensuring compliance with privacy regulations.

c) **Approval Process**: After incorporating the IRB's required adjustments to the survey and methodology, I obtained formal approval. I then distributed the survey to the broader participant pool.

This approach ensures ethical integrity and upholds participants' rights during NLP data collection.

The NLP analysis requires a systematic process for collecting, cleaning, and preparing text data for subsequent analysis. The text collection process involved the following steps:

a) **Web Scraping**: The NICCS website lists 14,696 certification programs derived from the NIST NICE framework. This data was collected using Python-based web scraping techniques, following the methodologies described by Mitchell (2015). This web scraping focused on collecting curriculum details from the NICCS database, which provides a comprehensive catalogue of over 14,696 training programs mapped to NICE Knowledge Statements (KSAs). This study categorized risk management content across these programs, with a specific focus on the 29 competency areas.

b) **Manual Retrieval**: Certification syllabi and inaccessible academic course descriptions were collected manually from publicly available websites through automated scraping. This process incorporates critical materials that web scraping missed, ensuring a comprehensive analysis by including training artifacts with limited web indexability in the corpus. This effort examined these documents using the competency taxonomy and NLP classification procedures described above to determine if the curriculum explicitly embedded formal risk



management constructs—particularly structured likelihood–impact reasoning, prioritization logic, and risk-based decision frameworks.

c) **Dataset Consolidation**: I aggregated all collected text data into a single database for preprocessing and analysis, attaching metadata to indicate the source framework, document type, and training context. This data set consolidation enabled easy categorization into the 29 competency areas, allowing consistent measurement of risk management principles across different training programs.

The text corpus generated by these procedures served as the foundation for the NLP pipeline, as detailed in the following Section. The NLP analysis assessed how risk management principles are integrated into the training content, aligning with the study's goal of evaluating risk management's effectiveness as a component of the taxonomy's 29 competency areas.

### 3.4.5    *Research Variables and Hypotheses*

Several key variables must be measured and analyzed to investigate the integration of risk management principles into cybersecurity training and its impact on professionals' competencies, as noted in Chapter 1. These variables include independent variables (IVs), dependent variables (DVs), control variables (CVs), and mediating variables (MeVs). The research questions and hypotheses developed in this study explore the relationships between these variables, as noted in Chapter 1.

This Section outlines how the variables discussed in Chapter 1 were operationalized and used in the measurement approach. It presents the connections between the research questions and hypotheses, demonstrating the role of each variable in addressing the study's key questions. The tables below map the independent variables (e.g., risk management content in training curricula) to the dependent variables (e.g., professionals' ability to apply risk management principles in practice) and



detail the measurement methods for each. This systematic presentation clarifies how these variables were analyzed and contribute to the research.

The following tables outline the relationships among the study's hypotheses, independent variables (IVs), dependent variables (DVs), and research questions (RQs), providing a clear roadmap for the data analysis and maintaining clarity throughout the study.

Table 6

*Research Variables and Hypotheses*

| Variable Type | Variable | Measurement Approach | Hypothesis | Measurement Tools |
|---|---|---|---|---|
| **Independent Variable (IV)** | Competencies in KSA Statements (IV1) | NLP analysis of competency frameworks (NICE, ECSF, SPARTA) to assess risk management competency (RISK) across all competencies, including CRYPT, PRDV, etc. | $H_{11}$: Cybersecurity training frameworks that integrate risk management principles exhibit a statistically significant representation of risk-related Knowledge, Skill, and Ability (KSA) statements (IV1) relative to other competency domains. | NLP analysis of frameworks like NICE, ECSF, SPARTA |
| **Dependent Variable (DV)** | Risk Management Competency in Real-World Scenarios (DV1 – DV4) | Survey and interview data assessing the ability to apply risk management in real-world scenarios. | $H_{12}$: If cybersecurity training integrates risk management competencies (IV1x), then professionals demonstrate higher risk-based behaviours (DV1–DV4), because training increases the conceptual salience and perceived relevance of risk management (MeV1), thereby strengthening real-world risk decision-making competence. | Surveys and interviews assessing competency in real-world situations. |
| **Dependent Variable (DV)** | Risk Management Competency in Real-World Scenarios (DV1 - DV4) | Survey and interview data assessing the ability to apply risk management in real-world scenarios. | $H_{13}$: If cybersecurity curricula inadequately address risk management, then the RISK competency (IV1) is underrepresented relative to other competencies, demonstrating a curricular gap. | Comparison of competency measures from surveys and NLP analysis of Interviews with competency measures from NICE, ECSF, and SPARTA frameworks |
| **Moderating Variable (MoV)** | The study tests leadership perspectives outside the SEM and excludes structural paths for | Survey data assessing leadership perspectives and views on the congruence of cybersecurity and risk management. | $H_{14}$: Organizations implicitly assign cybersecurity professionals responsibilities consistent with formal strategic risk management frameworks, assuming these professionals demonstrate skilled and differentiated risk management competencies aligned with likelihood–impact reasoning. | Surveys targeting leadership perspectives and their approach to risk management. |
| | these variables. | | | |
| **Independent Variable (IV)** | The SEM excludes structural paths for Professional Training in Risk; instead, a separate test evaluates this variable. | Survey data, competency-based assessments, and interview responses on formal and informal training experiences. | **H₁₅:** Cybersecurity professionals will demonstrate significantly higher risk management competence than non-cybersecurity professionals. | Surveys and interviews comparing professionals to non-professionals |

### 3.5 Data Analysis Methods

The data analysis process for this study applies Natural Language Processing (NLP) techniques to systematically evaluate the presence and coverage of risk management principles in cybersecurity training programs. This critical methodological choice aligns with research by Pandey et al. (2017a) and Mavrogiorgos et al. (2022b), who used NLP to derive scientifically valid metrics from large educational datasets in educational content analysis. NLP provided a scalable, data-driven approach to content analysis, enabling the study to move beyond manual reviews and subjectivity, yielding reliable, replicable findings. This Section outlines the analytical approach, detailing the NLP pipeline, procedures for analyzing survey data, and techniques to ensure the validity and reliability of the results.

The analysis comprises two main components:

1. NLP-based content analysis of cybersecurity training materials.

2. Statistical analysis of survey data collected from professionals and non-professionals.

Combining these methods provides quantitative insights and qualitative context regarding the integration of risk management principles in cybersecurity curricula. The results enabled the study to assess the alignment between training content and professional competency across the 29 competency areas, with a focus on risk management.



### 3.5.1  Natural Language Processing (NLP) Pipeline

The NLP pipeline consists of four stages: taxonomy construction, LLM-assisted NLP classification, structured classification, post-processing and cutoff application, and measurement and ranking. The following Sections describe each stage in detail.

#### 3.5.1.1  Step 1: Taxonomy Construction and Prompt Engineering

Before classification, a structured competency taxonomy was constructed for all 29 NICE competency categories as described in Section **Taxonomy of Cybersecurity Competencies and Variable Operationalization**. Four components defined each category: a label, a conceptual definition, a set of key terms, and a set of exclusion criteria: a plain-language description of what the competency covers; a discriminative keyword list; inclusion criteria that assign a TKS statement to that category based on specific conditions; and exclusion criteria specifying conditions under which apparent relevance should be rejected, with redirection to the more appropriate category. This taxonomy was serialized to JSON and fully injected into every classification prompt, ensuring the classifier applied consistent category definitions across all 2,111 TKS statements.

#### 3.5.1.2  Step 2: LLM-assisted NLP Structured Classification

Each TKS statement was classified using a large language model (GPT; OpenAI, 2025) via a structured output pipeline, rather than producing free-text labels. A Pydantic schema constrained the model to exactly 29 scored outputs per statement—one for each competency category. Each output comprised three fields: a relevance score (0.0–1.0) indicating the degree to which the statement addresses the competency; a confidence score (0.0–1.0) indicating the model's certainty in that relevance assessment; and a natural-language justification. Schema enforcement via OpenAI's structured output API guaranteed that every response was complete and type-valid, eliminating missing



or malformed classifications. The classification prompt provided explicit relevance anchors (0.0 = irrelevant, 0.25 = tangential, 0.50 = partial, 0.75 = strong, 1.00 = central) and decision rules for handling multi-domain statements and category disambiguation. The pipeline processed fifteen statements concurrently and persisted results after each statement to ensure crash recovery without data loss.

### 3.5.1.3  Step 3: Post-Processing and Cutoff Application

After classifying all 2,111 TKS statements, the pipeline applied a three-tier scoring structure. The post-processing script set relevance scores below 0.45 to 0, effectively excluding them from subsequent analysis; The analysis does not count these statements as belonging to any category. While scores in the 0.45–0.74 range constitute moderate matches, the study restricts primary distribution calculations to high-confidence results, retaining moderate matches only for completeness. Scores at or above 0.75 constitute strong matches — the high-confidence classification tier used for all primary analysis and reported findings, including the percentage of the total corpus attributed to each competency domain. A statement was classified as UNKN (unclassified) only when no category score survived the 0.45 minimum threshold.

The 0.45 minimum threshold was set on theoretical grounds to fall between the weak-match floor (0.25) and the strong-match threshold (0.75), excluding statements with only tangential relevance to a category while retaining those with at least moderate alignment. The 0.75 high-confidence threshold isolates statements where the classifier's relevance assessment remains unambiguous. Varying the strong-match threshold from 0.70 to 0.80 did not alter the rank ordering of competency categories, confirming the stability of the high-confidence statement counts.

Because the pipeline assigned 29 independent relevance scores per statement, it could classify a single TKS statement into multiple categories simultaneously. This multi-label design eliminates the need for tie-breaking rules. There is no forced single-category assignment, and apparent overlap



between categories reflects genuine multi-domain content rather than a classification conflict requiring resolution.

#### 3.5.1.4 Step 4: Measurement and Ranking

Post-processing computed the total number of TKS statements classified as relevant (relevance ≥ 0.45) and the subset classified as a strong match (relevance ≥ 0.75) for each of the 29 competency categories. Statement count determined category rank. The Risk Management (RISK) category was the primary focus of analysis: its rank among the 29 categories, its statement count, and the presence or absence of specific probabilistic reasoning terms (including "likelihood," "probability," and "impact") within classified statements were the key empirical outputs of the NLP analysis. Because the pipeline produced a natural-language justification for every classification decision, individual classifications were auditable: the recorded justification text allows for direct verification of any TKS statement's assignment to a category.

### 3.5.2   Survey Data Analysis

Survey responses were analyzed using descriptive and inferential statistics to evaluate participants' perceptions of their risk management training and the integration of risk management principles across the 29 competency areas. The analysis employed structured survey questions and unstructured responses (e.g., interviews or open-ended questions), with NLP used to identify and quantify risk management content.

a) **Data Cleaning and Screening**: A manual audit verified the completeness and consistency of the responses. Invalid or incomplete submissions were excluded from the analysis to ensure data quality and integrity. This step ensures that only valid responses are used in statistical procedures, thereby supporting accurate risk management measurements.



b) **Descriptive Statistics**: Descriptive statistics summarized participant characteristics and responses through means, medians, standard deviations, and frequency distributions. In particular, calculating descriptive statistics allowed for an assessment of the response distribution across the 29 competency areas. These statistics helped identify how participants perceive their risk management training and how it aligns with their professional experience.

c) **Inferential Statistics**: Analysis of Variance (ANOVA) and linear regression identified the relationships between risk management training and participants' perceived competency. ANOVA examined whether there are statistically significant differences in perceptions of risk management competency across groups (e.g., cybersecurity professionals, non-professionals, and management). Linear regression assessed the predictive relationship between the extent of risk management training and participants' perceived competency in specific competency areas, such as risk management, Incident Management, and Cybersecurity Governance.

d) **Competency-Specific Analysis**: Because the survey precisely measured competency categories (via structured questions and NLP-based analysis), we focused on responses to questions about risk management and other competencies. These responses were mapped to the 29 competency areas, ensuring the analysis captures the perceived strengths and gaps in training across risk management and related areas, such as Threats and vulnerabilities or Incident Management.

e) **Unstructured Data Processing**: NLP processed text and identified risk-management-related content in unstructured survey responses (e.g., open-ended questions or interviews). For example, in analyzing training materials, we utilized text classification to distinguish risk



management terms from other cybersecurity content. The analysis quantified the extent to which risk management content appears in professional discourse and participants' self-assessed competency by examining unstructured responses.

f) **Comparative Analysis**: The analysis compared responses from cybersecurity professionals and non-professionals to identify knowledge gaps stemming from risk management training. This analysis assessed disparities in risk management knowledge between individuals with formal cybersecurity training and those without, highlighting potential areas for improving training.

By analyzing structured and unstructured data, this study demonstrated the integration of risk management principles into cybersecurity training and measured their impact on professional practice. By triangulating quantitative and qualitative results, the study provided a comprehensive perspective on the current state of cybersecurity risk management training and its implications for workforce competency.

### 3.5.3   Leadership Interview Coding and Thematic Analysis

Leadership interviews were analyzed using a directed thematic coding analysis approach (Hsieh & Shannon, 2005) aligned with the study's conceptual framework, research questions, and structural equation model. The qualitative component serves to contextualize and triangulate quantitative findings rather than to generate grounded theory.

A master codebook of 28 deductively derived codes known as the Leadership Risk Cognition Codebook (LRCC) (Appendix C) guided the transcript analysis. To create the codebook, the study synthesizes enterprise risk management theory, the conceptual model of training exposure, and the four guiding research questions, ensuring construct-level alignment between qualitative coding and the SEM



architecture. The study categorizes codes into theoretically grounded domains that capture risk conceptualization, workforce competence expectations, and training architecture. Each code included operational definitions and inclusion/exclusion criteria to ensure consistent application.

All interviews were coded by a single researcher using a fixed codebook. This design decision is appropriate for this study's confirmatory purpose and is defensible on four grounds.

First, the coding approach is deductive rather than inductive. Enterprise risk management theory, the SEM construct architecture, and the four research questions informed the construction of the LRCC codebook. Because the code set was fixed and theory-driven, the research design structurally constrained coding discretion: the researcher applied predetermined categories rather than discovering them. Deductive coding from a pre-specified codebook reduces the interpretive latitude that makes inter-rater reliability most critical in inductive or grounded approaches (Hsieh & Shannon, 2005; Miles et al., 2014; Saldaña, 2016)

Second, the LRCC codebook provided operational definitions and explicit inclusion/exclusion criteria for each of the 28 codes, specifying the qualification requirements for each statement. These criteria function as decision rules that reduce coder discretion, as structured observation protocols do in quantitative behavioural research. By limiting interpretive range, the coding instrument minimizes the ambiguity typically captured by inter-rater reliability metrics.

Third, the study used a temporal consistency check as a proxy for reliability, involving a second independent pass of all transcripts two weeks after the initial coding phase. Resolving disagreements between the two passes ensured the integrity of the final coding. Temporal consistency — stability of a single coder's judgements across time — is a recognized reliability indicator for deductive coding in confirmatory contexts (Krippendorff, 2004; Miles et al., 2014).



Fourth, the qualitative component's role in this study is to triangulate and corroborate rather than to generate primary theory. The interview analysis does not generate the study's central findings; it contextualizes and extends findings established by the NLP analysis and the SEM. The evidential weight placed on any individual code assignment is therefore lower than it would be in a study where qualitative coding is the primary analytical instrument.

All interviews were segmented into meaningful idea units, each representing a coherent conceptual statement, and coded using only the predefined code set. The data fully aligned with the existing framework and required no additional emergent codes during analysis. The study employed descriptive statistics to calculate code frequency and salience, facilitating the identification of cross-case patterns. The study does not draw statistical inferences from qualitative code counts. This deductive strategy ensured direct traceability between interview findings and the constructs examined in the structural equation modelling phase, strengthening cross-method coherence.

Code frequency and salience were calculated descriptively to identify cross-case patterns, with particular attention to whether leaders articulated structured probabilistic reasoning (likelihood and impact), impact-only framing, adversarial threat framing, or undifferentiated assumptions of competence. Frequency counts enhance transparency and comparability across cases; the study avoids drawing statistical inferences from these qualitative code counts.

By employing a structured template, I mapped coded segments to specific research questions to facilitate a comprehensive thematic synthesis. Cross-case comparison assessed convergence and divergence in leadership expectations and observed reasoning patterns. This deductive coding approach enhanced construct alignment, case comparability, and methodological integration within the explanatory mixed-methods design.

### 3.5.4    Software and Tools



This research employed specific software tools for NLP classification, statistical analyses, and qualitative coding. These tools are well established in academic and industry research contexts, thereby ensuring a reliable and efficient analysis pipeline.

a) **Python (v3.12.10)**: The primary programming language for performing NLP tasks, such as text preprocessing, feature extraction, and content classification. Python's robust ecosystem of libraries enables scalable, flexible analysis.

b) **OpenAI Python SDK**: Provided an interface for submitting TKS statements to the GPT large language model and receiving structured classification responses. The SDK's structured output capability (responses parsed with Pydantic schema enforcement) ensured that every API response conformed to the required schema for 29 scored competency categories, each with a relevance score, a confidence score, and a natural-language justification.

c) **Scikit-learn**: This library supports statistical analysis of survey data, including descriptive summaries and inferential procedures. The NLP classification pipeline did not utilize this library.

d) **Jupyter**: A popular tool for interactive computing that allows for the development, testing, and documentation of the analysis pipeline in an efficient and reproducible way. By leveraging the iterative environment of Jupyter Notebooks, the NLP pipeline and statistical models were initially developed, optimized, and validated. I converted the Jupyter-based pipeline into linear, standalone Python scripts to ensure reproducibility and eliminate the hidden states common in notebook environments. This conversion ensured deterministic execution order, eliminated hidden dependencies, and enhanced full computational reproducibility. The flat Python implementation serves as the authoritative version of the



analytical pipeline and enables independent replication without relying on notebook execution history.

e) **Pandas & NumPy:** Essential Python libraries for data manipulation and statistical calculations. The Pandas library facilitated the handling and manipulation of survey data, enabling efficient cleaning and restructuring of the responses. In contrast, the NumPy library performed the numerical operations needed to compute summary statistics, including means, medians, and standard deviations.

f) **G*Power** is a statistical tool for determining sample size and calculating statistical power. It ensured that the sample size was sufficient to detect significant effects in the analysis of perceptions of risk management training and competency.

g) **SmartPLS**: SmartPLS is a software application for structural equation modelling (SEM) that explicitly uses partial least squares (PLS) path modelling. It facilitates SEM and PLS analyses, enabling users to evaluate model fit, estimate path coefficients, and generate comprehensive reports.

h) **JASP:** JASP is an open-source statistical software application that supports structural equation modelling using established SEM engines through a graphical interface. In this study, JASP was used to estimate confirmatory SEM models and evaluate model fit and parameter estimates.

i) **R** (with the lavaan package): R is an open-source statistical computing environment, and the lavaan package performs the structural equation modelling required for this research. R was used to specify and estimate measurement and structural models and to compute direct, indirect, and total effects.



j) **Chart.js:** After the Python-based NLP pipeline performs the underlying calculations, Chart.js visualizes the results in the presentation layer. Visualization scripts were linked directly to processed output files, ensuring that graphical representations reflected deterministic analytical results rather than manual manipulation.

All software (except SmartPLS) is open source and well-established, providing a reliable foundation for the analysis required to assess the integration of risk management principles into cybersecurity training programs.

### 3.5.5   *Post-Analysis Theoretical Interpretation*

This study aligns the insights from the NLP analysis with the theoretical frameworks of risk management and competency-based learning discussed in Chapter 2. Specifically, the analysis evaluates the frequency and distribution of semantically classified risk-related content across competency categories. It interprets those patterns in relation to the core principles of ISO 31000 and the NIST risk management Framework (RMF). These frameworks provide well-established guidelines for risk management, serving as the benchmark against which I assessed the training content's coverage of these critical areas.  This study employs Python-based analytical tools to execute quantitative and computational analyses, including text classification and model evaluation. This study used structural equation modelling to examine both direct and indirect relationships among the specified constructs.

Furthermore, competency-based learning theory examined discrepancies between the content of training materials and workforce needs. This theory emphasizes that effective learning outcomes should align with clearly defined competencies, including the 19 cybersecurity competencies in this study. By applying this framework, the study contextualized how gaps in risk management training



(identified through NLP analysis) might affect professional practice, particularly in terms of professionals' preparedness to address risk-based challenges in their roles.

These theoretical lenses enabled the study to generate actionable insights to enhance alignment between training programs and the competencies required in the cybersecurity workforce. The analysis evaluated whether the 29 competency areas placed adequate emphasis on risk management.

### 3.5.6    SEM Estimation and Model Refinement Strategy

While variance-based SEM tools such as SmartPLS were initially considered, the final analysis employed covariance-based SEM using Weighted Least Squares Mean and Variance adjusted (WLSMV) estimation in JASP and R (lavaan). Three factors mandated this approach: the ordinal nature of the survey indicators, the study's confirmatory orientation, and the necessity of robust global fit indices. WLSMV is widely recommended for SEM models involving ordered categorical indicators and provides unbiased parameter estimates and standard errors under these conditions.

Model refinement was conducted within a theory-constrained framework, focusing on validating the measurement structure of key constructs rather than altering hypothesized structural relationships. Initial testing of a second-order hierarchical specification for RM_Competence (with DV1 through DV4 as first-order factors) produced estimation problems that precluded its use as specified. The preliminary analysis dictated a unidimensional specification for the dependent variable. The empirical basis for this decision and its theoretical interpretation are addressed in Chapters 4 and 5, respectively. The underlying mediation structure and all hypothesized structural paths remained fixed throughout this process.

### 3.6    Ethical Considerations



Adherence to established research ethics principles and institutional guidelines is critical to maintaining the ethical integrity of this study. This Section outlines the ethical measures to protect participant welfare, maintain data confidentiality, and ensure responsible data handling throughout the research process.

The study complied with the ethical standards set forth by the Golden Gate University Institutional Review Board (IRB) and adhered to the principles outlined in the Belmont Report (National Commission for the Protection of Human Subjects of Biomedical and Behavioral Research, 1979), which includes respect for persons, beneficence, and justice. These principles guided the study's ethical practices. I committed to protecting participants' rights, handling data responsibly, and maintaining transparency throughout the research process.

### 3.6.1    Informed Consent

The study required informed consent from all participants before data collection. Research protocols ensured the anonymity of responses and the secure storage of data to prevent unauthorized access. All participants provided informed consent after receiving assurances that their involvement was voluntary and that they retained the right to withdraw at any time without penalty. The recruitment process included an explanation of data storage and usage policies to ensure compliance with ethical standards. All study participants provided informed consent before participating in the survey or interviews.  The American Psychological Association (APA7) Ethical Principles inform the study's ethical practices, specifically regarding research involving human subjects (American Psychological Association, 2016, 2017).

Participants received a detailed informed consent form that outlined the following:



1. The study's purpose, objectives, and research questions clearly explain the focus on risk management training and its integration into cybersecurity curricula.

2. The collected data types include survey responses and demographic information related to cybersecurity training.

3. The voluntary nature of participation and the right to withdraw at any time without penalty emphasize that participation is optional.

4. The study design incorporates specific safeguards to protect participant confidentiality and data privacy.

Participants must provide written or electronic consent before proceeding with the survey. The consent form also included contact information for the researcher and the IRB, allowing participants to seek clarification.

### 3.6.2 *Participant Confidentiality and Anonymity*

Confidentiality was maintained by anonymizing participant data and ensuring no identifiable information was published. To maintain security, I stored the data in protected locations accessible only to me. The study upheld strict confidentiality standards for all participants throughout the research process. This control aligns with standard ethical practices, as supported by texts such as those by Babbie (2013), who wrote on research ethics in the social sciences. The research design incorporates the following measures to ensure participant anonymity and data security:

a) **Data Anonymization**: All survey responses and interview transcripts were anonymized using unique participant codes. The research design excluded personally identifiable information (PII) to prevent participant re-identification.



b) **Data Encryption**: All data was stored on encrypted servers, utilizing encryption protocols that adhere to industry standards, such as AES-256. This safeguard ensures that sensitive information is securely stored and protected.

c) **Access Restrictions**: Raw data remained restricted to the researcher. To ensure privacy, the final report presents only aggregated and anonymized findings. To prevent re-identification, this study uses paraphrased versions of interview quotes when direct attribution would pose a privacy risk.

### 3.6.3    Survey Integrity and Participant Well-Being

The survey instrument minimizes participants' psychological discomfort by carefully sequencing questions and using neutral phrasing. Questions focused on participants' knowledge, perceptions, and training experiences in cybersecurity risk management. The research design omitted sensitive personal information, limiting collection to basic demographics to ensure participant privacy.

To further support participant well-being:

a) Participants could skip any questions they found uncomfortable and quit the survey.

b) A debriefing statement was included at the end of the survey, summarizing the study's goals and the potential applications of the research findings. This statement provided participants with a clear understanding of how their responses contribute to improving cybersecurity training and risk management education.

These ethical considerations ensured that the research process respected participants' rights and privacy, adhered to industry standards, and maintained the integrity and confidentiality of the data.

### 3.6.4    Ethical Approval and Compliance



The research plan, including survey instruments and interview protocols, was submitted to the Golden Gate University IRB for review and approval before survey distribution. The study waited for written IRB approval before starting data collection. These research activities followed the ethical guidelines of the 7th Edition of the American Psychological Association (2020) and established cybersecurity research standards.

### 3.6.5    Data Handling and Retention

Data-handling procedures prioritized the confidentiality, accuracy, and security of information collected throughout the study. The study employs the following practices to safeguard participant data and uphold responsible research standards:

**Data Retention Period**: All research data, including survey responses and NLP-analyzed text, was retained for 3 years following the study's conclusion or as required by Golden Gate University (GGU) institutional policies. This period allowed for further analysis or potential re-examination of findings, ensuring that the data remains accessible for academic or regulatory purposes.

**Secure Storage**: To protect the dataset, I hosted all files on institutional servers and implemented multi-factor authentication (MFA) to ensure exclusive access for researchers. This safeguard ensures that access to the data is highly restricted and protected from unauthorized use. The study uses encrypted storage to protect participant information and risk-management-related content. The data collection process excludes sensitive personal information.

**Data Disposal**: The study employed secure data-wiping techniques (e.g., DoD 5220.22-M) to permanently delete all datasets after the retention period. These protocols prevented sensitive data from being stored beyond the necessary timeframe. This procedure ensures strict adherence to privacy standards and mandates the responsible handling of participant data.



These data handling measures ensure data integrity and compliance with ethical guidelines for the safe storage and disposal of research data.

### 3.6.6    Ethical Risks and Mitigation Strategies

The primary ethical risks associated with this research include potential breaches of participant confidentiality and biases in survey responses. The study employs the following mitigation strategies to minimize risks and maintain strict ethical standards:

**Bias Mitigation**: The survey instrument used neutral language and avoided leading questions to reduce the risk of response bias. This approach helped ensure that participants' responses accurately reflected their perceptions and experiences regarding risk management training and competency. A pilot test identified unintentional biases in the survey design, allowing me to refine the instrument before full implementation. To prioritize neutrality, I adjusted the survey instrument to eliminate identified biases before full implementation.

**Data Breach Prevention**: All digital data transfers (from survey responses to analysis) used end-to-end encryption, ensuring data integrity and confidentiality from collection through analysis. This measure ensures that any personal or sensitive information related to risk management training is securely transmitted and stored.

**Transparency**: The study provided participants with detailed information regarding data usage, storage, and protection. Before beginning, every participant signed an informed consent form, confirming their understanding of the research process. Upon request, a summary of the findings was provided to participants, demonstrating transparency regarding how their participation contributes to the research.



These strategies aimed to minimize ethical risks, ensure the highest level of study integrity, protect participants' privacy, and yield valuable insights into risk management in cybersecurity training.

**3.7     Limitations of the Methodology**

All research endeavours involve inherent limitations that may impact the interpretation and generalizability of their findings. This study is no exception. Building on the limitations identified in Chapter 1, this Section evaluates methodological constraints and discusses the mitigation strategies used to maintain research integrity.

1. **Dependence on NICE, ECSF, and SPARTA Frameworks as Proxies for Cybersecurity Training Content**

This study relies heavily on the NIST NICE Cybersecurity Workforce Framework (U.S.), the SPARTA Cybersecurity Skills Framework (Europe), and the ECSF to represent the training content delivered across more than 14,696 academic and certification-based cybersecurity programs. While these frameworks provide a standardized structure for defining cybersecurity competencies, they may not fully capture content from programs that do not align with NICE or SPARTA guidelines. Additionally, in-house training programs developed by organizations for their specific needs are typically not documented in these frameworks, potentially excluding relevant risk management training content. This limitation is the challenge of ensuring that the training programs included in the analysis are representative of the broader cybersecurity education landscape.  This limitation is related to the challenges of generalizing from specific case studies, as discussed in general methodological texts, such as Creswell (2014).

**Mitigation Strategy:**



The study collected supplementary training materials from publicly accessible syllabi and descriptions of major certifications (e.g., CISSP, Security+, CISM) to broaden the representation of risk management principles in training content, thereby mitigating this limitation. Survey responses from cybersecurity professionals also provided qualitative insights into training experiences that might not be reflected in NICE or SPARTA documentation, further enhancing the study's comprehensiveness.

2. **NLP Model Limitations and Context Misclassification**

Ambiguous terms and domain-specific language often hinder an NLP model's ability to categorize risk management data accurately. For example, "vulnerability" can refer to technical system weaknesses and broader risk management considerations. Context misclassification could lead to biased estimates of the prevalence of risk management concepts in training materials.

**Mitigation Strategy:**

The pipeline addressed this concern through the category taxonomy's explicit include/exclude rules and the large language model's context-sensitive reasoning. For terms such as "vulnerability," which may appear in technical system contexts (appropriately classified under Threats and Vulnerabilities) or in risk exposure contexts (appropriately classified under Risk Management), the classification prompt provided explicit disambiguation criteria directing the model to prefer the category most consistent with the statement's primary analytical purpose. Because the model produced a natural-language justification for every classification decision, ambiguous cases were auditable: any individual TKS statement's category assignment could be verified by inspecting the recorded justification text, providing a transparency mechanism absent from black-box classifiers.

3. **Potential Response Bias in Survey Data**



Survey-based research is inherently prone to response bias, particularly when respondents over- or underestimate their knowledge or training experience. Social desirability bias may also lead professionals to overstate their familiarity with risk management principles.

**Mitigation Strategy:**

The survey utilized neutral, behaviorally anchored questions to minimize the risk of subjective interpretations and reduce response bias. The pre-test pilot helped identify any unintentional bias-inducing questions. The analysis phase included comparisons between survey responses and NLP-derived findings to identify discrepancies indicative of response bias, such as overreporting the prevalence of risk management concepts in training programs.

4. **Common Method Bias in Self-Report Survey Data**

A structural threat inherent in this study's survey design is the risk of Common Method Bias (CMB), also referred to as common method variance. CMB arises when the independent and dependent variables of interest are collected from the same respondents, using the same instrument, at the same point in time; under these conditions, observed correlations between constructs may be partially inflated by measurement artifact rather than by the substantive relationships of theoretical interest (Podsakoff et al., 2003). In the present study, IV1x — self-reported exposure to risk-embedded training content — and DV1–DV4 — the four Risk Management Competence indicators — are both operationalized as Likert-scale items within a single Qualtrics survey instrument administered in a single session. This design satisfies the defining conditions for CMB risk as described in the organizational behaviour and applied research methodology literature (Chang et al., 2010; Harman, 1967; Podsakoff et al., 2003). The concern is not merely theoretical: in a SEM context, CMB can produce inflated path coefficients on the IV1x → DV1–DV4 pathways, increasing the risk of falsely supporting $H_{12}$ or overstating the magnitude of a genuine effect.



**Mitigation Strategy:**

The analysis used Harman's single-factor test to assess all survey scale items and provide a preliminary diagnosis of CMB severity. This procedure evaluates whether a single unrotated factor in an exploratory factor analysis accounts for the majority of the total variance. This result would indicate pervasive method variance as the primary source of covariance in the data. In this study, the first unrotated factor accounted for 46.3% of the total variance across all survey scale items (below the conventional 50% threshold), providing preliminary evidence that a single common method factor does not account for the majority of observed covariance, and that CMB alone is unlikely to explain the inter-construct relationships estimated in the SEM (Podsakoff et al., 2003). Scholars widely acknowledge Harman's test as a conservative and imperfect diagnostic for common method bias; it identifies whether method variance is the sole explanation for the data structure, not whether it contributes at a level that may influence path estimates.

Three additional design features further reduce the CMB risk. First, the survey instrument employed behaviourally anchored response items targeting distinct referents: IV1x items addressed the content of prior training programs, while DV1–DV4 items addressed current professional work behaviour. This temporal and conceptual separation reduces the shared-evaluation process, which is most responsible for inflating cross-construct correlations. Second, neutral item wording and the absence of leading or evaluative language — verified during the pre-test pilot — reduced the shared affective frame that amplifies method effects. Third, and most consequentially for this study's overall evidentiary structure, the self-report SEM strand operates alongside an independent, non-self-report measure of training architecture: the NLP-derived content analysis of the NICE framework. This NLP measure is entirely free of common method variance by construction, as it is derived from objective computational classification of training text rather than from respondent perception. Where the NLP



findings and the SEM results converge (specifically, where domains identified as risk-deficient by the NLP analysis are also associated with lower self-reported competence in the SEM), the convergent pattern constitutes cross-method corroboration that is structurally independent of CMB. Such convergence, when observed, substantially strengthens confidence that the SEM path estimates reflect genuine training-competence relationships rather than shared measurement artifact.

Notwithstanding these mitigating features, the study maintains a conservative interpretation of the path coefficients on the IV1x → DV1–DV4 pathways. Some portion of the observed covariance may reflect a consistent internal frame of reference applied by respondents across training and competence items. The multi-method design provides the primary structural defence against over-reliance on single-source inference.

**5. Limited Generalizability Across Regions and Sectors**

The study primarily focuses on training materials associated with the NICE, ECSF, and SPARTA frameworks, which represent training models from North America and Europe. As a result, the findings may not be directly generalizable to other regions of the world, such as Asia, Africa, or South America, where cybersecurity training frameworks may emphasize different competencies.

**Mitigation Strategy:**

The study contextualized its findings by comparing coverage patterns across NICE, ECSF, and SPARTA and by identifying training content patterns that may apply to other frameworks. This strategy improved the study's external validity. Future research could extend the analysis to additional regions, offering a more global perspective on cybersecurity risk management training.

**6. Constraints on Data Access and Completeness**

The study's ability to analyze training content depends on publicly available materials, which may not provide complete or up-to-date representations of all training programs. Proprietary or



confidential training materials, such as those developed internally by organizations, are generally not accessible for analysis.

**Mitigation Strategy:**

The study prioritized training programs in the NICCS database, which maps certification courses to NICE Knowledge Statements (KSAs). Survey respondents described the nature of their training experiences, offering qualitative context that helped identify potential content gaps. These responses allowed the study to gather data on risk management training content that the publicly available frameworks may not capture.

7. **Temporal Limitations and Static Content Analysis**

Cybersecurity training programs continually evolve to address emerging threats and evolving regulatory requirements. This study's analysis is based on training materials collected at a specific point in time and may not reflect subsequent curricular changes.

**Mitigation Strategy:**

The study documented the data collection timeline and acknowledged the potential for content evolution. The NLP methodology, including the pipeline and classification models, was thoroughly documented to facilitate replication and longitudinal comparisons in future research. This transparency enabled future studies to revisit and build upon the research as training programs evolve.

8. **Limitations in the Measurement of Professional Competency**

Competency in cybersecurity risk management extends beyond theoretical knowledge and into practical application. While this study focuses on measuring the presence of risk management content in training materials and self-reported competency among professionals, it does not directly measure task performance in live cybersecurity scenarios.

**Mitigation Strategy:**



While direct performance testing is beyond the scope of this study, the survey included scenario-based questions designed to assess participants' ability to demonstrate risk management competencies in realistic situations. Additionally, using NLP to analyze professional discourse in interviews helped identify practical applications of theoretical knowledge, bridging the gap between training content and real-world practice.

9. **Single-Coder Qualitative Analysis**

I coded all seven leadership interviews. In qualitative research traditions that use inductive coding to discover emergent themes, single-coder designs are a recognized limitation because they risk leading the resulting codes to reflect the researcher's interpretive frame rather than patterns independently present in the data. Quantification of that risk typically requires inter-rater reliability testing with a second independent coder.

**Mitigation Strategy:**

Four design features reduce the reliability risk inherent in single-coder analysis in this study. The coding approach is deductive: specifying the LRCC codebook in full preceded the coding of any transcripts, thereby fixing the code set and eliminating emergent coding discretion. Each code was defined with operational inclusion and exclusion criteria, functioning as decision rules that constrain interpretive variation. A temporal consistency check — a second independent coding pass conducted two weeks after initial coding — served as a within-coder reliability proxy; the reconciliation of discrepancies between passes ensured the accuracy of the final codes. Finally, the qualitative component's function is triangulation and corroboration rather than primary evidence generation; the study's central findings are established by the NLP analysis and the SEM, with interview analysis providing contextual depth and convergent validation. The single-coder limitation is therefore real but



bounded: it affects the precision of individual code assignments within a fixed theoretical scheme, not the validity of the study's primary empirical claims.

### 3.7.1    Conclusion about Limitations

These limitations present challenges; however, the research design incorporates specific mitigation strategies to lessen their impact on the study's findings. The risk of common method bias inherent in the single-instrument self-report design is partially offset by Harman's single-factor diagnostic and, more substantially, by the structural independence of the NLP-derived measures, which provide a cross-method verification anchor for the SEM findings. The study's combination of NLP analysis and survey insights establishes a robust foundation for examining the integration of risk management principles into cybersecurity training programs and measuring professional competency. This approach enables the study to address the complexities of cybersecurity education, ensuring that the findings are valid and applicable across diverse contexts and regions.

## 3.8    Summary

This chapter outlined the research design and methodological approach used to investigate the extent to which risk management principles are integrated into cybersecurity training and reflected in professional practice. This methodology targets the study's central research question by evaluating the effectiveness of current cybersecurity training programs in developing the functional risk management skills required to identify, assess, prioritize, mitigate, and communicate cyber risks. A mixed-methods design was employed to support a comprehensive and empirically grounded examination of this question.

The chapter introduced the study's overall methodological approach, which integrated NLP-based content analysis with survey-based latent-variable modelling, cross-group comparisons, and



leadership interviews. This study uses NLP techniques to examine publicly available cybersecurity training frameworks, such as NIST NICE, ECSF, and SPARTA, which standardize competency structures used in the NICCS database and industry training. This analysis quantified the prevalence and distribution of risk management concepts across cybersecurity competency areas.

Survey data collected from cybersecurity professionals provided individual-level measures of exposure to risk-embedded training, perceived relevance of risk management, and risk-based professional behaviours, enabling examination of how training emphasis relates to applied competence in practice.

The research design Section detailed the rationale for selecting NLP as a systematic and scalable method for evaluating training content and for employing survey-based structural equation modelling to test theoretically grounded relationships between training exposure and professional competence. The target population included cybersecurity professionals across roles and levels, with a non-cybersecurity comparison group used to contextualize baseline understanding of risk management concepts.

Data collection procedures were described, including the development, pilot testing, and deployment of the survey instrument. The instrument measured risk management exposure, perceived relevance, and applied competence, along with relevant demographic and experiential variables. The study complies with Golden Gate University's Institutional Review Board requirements. I established ethical safeguards to guarantee informed consent and protect participant confidentiality.

The data analysis Section outlined the multi-stage NLP pipeline used to preprocess, classify, and measure risk-management content in training materials. It specified a theory-driven structural equation model grounded in competency-based learning theory. The research design specifies all constructs and relationships a priori, enabling confirmatory hypothesis testing and minimizing the risks associated with exploratory analysis. Methodological limitations—such as potential NLP misclassification, evolving aining



content, and survey response bias—were acknowledged, along with mitigation strategies designed to preserve validity and reliability.

In summary, the methodology presented in this chapter provides a systematic, replicable approach to evaluating both the integration of risk management principles into cybersecurity training and their translation into professional competency in practice. Chapter 4 presents the empirical results of these analyses, including findings from the NLP content analysis, measurement model evaluation, and structural equation modelling. This chapter presents the results descriptively and avoids interpretation. Chapter 5 evaluates the findings, discussing their theoretical implications and practical significance.



**Chapter 4 – Results and Data Analysis**

The following chapter presents the empirical results of all four analytical strands (NLP content analysis, structural equation modelling, cross-group comparison, and leadership interview analysis), organized to directly address the research questions and hypotheses established in Chapter 1.

## 4.1    Introduction to Results

This chapter presents the results of quantitative and qualitative analyses examining whether cybersecurity training programs meaningfully embed risk management principles and whether such training, directly or indirectly, influences professionals' risk management competence through perceived relevance and cognitive framing. Consistent with this dissertation's mixed-methods design, the chapter reports findings from Structural Equation Modelling (SEM) based on the survey data, followed by supplementary analyses and qualitative interview results presented in later Sections.

Chapter 4 presents the study's empirical findings in four stages. First, the Natural Language Processing (NLP) analysis evaluates the structural representation of cybersecurity competencies across the NICE framework. Second, this chapter evaluates the measurement properties of the survey constructs, with a focus on reliability and validity. Third, the structural equation modelling (SEM) results are presented, including model comparison and mediation analysis. Finally, this chapter reports supplementary descriptive and group-level analyses to provide additional context for the findings. This chapter presents the results, while Chapter 5 addresses the interpretive and theoretical implications.

The primary quantitative analysis employed Structural Equation Modelling using ordinal Likert-type indicators and was estimated using the weighted least squares mean- and variance-adjusted (WLSMV) estimator in lavaan (R). This estimator was selected to accommodate the ordinal measurement scale of the survey items and to provide robust parameter estimates in the presence of



non-normality. The SEM examines direct and indirect relationships between training exposure and risk management competence, using perceived relevance as a mediating construct.

The SEM examines direct and indirect relationships between training exposure and risk management competence, using perceived relevance as a mediating construct. This chapter starts by summarizing the preparation of survey data and the descriptive statistics. The measurement model results are then presented, including factor loadings, reliability estimates, and assessments of convergent and discriminant validity. Following these measurements, the chapter presents the structural model results, including standardized path coefficients, statistical significance, and overall model fit indices. This chapter also reports direct, indirect, and total effects to evaluate the hypothesized mediation relationships. A separate Section summarizes the influence of control variables, while later parts of the chapter integrate supplementary quantitative data and qualitative interview findings.

Throughout this chapter, the results are reported objectively and descriptively, without interpretation, explanation, or linkage to theory or practice. This chapter summarizes the outcomes of hypothesis testing based solely on statistical evidence, indicating which hypotheses the data support. Chapter 5 reserves the interpretation and theoretical implications of these findings.

## 4.2    Cybersecurity Training Frameworks (NLP)

This analysis focused on the presence and relative emphasis of risk management concepts within training artifacts and does not evaluate instructional quality or learning outcomes.

### 4.2.1    Corpus Description and Analytical Approach

The NLP corpus consisted of 2111 Training, Knowledge, and Skills (TKS) statements extracted from the NIST NICE framework (631 Knowledge Statements, 538 Skills, 942 Tasks). This NLP approach



treated each TKS statement as a distinct unit of analysis and matched risk-management keywords semantically. The NLP techniques described in Chapter 3 identified and classified the risk-related content within the corpus. This study uses the NIST NICE framework as the primary NLP analysis framework. Its comprehensive structure and well-documented competency framework make it the ideal candidate for systematic content analysis.

Instrument development included comparing the ECSF and SPARTA frameworks with the NIST NICE framework. Both European taxonomies mirror the NICE structure by categorizing governance, technical operations, and risk management as distinct competency domains. The NICE findings are therefore treated as representative of the broader framework family for this study, with the caveat that framework-specific variations in competency labelling and scope may limit direct comparability. While the primary analysis uses the NICE v2.0.0 corpus, the ECSF and SPARTA frameworks corroborate the key probabilistic vocabulary finding reported in Section 4.2.6 via lexical analysis.

### 4.2.2    Frequency and Distribution of Risk Management Content

The analysis calculates the frequency and distribution of competency activation across the full corpus of NIST NICE Task, Knowledge, and Skill (TKS) statements. The analysis examined both how often risk-related statements appeared (absolute counts) and what proportion of the overall framework they represented (relative proportion). In addition to binary classification of risk-related versus non-risk-related content, semantic similarity scoring enabled categorization by competency domain and by confidence threshold. These distributions provide an empirical overview of how the NICE framework incorporates formal risk management constructs and other cybersecurity competencies.

### 4.2.3    Comparative Distribution Across Competency Categories



To determine the prevalence of risk management, this Section compares the distribution of risk-related content with that of other cybersecurity domains within the NICE framework. Table 7 presents the NLP-based distribution of competency activation across all NICE Knowledge and Skill statements. As shown in Table 7, the risk management competency (RISK) accounted for 4.5% of high-confidence statements, ranking 18th of 29 competencies. In contrast, several technical domains exceeded 8–10% representation. This distribution indicates that risk-related competencies are structurally present within the framework but comparatively underrepresented relative to operational and technical categories.

**Table 7**

*NLP Distribution of Competency Activation Across NICE K / S Statements*

| Competency Area | Code | Total Active (≥0.45) | High (≥0.75) | Moderate (0.45–0.74) | % of Total Corpus (≥0.75) |
|---|---|---|---|---|---|
| Cybersecurity Foundations | CYBR | 693 | 137 | 556 | 6.5% |
| Cybersecurity Governance | CGOV | 672 | 396 | 276 | 18.8% |
| Security Operations | SOPS | 611 | 248 | 363 | 11.7% |
| Systems Security Architecture | SYSA | 556 | 183 | 373 | 8.7% |
| Detection Operations | DEOP | 453 | 189 | 264 | 9.0% |
| Threat Management | THMA | 421 | 116 | 305 | 5.5% |
| Cyber Threat Intelligence | THRT | 316 | 144 | 172 | 6.8% |
| Digital Forensics | DFOR | 285 | 129 | 156 | 6.1% |
| Data Security & Protection | DATA | 258 | 125 | 133 | 5.9% |
| Security Testing | TEST | 244 | 109 | 135 | 5.2% |
| Secure Software Development | SESO | 238 | 91 | 147 | 4.3% |
| Asset Management | ASST | 194 | 77 | 117 | 3.6% |
| Cyber Resilience | RESI | 194 | 75 | 119 | 3.6% |
| Communication Security | COMM | 191 | 82 | 109 | 3.9% |
| Security Awareness | AWAR | 163 | 76 | 87 | 3.6% |
| Operating System Security | OPSC | 155 | 60 | 95 | 2.8% |
| DevSecOps | DEVS | 149 | 20 | 129 | 0.9% |
| **Cyber Risk Management** | **RISK** | **137** | **96** | **41** | **4.5%** |
| Offensive Security | PENN | 130 | 50 | 80 | 2.4% |
| Access Controls | ACCO | 102 | 43 | 59 | 2.0% |
| Supply Chain Security | SCSY | 92 | 65 | 27 | 3.1% |
| Malware & Software Analysis | MASO | 86 | 33 | 53 | 1.6% |
| Cryptography | CRYP | 54 | 30 | 24 | 1.4% |



| Operational Technology Security | OTSC | 48 | 43 | 5 | 2.0% |
|---|---|---|---|---|---|
| Secure Hardware & Embedded Systems | SEHA | 43 | 15 | 28 | 0.7% |
| Platform & Virtualization Security | PLVI | 20 | 4 | 16 | 0.2% |
| Emerging Technology Risk Management | EMTE | 17 | 9 | 8 | 0.4% |
| Cloud Security | CLOU | 8 | 2 | 6 | 0.1% |
| Artificial Intelligence Security | AISC | 7 | 4 | 3 | 0.2% |

Cybersecurity Foundations (CYBR) and Cybersecurity Governance (CGOV) demonstrated the highest total activation counts, followed by operational domains such as Security Operations (SOPS) and Systems Security Architecture (SYSA). These domains collectively accounted for the largest proportion of competency activation within the framework.

Cyber risk management (RISK) demonstrated lower total activation counts than the foundational and operational domains. However, the analysis identified a substantial proportion of RISK matches that exceeded the high-confidence threshold (≥0.75), demonstrating strong semantic alignment.

The relative ordering of competency domains remained stable during threshold refinement testing, suggesting that the observed distribution patterns were not materially sensitive to changes in similarity thresholds.

A cross-category contrast is a comparison of different competency areas to determine which are emphasized more or less within the framework. The analysis computes cross-category contrasts by comparing absolute activation counts, proportional representation, and high-confidence match density to identify areas of concentration and absence within the framework.

### 4.2.4   Summary of NLP Content Analysis Results

We restricted primary distribution calculations to strong semantic matches (≥ 0.75) to ensure high-confidence classification. The model also generated weaker semantic matches (0.25–0.74), which we retained and reported for completeness.  Risk management (RISK) comprised 4.5% of high-



confidence training content (96 of 2,111 TKS statements), ranking 18th among 29 competency areas. In comparison, Cybersecurity Governance (18.8%), Security Operations (11.7%), and Systems Security Architecture (8.7%) demonstrated substantially higher representation. An equal distribution across 29 competency categories would yield approximately 3.4% representation per category; the observed 4.5% representation for RISK is modestly higher than this baseline but remains substantially lower than several operational and governance domains.

Notably, when the model identified risk-related content, the data exhibited high semantic confidence; indeed, 70.1% of RISK-classified statements met the $\geq$

**Figure 3**

*Semantic Activation Frequency of RISK Competency Category*

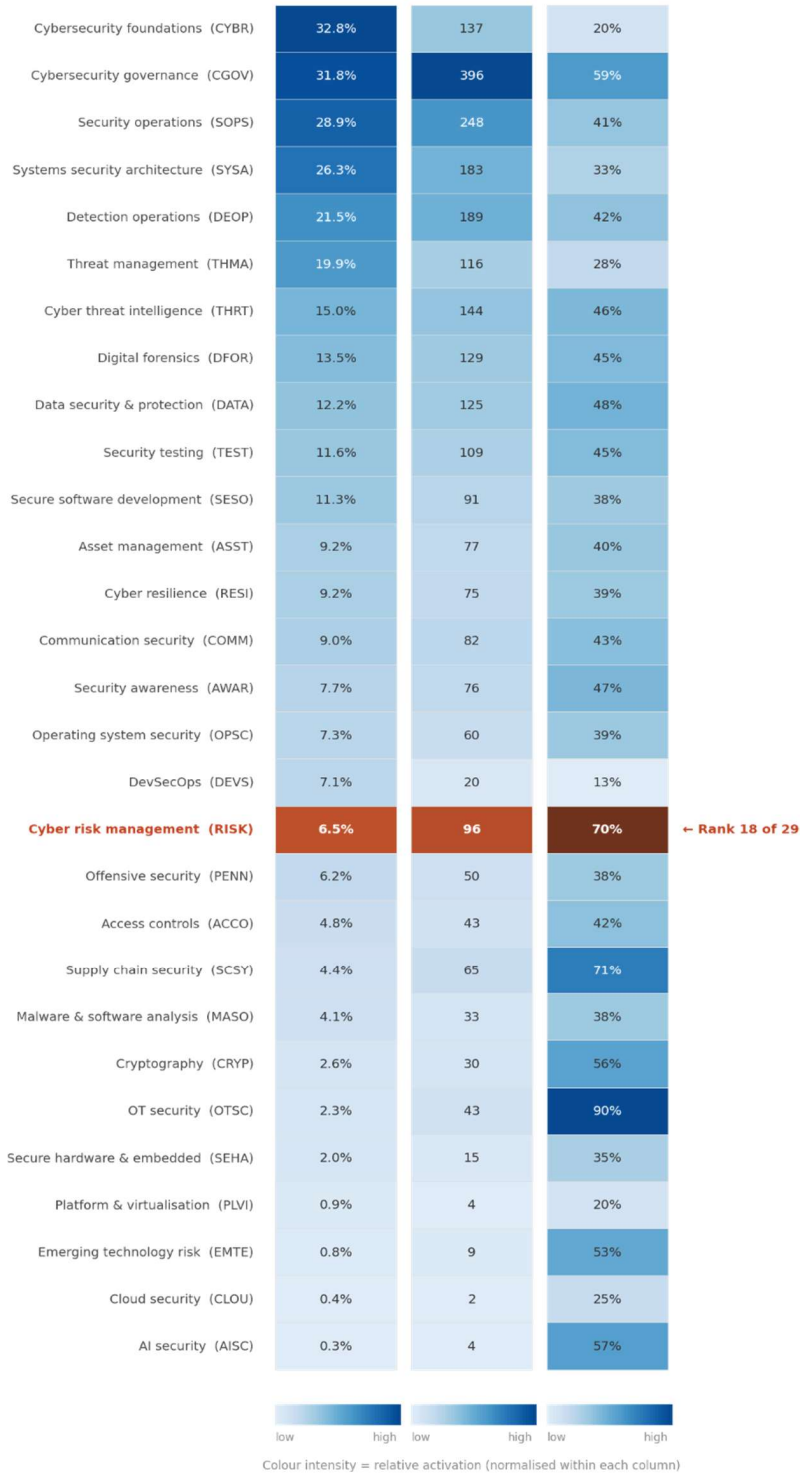



0.75$ threshold, showing that authors articulate risk concepts clearly when they include them. Still, their overall prevalence within the training corpus is comparatively limited.

The NLP analysis provides a descriptive account of how competency domains—including risk management—are represented within the NICE Knowledge and Skill framework. These findings reflect the structural composition of formal training artifacts; they do not assess how cybersecurity professionals perceive, interpret, or apply this content. Rather, the analysis evaluates whether the framework explicitly includes elements of formal risk reasoning, such as likelihood estimation, likelihood–impact risk calculation, structured risk-based prioritization, and control selection grounded in risk-reduction logic (reduction of impact, likelihood, or both). Section 4.2.5 details the results of this construct-level analysis.

The findings reported in this Section directly address $H_{13}$ by evaluating the relative representation of risk management constructs within formal competency frameworks. The SEM analysis reported later in this chapter does not test $H_{13}$ directly; instead, it examines whether variation in exposure to risk-embedded training content is associated with differences in professional risk management competence. In this sense, the SEM results provide a complementary assessment of the practical consequences of the structural patterns identified in the NLP analysis.

If training frameworks unevenly represent risk management principles, professionals who engage with explicitly risk-framed content should demonstrate higher levels of applied risk competence. The SEM therefore evaluates whether the structural characteristics of training artifacts predict respondent-level competence outcomes.



**Figure 4**

*Distribution of high-confidence TKS statements across competencies*

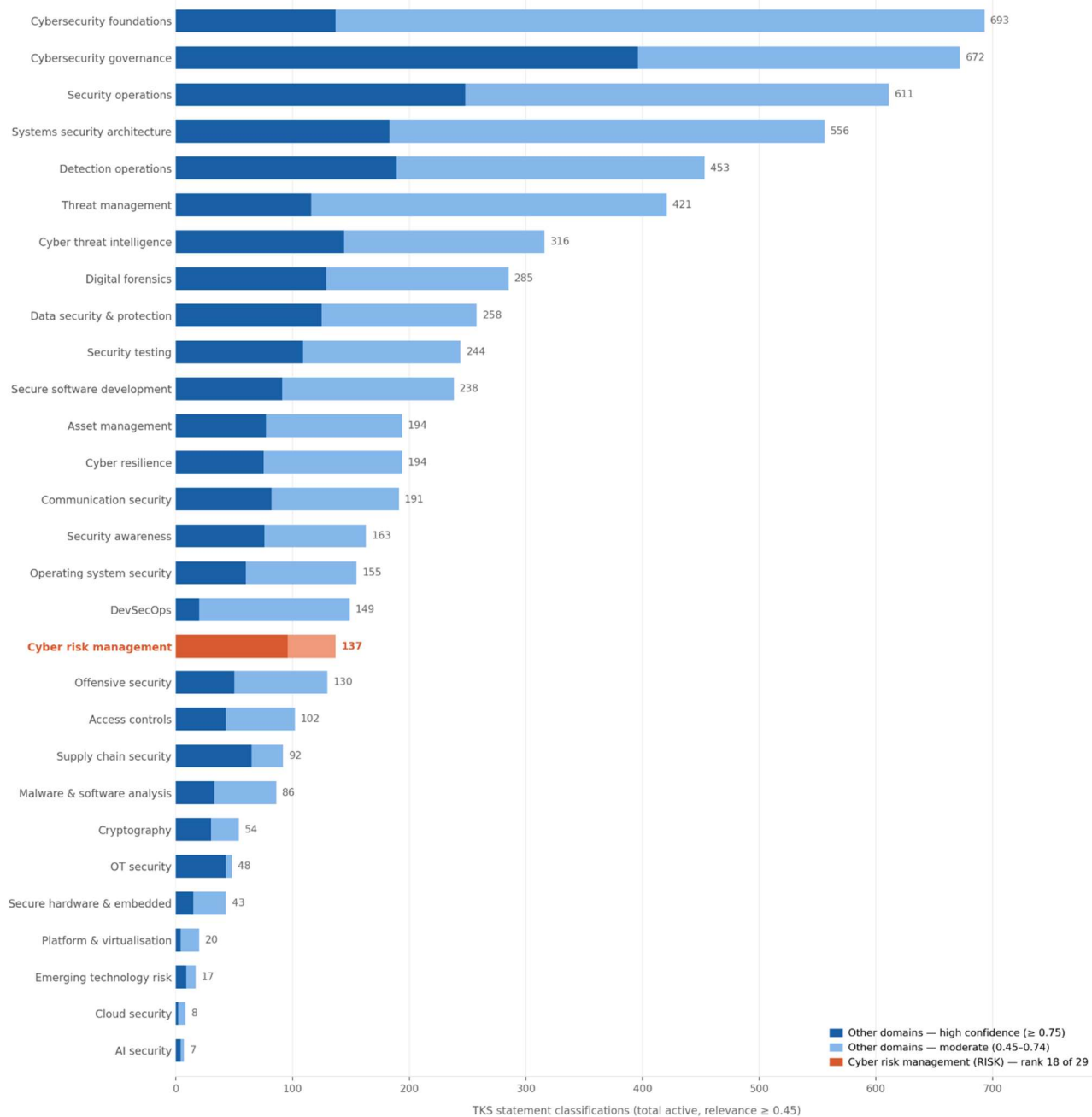





Having established the distribution of competency activation within training frameworks, the analysis now shifts to the survey dataset. Whereas this Section examined framework-level content, the next Section introduces respondent-level data preparation procedures for the survey instrument used to assess perceived training exposure, conceptual salience of risk management, and self-reported risk management competence.

Taken together, the NLP findings describe the structural distribution of risk-related content within formal competency frameworks. Subsequent Sections examine whether this structural representation corresponds to measurable differentiation in practitioner-level risk management competence (survey and SEM results) and to patterns of expectation articulated within leadership discourse (interview analysis). These results are reported separately below before interpretive integration in Chapter 5.

### 4.2.5    Construct-Level Analysis: Risk Vocabulary and Risk Reasoning

A secondary analysis examined whether risk-related TKS statements encode the reasoning construct that the dissertation's theoretical framework defines as constitutive of risk management: the joint estimation of likelihood and impact and their combination into a risk judgment ($L \times I$). The study evaluates all 2,111 TKS statements across the Task (T), Knowledge (K), and Skill (S) layers of the framework. The terms likelihood and probability do not appear in any of the 2,111 TKS statements. The likelihood × impact construct, which is definitional to risk as a computational object in enterprise risk management frameworks such as ISO 31000 and NIST SP 800-30, is absent at the lexical level. Impact language is present but consistently appears as a standalone consequence descriptor without a probability counterpart.



The next stage of the analysis evaluated each of the four dependent variable constructs for explicit **L x I** risk-calculation reasoning requirements. The analysis categorized TKS statements as strict **L x I**, method-agnostic strong, or method-agnostic weak, depending on how explicitly they require probabilistic reasoning. Table 8 presents the results across T, K, and S statement types.

**Table 8**

*NICE Framework v2.0.0: Complete DV Classification: TKS Statements*

| | | | | | | | | | | | | |
|---|---|---|---|---|---|---|---|---|---|---|---|---|
| *Total corpus: 2,111 (T=942 \| K=631 \| S=538) \| Strict L × I = 0 across ALL statement types \| Likelihood / Probability: 0 occurrences in full corpus* | | | | | | | | | | | | |
| | **T Statements (n=942)** | | | **K Statements (n=631)** | | | **S Statements (n=538)** | | | **All TKS (n=2,111)** | | |
| **Variable** | **Str** | **Wk** | **Tot** | **Str** | **Wk** | **Tot** | **Str** | **Wk** | **Tot** | **Str** | **Wk** | **Tot** |
| **DV1** | 0 | 0 | 0 | 0 | 0 | 0 | 0 | 0 | 0 | 0 | 0 | 0 |
| **DV1-MA** | 12 | 3 | **15** | 8 | 3 | **11** | 4 | 1 | **5** | 24 | 7 | **31** |
| **DV2** | 0 | 0 | 0 | 0 | 0 | 0 | 0 | 0 | 0 | 0 | 0 | 0 |
| **DV2-MA** | 0 | 0 | 0 | 0 | 0 | 0 | 0 | 0 | 0 | 0 | 0 | 0 |
| **DV3** | 0 | 0 | 0 | 0 | 0 | 0 | 0 | 0 | 0 | 0 | 0 | 0 |
| **DV3-MA** | 7 | 1 | **8** | 8 | 6 | **14** | 2 | 0 | **2** | 17 | 7 | **24** |
| **DV4** | 0 | 0 | 0 | 0 | 0 | 0 | 0 | 0 | 0 | 0 | 0 | 0 |
| **DV4-MA** | 2 | 0 | **2** | 0 | 0 | 0 | 0 | 0 | 0 | 2 | 0 | **2** |
| **TOTAL MA** | **21** | **4** | **25** | **16** | **9** | **25** | **6** | **1** | **7** | **43** | **14** | **57** |
| **% of corpus** | 2.7% | | | 4.0% | | | 1.3% | | | 2.7% | | |

**KEY:** Str = Strong (risk is primary object) | Wk = Weak (risk qualifies threat/compliance frame) | Tot = Strong + Weak | MA = Method Agnostic | Strict **L × I** = 0 across all 2,111 TKS entries

A comprehensive review of the framework reveals a total absence of TKS statements that require Likelihood x Impact risk reasoning. The 57 statements (2.7% of the full corpus) that contain risk vocabulary assignable to the four study constructs are entirely method-agnostic: risk language designates the subject domain of the activity. Still, it does not specify the reasoning method required to perform it.



Two findings warrant specific notation. First, DV2 (Risk Prioritization) returned zero matches at all levels —Task, Knowledge, and Skill. The concept of ranking risks by computed likelihood × impact magnitude is absent from the framework, not merely as a specified method but as a named activity. Second, no Knowledge or Skill statement addresses the communication of risk judgments to governance stakeholders (DV4), indicating that the framework does not define a knowledge base or skill set for the translational function that risk communication between technical and governance roles requires.

These findings supplement the distributional results reported in Sections 4.2.2 and 4.2.3. While the previous Sections established that, within the framework, risk management content is proportionally scarce, this Section demonstrates that the existing content fails to encode the probabilistic reasoning necessary to distinguish risk management from threat management as a cognitive practice.

### 4.2.6    Cross-Framework Lexical Confirmation: ECSF and SPARTA

The full NLP classification pipeline described in Chapter 3 was applied to NICE v2.0.0 because its 2,111 discrete TKS statements provide the structured, statement-level corpus required for semantic classification. The European Cybersecurity Skills Framework (European Union Agency for Cybersecurity, 2022) and the SPARTA Cybersecurity Skills Framework (Piesarskas et al., 2019) organize their content as narrative role profiles and sector-level competency descriptions, respectively — formats that do not present the volume of consistently structured discrete statements required for the same pipeline.

However, the primary finding reported in Section 4.2.5 (the complete absence of probabilistic reasoning vocabulary from the NICE TKS corpus) is directly testable in ECSF and SPARTA through targeted lexical analysis. A comparative analysis of the two European frameworks reveals four terms whose absence in NICE constitutes the finding: likelihood, probability, probabilistic, and expected loss.



The results were uniform across all three frameworks. Table 9 reports the term frequency across NICE v2.0.0, the ECSF Role Profiles document, and the SPARTA D9.1 document.

**Table 9**

*Cross-Framework Frequency of Probabilistic Risk Constructs*

| Term | NICE v2.0.0 (2,111 TKS statements) | ECSF Role Profiles (12 role profiles) | SPARTA D9.1 |
|---|---|---|---|
| "likelihood" | 0 | 0 | 0 |
| "probability" | 0 | 0 | 0 |
| "probabilistic" | 0 | 0 | 0 |
| "expected loss" | 0 | 0 | 0 |
| "risk" (for contrast) | 86 | 58 | 49 |

**Note**. Counts reflect occurrences in the full published text of each framework document. The v2_0_0_TKS_Statements.csv corpus provides the counts for the NIST NICE TKS statements. The European Cybersecurity Skills Framework Role Profiles (European Union Agency for Cybersecurity, 2022) provide the ECSF counts utilized in this cross-framework analysis. The SPARTA Deliverable D9.1 (Piesarskas et al., 2019) provided the SPARTA counts.

The absence of vocabulary for probabilistic reasoning holds across all three frameworks. "Risk" appears with moderate to high frequency in each (86 occurrences in NICE, 58 in ECSF, and 49 in SPARTA), confirming that risk vocabulary is structurally present across the international framework family. The terms that operationalize risk as a computational construct — specifically the probabilistic and expected-value reasoning that enterprise risk doctrine requires — are absent from all three

This cross-framework result confirms that the finding reported for NICE in Section 4.2.5 is not an artifact of NICE's TKS statement architecture, the classification pipeline, or the threshold decisions applied in the primary analysis. The identical structural absence holds in frameworks with different organizational formats and distinct jurisdictional origins, suggesting the finding reflects a characteristic of the profession's dominant international training architecture rather than a property of any single national framework.

**4.3     Survey Data Preparation and Descriptive Statistics**



This Section describes the preparation of the survey data and presents descriptive statistics for the study sample and observed variables used in the Structural Equation Modelling analysis. The purpose of this Section is to document data screening and cleaning, and to present the dataset's preliminary characteristics before estimating the measurement and structural models.

### 4.3.1 Data Screening and Preparation

Survey responses were exported from Qualtrics and prepared for analysis in JASP and R. Preliminary data screening ensured the dataset's completeness, response quality, and alignment with the assumptions of ordinal SEM before model estimation. Responses that did not meet predefined inclusion criteria—such as incomplete surveys or responses outside the defined sampling frame—were removed before analysis.

Table 10 summarizes survey participation and data screening outcomes. The recruitment process yielded 136 responses to 142 invitations. Pre-screening resulted in the exclusion of seven incomplete or non-consenting responses. The study's geographic restrictions required the removal of an additional three cases that originated outside the defined sampling frame (Europe and North America). After applying these exclusion criteria, 126 complete survey responses were retained for analysis, yielding an overall usable response rate of 89%. All retained responses contained complete item-level data, as survey logic required responses to all items before submission.

#### 4.3.1.1 Response Rate

**Table 10**

*Survey Response and Inclusion Summary*

| Description | Count |
| --- | --- |
| Survey invitations distributed | 142 |
| Surveys initiated | 136 |
| Incomplete or non-consented responses | 7 |
| Responses excluded (outside sampling frame) | 3 |
| **Final analytic sample** | **126** |
| **Usable response rate** | **89%** |

All retained survey items were measured using ordinal Likert-type response scales. Consequently, the analysis handles these items as ordered categorical indicators to ensure statistical accuracy in the SEM. The survey design utilized Qualtrics' mandatory response settings for all questions, thereby eliminating the possibility of missing data in the retained cases. Accordingly, all 126 retained surveys contained complete responses across all observed indicators used in the Structural Equation Model.

The analysis, therefore, relies solely on the case-level exclusions detailed above to address missing data. The retained sample of 126 cases contained complete item-level data across all observed indicators, confirmed by the single missing-data pattern reported in the lavaan output. I specified pairwise deletion in conjunction with the weighted least squares mean- and variance-adjusted (WLSMV) estimator, consistent with recommended practice for ordinal SEM estimation in lavaan. Because the retained sample contained complete item-level data, the pairwise-deletion specification did not affect parameter estimation; the model computed all polychoric correlations for the WLSMV weight matrix using the full 126-case sample. Preliminary data cleaning involved inspecting variables to verify coding accuracy and identify anomalous response patterns. Maintaining consistent directional interpretation precluded the need for recoding, as all items followed a uniform positive scale. Preliminary diagnostics verified that the data characteristics met the requirements for the WLSMV estimator. This process evaluated category frequencies to detect potential empty or near-empty response cells that might destabilize the WLSMV estimator. The dataset met all requirements for estimating the specified measurement and structural models, with no problematic characteristics emerging during screening.

**4.3.2    Sample Characteristics and Participant Profile**



The final analytic sample consisted of 126 respondents whose survey responses met all the inclusion criteria described in Section 4.3.1. The final analysis includes only participants who consented to the study, finished the entire survey, and confirmed a geographic location in North America or Europe. This final sample provides the basis for all descriptive statistics reported in this Section.

This Section outlines participant characteristics to contextualize the survey sample and describes respondents' professional backgrounds. These data are reported descriptively and are not interpreted or linked to hypotheses in this Section.

### 4.3.2.1 Geographic Distribution

Respondents were primarily based in North America (73%) and Europe (27%), reflecting the study's intended geographic scope. Figure 5 illustrates the distribution of respondents across these regions.

**Figure 5**

*Geographic Distribution of Survey Respondents*

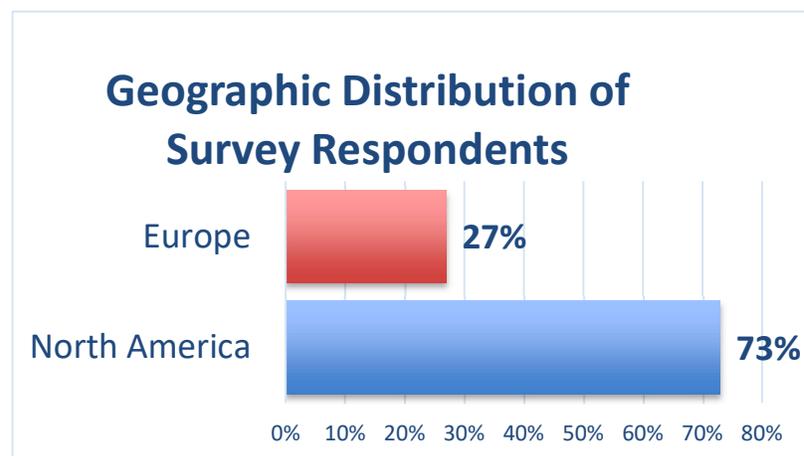



#### 4.3.2.2 Industry Sector Representation

Participants reported employment across a range of industry sectors, including technology, finance, healthcare, government, and other sectors with cybersecurity-relevant functions. Figure 6 illustrates the distribution of respondents by industry sector. This diversity reflects the cross-sector nature of cybersecurity practice.

**Figure 6**

*Survey Industry Sector Representation*

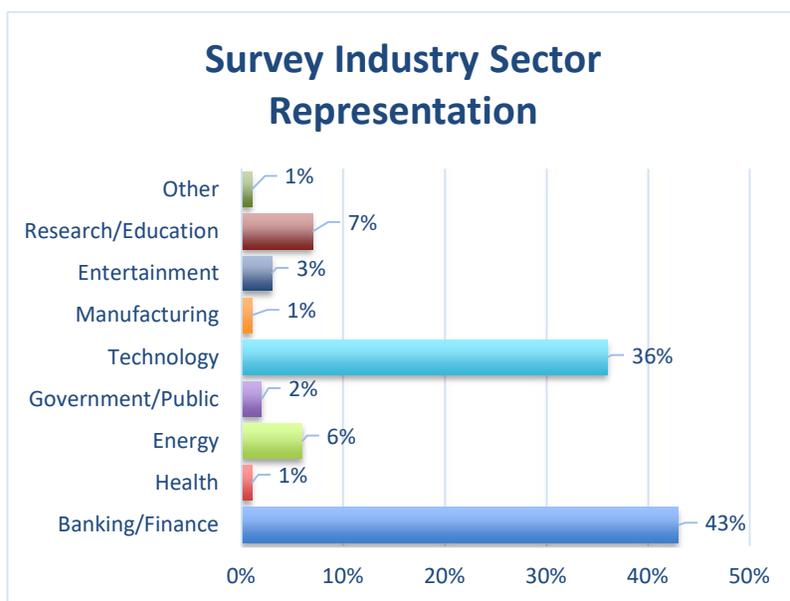

#### 4.3.2.3 Size of Organization

Respondents reported employment across organizations of varying sizes. The majority of participants (68%) reported working in organizations with more than 1,000 employees. An additional 25% reported working in organizations with 100–499

**Figure 7**

*Size of Organization*

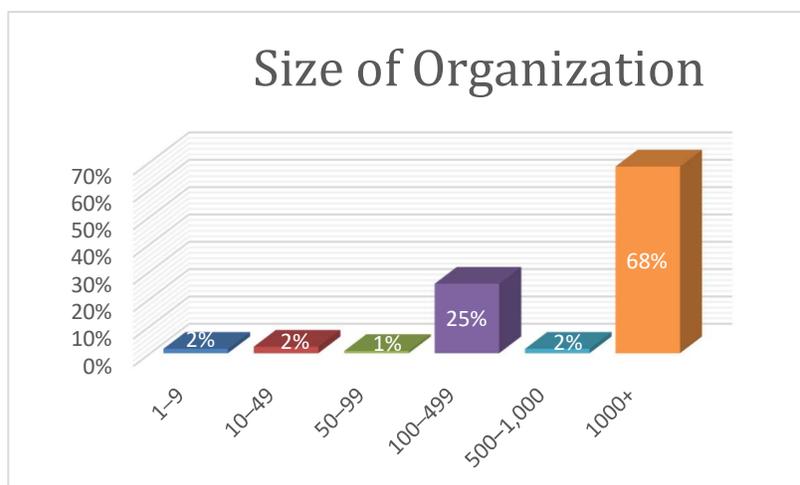

employees. Smaller proportions of respondents were employed in organizations with 500–1,000 employees (2%), 50–99 employees (1%), 10–49 employees (2%), and 1–9 employees (2%). These data



indicate that the sample is primarily composed of professionals working within large organizational environments, with additional representation from small and medium-sized enterprises.

**4.3.2.4  Primary Professional Role**

Respondents reported a diverse range of primary professional roles within cybersecurity and related functions. Engineering and Architecture roles comprised the largest category (25%), followed by Security Operations and Monitoring (21%). Leadership and Management roles represented 17% of the sample, while Governance, Risk, and Compliance (GRC) roles accounted for 16%. Smaller proportions of respondents identified as holding Generic Cybersecurity roles (10%), Offensive and Assurance roles (3%), Development and Integration roles (2%), Advisory and Support roles (2%), or Other roles (2%). This distribution reflects participation across technical, operational, governance, and leadership domains within cybersecurity practice.

**Figure 8**
*Primary Professional Role*

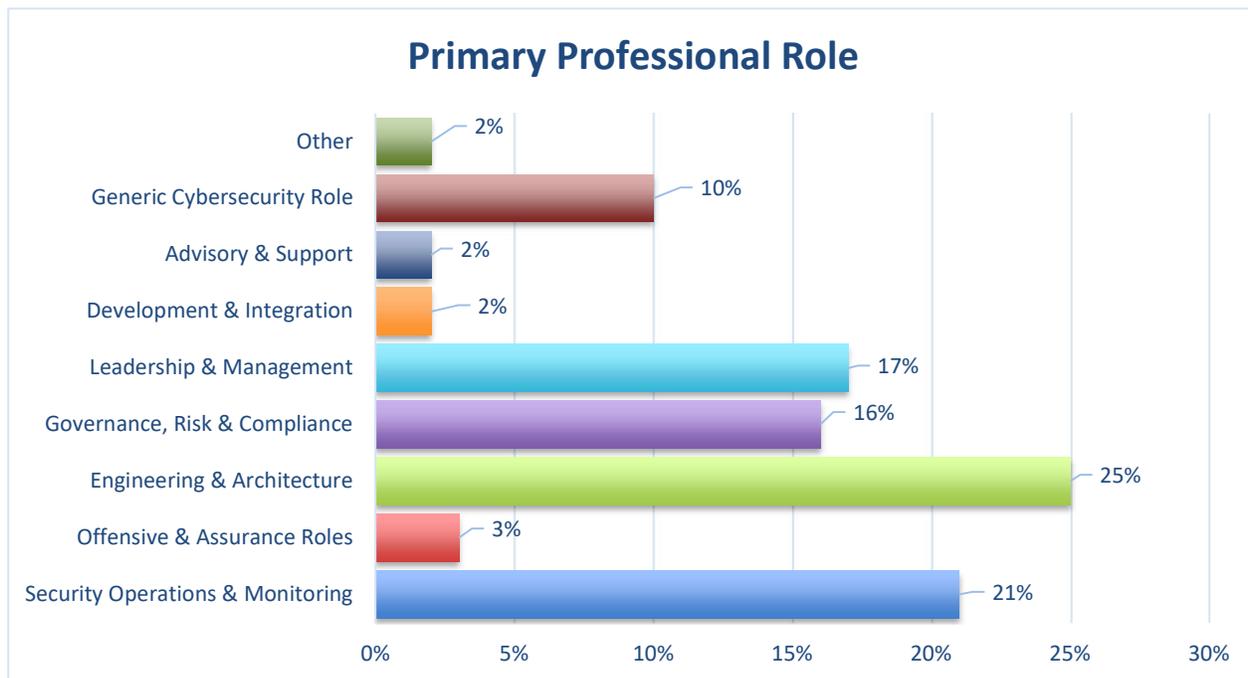



**4.3.2.5  Highest Education Level**

Respondents reported a range of formal educational attainment levels. The largest proportion of participants held a master's degree (41%), followed by those with a bachelor's degree (37%). A smaller proportion reported a college diploma (12%) or high school as their highest level of completed education (5%). Respondents with doctoral-level education accounted for 3% of the total sample. These data indicate that the sample includes participants with varied educational backgrounds at both the undergraduate and graduate levels.

**Figure 9**
*Highest Level of Education*

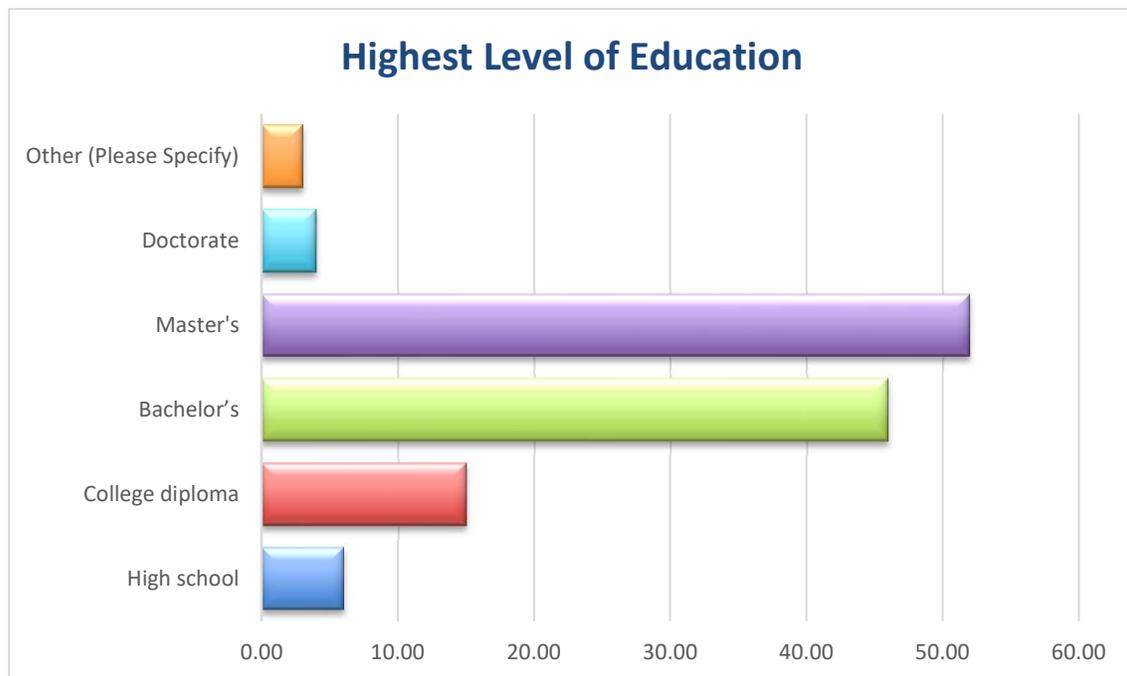

**4.3.2.6  Leadership Responsibility (Seniority)**

The survey asked participants whether their current role included formal leadership or managerial responsibilities. Responses indicate that the sample included both individuals in leadership



positions and those in non-leadership roles. This distinction is relevant for later analyses of leadership perspectives. **Error! Reference source not found.** shows the distribution of respondents by leadership responsibility.

Respondents reported a range of organizational seniority levels. Individual contributors represented the largest group (38%), followed by team leads (21%), directors (16%), managers (13%), and executives (12%). This distribution reflects participation from both non-managerial and leadership roles within cybersecurity functions.

**Figure 10**
*Leadership Responsibility (Seniority)*

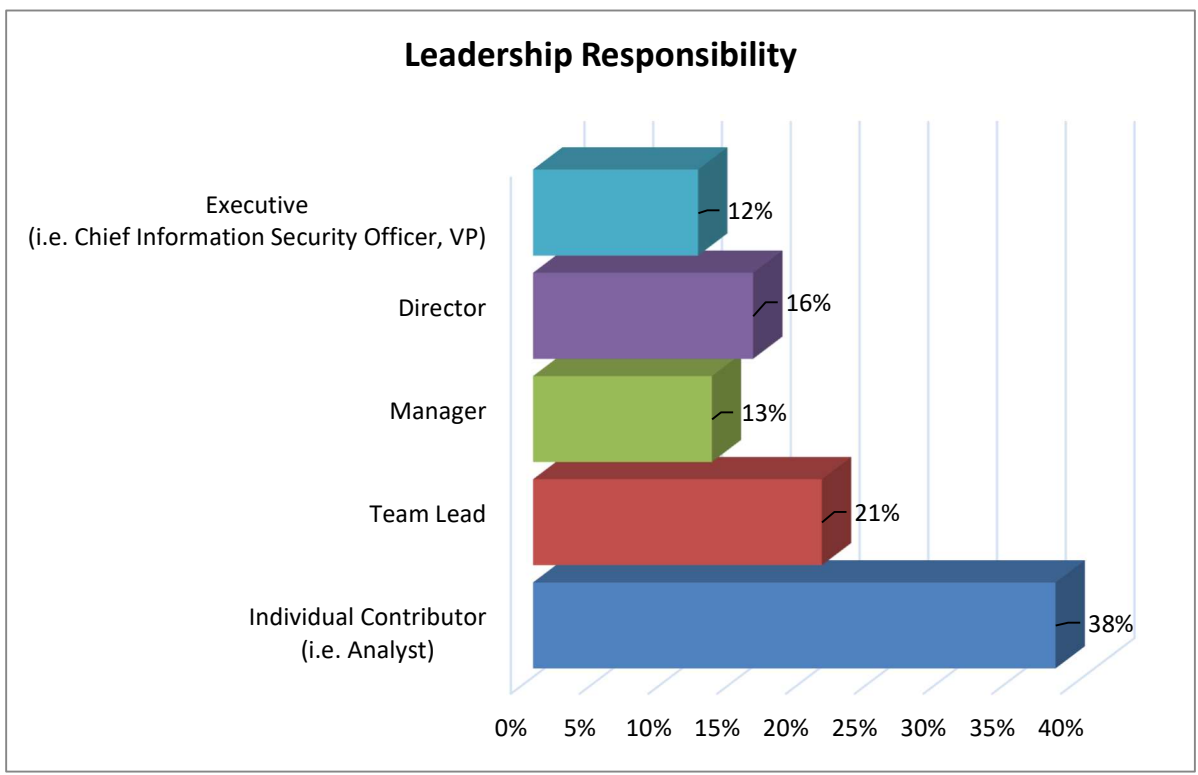

**4.3.2.7 Years in Cybersecurity**

Participants reported varying levels of professional experience in cybersecurity-related roles. The largest segment of the sample reported between 5–9 years of experience (29%), followed by 10–14



years (21%) and 2–4 years (21%). A smaller proportion reported 20 or more years of experience (14%) and 15–19 years (10%). Five respondents (4%) reported 0–1 year of experience. These distributions

**Figure 11**
*Years of Experience*

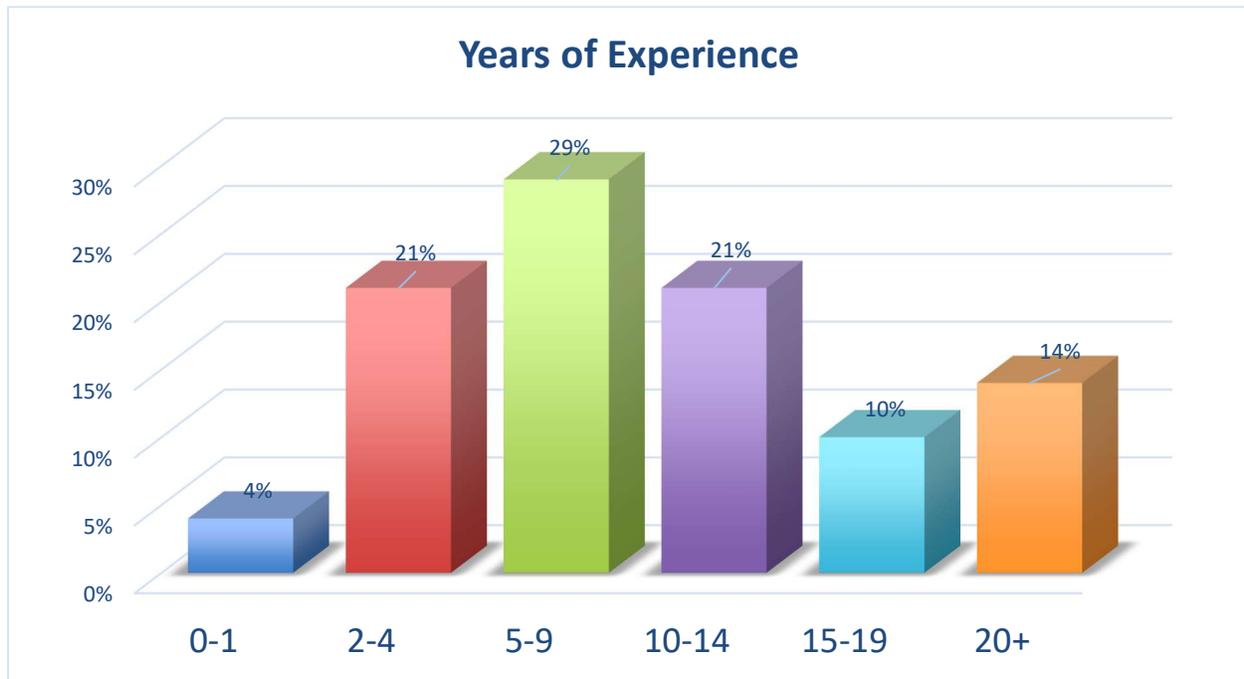

indicate representation across early-, mid-, and senior-career professionals.

#### 4.3.2.8 Exposure to risk management Learning

The survey prompted participants to identify sources of their risk management knowledge. On-the-job training (OJT) emerged as the most common source, with 43% of respondents selecting it. While 17% of participants identified certifications and 14% reported self-directed learning, formal coursework and standards-based frameworks (ISO/NIST) attracted the lowest engagement, at 8% each. Participants who indicated that none of the listed sources contributed to their understanding of risk management was 10%.



These responses indicate variation in how participants reported acquiring risk management knowledge, with experiential workplace exposure representing the most commonly identified source.

**Figure 12**
*Exposure to risk management Learning*

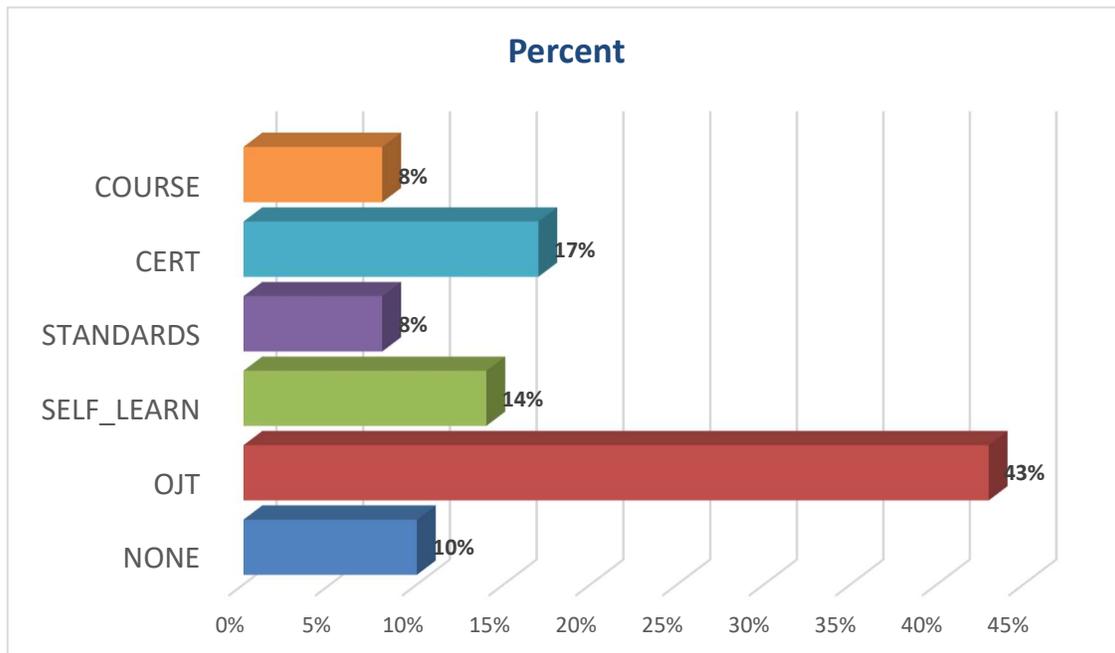

**4.3.2.9  Summary of Participant Characteristics**

Taken together, these descriptive statistics indicate that the survey sample comprised professionals with varied backgrounds, levels of experience, and organizational roles, drawn from multiple regions and industry sectors. Table 11 summarizes respondent demographics and professional characteristics.

Table 11 presents the demographic and professional characteristics of the final analytic sample (N = 126). Respondents were predominantly based in North America (73%) and represented a range of industry sectors, with the largest concentrations in banking/finance (43%) and technology (36%). Most participants worked in large organizations with more than 1,000 employees (68%). Professional roles



were diverse, with engineering and architecture (25%) and security operations (21%) most frequently

reported. Respondents were generally experienced professionals, with 71% reporting 5 or more years of

cybersecurity experience. Seniority levels ranged from individual contributors to executive leadership

roles.

**Table 11**

*Demographic & Professional Profile of Survey Respondents (n=126)*

| Characteristic | Category | n | % |
|---|---|---|---|
| **Geographic Distribution** | North America | 92 | 73 |
| | Europe | 34 | 27 |
| **Industry Sector Representation** | Banking/Finance | 54 | 43 |
| | Technology | 45 | 36 |
| | Research/Education | 9 | 7 |
| | Energy | 8 | 6 |
| | Entertainment | 4 | 3 |
| | Government/Public | 3 | 2 |
| | Health | 1 | 1 |
| | Manufacturing | 1 | 1 |
| | Other | 1 | 1 |
| **Organization Size (Employees)** | 1000+ | 86 | 68 |
| | 100–499 | 32 | 25 |
| | 10–49 | 3 | 2 |
| | 500–1,000 | 2 | 2 |
| | 1–9 | 2 | 2 |
| | 50–99 | 1 | 1 |
| **Primary Professional Role** | Engineering & Architecture | 31 | 25 |
| | Security Operations & Monitoring | 27 | 21 |
| | Leadership & Management | 22 | 17 |
| | Governance, Risk & Compliance | 20 | 16 |
| | Generic Cybersecurity Role | 13 | 10 |
| | Offensive & Assurance Roles | 4 | 3 |
| | Development & Integration | 3 | 2 |
| | Advisory & Support | 3 | 2 |
| | Other | 3 | 2 |
| **Highest Education Level** | Master's | 52 | 41 |
| | Bachelor's | 46 | 37 |
| | College diploma | 15 | 12 |
| | High school | 6 | 5 |
| | Doctorate | 4 | 3 |
| | Other | 3 | 2 |
| **Years in Cybersecurity** | 5–9 years | 37 | 29 |
| | 10–14 years | 27 | 21 |
| | 2–4 years | 26 | 21 |
| | 20+ years | 18 | 14 |
| | 15–19 years | 13 | 10 |
| | 0–1 year | 5 | 4 |
| **Leadership (Seniority Level)** | Individual Contributor | 48 | 38 |
| | Team Lead | 27 | 21 |
| | Director | 20 | 16 |
| | Manager | 16 | 13 |
| | Executive | 15 | 12 |

Providing these characteristics establishes the necessary context for the measurement and structural analyses that follow.

### 4.3.3   Descriptive Statistics for Observed Variables

The preliminary analysis calculated descriptive statistics for all observed indicators, spanning training exposure (IV1x), perceived relevance (MeV1), and the four competence dimensions (DV1–DV4). The preliminary analysis involved examining response distributions and calculating summary statistics for each item. Although the indicators were measured using Likert-type ordinal scales, subsequent structural analyses employed robust weighted least squares estimation (WLSMV), which models polychoric correlations and does not assume interval-level measurement.

The screening process evaluated these descriptive statistics for potential issues, including extreme skewness, restricted variance, and floor or ceiling effects.

Across all observed indicators, mean responses ranged from 3.06 to 4.17 on the five-point Likert scale. Training exposure items (IV1x) had means of 3.70-3.91, while conceptual salience items (MeV1) ranged from 3.06 to 3.92. Risk competence indicators (DV1–DV4) generally showed higher means, with



most items ranging from 3.66 to 4.17. These values indicate that respondents tended to rate both their exposure to risk-related training and their applied risk competence above the midpoint of the scale.

Skewness values were predominantly negative across competence indicators, indicating a modest clustering of responses toward the higher end of the scale. Skewness values ranged approximately from −0.37 to −1.27 across items, with no indicators exhibiting extreme skewness or kurtosis that would threaten model estimation.

Across indicators, mean responses were generally above the midpoint of the 5-point scale, and several risk competence items showed moderate negative skewness. These distributions suggest a tendency toward elevated self-assessment but did not violate assumptions for ordered categorical modelling under WLSMV estimation. Examination of skewness and kurtosis values did not reveal extreme deviations from normality. No items exhibited restricted variance or evidence of floor or ceiling effects. Accordingly, the study retained all indicators for evaluation of the measurement model. The study retained all items at this stage, as descriptive statistics alone did not warrant removal. The measurement model evaluation governed decisions on item retention and variance constraints; the following Section reports these specific adjustments.

### 4.3.4    Distributional Characteristics of Competence Indicators

Table 12 summarizes the descriptive statistics for the four competence dimensions (DV1–DV4). Observed item means ranged from 3.66 to 4.17 on the five-point Likert response scale, uniformly exceeding the scale midpoint of 3.0. This pattern indicates above-midpoint endorsement across all competence indicators.



**Table 12**

*Descriptive Statistics for SEM Observed Indicators (N = 126)*

| Construct | Item | Mean | SD | Min | Max | Skew | Kurtosis |
|---|---|---|---|---|---|---|---|
| **Training Exposure (IV1x)** | IV1x_1 | 3.7 | 0.92 | 1 | 5 | -0.85 | 0.51 |
| | IV1x_2 | 3.91 | 1.03 | 1 | 5 | -1.27 | 1.44 |
| | IV1x_3 | 3.76 | 1.02 | 1 | 5 | -0.87 | 0.41 |
| | IV1x_4 | 3.75 | 1.06 | 1 | 5 | -0.79 | -0.09 |
| **Conceptual Salience (MeV1)** | MeV1_1 | 3.11 | 0.77 | 1 | 4 | -0.71 | 0.3 |
| | MeV1_2 | 3.92 | 0.82 | 1 | 5 | -0.81 | 1.32 |
| | MeV1_3 | 3.06 | 0.79 | 1 | 4 | -0.69 | 0.29 |
| | MeV1_4 | 3.82 | 0.98 | 1 | 5 | -0.91 | 0.63 |
| | MeV1_5 | 3.65 | 0.84 | 1 | 5 | -0.68 | 0.59 |
| | MeV1_6 | 3.78 | 0.88 | 1 | 5 | -0.43 | -0.14 |
| **Risk-Based Assessment (DV1)** | DV1_1 | 3.95 | 0.96 | 1 | 5 | -0.83 | 0.3 |
| | DV1_2 | 4.04 | 0.85 | 1 | 5 | -1.26 | 2.54 |
| | DV1_3 | 3.86 | 0.83 | 1 | 5 | -1.02 | 1.98 |
| | DV1_4 | 3.94 | 1.15 | 1 | 5 | -1.05 | 0.31 |
| | DV1_5 | 3.67 | 1.04 | 1 | 5 | -0.82 | 0.42 |
| **Risk-Based Prioritization (DV2)** | DV2_1 | 3.74 | 1.13 | 1 | 5 | -0.77 | -0.23 |
| | DV2_2 | 3.66 | 0.93 | 1 | 5 | -1.03 | 1.05 |
| | DV2_3 | 3.73 | 0.93 | 1 | 5 | -0.89 | 0.58 |
| | DV2_4 | 3.21 | 1.02 | 1 | 5 | -0.37 | -0.25 |
| | DV2_5 | 3.95 | 0.95 | 1 | 5 | -0.8 | 0.53 |
| **Risk-Based Control Selection (DV3)** | DV3_1 | 3.74 | 0.98 | 1 | 5 | -0.99 | 0.77 |
| | DV3_2 | 3.97 | 1.08 | 1 | 5 | -1.17 | 0.85 |
| | DV3_3 | 3.88 | 0.94 | 1 | 5 | -0.4 | -0.55 |
| | DV3_4 | 3.74 | 1.01 | 1 | 5 | -0.84 | 0.44 |
| | DV3_5 | 3.91 | 1.07 | 1 | 5 | -0.65 | -0.55 |
| **Risk Communication & Rationale (DV4)** | DV4_1 | 4.03 | 0.87 | 1 | 5 | -1.01 | 0.97 |
| | DV4_2 | 4.17 | 0.86 | 1 | 5 | -1.01 | 0.9 |
| | DV4_3 | 3.75 | 1.01 | 1 | 5 | -0.58 | -0.17 |
| | DV4_4 | 3.97 | 0.89 | 1 | 5 | -0.55 | -0.13 |
| | DV4_5 | 3.91 | 1.06 | 1 | 5 | -0.64 | -0.34 |

**Note**. The study employed a 5-point Likert ordinal scale to measure indicators; Means and standard deviations are reported for descriptive purposes only. Structural analyses employed robust weighted least squares (WLSMV), which is appropriate for ordinal indicators.

Skewness values were negative across all items, ranging from −0.37 to −1.27, indicating clustering of responses toward the higher response categories. Table 12 reports the Kurtosis values.



None of the observed distributional properties precluded the use of the WLSMV estimator for ordered categorical modelling.

Despite elevated means and negative skewness, all indicators retained sufficient variance for further analysis. No item exhibited a response concentration consistent with a functional ceiling effect (defined as ≥85% of responses within the highest one or two categories). Accordingly, the competence indicators retained sufficient distributional spread for latent factor estimation.

Taken together, the observed distributional properties indicate generally positive self-assessments of risk-related behaviour, alongside measurable variation at the respondent level.

The next Section presents the results of the measurement model, including factor loadings, reliability estimates, and assessments of convergent and discriminant validity for the latent constructs used in the structural model.

### 4.3.5    Assessment of Common Method Bias

Because all survey constructs —training exposure to risk-embedded content (IV1x), perceived relevance of risk management (MeV1), and risk management competence (*RM_Competence*) — were collected from the same respondents using the same self-report instrument administered at a single time point, the potential for common method bias (CMB) was evaluated (Podsakoff et al., 2003). CMB refers to artifactual covariance that may arise when predictor and criterion variables share a common measurement source, thereby inflating observed associations and threatening the validity of structural inferences.

The instrument design incorporates several procedural safeguards to mitigate CMB susceptibility. The independent construct (IV1x) and the outcome constructs (*RM_Competence*, DV1–DV4) were separated within the survey by an intervening Section measuring the mediating construct



(MeV1) and a block of demographic and covariate items, thereby introducing psychological distance between antecedent and outcome measurement (Podsakoff et al., 2003). Item stems were varied across construct blocks to reduce acquiescence and consistency-motive responding. The survey protocol ensured respondent anonymity to minimize social desirability bias. Control variables were measured on distinct response formats (numeric entry and categorical selection) rather than on the Likert scale used for the substantive constructs, further differentiating the measurement context across construct types.

Harman's (1976) single-factor test provided the statistical assessment for common method bias. The analysis constrained all 30 SEM indicators to a single factor, using the WLSMV estimator and the theta parameterization to account for the ordinal nature of the data. The single-factor solution accounted for a mean of 50.7% of total item variance across indicators (mean $R^2 = 0.507$; individual item $R^2$ values ranged from 0.301 to 0.711). This value marginally exceeds the conventional 50% screening threshold (Podsakoff et al., 2003).

This marginal exceedance warrants transparent acknowledgement and careful interpretation, though it does not, in isolation, constitute evidence of method-driven inflation of the structural paths. Three considerations inform this assessment. First, in the theta parameterization used throughout this study, each indicator's unique variance is constrained to 1.0, meaning that a single-factor solution in this parameterization will tend to concentrate explained variance in the common factor more readily than in an unconstrained solution; the resulting mean $R^2$ is therefore not directly comparable to the conventional threshold derived from unstandardized exploratory factor analysis. Second, the marginal exceedance is analytically consistent with the holistic structure of *RM_Competence* documented in Section 4.4.5, in which all four theoretically specified behavioural dimensions loaded at near-unity on a single second-order factor and failed to demonstrate empirical separation. A strong general factor across the 30 items reflects more than a mere method artifact; it represents the statistical signature of a



sample that fails to cognitively differentiate risk management competence into distinct sub-dimensions. Third, while Harman's test functions as a common screening procedure, its conservative nature limits its utility; it fails to distinguish method-driven covariance from the theoretically coherent interrelatedness of the items (Richardson et al., 2009; Williams et al., 2010). Its presence here is therefore a transparency measure rather than a conclusive diagnostic measure.

Two additional features of the study's design further limit the plausibility of a CMB-driven explanation for the observed structural associations. First, the three primary constructs are conceptually and operationally distinct across temporal and referential frames: IV1x measures retrospective exposure to training content, MeV1 measures present-state appraisal of risk management relevance, and *RM_Competence* measures self-reported behavioural enactment in professional practice. These referential differences attenuate the likelihood of a unified method distortion accounting for the observed inter-construct covariance. Second, convergent evidence across three independent data sources — NLP content analysis of training framework artifacts (Section 4.2), SEM structural path estimates (Section 4.5), and qualitative leadership interview findings (Section 4.8) — yields consistent, theoretically coherent patterns. A CMB account, which operates exclusively at the level of self-report covariance, cannot explain the convergence of findings across sources that include both objective text analysis and independently coded leadership discourse. Accordingly, while a single-method design precludes the absolute exclusion of CMB, the procedural, statistical, and cross-method evidence collectively indicate that method artifact is unlikely to account for the primary structural findings reported in this chapter.

**4.4    Measurement Model Results (Survey Data)**



Enterprise risk theory provides the foundation for specifying risk management competence as four differentiated behavioral dimensions: (1) risk-based assessment (using the likelihood x impact calculation), (2) risk-calculation-based prioritization, (3) control selection justified through risk reduction logic (reduction of impact, or likelihood or both), and (4) risk communication and decision rationale (i.e., the ability to explain and defend risk assessments based on explicit evaluation of likelihood and impact, and to explain how those assessments inform subsequent decisions). This normative structure reflects established definitions of risk as likelihood × impact, as well as the associated decision processes that guide mitigation and governance. The measurement analysis tests the alignment between the theoretical structure and respondents' self-assessments.

This Section reports the results of the measurement model evaluation conducted before testing the structural relationships. The purpose of the analysis was to assess the reliability and validity of the latent constructs used in the Structural Equation Model. Evaluation criteria included standardized factor loadings, internal consistency reliability, convergent validity, discriminant validity, and overall model fit.

### 4.4.1 Confirmatory Factor Analysis Results

I conducted a Confirmatory Factor Analysis (CFA) to estimate the measurement model for the observed survey indicators. All latent constructs were estimated using ordinal indicators and the WLSMV estimator. The fixed-factor variance approach (std.lv = TRUE) scales the model's latent variables.

Standardized factor loadings were statistically significant for all observed indicators (p < .05). Loading magnitudes ranged from 0.770 to 0.815 for IV1x, from 0.719 to 0.896 for MeV1, and from 0.518 to 0.827 for *RM_Competence*.



**Table 13**

*IV1x: Standardized Factor Loadings (SEM Measurement Model)*

| Indicator | Loading (b) | SE | z | p | Std. loading (β) |
|-----------|-------------|-------|-------|-------|------------------|
| IV1x_1 | 1.253 | 0.259 | 4.834 | <.001 | 0.782 |
| IV1x_2 | 1.289 | 0.221 | 5.825 | <.001 | 0.79 |
| IV1x_3 | 1.224 | 0.213 | 5.75 | <.001 | 0.774 |
| IV1x_4 | 1.456 | 0.327 | 4.448 | <.001 | 0.824 |

**Table 14**

*MeV1: Standardized Factor Loadings (SEM Measurement Model)*

| Indicator | Loading (b) | SE | z | p | Std. loading (β) |
|-----------|-------------|-------|-------|-------|------------------|
| MeV1_1 | 1.177 | 0.222 | 5.304 | <.001 | 0.88 |
| MeV1_2 | 0.904 | 0.134 | 6.733 | <.001 | 0.819 |
| MeV1_3 | 1.328 | 0.232 | 5.726 | <.001 | 0.902 |
| MeV1_4 | 0.987 | 0.141 | 7.011 | <.001 | 0.841 |
| MeV1_5 | 1.229 | 0.199 | 6.177 | <.001 | 0.889 |
| MeV1_6 | 0.662 | 0.09 | 7.316 | <.001 | 0.722 |

**Table 15**

*RM_Competence: Standardized Factor Loadings (SEM)*

| Indicator | Loading (b) | SE | z | p | Std. loading (β) |
|-----------|-------------|-------|-------|-------|------------------|
| DV1_1 | 0.575 | 0.084 | 6.828 | <.001 | 0.667 |
| DV1_2 | 0.817 | 0.113 | 7.219 | <.001 | 0.786 |
| DV1_3 | 0.794 | 0.115 | 6.903 | <.001 | 0.777 |
| DV1_4 | 0.933 | 0.131 | 7.134 | <.001 | 0.824 |
| DV1_5 | 0.529 | 0.08 | 6.617 | <.001 | 0.636 |
| DV2_1 | 0.401 | 0.064 | 6.308 | <.001 | 0.529 |
| DV2_2 | 0.704 | 0.105 | 6.684 | <.001 | 0.739 |
| DV2_3 | 0.729 | 0.093 | 7.818 | <.001 | 0.75 |
| DV2_4 | 0.477 | 0.066 | 7.251 | <.001 | 0.596 |
| DV2_5 | 0.547 | 0.076 | 7.178 | <.001 | 0.648 |
| DV3_1 | 0.893 | 0.123 | 7.266 | <.001 | 0.812 |
| DV3_2 | 0.816 | 0.115 | 7.102 | <.001 | 0.786 |
| DV3_3 | 0.524 | 0.074 | 7.066 | <.001 | 0.632 |
| DV3_4 | 0.716 | 0.106 | 6.767 | <.001 | 0.744 |
| DV3_5 | 0.534 | 0.077 | 6.929 | <.001 | 0.639 |
| DV4_1 | 0.939 | 0.143 | 6.586 | <.001 | 0.825 |
| DV4_2 | 0.584 | 0.089 | 6.589 | <.001 | 0.672 |
| DV4_3 | 0.951 | 0.129 | 7.4 | <.001 | 0.829 |
| DV4_4 | 0.654 | 0.106 | 6.181 | <.001 | 0.713 |
| DV4_5 | 0.554 | 0.077 | 7.169 | <.001 | 0.653 |

Global model fit indices for the measurement model are reported alongside those for the structural model to ensure consistency in model evaluation (see Section 4.5). The survey design and theoretical construct definitions guided the a priori specification of all indicators.

### 4.4.2    Composite Reliability (CR)

**Composite reliability (CR) estimates, derived from the standardized factor loadings in Table 13 and**

Table 15 verifies the internal consistency of each latent construct. The analysis yielded composite reliability values for training exposure (IV1x), perceived relevance (MeV1), and overall risk management competence (*RM_Competence*). See Table 16.

Table 16
*Composite Reliability (CR) Values by Construct*

| Construct | Indicators (k) | Std. loading range | Composite Reliability (CR) |
|---|---|---|---|
| IV1x | 4 | 0.774–0.824 | 0.934 |
| MeV1 | 6 | 0.722–0.902 | 0.935 |
| *RM_Competence* | 20 | 0.529–0.829 | 0.972 |

I retained all indicators in the measurement model, as none failed to meet the established reliability or internal consistency criteria.

### 4.4.3    Convergent Validity (AVE)



I evaluated convergent validity by calculating the average variance extracted (AVE), which measures the proportion of variance captured by a construct relative to measurement error. See Table 17.

**Table 17**
*AVE Values by Construct*

| Construct | Indicators (k) | Std. loading range | AVE |
|-----------|----------------|--------------------|-----|
| IV1x | 4 | 0.774–0.824 | 0.613 |
| MeV1 | 6 | 0.722–0.902 | 0.708 |
| RM_Competence | 20 | 0.529–0.829 | 0.492 |

The measurement analysis evaluated the convergent validity of all latent constructs retained in the structural model.

### 4.4.4   Second-Order Risk Management Competence Construct

Risk management competence was modelled as a second-order latent construct, measured by four first-order latent dimensions: risk-based assessment behaviour (DV1), risk-based prioritization behaviour (DV2), control selection justified by risk reduction (DV3), and risk communication and decision rationale (DV4).

Multiple observed indicators measure each first-order dimension, while a second-order factor captures the shared variance among the four behavioural dimensions. See Table 18.

**Table 18**
*Second-Order Standardized Loadings for RM_Competence (WLSMV)*

| First-order factor | Standardized loading (β) |
|--------------------|--------------------------|
| DV1 (Risk-based assessment behaviour) | 1 |
| DV2 (Risk-based prioritization behaviour) | 0.978 |
| DV3 (Control selection justified by risk reduction) | 1 |
| DV4 (Risk communication and decision rationale) | 0.935 |



The second-order specification was estimated to evaluate dimensional structure; the structural model retained the higher-order latent factor *RM_Competence*. The second-order specification (Model 2) was estimated to evaluate the dimensional structure of the risk management competence construct and to assess the distinctiveness of the four behavioural domains relative to a single-factor representation.

### 4.4.5    Evaluation of Second-Order Specification (Model 2)

Model 2 specified risk management Competence as a second-order construct reflected by four first-order latent dimensions: risk assessment (DV1), prioritization (DV2), control selection (DV3), and risk communication (DV4). This specification tested whether the four theoretically distinct behavioural domains demonstrated empirical separation while loading on a common higher-order factor.

The standardized second-order loadings were uniformly high (DV1 = 1.000, DV2 = 0.978, DV3 = 1.000, DV4 = 0.935). Residual variances for the first-order factors were correspondingly small (DV1 = 0.000, DV2 = 0.044, DV3 = 0.001, DV4 = 0.125), indicating minimal unique variance at the first-order level after accounting for the higher-order factor.

The implied correlations among the 4 first-order dimensions ranged from 0.914 to 1.000 (see Table 19):

**Table 19**
*Second-Order Loadings & Inter-Factor Correlations (Model 2)*

Section 1: Second-Order standardized loadings:

| First-order factor | Std. loading on *RM_Competence* |
|---|---|
| DV1 | 1.000 |
| DV2 | 0.978 |
| DV3 | 1.000 |
| DV4 | 0.935 |



Section 2: implied correlations (can be a small matrix or a 6-row list):

| Pair | Implied correlation |
|---|---|
| DV1–DV2 | 0.978 |
| DV1–DV3 | 1.000 |
| DV1–DV4 | 0.935 |
| DV2–DV3 | 0.978 |
| DV2–DV4 | 0.914 |
| DV3–DV4 | 0.935 |

The near-unity of the implied correlations and the minimal residual variance indicate that the first-order factors were not empirically distinguishable within this sample. Several correlations approached or reached unity. Two second-order loadings were fixed at 1.000 for scaling purposes because the associated first-order factors had near-zero residual variances. This pattern indicates that respondents evaluated the four behavioural domains as closely interconnected aspects of a single applied decision process rather than as distinct analytical stages.

The second-order model also produced an estimation warning indicating a non-positive-definite variance–covariance matrix, consistent with the near-unity inter-factor correlations. Although global fit indices for Model 2 were comparable to those of the single-factor model, the combination of:

- Near-perfect inter-factor correlations,

- Minimal residual variance at the first-order level, and

- Variance–covariance instability

indicated that the four behavioural dimensions were not empirically distinguishable in this sample.

The lack of empirical separation among first-order factors in Model 2 precluded its retention as a higher-order specification for structural analysis. Model 1 yielded stable parameter estimates without identification warnings; consequently, this study adopts the single-factor representation as the measurement model for structural testing.



### 4.4.6   Discriminant Validity (Model 1)

The Fornell–Larcker criterion (1981) was applied to assess discriminant validity for the retained structural model (Model 1). This criterion requires that the square root of the average variance extracted (VAVE) for each latent construct exceed its correlations with all other constructs in the model.

Table 20 presents the completed discriminant validity matrix for the three constructs retained in the structural model: IV1x (Training Exposure), MeV1 (Conceptual Salience), and *RM_Competence* (Risk Management Competence). Diagonal elements report VAVE for each construct; off-diagonal elements report model-implied latent correlations extracted from the standardized WLSMV solution.

**Table 20**

*Fornell–Larcker Discriminant Validity Matrix (VAVE on diagonal)*

|  | IV1x | MeV1 | *RM_Competence* |
|---|---|---|---|
| IV1x | **0.783** | | |
| MeV1 | 0.453 | **0.841** | |
| *RM_Competence* | 0.629 | 0.675 | **0.701** |

**Notes**:
1. The Diagonal elements (bold) are the VAVE for each construct (IV1x: $\sqrt{0.613} = 0.783$; MeV1: $\sqrt{0.708} = 0.841$; *RM_Competence*: $\sqrt{0.492} = 0.701$).
2. Off-diagonal elements model-implied latent correlations from the standardized WLSMV solution (theta parameterization).

The Fornell–Larcker criterion is satisfied across all construct pairs. For IV1x, VAVE = 0.783 exceeds r(IV1x, MeV1) = 0.453 and r(IV1x, *RM_Competence*) = 0.629 by margins of 0.330 and 0.154, respectively. For MeV1, VAVE = 0.841 exceeds r(MeV1, IV1x) = 0.453 and r(MeV1, *RM_Competence*) = 0.675 by margins of 0.388 and 0.166, respectively. For *RM_Competence*, VAVE = 0.701 exceeds r(*RM_Competence*, IV1x) = 0.629 and r(*RM_Competence*, MeV1) = 0.675, though by narrower margins of 0.072 and 0.026, respectively.

The narrower margins for *RM_Competence* are consistent with the AVE value of 0.492 reported in Section 4.4.3, which approaches but does not exceed the conventional 0.50 threshold. However,



composite reliability for *RM_Competence* (CR = 0.972) substantially exceeds the 0.70 benchmark, indicating strong internal consistency. The observed correlations reflect theoretically expected relationships among training exposure, conceptual salience, and competence; at no point does a latent correlation equal or exceed the VAVE of the associated construct.

Accordingly, the results support discriminant validity for the three-construct structural model retained for hypothesis testing.

### 4.4.7    *Measurement Model Summary*

The measurement model was successfully estimated using ordinal indicators for training exposure (IV1x_1–IV1x_4), perceived relevance of risk management (MeV1_1–MeV1_6), and the 4 first-order risk management competence dimensions (DV1_1–DV1_5, DV2_1–DV2_5, DV3_1–DV3_5, DV4_1–DV4_5). I conducted the estimation using the WLSMV estimator in lavaan and scaled the latent variables via the fixed-factor variance (std.lv = TRUE) approach. The measurement model results supported the inclusion of all latent constructs in the structural model without further modification.

Taken together, the measurement model results establish a key empirical finding with direct implications for the structural analysis and for Chapter 5's theoretical interpretation: the four theoretically specified dimensions of risk management competence — risk-based assessment (DV1), risk-based prioritization (DV2), control selection justified by risk reduction (DV3), and risk communication and decision rationale (DV4) — did not demonstrate empirical separation in this sample. The evidence for this conclusion converges across four independent indicators: (1) near-unity implied correlations among the first-order factors under the second-order specification, ranging from 0.914 to 1.000 (Section 4.4.5); (2) near-zero residual variances at the first-order level (DV1 = 0.000, DV3 = 0.001), indicating that the higher-order factor accounted for virtually all variance in those dimensions; (3)



identification instability in the second-order model, evidenced by a non-positive-definite variance–covariance matrix (smallest eigenvalue = −0.016), which rendered Model 2 inferentially unusable (Section 4.4.5); and (4) narrow Fornell–Larcker discriminant validity margins for *RM_Competence* relative to its predictors (Section 4.4.6), consistent with a construct that responds to training exposure and salience as an integrated whole rather than as four separable behavioural capacities. The single-factor specification (Model 1) is therefore empirically justified: it is not a simplifying assumption imposed on the data, but the representation that the data support. Chapter 5 examines the implications of this empirical result, exploring its effects on practitioner cognition.

## 4.5   Structural Equation Model Results

This Section reports results from the Structural Equation Model (SEM) estimated to examine relationships among training exposure to risk-embedded content (IV1x), perceived relevance of risk management (MeV1), and risk management competence (*RM_Competence*). The lavaan package estimated the model using the WLSMV estimator and theta parameterization. At the same time, pairwise methods addressed the missing data. Reported results include estimated structural paths, mediation effects, and global model fit indices.

### *4.5.1   Structural Model Specification*

The structural model specified training exposure to risk-embedded content (IV1x) as an exogenous latent variable predicting perceived relevance of risk management (MeV1) and risk management competence (*RM_Competence*). The structural model posits that perceived relevance (MeV1) mediates the relationship between IV1x and *RM_Competence*. For structural testing, *RM_Competence* was modelled as a single latent construct, directly measured by the retained set of



observed indicators, consistent with the Model 1 specification adopted following evaluation of the
second-order alternative, as described in Section 4.4.

Observed control variables—including years of cybersecurity experience (CV1_YEARS), seniority
level (CV1_SENIORITY), and indicators of prior risk management learning (CV2_RM_LRN1,
CV2_RM_LRN2, and CV2_RM_LRN_CHK1) —were specified as predictors of perceived relevance of risk
management. The theoretical rationale for restricting covariates to the MeV1 equation is that
background experience and prior learning operate on risk management competence indirectly —
through their influence on how professionals perceive and appraise training relevance — rather than by
directly generating competence outcomes independent of that appraisal process. Section 4.5.5 reports a
sensitivity analysis that evaluates the robustness of the primary structural paths under an alternative
specification in which covariates also enter the *RM_Competence* equation directly. Covariances among
selected control variables were specified where consistent with the model design. The lavaan package
estimated the structural model using the WLSMV estimator, while the fixed-factor variance (std.lv =
TRUE) approach provided the necessary scaling for the latent variables.

### 4.5.2   Direct Effects

The structural model produced standardized path coefficients (std.all) to evaluate the
hypothesized direct effects (Table 21).

**Table 21**
*Structural path estimates (Model 1; n = 126)*

| Path | b | SE | z | p | Std. β |
|---|---|---|---|---|---|
| IV1x → MeV1 | 0.715 | 0.098 | 7.29 | <.001 | 0.453 |
| MeV1 → *RM_Competence* | 0.484 | 0.11 | 4.422 | <.001 | 0.491 |
| IV1x → *RM_Competence* (c') | 0.632 | 0.129 | 4.888 | <.001 | 0.406 |



The structural model evaluated the direct effect of training exposure (IV1x) on perceived relevance (MeV1).

- **IV1x → MeV1**: unstandardized b = 0.715, SE = 0.098, z = 7.290, p < .001, 95% CI [0.523, 0.907], standardized β = 0.453.

Perceived relevance of risk management (MeV1) significantly predicted risk management competence (*RM_Competence*).

- **MeV1 → *RM_Competence***: unstandardized b = 0.484, SE = 0.110, z = 4.422, p < .001, 95% CI [0.270, 0.699], standardized β = 0.491.

The structural model also evaluated the direct path from training exposure (IV1x) to risk management competence (*RM_Competence*).

**Figure 13**
*SEM Path Diagram*

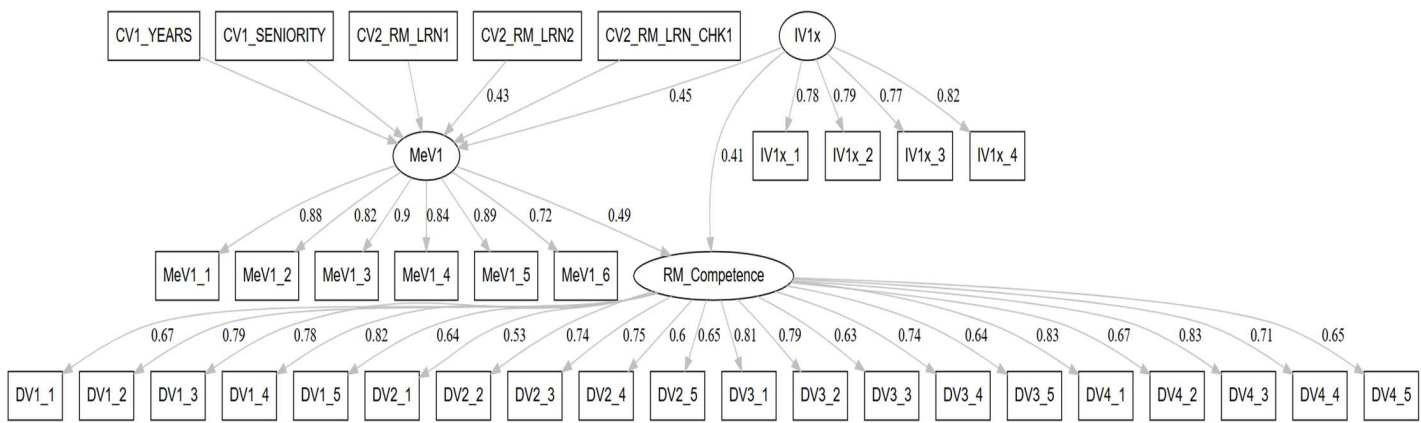

**Note**: *This represents the final estimated model, not the a priori specification*

- **IV1x → *RM_Competence* (direct effect; c')**: unstandardized b = 0.632, SE = 0.129, z = 4.888, p < .001, 95% CI [0.379, 0.885], standardized β = 0.406.

### 4.5.3   Mediation Analysis



To test for mediation, I defined a new parameter, indirect (a*b), to capture the indirect effect of IV1x on *RM_Competence* via MeV1. The following Sections detail the direct, indirect, and total effects:

- **Indirect effect (a × b)**: unstandardized = 0.346, SE = 0.074, z = 4.670, p < .001, 95% CI [0.201, 0.492], standardized β = 0.223.

- **Direct effect (c')**: unstandardized = 0.632, SE = 0.129, z = 4.888, p < .001, 95% CI [0.379, 0.885], standardized β = 0.406.

- **Total effect (a × b + c')**: unstandardized = 0.978, SE = 0.137, z = 7.136, p < .001, 95% CI [0.710, 1.247], standardized β = 0.629

**Table 22**
*Direct, indirect, and total effects (Model 1; n = 126)*

| Effect | Estimate | SE | z | p | Std. β |
|---|---|---|---|---|---|
| Indirect (a×b) | 0.346 | 0.074 | 4.670 | <.001 | 0.223 |
| Direct (c') | 0.632 | 0.129 | 4.888 | <.001 | 0.406 |
| Total | 0.978 | 0.137 | 7.136 | <.001 | 0.629 |

Both the direct and indirect effects were statistically significant. The standardized direct effect (β = 0.406) exceeded the indirect effect through conceptual salience (β = 0.223), indicating complementary (partial) mediation rather than full mediation.

### 4.5.4   Explained Variance and Proportion Mediated

The structural model explained approximately 58.7% of the variance in risk management competence ($R^2$ = 0.587), indicating that training exposure to risk-embedded content and perceived relevance of risk management, estimated jointly as direct and indirect predictors, account for a substantial portion of the observed variation in *RM_Competence* across the sample. The proportion of *RM_Competence* variance not accounted for by the model reflects the contribution of unmeasured



antecedents — such as informal developmental exposure, organizational role demands, and individual cognitive factors — acknowledged as boundary conditions of this study and discussed in Chapter 5.

The structural model also explained 59.8% of the variance in perceived relevance of risk management ($R^2$(MeV1) = 0.598). This estimate reflects the combined contribution of training exposure (IV1x, β = 0.453) and the control variables, particularly prior formal risk management learning (CV2_RM_LRN2, β = 0.432).

**Table 23**

*Explained Variance ($R^2$) for Endogenous Latent Constructs (Model 1)*

| Endogenous Construct | Std. Residual Variance | R² |
|---|---|---|
| MeV1 (Perceived Relevance) | 0.402 | 0.598 |
| *RM_Competence* | 0.413 | 0.587 |

**Note.** $R^2$ = 1 − standardized residual variance (Std.all), derived from WLSMV theta-parameterized solution (lavaan 0.6-21). $R^2$(MeV1) reflects the prediction from IV1x and all five control variables. $R^2$(*RM_Competence*) reflects prediction from MeV1 and IV1x only, with covariates restricted to the MeV1 equation in the primary specification.

Of the total standardized effect of training exposure on risk management competence (β_total = 0.629), the indirect pathway through perceived relevance accounted for β_indirect = 0.223, representing a proportion mediated of approximately 35.5% (PM = 0.223/0.629). The remaining 64.5% of the total effect was transmitted directly, independent of the mediating mechanism. These figures are consistent with complementary partial mediation (Baron & Kenny, n.d.; Zhao et al., 2010): perceived relevance of risk management enhances the training–competence relationship but does not fully account for it. Chapter 5 provides the practical interpretation of this decomposition, translating the mathematical findings into professional implications.

### 4.5.5    Sensitivity Analysis: Covariate Placement on RM_Competence

A sensitivity model assessed the robustness of the primary structural paths by incorporating five covariates—years of experience (CV1_YEARS), seniority level (CV1_SENIORITY),  and three prior learning



variables (CV2_RM_LRN1, CV2_RM_LRN2, CV2_RM_LRN_CHK1) - into both the MeV1 and

*RM_Competence* equations simultaneously.

    The primary model (Model 1) restricts covariates to the MeV1 equation, reflecting the

theoretical premise that background experience and prior learning operate on *RM_Competence* through

the mediating role of perceived relevance. The sensitivity model relaxed this constraint to evaluate

whether the primary paths were stable across specifications.

    The direct effect of training exposure on *RM_Competence* (c') demonstrated stability across

both specifications, shifting from $\beta = 0.406$ in the primary model to $\beta = 0.434$ in the sensitivity model ($\Delta\beta$

$= 0.028$), well within conventional robustness thresholds. Both estimates remained statistically

significant ($p < .001$). This result confirms that the direct relationship between training exposure and risk

management competence is not an artifact of covariate placement.

    The mediating path from perceived relevance to *RM_Competence* (b) attenuated from $\beta = 0.491$

to $\beta = 0.295$ ($\Delta\beta = -0.196$) when covariates were permitted to explain variance directly in

*RM_Competence*. This attenuation is attributable to shared variance between the covariates —

particularly prior risk management learning (CV2_RM_LRN2, $\beta = 0.432$ on MeV1 in the primary model)

— and *RM_Competence*. The b path remained statistically significant in the sensitivity model ($p = .004$),

indicating that perceived relevance continues to have a meaningful predictive relationship with

*RM_Competence* after covariate adjustment. Consequently, the indirect effect attenuated from $\beta =$

$0.223$ to $\beta = 0.162$, while remaining statistically significant ($p = .001$).

    The direct effect of training exposure on competence withstands different covariate

specifications, though the mediated pathway varies across prior-learning-experience models. All

subsequent analyses utilize the primary model specification, which positions covariates as predictors of



the mediator's antecedent. This section presents the sensitivity analysis results, demonstrating methodological transparency and defining the upper and lower bounds of the mediation estimate.

### 4.5.6 Model Fit Indices

The Comparative Fit Index (CFI), Tucker–Lewis Index (TLI), RMSEA, and SRMR were used to evaluate model fit, as they accommodate WLSMV estimation for ordinal data. The chi-square test statistic was $\chi^2(547) = 709.114$; p-value NA (and scaled $\chi^2 = 745.984$, df = 547, p = 0.000). The WLSMV estimation approach omits a corresponding p-value for the robust chi-square statistic and instead relies on alternative fit indices. Approximate fit indices (CFI, TLI, RMSEA) served as the primary criteria for evaluating model fit (See Table 24). Approximate fit indices provide a more reliable assessment of model utility than the robust Chi-square p-value, which is highly sensitive to sample size in complex structural models. Robust Chi-square p-values often produce inflated Type I error rates in large samples and complex structural models.

Additional fit indices indicated the following values: Comparative Fit Index (CFI) = 0.990, Tucker–Lewis Index (TLI) = 0.992, Root Mean Square Error of Approximation (RMSEA) = 0.049; RMSEA 90% CI [0.038, 0.059] (scaled CI [0.044, 0.063]), and Standardized Root Mean Square Residual (SRMR) = 0.079.

**Table 24**
*Model fit indices (Model 1; n = 126)*

| $\chi^2$ | df | Scaled $\chi^2$ | df | CFI | TLI | RMSEA | RMSEA 90% CI | SRMR |
|---|---|---|---|---|---|---|---|---|
| 709.114 | 547 | 745.984 | 547 | 0.99 | 0.992 | 0.049 | [0.038, 0.059] | 0.079 |

These indices are reported descriptively and are used to characterize overall model fit in conjunction with the structural path estimates reported above. The following Section summarizes the hypothesis-testing outcomes based solely on the structural model estimates reported here.



### 4.5.7 Model Comparison Results (Single-Factor vs. Second-Order Specification)

To document alternative representations of risk management competence, two structurally equivalent models were estimated and compared. Model 1 specified risk management competence as a single latent construct directly measured by all observed indicators (DV1_1–DV4_5). Model 2 specified risk management competence as a second-order latent construct measured by four first-order dimensions (DV1–DV4), as described in Section 4.4.4. Both models included identical structural paths, mediating relationships, control variables, estimation settings, and data treatment.

Latent correlations among the 4 first-order dimensions in the second-order specification were uniformly high, ranging from 0.914 to 0.999, with several correlations approaching unity. Residual variances for certain first-order factors were near zero, indicating limited empirical separation among the behavioural domains. To address near-Heywood conditions during estimation of the second-order model, residual variances for all four first-order factors (DV1–DV4) were constrained to 0.001 as numerical stabilization constraints. These constraints were applied symmetrically across all four factors and represent a standard computational remedy rather than a theoretically motivated modification; they do not alter the substantive interpretation of the second-order structure.

The variance-covariance matrix of estimated parameters (vcov) for Model 2 produced a smallest eigenvalue of −0.016, indicating a non-positive-definite solution. Although this value represents a substantial improvement over the unconstrained second-order solution, it nonetheless signals residual identification instability. This value confirms that the standard errors for the Model 2 parameter estimates lack sufficient reliability for inferential use. The variance-covariance matrix for Model 1 produced a smallest eigenvalue of $-3.98 \times 10^{-15}$, which is within machine-precision rounding error and does not indicate any identification concern. Together, the near-unity latent correlations, near-zero residual variances, and vcov instability in Model 2 indicate that the four theoretically specified



behavioural dimensions were not empirically separable in this sample. This pattern reflects respondents' failure to cognitively differentiate among the normative conceptual distinctions embedded in the survey instrument rather than a model specification error. The vcov instability in Model 2 further reinforces — rather than merely coincides with — the decision to adopt the single-factor representation for all structural inference.

Model fit indices were examined descriptively for both specifications using the Comparative Fit Index (CFI), Tucker–Lewis Index (TLI), Root Mean Square Error of Approximation (RMSEA), and Standardized Root Mean Square Residual (SRMR). Fit indices were highly similar across the two models.

For Model 1, fit indices were CFI = 0.990, TLI = 0.992, RMSEA = 0.049, and SRMR = 0.079.

For Model 2, fit indices were CFI = 0.990, TLI = 0.992, RMSEA = 0.048, and SRMR = 0.078. Table 25 shows these values.

**Table 25**

*Model Fit Comparison: Single-Factor vs. Second-Order Models*

| Model | CFI | TLI | RMSEA | SRMR |
|---|---|---|---|---|
| Model 1: Single-Factor *RM_Competence* | 0.990 | 0.992 | 0.049 | 0.079 |
| Model 2: Second-Order *RM_Competence* | 0.990 | 0.992 | 0.048 | 0.078 |

**Note.** Fit indices are from WLSMV/DWLS estimation (theta parameterization). Models were estimated using identical structural paths and control variables.

### 4.5.8    Distribution of Standardized Risk Management Competence Scores

To characterize respondent-level variation in the retained single-factor risk management competence construct (*RM_Competence*), standardized latent factor scores were extracted and examined for distributional properties. Factor scores were derived under fixed-factor variance scaling (std.lv = TRUE), producing a distribution with a theoretical mean of zero and unit standard deviation, thereby enabling comparison of relative standing across respondents independently of the raw indicator scale.



As anticipated under this scaling convention, standardized *RM_Competence* scores were approximately normally distributed (M = 0.00, SD = 1.00). This result is consistent with the psychometric properties of a well-specified single-factor model and, in itself, does not confirm the normality of the underlying latent trait in the population; rather, it reflects the standardization imposed by the estimation approach. This distribution warrants comparison with Section 4.3.3; in both cases, the data reveal indicator means consistently above the midpoint of the scale. The convergence of elevated raw means with a symmetrically distributed latent score variable indicates that, while the sample as a whole endorsed competence items at above-average levels, within-sample variance remained present across respondents.

Table 26 presents the distribution of respondents across four standardized competence tiers, partitioned at conventional ±1 SD boundaries.

**Table 26**

*Distribution of Standardized RM_Competence Scores by Competence Tier*

| Competence Tier (Standardized Score) | n | % |
|---|---|---|
| ≥ +1.00 SD — High Differentiation | 15 | 11.9% |
| 0.00 to +0.99 SD — Above Average | 52 | 41.3% |
| −0.99 to 0.00 SD — Below Average | 46 | 36.5% |
| ≤ −1.00 SD — Low Differentiation | 13 | 10.3% |

**Note**. Standardized factor scores derived from Model 1 (retained single-factor specification, std.lv = TRUE). Conventional standard deviation thresholds establish the tier boundaries relative to the sample mean. These boundaries function as internal benchmarks rather than externally validated competence cut scores.

Approximately 50.8% of respondents scored above the sample mean (tiers ≥ 0 SD), while 49.2% scored below, consistent with the expected symmetry of standardized scores. Notably, the high-differentiation tier (≥ +1 SD) comprised only 11.9% of respondents, and the low-differentiation tier (≤ −1 SD) comprised 10.3%, together accounting for approximately one-fifth of the sample (22.2%). The concentration of respondents within the central bands (±1 SD) suggests that, while extreme scores are



present in both directions, the sample does not exhibit pronounced bimodality or clustering at either pole, a distribution that does not exhibit clustering at either extreme of the scale, as it reduces the risk of floor or ceiling effects attenuating path coefficients in subsequent analyses.

It bears emphasis that standardized competence tiers are sample-referential constructs. While these results reflect a participant's relative standing, they do not constitute an absolute measure of adequate or inadequate risk-management competence. The absence of externally validated cut scores for *RM_Competence* is a recognized boundary condition of this study. The study did not establish externally validated criterion cut-scores for *RM_Competence*. Nonetheless, the observed within-sample variance is sufficient to detect meaningful structural relationships and to characterize the heterogeneity of risk management competence among the sample's practicing cybersecurity professionals.

### 4.6      Effects of Control Variables

To isolate the primary effects, the structural model incorporates control variables for professional experience, prior risk management learning, and leadership status. This Section presents the estimates of the control variables. The CV2_RM_LRN2 finding warrants interpretive attention beyond standard covariate reporting and is discussed in Chapter 5.

Estimating the effects of the control variables concurrently with the primary paths accounts for potential confounding variance throughout the structural model.

**The effect of years of cybersecurity experience (CV1_YEARS)** on perceived relevance of risk management was estimated (b = −0.087, SE = 0.118, z = −0.742, p = .458, standardized β = −0.078).

**The effect of seniority level (CV1_SENIORITY)** was estimated (b = 0.194, SE = 0.125, z = 1.553, p = .120, standardized β = 0.176).

The analysis also incorporates two indicators of prior risk management learning as covariates.



**The effect of CV2_RM_LRN1** was estimated (b = 0.169, SE = 0.176, z = 0.961, p = .337, standardized β = 0.127); and

The effect of **CV2_RM_LRN2** was estimated (b = 0.645, SE = 0.234, z = 2.756, p = .006, standardized β = 0.432).

The standardized effect of CV2_RM_LRN2 on perceived relevance (β = 0.432) approached the magnitude of the primary training exposure path (IV1x → MeV1; β = 0.453), indicating that prior formal risk management learning is a substantively important antecedent of perceived relevance. Section 4.5.5 addresses the implications of this finding for interpreting the mediation pathway, and Chapter 5 discusses the practical significance of this result for training design and participant prerequisite considerations.

In addition, the **effect of the prior risk management learning check indicator (CV2_RM_LRN_CHK1)** was estimated (b = 0.205, SE = 0.111, z = 1.853, p = .064, standardized β = 0.199 (scaled CI [−0.012, 0.422])).

### 4.7    Group Comparison Analyses (Quantitative Supplementary Analyses)

This Section presents supplementary quantitative analyses examining group-level differences among survey respondents. The analyses include a comparison between cybersecurity professionals and a non-cybersecurity control group using a reduced foundational risk-reasoning scale, as well as within-sample comparisons between leadership and non-leadership participants. These analyses are descriptive and intended to provide additional context for the primary Structural Equation Modelling results reported earlier in this chapter, including the observed measurement behaviour of the risk management competence construct. The primary hypothesis tests rely on the full-sample structural model; consequently, this study omits group comparisons except where explicitly noted. The primary



hypothesis tests rely on the full-sample structural model; consequently, this study reports group comparisons only where they provide relevant contextual insight into the observed measurement behaviour of the risk management competence construct reported earlier in this chapter.

### 4.7.1    Cybersecurity Professionals vs Non-Cybersecurity Professionals

To provide an external reference point for the foundational risk-reasoning levels observed within the cybersecurity sample, this study compared responses from cybersecurity professionals with those from a control group of non-cybersecurity professionals. The comparison focused on a reduced eight-item foundational risk-reasoning scale drawn from three of the four competence dimensions assessed in the primary SEM instrument: risk-based assessment (ASMT_1–3, corresponding to DV1_1– DV1_3), risk-based prioritization (PRIO_1–3, corresponding to DV2_1–DV2_3), and risk-based control selection (ACTN_1–2, corresponding to DV3_1–DV3_2). Table 27 presents the item-level descriptions for all eight scale indicators. Items from the fourth competence dimension — Risk Communication and Rationale (DV4) — were administered in the control group survey as RSID_1 and RSID_2 but were excluded from the foundational scale because those items presuppose an organizational decision-making context tied specifically to cybersecurity practice and would not function as domain-neutral indicators in a general professional sample. The study retains eight items featuring context-independent language, thereby ensuring a fair cross-group comparison.

The analytical sequence included confirmatory factor analysis, cross-group measurement invariance testing, and distributional comparison. The analysis addresses a single evaluative question: Does cybersecurity training impart a substantive, measurable advantage in foundational risk reasoning compared with professionals without cybersecurity-specific training? The Section proceeds in three stages: the first establishes measurement validity and cross-group comparability; the second compares



central tendency; and the third examines distributional dispersion and the nature of variability differences between groups.

**Table 27**

*Foundational Risk-Reasoning Scale: Item Structure & Sub-Dimensions*

| Item Code | SEM Indicator | Sub-Dimension | Item Description |
|---|---|---|---|
| ASMT_1 | DV1_1 | Risk-Based Assessment | Changes in how often an issue occurs or how disruptive it appears influence how practitioners prioritize that issue in practice. |
| ASMT_2 | DV1_2 | Risk-Based Assessment | When new information changes the estimated likelihood of a problem, practitioners reassess the priority they assign to that risk. |
| ASMT_3 | DV1_3 | Risk-Based Assessment | Changes in operating conditions influence how practitioners evaluate the severity of potential consequences. |
| PRIO_1 | DV2_1 | Risk-Based Prioritization | When an issue is urgent but has a lower business impact, practitioners prioritize it based on its impact rather than its urgency. |
| PRIO_2 | DV2_2 | Risk-Based Prioritization | When expected operational effects change, practitioners revisit earlier prioritization decisions to reflect the new context. |
| PRIO_3 | DV2_3 | Risk-Based Prioritization | Practitioners use differences in expected exposures to justify prioritizing one item over another. |
| ACTN_1 | DV3_1 | Risk-Based Control Selection | Differences in how safeguards affect likelihood or impact influence the controls that practitioners recommend. |
| ACTN_2 | DV3_2 | Risk-Based Control Selection | Control recommendations differ depending on whether the selected control reduces likelihood, reduces impact, or both. |

*Note.* Item codes ASMT_1–3, PRIO_1–3, and ACTN_1–2 correspond to items DV1_1–DV1_3, DV2_1–DV2_3, and DV3_1–DV3_2 respectively within the primary SEM instrument. This study utilizes paraphrased item descriptions to improve cross-group clarity; however, the underlying constructs remain identical to those tested in the cybersecurity professional sample. Two additional items administered in the control group survey (RSID_1: residual risk consideration after control selection; RSID_2: communication of risk trade-offs to non-technical colleagues) correspond to the DV4 Risk Communication and Rationale dimension and were excluded from the foundational scale because their content is substantively context-dependent on cybersecurity professional practice and does not constitute domain-neutral foundational reasoning assessable across professional populations.

#### 4.7.1.1 Stage 1: Measurement Validity and Cross-Group Portability

Before any group comparison, the analysis evaluated whether the eight-item foundational scale demonstrated acceptable measurement properties within each population independently and whether it functioned equivalently across groups. Single-group confirmatory factor analyses were conducted



separately for cybersecurity professionals and the control group, using the WLSMV estimator with ordered indicators and theta parameterization. Fit indices for the cybersecurity professional group indicated strong model fit: CFI = 0.998, RMSEA = 0.040, SRMR = 0.034. Fit indices for the control group indicated acceptable model fit: CFI = 0.945, RMSEA = 0.049, SRMR = 0.062. The slightly attenuated fit in the control group is consistent with the instrument's adaptation to that population and does not indicate structural misspecification. These results confirm that the eight-item construct functioned as a coherent unidimensional measure in both populations.

Reliability analysis confirmed internal consistency across the combined dataset. Cronbach's α = 0.83, with satisfactory item–total correlations across all eight indicators and no item identified as degrading overall scale reliability. The absence of weak indicators confirms that subsequent group comparisons reflect genuine between-group differences rather than measurement instability within the scale. A configural-to-metric invariance sequence validated cross-group measurement portability within the multi-group confirmatory factor analysis framework. The configural model, which freely estimated factor loadings within each group, yielded a CFI of 0.995. The metric model, which constrained factor loadings to equality across groups, produced CFI = 0.989. The resulting ΔCFI of −0.006 falls well within the conventionally accepted threshold of |ΔCFI| ≤ 0.010 (Cheung & Rensvold, 2002), indicating that the factor loadings were statistically equivalent across groups. These results confirm metric invariance and establish the minimum condition for legitimate comparison of observed composite scores between groups. Taken together, the results of Stage 1 confirm that the eight-item foundational risk-reasoning scale is internally reliable, structurally coherent in both populations, and metrically portable across the cybersecurity professional and control groups. A validated measurement framework underpins the group comparisons reported in Stages 2 and 3.



**4.7.1.2 Stage 2: Comparison of Central Tendency**

Having established measurement equivalence, the analysis compared mean levels of foundational risk reasoning between the two groups using observed eight-item composite scores. Table 28 presents the descriptive statistics.

Table 28

*Comparison of Central Tendency*

|  | Cybersecurity Professionals | Control Group |
| --- | --- | --- |
| **n** | 126 | 133 |
| **Mean** | 3.82 | 3.72 |
| **SD** | 0.777 | 0.402 |
| **Median** | 4 | 3.75 |
| **Min** | 1 | 2.5 |
| **Max** | 5 | 4.88 |
| **Welch t** | 1.271 | |
| **p** | 0.205 | |
| **Cohen's d** | 0.16 | |

Cybersecurity professionals exhibited a mean composite score of 3.82 (SD = 0.777, Median = 4.00) on the five-point response scale. The control group exhibited a mean composite score of 3.72 (SD = 0.402, Median = 3.75). Both groups thus demonstrate mean-level responses in the upper region of the scale, with median scores reinforcing the proximity of group-level central tendency. Given the non-normal distribution of the composite scores — with responses concentrated at 4 and 5 across both groups — median values are reported alongside means as complementary indicators of central tendency.

A Welch independent-samples t-test was conducted (appropriate given the unequal group variances confirmed in Stage 3) to compare mean foundational risk reasoning scores between cybersecurity professionals (*M* = 3.82, *SD* = 0.777) and the control group (*M* = 3.72, *SD* = 0.402). The test yielded a non-significant result, $t(185.16) = 1.271$, *p* = .205 (Cohen, 1988). Cohen's d for the observed



difference was 0.16, 95% CI [−0.08, 0.40]. The CI crossed zero and was bounded at both ends within the trivial-to-small effect range, providing precision-based support for the null inference beyond the p-value alone. Cohen's *d* for the observed difference was 0.16, 95% CI [−0.08, 0.40]. The CI crossed zero and was bounded at both ends within the trivial-to-small effect range, providing precision-based support for the null inference beyond the *p*-value alone. These results indicate that cybersecurity professionals did not demonstrate a substantive incremental advantage over the control group in mean-level foundational risk reasoning. Item-level testing provided additional resolution. Independent-samples t-tests showed no significant differences between groups for six of the eight indicators.

Only two items emerged as exceptions to this trend. ACTN_2, reflecting the behavioural response dimension of risk practice, demonstrated a statistically significant group difference (t = 2.468, p = .014). ASMT_2, reflecting evaluative risk assessment, approached but did not reach statistical significance (p = .053). The dominant pattern across items was one of cross-group equivalence, with localized differentiation concentrated in the action and assessment dimensions rather than distributed systematically across the full construct. Table 29 summarizes these results.

**Table 29**

*Item-Level Mean Comparison: Cybersecurity vs Control Group*

| Item Code | Indicator Description | Cyber Mean | Control Mean | t | p |
|---|---|---|---|---|---|
| ASMT_1 | Changes in likelihood or disruption influence the decision assessment | 3.94 | 3.79 | 1.360 | 0.175 |
| ASMT_2 | New likelihood information affects priority assessment | 4.02 | 3.81 | 1.946 | 0.053 |
| ASMT_3 | Operating conditions influence perceived consequence severity | 3.85 | 3.70 | 1.404 | 0.162 |
| PRIO_1 | Higher business impact drives prioritization decisions | 3.72 | 3.62 | 0.844 | 0.400 |
| PRIO_2 | Impact outweighs urgency when assigning priority | 3.63 | 3.74 | -0.902 | 0.638 |
| PRIO_3 | Risk prioritization considers consequence magnitude | 3.71 | 3.73 | -0.135 | 0.892 |
| ACTN_1 | Response selection reflects impact reduction considerations | 3.72 | 3.73 | -0.064 | 0.949 |
| ACTN_2 | Mitigation choice reflects the likelihood or consequence reduction | 3.94 | 3.65 | 2.484 | 0.014 |

 **Note.** Welch two-sample t-tests were used for all item comparisons because the variances were unequal between groups. Cybersecurity sample n = 126; control group n ≈ 90. Only ACTN_2 reached conventional statistical significance (p < .05). ASMT_2 approached significance (p = .053). All remaining items showed no statistically significant group differences.

#### 4.7.1.3  Stage 3: Comparison of Distributional Dispersion

While the group means were substantially equivalent, the two populations differed markedly in score dispersion. This divergence is the most distinctive distributional characteristic to emerge from the control-group comparison. The cybersecurity professional group exhibited a standard deviation of 0.777, compared to 0.402 for the control group, yielding a variance ratio of 3.74. An independent variance test confirmed that this difference was statistically significant (p < .001). The magnitude of this difference — a nearly fourfold difference in variance — is substantively large and is not attributable to sampling fluctuation. The control group showed a relatively compact distribution, with scores clustered closely around the group mean. The cybersecurity professional group, in contrast, exhibited a substantially wider spread, with greater representation at both the upper and lower regions of the composite score distribution. This dispersion asymmetry remained independent of the non-significant difference in central tendency: while the groups shared comparable average levels of foundational risk reasoning, individual practitioners demonstrated varying degrees of consistency in using that reasoning.

#### 4.7.1.4  Summary

The three-stage analysis yields a coherent and internally consistent pattern of results. Stage 1 confirmed that the eight-item foundational risk-reasoning scale is reliable, structurally valid, and metrically equivalent across the cybersecurity professional and control groups, establishing a credible basis for comparison. Stage 2 established that group means are substantially equivalent, with no statistically significant or practically meaningful difference in central tendency (t = 1.271, p = .205;



Cohen's d = 0.16). Stage 3 revealed a striking contrast in distributional dispersion: while average levels of foundational risk reasoning were comparable, the cybersecurity professional group exhibited substantially greater variability (SD = 0.777 vs. 0.402; variance ratio = 3.74, p < .001). These findings collectively indicate that cybersecurity professionals do not demonstrate a systematic, incremental advantage over non-cybersecurity professionals in foundational risk reasoning, as measured by this instrument. That cybersecurity training does not appear to produce consistent or standardized outcomes on this construct across the practitioner population. Chapter 5 examines the theoretical implications of these results.

### 4.7.2    Leadership vs Non-Leadership Participants

This study employs group comparisons to identify differences based on formal leadership or managerial responsibilities. Leadership status derives from self-reported role characteristics collected during the demographic portion of the survey.

I computed descriptive statistics separately for leadership and non-leadership participants across selected variables to identify baseline differences. These comparisons characterize potential differences in professional perspectives and offer supplementary context for the subsequent qualitative leadership interview findings.

**Table 30**
*Leadership Classification Based on Seniority Level (n = 126)*

| Seniority Level | Non-Leadership n (%) | Leadership n (%) |
|---|---|---|
| Individual Contributor | 48 (53%) | — |
| Team Lead | 27 (30%) | — |
| Manager | 16 (18%) | — |
| Director | — | 20 (57%) |
| Executive | — | 15 (43%) |
| **Total** | **91 (72%)** | **35 (28%)** |





Where applicable, group-level contrasts are reported descriptively without inferential testing. This Section avoids drawing causal inferences from group comparisons and focuses instead on descriptive representation.

Additional comparisons examined differences between participants in leadership and non-leadership roles. This Section presents group-level contrasts without interpretation, focusing strictly on the statistical output.

#### 4.7.2.1 Notes on Supplementary Analyses

The group comparison analyses in this Section complement the primary SEM results without replacing them. Findings from these analyses are reported descriptively and are not used to infer causal relationships. Chapter 5 provides a detailed interpretation of these observed differences.

### 4.7.3 Structured Risk Practice and Competence Differentiation

A supplementary analysis evaluates behavioural differentiation by testing whether consistent use of structured risk decision practices correlates with higher latent risk management competence. Four survey items reflecting risk-reasoning behaviours (DV2_3, DV2_4, DV3_1, and DV3_4) serve as a proxy for structured risk practice. The structured practice classification required respondents to endorse at least 3 of the 4 items at the Agree or Strongly Agree level (endorsement at level 4 or 5 on the Likert scale (Agree or Strongly Agree)).

Of the 126 respondents, 68 (54.0%) met the structured-practice threshold.

A Welch two-sample t-test compared the latent *RM_Competence* factor scores (standardized z-scores) of structured and non-structured respondents. Respondents classified as structured practitioners demonstrated substantially higher competence scores (M = 0.528) than non-structured



respondents (M = −0.619), representing a mean difference of 1.15 standard deviations (95% CI [0.84, 1.45], p < .001).

Effect size estimation using Cohen's d indicated a large effect size (d = 1.39, 95% CI [1.00, 1.78]). Competence tier analysis further showed:

28.6% of respondents fell at or above +0.5 SD (Elevated Competence Tier)

11.9% fell at or above +1 SD (High Competence Tier)

These findings are presented as supplementary differentiation analysis and are not used to test the primary SEM hypotheses directly. Chapter 5 interprets the implications of these findings in greater detail.

### 4.8    Leadership Interviews

This Section reports the results of the study's qualitative component. Seven leadership interviews provided the primary data source for this phase of analysis. The unit of analysis is leadership discourse: the spoken accounts, evaluative judgments, and normative prescriptions of senior cybersecurity leaders regarding professional cognitive work, enterprise risk governance, and the developmental conditions under which risk judgment forms. Section 4.8.1 describes the analytical framework and coding approach. Section 4.8.2 presents individual case findings for each participant. Section 4.8.3 reports the cross-case analysis.

#### *4.8.1    Analytical Framework and Coding Approach*

The study analyzed interview transcripts through directed content analysis, using a fixed codebook developed deductively from the theoretical constructs examined in the quantitative model. Leadership Risk Cognition Codebook (LRCC) organizes 28 codes across five domains: Enterprise Risk Framing (ERF), Risk Governance and Process (RGP), Expected Cognitive Work (COG), Observed



Professional Gaps (GAP), and Development of Risk Judgment (DEV). Appendix C presents the full codebook, including code definitions, behavioural anchors, and salience scoring rules.

The analysis applied the codebook to segmented transcripts using meaningful idea-unit segmentation, targeting 18–30 units per transcript. Each idea unit received one or more primary domain codes. This analyst preserved leadership-specific language during segmentation and applied secondary codes only when the evidence warranted them. The analyst assigned a salience score of 0–4 to each code based on frequency, elaboration, and the prompted/unprompted character of evidence — behavioural anchors for each score level appear in the codebook. Table 31 presents the five salience levels and their defining criteria.

**Table 31**
*Salience Scoring Scale*

| 0 | Absent | The code concept does not appear in the transcript in any form; No evidence, direct or indirect. |
|---|---|---|
| 1 | Present | The concept appears once or twice in direct response to a prompted question. The evidence is brief, not elaborated, and the participant does not return to the theme independently. |
| 2 | Moderate | The concept appears in two or more units across at least two distinct question contexts or appears once with substantive elaboration beyond the minimum required by the prompt. The evidence is consistent and clear, but remains primarily within prompted responses. |
| 3 | High | Multiple units within the text feature and expand upon this concept frequently. At least one instance involves an unprompted extension beyond the scope of the question, or the participant returns to the theme independently in a different question context. Evidence is sustained and analytically developed. |
| 4 | Emphatic | The concept appears repeatedly, is elaborated at length, and includes multiple instances that are partially or fully unprompted. The participant foregrounds the theme as a central organizing idea. Evidence is emotionally weighted, analytically specific, or accompanied by concrete illustrative examples that exceed what the question required. |

Note: The full behavioural anchors, including prompted/unprompted distinction criteria, appear in Appendix C

Following individual case coding, the analyst produced a Case Abstraction Profile (CAP) for each interview. Each CAP includes a coding worksheet, a code frequency table with salience scores for all 28



codes, an Expectation–Gap Tension Index (calculated as the sum of salience scores across seven COG and seven GAP codes, with a maximum of 28 per domain), and a thematic synthesis organized by domain. The CAPs served as the primary data source for the cross-case analysis reported in Section 4.8.3. No re-coding or reinterpretation of transcript data occurs during the cross-case stage.

The coding approach was deductive rather than emergent. The codebook constructs derive from the study's theoretical framework — Competency-Based Learning Theory and enterprise risk management principles — and map directly onto the constructs examined in the quantitative model. The framework allowed flexibility in the weight and elaboration of evidence across codes. Still, it introduced no emergent codes and admitted no reinterpretation of previously assigned codes at the cross-case stage. This design facilitated alignment of constructs across the qualitative and quantitative phases of the mixed-methods analysis.

### 4.8.2    Individual Case Findings

The following Sections present analytical findings for each interview, drawn directly from the finalized CAP for that case. Each case report states the participant's organizational context, summarizes the dominant patterns across the five codebook domains, and records the Expectation–Gap Tension Index scores and typological classification. Salience scores appear in parentheses throughout. Findings are presented descriptively and without cross-case interpretation; cross-case comparison appears in Section 4.8.3

### 4.8.2.1 INT-01 — Chief Information Security Officer, iGaming / Online Gambling

| COG  13/28 | GAP  13/28 | High Expectation / High Gap  |  Symmetric |
|---|---|---|
| Role | Chief Information Security Officer (CISO) | |
| Sector | Online Gambling / iGaming | |
| Size | ~290 employees | |
| Location | Malta; additional European sites | |



INT-01 brings a dual accountability — cybersecurity leadership and regulatory compliance — that shapes every dimension of his analytical orientation. In a sector where licensing exposure is existential, cybersecurity and regulatory risks converge structurally: virtually any security breach simultaneously threatens the organization's compliance standing, giving cyber incidents board-level visibility through a regulatory channel that most organizations access only through financial magnitude.

**Enterprise Risk Framing.** INT-01 frames risk across three co-equal dimensions: financial impact (ERF-FIN, Sal. 2), reputational consequence (ERF-REP, Sal. 3), and regulatory and compliance exposure (ERF-REG, Sal. 3). Regulatory framing carries the greatest organizational salience — the participant describes board-level cyber risk attention as driven primarily by licensing obligation. Reputational framing reaches the interview's strongest emotional weight: the participant asserts that even a minor breach entering public discourse could render the organization insolvent. The financial impact is consistently present but positioned as one dimension among several rather than as the primary organizing frame.

**Risk Governance and Process.** INT-01 describes the most formalized governance structure among the seven cases: a two-tier risk committee architecture in which a cybersecurity-specific committee conducts an initial likelihood × impact rating before escalating high-risk items to an enterprise risk committee that includes the COO, CTO, and Chief Compliance Officer (RGP-LXI, Sal. 3; RGP-COLL, Sal. 3; RGP-ESC, Sal. 2). The process explicitly rejects arbitrary risk-level assignment. A gold/silver/bronze-tiered remediation framework governs residual risk options, and acceptance of any option below gold requires the accepting party to provide documented accountability.

**Expected Cognitive Work.** The participant expects professionals to take a holistic, enterprise-wide perspective (COG-HOL, Sal. 3), bring likelihood estimation as their primary technical contribution to



collaborative governance processes (COG-LXI, Sal. 2), and understand business impact across regulatory, financial, and reputational dimensions (COG-IMP, Sal. 2). The COG profile reflects a collaborative standard; the governance structure situates professional competence within a system that distributes **L x I** reasoning across multiple functions instead of mandating it solely for the cybersecurity team.

**Observed Professional Gaps.** INT-01 reports a distinctive dual technocentric pattern. Technical problem framing (TFR-PROB, Sal. 3) and tool-centric solution orientation (TFR-SOLN, Sal. 3) co-occur at equal salience — the only case in the dataset where both reach salience 3 simultaneously. Professionals not only define risk events in technical terms but also reflexively default to acquiring and deploying tools as the remediation response, without examining root causes or process-level contributors. Business-value blindness (GAP-BUS, Sal. 3) and myopic thinking (GAP-MYO, Sal. 2) reinforce the pattern. Notably, structured **L × I** reasoning (GAP-LXI) and entry-level immaturity (GAP-ENTRY) both record salience 0; the two-tier committee validation process distributes and compensates for individual professional deficits in these areas.

**Developmental Patterns.** INT-01 endorses a multi-channel developmental model. Formal training in regulatory literacy and business context carries the highest priority (DEV-FORM, Sal. 3), with experiential learning (DEV-EXP, Sal. 2), cross-domain exposure (DEV-MULTI, Sal. 2), and executive mandate (DEV-EXEC, Sal. 2) endorsed as complementary accelerants. INT-01 retains all pathways as viable.

**Tension Index.** Symmetric COG and GAP scores (13/28 each) classify this case as High Expectation / High Gap — symmetric sub-pattern. The participant articulates cognitive expectations and reports gaps with equal elaboration and salience, indicating a leader who defines the desired



professional cognition as clearly as he diagnoses its failure. The full case analysis appears in the Appendices.

**4.8.2.2 INT-02 — Director of Security Services, IT/Cyber Managed Services**

| COG  9/28 | GAP  18/28 | Low Expectation / High Gap |
|---|---|---|
| Role | Director of Security Services | |
| Sector | IT and Cybersecurity Managed Services | |
| Size | ~150 employees | |
| Location | Toronto, Canada | |

INT-02 reviews practitioner-produced risk outputs on a high-frequency, first-hand basis. Her evidence base is not reflective or historical — it is current, direct, and accumulated across hundreds of professional interactions. This experience gives her an empirical sense of gap diagnosis, distinguishing her from leaders who observe professional performance at one remove. Her dominant analytical orientation is business survival: risk becomes consequential when it threatens the organization's capacity to maintain client relationships and deliver contracted services, and her frustration with practitioners stems from their failure to reason within that logic.

***Enterprise Risk Framing.***  INT-02 frames risk primarily through business continuity and survival (ERF-SURV, Sal. 3), with financial impact (ERF-FIN, Sal. 2), reputational exposure (ERF-REP, Sal. 2), and legal liability (ERF-REG, Sal. 1) as co-present dimensions. The transcript's most emphatic formulation, 'the company might not even survive if those cases are not addressed properly (by cybersecurity professionals)'—dictates the implicit benchmark leaders apply when evaluating all practitioner outputs. The participant positions cybersecurity as one domain within a broader enterprise risk architecture rather than as an autonomous technical function.



*Risk Governance and Process.*  Governance structures exist and are named, but the participant qualifies both immediately and persistently: the **L × I** evaluation procedure is undocumented, inconsistently applied, and too informal to supply quantitatively defensible outputs at the leadership level (RGP-LXI, Sal. 2; RGP-ESC, Sal. 2). INT-02 records the lowest governance formalization level in the dataset. The participant frames this not as an organizational design failure but as a direct consequence of practitioner outputs that lack the analytical foundation needed for structured governance to work.

*Expected Cognitive Work.*  The participant expects professionals to understand and communicate business impact in financial terms (COG-IMP, Sal. 2), estimate likelihood as an analytical contribution (COG-LXI, Sal. 2), and think holistically about the organizational scope of identified risks (COG-HOL, Sal. 3). The holistic expectation reaches its most emphatic expression in the closing statement of the interview, where the participant frames enterprise-oriented risk thinking as the defining differentiator between a technical practitioner and a risk-aware professional.

*Observed Professional Gaps.*  INT-02 produces the highest GAP score in the dataset (18/28) and the highest-saturation gap profile across all seven cases — six of seven GAP codes register above zero, with four reaching salience 3: technical problem framing (TFR-PROB, Sal. 3), business-value blindness (GAP-BUS, Sal. 3), scale miscalibration (GAP-SCAL, Sal. 3), and absence of structured **L × I** reasoning (GAP-LXI, Sal. 3). The participant estimates that 90% or more of practitioners require leadership to perform risk translation on their behalf and cannot recall a



single instance of a professionally adequate risk submission. She provides multiple concrete examples of scale miscalibration — open FTP ports, unpatched RDP servers, missing endpoint-detection coverage — all assigned high severity based solely on the presence of a control gap, without enterprise-scale contextualization of likelihood and impact.

*Developmental Patterns.* Formal training is the primary and explicitly non-negotiable corrective prescription (DEV-FORM, Sal. 3), delivered at the point of professional entry. The interviewee specifies a curriculum: business goal comprehension, financial literacy, risk framework literacy, and cost-benefit analysis. She explicitly subordinates experiential learning and coaching (DEV-EXP, Sal. 1), labelling them insufficient on their own.

*Tension Index.* The COG-GAP asymmetry (9 vs. 18) classifies this case as Low Expectation / High Gap — the only case in the dataset to fall outside the High Expectation quadrant. The nine-point asymmetry represents the widest COG-GAP spread in the dataset and reflects a leader whose accumulated practitioner-facing experience has produced a highly elaborated gap diagnosis. At the same time, the articulation of an idealized cognitive standard remains comparatively compressed. The gap is not absent from the expectation domain, but it is less elaborated than in the six High Expectation cases. The full case analysis appears in Appendix D.

### 4.8.2.3 INT-03 — Senior Vice President, Cybersecurity, Financial Services

| COG  13/28 | GAP  16/28 | High Expectation / High Gap  |  Gap-dominant |
|---|---|---|
| Role | Senior Vice President, Cybersecurity | |
| Sector | Financial Services / Multinational Banking | |
| Size | ~80,000 employees | |



| Location | New York / global |
|---|---|

INT-03 operates within the most institutionally sophisticated governance environment in the dataset. A Senior Vice President with direct executive oversight of a large cybersecurity function embedded within a multinational financial institution, he speaks with authority. Likewise, he diagnoses professional shortcomings with specificity and consistency. His analytical orientation is financial and diagnostic: risk is real when money is at stake, and the professional failure he identifies most persistently is the inability — or unwillingness — of cybersecurity practitioners to reason within that logic.

***Enterprise Risk Framing.*** INT-03 frames risk primarily through financial magnitude (ERF-FIN, Sal. 3), with regulatory and audit exposure (ERF-REG, Sal. 2), business disruption (ERF-SURV, Sal. 2), and strategic positioning (ERF-STRAT, Sal. 2) as secondary dimensions. The participant's most emphatic formulation — 'it's always business versus technology — it's always money that wins the game' — establishes financial logic as the operative referent for all risk prioritizations. Regulatory framing functions as an automatic attention trigger: risks surfaced through audit examination receive immediate escalation regardless of financial magnitude. Reputational framing is absent (ERF-REP, Sal. 0), a notable omission for an institution where public trust is a primary governance concern.

***Risk Governance and Process.*** The governance architecture is the most formally institutionalized in the dataset. Dedicated risk officers hold quantitative **L × I** responsibility, supplying financial frequency data and probability estimates that the cybersecurity team does



not independently generate (RGP-LXI, Sal. 2). Cybersecurity professionals contribute technical and qualitative context; the quantitative risk calculus sits structurally outside their function. A formal four-option decision flow — mitigate, decommission, remediate, or accept — governs residual risk handling (RGP-ACC, Sal. 2). Cross-functional committees operate at scaled severity thresholds. All RGP codes record moderate salience (Sal. 2), reflecting a mature and functionally distributed system operating with institutional regularity.

**Expected Cognitive Work.**  The participant expects professionals to approach risk with holistic, architecturally informed enterprise perspective (COG-HOL, Sal. 3) and to estimate likelihood as a component of their analytical contribution (COG-LXI, Sal. 2). Business impact reasoning records the lowest COG salience in this case (COG-IMP, Sal. 1) — the governance structure assigns financial impact quantification to risk officers, softening the individual professional expectation accordingly. The participant endorses tiered options reasoning (COG-OPT, Sal. 2) and residual risk evaluation (COG-RES, Sal. 2), both of which the four-option institutional framework operationalizes.

**Observed Professional Gaps.**  The participant treats business-value blindness as the primary explanatory construct for all professional failure: practitioners do not understand, access, or reason within the monetary logic that governs executive decisions. TFR-PROB (Sal. 3) and GAP-MYO (Sal. 3) reinforce this diagnosis. GAP-LXI (Sal. 3) records high salience despite the governance delegation of quantitative **L × I** to risk officers — professionals neither develop nor apply independent **L × I** reasoning. In a striking unsolicited observation, the participant cites



CISSP credential examination failure rates as profession-wide evidence of technology-oriented reasoning bias, repositioning the credential not as a competency standard but as a diagnostic instrument.

*Developmental Patterns.* Formal training earns moderate endorsement (DEV-FORM, Sal. 2), but the participant assesses it as generic and insufficient for developing contextual judgment. Experiential learning — on-the-job participation in institutional risk processes — carries the primary developmental weight (DEV-EXP, Sal. 2). Risk officers' dedicated training is cited as the functional model cybersecurity professionals should emulate.

*Tension Index.* The COG-GAP asymmetry (13 vs. 16) classifies this case as High Expectation / High Gap — gap-dominant sub-pattern. Gap observations are more emphatic and sustained than expectation articulations, consistent with an interview style that leads with diagnosis rather than prescription. The full case analysis appears in Appendix D.

### 4.8.2.4 INT-04 — Executive Director, Customer Success, Cybersecurity SaaS

| COG 12/28 | GAP 12/28 | High Expectation / High Gap  |  Symmetric |
|---|---|---|
| Role | | Executive Director, Customer Success |
| Sector | | Cybersecurity SaaS |
| Size | | ~47 employees |
| Location | | Chicago / global |

INT-04 occupies an analytically distinctive position in this sample: an executive in a small, lightly regulated organization with cross-functional visibility into both cybersecurity practice and product delivery, who approaches risk through a lens of parsimonious financial



survival. His analytical judgments about the profession are long-held and broadly applied — the phrase 'forest from the trees,' which he uses to characterize specialization-induced myopia, was well established before the interview, signalling a conviction formed well beyond the scope of a single organizational experience.

***Enterprise Risk Framing.***  INT-04 frames risk through two reinforcing lenses — financial survival (ERF-FIN, Sal. 3) and organizational continuity (ERF-SURV, Sal. 2). The operative executive question is whether a risk jeopardizes the survival of the company; financial quantifiability is the primary admissibility criterion for executive attention. Regulatory framing is structurally absent (ERF-REG, Sal. 0) in a lightly regulated SaaS environment. Cybersecurity sits within the engineering function and connects directly to the revenue-generating product platform — there is no separate cybersecurity organizational structure (ERF-STRAT, Sal. 2).

***Risk Governance and Process.***  INT-04 describes a compressed but coherent governance model appropriate to a small-organization scale. A formal triage process gates escalation: teams evaluate impact and severity across multiple perspectives before escalation is considered (RGP-ESC, Sal. 2; RGP-LXI, Sal. 2). At high-severity thresholds — those that implicate organizational survival — CEO-level involvement is standard, and cancelled vacations signal genuine criticality. The same cross-functional leadership group engages across all risk types (RGP-COLL, Sal. 2), providing structural consistency without layered formality.

***Expected Cognitive Work.***  COG-HOL reaches maximum salience (4) — the highest of any code across all seven cases. The participant's defining cognitive standard is that



professionals synthesize across technical, organizational, and risk domains simultaneously, producing an analysis consumable by an executive awakened at three o'clock in the morning. This expectation is framed not as aspirational but as a baseline for professional credibility at the leadership interface. Business impact reasoning (COG-IMP, Sal. 3) and likelihood estimation (COG-LXI, Sal. 2) appear at substantive salience. The participant explicitly exempts entry-level professionals from this expectation, positioning it as a mid-career milestone rather than a hiring standard.

*Observed Professional Gaps.* Specialization-induced myopia (GAP-MYO, Sal. 3) is the central diagnostic construct — the pre-formed 'forest from the trees' language functions as the interview's organizing metaphor. The participant describes this as an industry-wide structural pattern produced by deep specialization: professionals see individual technical vulnerabilities with precision while missing the organizational risk event entirely. Technical problem framing (TFR-PROB, Sal. 2), business-value blindness (GAP-BUS, Sal. 2), absence of **L × I** reasoning (GAP-LXI, Sal. 2), and entry-level immaturity (GAP-ENTRY, Sal. 2) all appear at moderate salience. Tool-centric remediation (TFR-SOLN) is absent; the primary gap is myopic framing, not tool-centrism.

*Developmental Patterns.* Experiential learning is the primary developmental mechanism (DEV-EXP, Sal. 3) — the transition from technical to risk-based cognition is a mid-career milestone achieved through organizational exposure, not coursework. Executive mandate (DEV-EXEC, Sal. 2) is the primary organizational driver of risk-language adoption:



professionals adopt risk-oriented reasoning because the organization demands it through escalation of expectations, not through intrinsic professional motivation. Formal training is acknowledged but not observed as a systematically deployed mechanism (DEV-FORM, Sal. 2).

*Tension Index.* Symmetric COG and GAP scores (12/28 each) classify this case as High Expectation / High Gap — symmetric sub-pattern. The participant defines desired professional cognition with equal clarity and precision to his gap diagnosis. This configuration reflects a leader with high, well-specified expectations who observes an industry-wide failure to meet them. The full case analysis appears in Appendix D.

### 4.8.2.5 INT-05 — Director / CISO, Post-Secondary Education & Public Service

| COG 12/28 | GAP 13/28 | High Expectation / High Gap  \|  Near-symmetric |
|---|---|---|
| Role | Director / Chief Information Security Officer (CISO) | |
| Sector | Post-Secondary Education and Public Service | |
| Size | ~500–600 employees | |
| Location | British Columbia, Canada | |

INT-05 is the only participant in this sample whose primary analytical frame is methodological rather than categorical. Where other leaders organize risk through business survival, financial magnitude, or regulatory exposure, INT-05 organizes risk through a formula: Impact × Likelihood − effective controls − organizational risk tolerance. This formula represents more than conceptual shorthand; it serves as the participant's operational definition of a legitimate risk statement, and the participant measures every gap diagnosis against it. His professional frustration is not that practitioners fail to communicate well, but that they fail to think precisely, and he brings sector-forum-level evidence to support that judgment.



***Enterprise Risk Framing.***  ERF aggregate salience is the lowest in the dataset. Financial (ERF-FIN, Sal. 1) and regulatory (ERF-REG, Sal. 1) dimensions appear descriptively — as observations about what drives organizational urgency — not as the participant's own risk-organizing lens. Survival, reputational, and strategic framings are absent. The low salience does not reflect a lower standard: it reflects a cognitive architecture in which the formula renders categorical labelling redundant. Risk type is analytically subordinate to methodological discipline.

***Risk Governance and Process.***  INT-05 describes the most structurally elaborate governance architecture in the dataset. Three concurrent escalation routes operate: upward reporting from the Information Security department, reactive escalation from security events, and — increasingly — board-initiated risk inquiry in which the board poses direct questions about specific risks on the registry (RGP-ESC, Sal. 4). A board-level risk registry integrates cyber risk with all other organizational risk categories, enabling comparative cross-domain prioritization (RGP-COMP, Sal. 2). The participant frames governance quality not by structural existence but by whether the architecture produces proportional, calibrated, evidence-defensible inputs to decision-makers.

***Expected Cognitive Work.***  Likelihood estimation (COG-LXI, Sal. 3) and business impact reasoning (COG-IMP, Sal. 3) co-dominate — the only case in the dataset where this pair leads over COG-HOL. The expectation is evidence-based and precise: professionals must defend their likelihood placement with an explicit account of the evidence that supports it, not merely state



the placement. Comparative risk reasoning against the registry baseline (COG-COMP, Sal. 2) and residual risk evaluation after controls (COG-RES, Sal. 2) are also registered at substantive salience. COG-HOL records its lowest salience here (Sal. 1); the participant's professional standard is methodological precision rather than architectural scope.

***Observed Professional Gaps.***  Three codes co-dominate at equal salience: technical problem framing (TFR-PROB, Sal. 3), absence of structured **L × I** reasoning (GAP-LXI, Sal. 3), and entry-level immaturity (GAP-ENTRY, Sal. 3). The **L × I** gap is the direct inverse of the participant's own standard: practitioners assert 'high risk' without proportional likelihood and impact calibration — 'loosey-goosey' in his characterization. Entry-level immaturity receives the most extensive evidential support in the dataset: national post-secondary and government information security forum observations document low professional community interest in risk formalization, positioning the gap as a profession-wide socialization failure rather than an individual performance shortfall.

***Developmental Patterns.***  DEV-FORM reaches the highest salience in the dataset (4); INT-05 identifies professional certification in risk, privacy, and audit as a non-negotiable corrective prescription. This leader assesses current formal training as 'fairly weak'. Cross-domain exposure to risk and privacy functions provides a compensatory informal pathway (DEV-MULTI, Sal. 2). Experiential learning is acknowledged as the prevailing reality but assessed as insufficient without structured credentialing (DEV-EXP, Sal. 1). Executive mandate is absent



(DEV-EXEC, Sal. 0) — professional credentialing rather than organizational culture is the theory of development here.

   *Tension Index.* Near-symmetric COG and GAP scores (12 and 13/28, respectively) classify this case as High Expectation / High Gap — near-symmetric sub-pattern with slight gap dominance. The participant's methodological precision applies symmetrically to both expectation and gap: he defines the standard with the same exactness he applies to diagnosing its absence. The full case analysis appears in Appendix D.

**4.8.2.6 INT-06 — CEO, Cybersecurity Awareness, Training, Behaviour & Culture SaaS Company**

| COG 12/28 | GAP 17/28 | High Expectation / High Gap  |  Gap-dominant |
|---|---|---|
| Role | Founder / Chief Executive Officer (CEO) | |
| Sector | ICT — Cybersecurity Awareness, Training, Behaviour & Culture SaaS | |
| Size | ~46 FTEs; ~1,500 clients; ~1M users | |
| Location | New Brunswick, Canada | |

   INT-06 occupies a singular position in this sample: the only participant who is simultaneously a practicing cybersecurity professional, a product vendor, and an organizational CEO. Leading a SaaS company whose product is cybersecurity awareness, training, behaviour, and culture for approximately 1,500 client organizations and one million end users, he evaluates risk not from behind a governance desk but from within the operational centre of an organization that treats cyber risk as both its primary threat environment and its market proposition. His analytical orientation is strategic and entrepreneurial: risk is the cost dimension of a risk–reward equation, always contextual, always calibrated to what is at stake, and persistently distorted by cognitive biases that he names precisely and corrects deliberately. His gap diagnosis is the most theoretically elaborated in the dataset, grounded in



cognitive science, evidence on organizational failure, and an original distinction between IT security and cybersecurity as professional identities.

*Enterprise Risk Framing.* INT-06 frames risk through an explicitly strategic lens (ERF-STRAT, Sal. 3), defining it as the cost side of a risk–reward trade-off; here, the practitioner must evaluate every risk against the value it could compromise. Business continuity and survival framing (ERF-SURV, Sal. 3) appears across three units — a cascading systemic failure metaphor, the generative AI phishing escalation, and the adversarial threat environment — all of which position risk as a potential organizational viability event. In this instance of ERF-REG (Sal. 1), the participant characterizes the Canadian regulatory environment as non-operative. Instead, industrial standards in banking and data integrity dictate the organization's meaningful risk exposure. Financial framing (ERF-FIN, Sal. 1) appears only in a client-organization case example, not as a primary self-referencing frame. The total absence of reputational framing (ERF-REP, Sal. 0) signals a specific cognitive architecture. In this model, strategic consequence supersedes stakeholder perception as the primary organizing principle for risk.

*Risk Governance and Process.* The participant describes a pragmatic, informally structured governance architecture appropriate to a ~46-person organization. Risk escalation runs directly to the CEO and the two most senior technical leaders — the Chief Security Officer and CTO — through a consistently applied but undocumented process (RGP-ESC, Sal. 2). The participant acknowledges explicitly that no formal likelihood methodology exists; the operative default is to treat a risk as likely until it can be proven otherwise (RGP-LXI, Sal. 2; RGP-ACC, Sal. 2). The culture embeds cross-functional risk identification by requiring all team members to escalate risk signals. To validate this model (RGP-COLL, Sal. 2), the participant presents a concrete exemplar: a help desk ticket that successfully traced a criminal account compromise. ISO 27001 quarterly review meetings and a revised software development lifecycle provide structural governance anchors, endorsed for substantive process value



rather than certification status. Compared to the more institutionalized governance environments of INT-01, INT-03, and INT-05, INT-06 describes a lean but functionally coherent system in which cultural commitment compensates for procedural informality.

***Expected Cognitive Work.*** The participant articulates four active cognitive expectations for cybersecurity professionals, all with a salience rating of 3. Likelihood estimation (COG-LXI, Sal. 3) is the most theoretically grounded expectation in the dataset: the participant invokes Protection Motivation Theory to argue that effective risk communication must first establish personal likelihood credibility before protective action is motivated, and identifies optimism bias and novelty bias as cognitive mechanisms that systematically distort likelihood calibration. This invocation positions the expectation beyond mechanical risk calculation and into metacognitive self-correction. In this instance of COG-IMP (Sal. 3), the rhetoric moves beyond technical metrics. Instead, it compels professionals to articulate risk in terms of 'loss aversion' to influence executive decision-makers. Holistic thinking (COG-HOL, Sal. 3) is the most frequently coded expectation (four units) and centres on the capacity for systemic imagination — anticipating the cascading, combinatorial failure modes that adversaries exploit and that professionals with limited imagination cannot see. Strategic cognition (COG-STRAT, Sal. 3) is captured in the participant's prescription of a 'department of know-how' role: professionals must reason in outcome-driven terms — if we do X, reduce Y, enabling Z — and position cybersecurity as an enabler of organizational mission rather than a constraint upon it.

***Observed Professional Gaps.*** INT-06 produces the second-highest GAP score in the dataset (17/28), exceeded only by INT-02 (18/28), and generates the most theoretically elaborated gap diagnosis across all seven cases. GAP-BUS (Sal. 4) is the sole salience-4 code in this case and the organizing argument of the entire gap analysis: the field produces IT security professionals, not cybersecurity professionals, and the difference is definitional. Where 'cyber' encompasses people,



technology, and control — in that order — IT security training produces professionals anchored exclusively to the technology dimension, without grounding in business management, liberal arts, or the organizational empathy required to translate technical findings into executive-actionable intelligence. The $2M firewall case anchors this diagnosis as a concrete negative exemplar. Technical problem framing (TFR-PROB, Sal. 3) is the structural mechanism of GAP-BUS: the IT security professional identity produces business blindness as a predictable output. GAP-MYO (Sal. 3) extends the diagnosis to include the misanthropic user-blaming syndrome — 'users are stupid,' 'if I didn't have users, we'd have a perfect business' — and the CISO scapegoat dynamic, in which organizational positioning as blame-absorber perpetuates the field's inability to develop a genuine risk management orientation. GAP-LXI (Sal. 3) receives its most theoretically elaborated treatment in this case: the participant applies cognitive science — optimism bias, novelty bias, and emergent systemic complexity — to explain why likelihood estimation fails structurally, and proposes a corrective heuristic: treat as likely until proven otherwise. The formulation 'it's the likelihood where we die' is the most emphatic likelihood diagnosis in the dataset. The participant expresses GAP-ENTRY (Sal. 2) comparatively by positioning his own organizational risk culture at the extreme edge of the professional bell curve, thereby implying a field-wide normative gap.

*Developmental Patterns.* The participant's developmental prescription is bifocal and analytically specific. Formal training (DEV-FORM, Sal. 3) receives the most philosophically grounded prescription in the dataset: the curriculum must begin with the etymological foundations of 'cyber' — people, technology, and control — and of 'technology' — techne as the art and skill of building, and logos as the careful consideration of what you gain or lose — treating these as the conceptual foundations of risk management rather than marketing terms. Accreditation bodies mandate ethics and business management courses across the field. Mentored experiential learning (DEV-EXP, Sal. 2) is the



primary mechanism for developing professional judgment: mentorship is framed not as skill transfer but as safe failure management, providing conditions in which professionals can make consequential mistakes without career-ending consequences. The participant introduces an unprompted structural concern: AI's potential to eliminate the human guide who contextualizes safe failure may degrade professional judgment development across the field. DEV-MULTI is absent — a significant omission given that the diagnosed gap would logically prescribe cross-domain exposure. The prescribed remedy is formal curricular reform, not rotation. The code DEV-EXEC (Sal. 2) reveals a dual rhetoric: it positions organizational culture as a development asset while exposing the 'CISO scapegoat' role as a structural barrier to field-level progress.

*Tension Index.* The COG-GAP asymmetry (12 vs. 17) classifies this case as High Expectation / High Gap — gap-dominant sub-pattern. Among the High Expectation cases, INT-06 records the widest COG-GAP spread at five points, exceeded across the full dataset only by INT-02, whose Low Expectation classification reflects a nine-point asymmetry of a structurally different character. The asymmetry reflects a leader who articulates four active cognitive expectations at consistently high levels of elaboration, while producing a gap diagnosis that substantially exceeds those expectations in both frequency and theoretical depth. Unlike INT-02, where the asymmetry reflects frustration accumulated through practitioner-facing experience, INT-06's asymmetry reflects the deliberate application of cognitive science, evidence of organizational failure, and professional identity theory to a diagnostic argument that the participant has developed well beyond the scope of this interview. The full case analysis appears in Appendix D.

### 4.8.2.7 INT-07 — CISO, Higher Education

| COG 11/28 | GAP 15/28 | High Expectation / High Gap  |  Gap-dominant |
|---|---|---|
| Role | | Director of Information Security Services (functionally CISO) |
| Sector | | Higher Education |



| Size | ~9000 FTE |
| --- | --- |
| Location | Ontario, Canada |

INT-07 describes a higher education environment in which cybersecurity governance matured reactively following a realized breach rather than through anticipatory enterprise risk reasoning. The participant articulates a definition of risk grounded in likelihood × impact and reports that this formulation is generally accepted on campus when formally stated. Governance evidence is substantive but moderately formalized, including tiered severity escalation, cross-functional alignment with safety and enterprise risk functions, and post-breach elevation of cyber risk to parity with physical threats. The dominant professional gap is business-value blindness, reinforced by technocentric problem framing—a pattern the participant attributes to an autodidactic professional culture, deep specialization, and limited engagement with business-facing analytical frameworks.

Enterprise Risk Framing. INT-07 frames enterprise risk through a mixed rather than singular configuration. Financial impact (ERF-FIN, Sal. 2) functions as a consistent referent for executive attention, while reputational consequence (ERF-REP, Sal. 2) gained substantially elevated salience following the 2023 breach and associated media coverage. Business continuity and organizational survival concerns are present in the post-breach environment (ERF-SURV, Sal. 2), where disruption to institutional operations serves as a sustained organizational reference point. Regulatory exposure contributes descriptively to the risk environment in an accreditation-sensitive sector (ERF-REG, Sal. 1), though it does not function as the primary organizing frame. Strategic positioning is absent as an explicit framing dimension (ERF-STRAT, Sal. 0). The ERF profile reflects an institution in which realized exposure—not strategic abstraction—shaped risk salience. In this environment, the post-breach experience continues to dictate how leadership articulates cybersecurity risk.



Risk Governance and Process. Governance is moderately formalized and substantively operational, though it reflects a maturity trajectory shaped more by reactive intensification than by anticipatory design. Tiered severity escalation governs upward reporting (RGP-ESC, Sal. 3), and the participant describes clear threshold-based routing of incidents and risks through institutional channels. Likelihood × impact evaluation aligns with the participant's personal risk formulation (RGP-LXI, Sal. 2). However, the participant does not describe the operationalization of this formulation within the university as a quantitatively codified process. Cross-functional collaboration with safety and enterprise risk functions provides structural integration (RGP-COLL, Sal. 2), and post-breach elevation of cyber risk to parity with physical threats reflects a meaningful expansion of the governance architecture's scope. Risk acceptance logic is present but not formally codified (RGP-ACC, Sal. 1). Comparative registry-level risk reasoning is not described (RGP-COMP, Sal. 0). The governance profile positions INT-07 in the moderate-formalization range, above INT-02 but below the institutionally mature architectures of INT-01, INT-03, and INT-05.

Expected Cognitive Work. The participant articulates a pragmatic and seniority-stratified cognitive standard. Likelihood estimation (COG-LXI, Sal. 3) is the most salient expectation: the participant endorses likelihood × impact as the operative definition of a legitimate risk statement and expects senior professionals to apply and defend that formulation in institutional contexts. Business impact reasoning (COG-IMP, Sal. 2) reaches substantive salience; the organization requires senior staff to produce business cases, quantify time and cost implications, and justify risk trade-offs in organizational terms. Holistic enterprise-wide reasoning (COG-HOL, Sal. 2) is present but does not reach the maximum salience observed in INT-04; the expectation applies differentially, with senior staff held to a broader analytical standard and junior staff not yet expected to perform formal risk analysis independently. Strategic alignment reasoning (COG-STRAT, Sal. 2) is implicit in the expectation that



senior staff produce decision-useful framing for executive audiences. Tiered options reasoning (COG-OPT, Sal. 1) and residual risk evaluation (COG-RES, Sal. 1) appear at low salience. The seniority-stratified framing distinguishes INT-07 from cases in which a uniform cognitive standard applies across role levels.

Observed Professional Gaps. Business-value blindness (GAP-BUS, Sal. 4) is the dominant gap and reaches maximum salience, co-occurring with INT-03 and INT-06 as the highest single-code gap observations in the dataset. The participant describes professionals as overproducing technical detail while underproducing business rationale and strategic context — a pattern in which depth of analysis does not translate into decision-useful institutional framing. Technical problem framing (TFR-PROB, Sal. 3) reinforces and structurally underpins this diagnosis: cybersecurity professionals define risk events in technical rather than organizational terms, rendering their outputs incomplete for governance purposes. Myopic, siloed reasoning (GAP-MYO, Sal. 3) extends the diagnosis — practitioners focus with precision on technical vulnerabilities while missing the organizational risk event that those vulnerabilities constitute. The participant attributes this pattern partly to autodidactic professional culture and self-reliance, and partly to underutilization of available business-analysis resources. Absence of structured **L × I** reasoning (GAP-LXI, Sal. 2) appears at moderate salience, consistent with the observation that practitioners assert risk significance without proportional likelihood and impact calibration. Entry-level professional immaturity (GAP-ENTRY, Sal. 2) is present, linked to the autodidactic trajectory through which many practitioners develop without structured engagement with risk formalization. Scale miscalibration (GAP-SCAL, Sal. 1) appears at low salience. Tool-centric remediation orientation (TFR-SOLN, Sal. 0) is absent.

Developmental Patterns. The participant endorses a mixed developmental model with experiential learning as the primary mechanism. Experiential learning and cross-domain mentorship (DEV-EXP, Sal. 3) carry the greatest developmental weight — the transition from technical to risk-



oriented cognition is described as an organizational socialization process, supported by exposure to business-facing decision contexts and mentorship from professionals who operate across institutional domains. Cross-domain exposure (DEV-MULTI, Sal. 2) functions as a complementary pathway, with engagement across risk, safety, and enterprise functions identified as a substantive corrective to siloed professional development. Formal training is acknowledged and assessed as growing in relevance (DEV-FORM, Sal. 2), though the participant does not prescribe it as the primary or non-negotiable corrective mechanism. Executive mandate (DEV-EXEC, Sal. 1) appears descriptively in the participant's observation that institutional expectations shifted following the breach; however, the participant does not frame this shift as a developmental theory. The developmental prescription is therefore not simply more training, but structured exposure to business-facing reasoning and institutional decision contexts, with formal training as a supporting rather than anchoring mechanism.

Tension Index. The Tension Index score (COG 11/28; GAP 15/28) classifies this case as High Expectation / High Gap — gap-dominant sub-pattern. Gap observations exceed expectation articulations by four points, with GAP-BUS at maximum salience constituting the dominant analytical weight of the case. The participant devotes greater elaboration to diagnosing practitioner underperformance than to specifying the full idealized model of risk cognition, consistent with a leader whose institutional experience centres on the consequences of professional shortfall rather than on prescribing cognitive standards. The full case analysis appears in Appendix D.

### 4.8.3    Cross-Case Analysis

The following Sections report cross-case results derived from seven finalized Case Abstraction Profiles (CAPs), produced through directed content analysis of leadership interview transcripts using the Leadership Risk Cognition Codebook (LRCC). Analysis draws on code frequency counts, salience scores,



and domain-level summaries generated at the individual case level. No recoding or reinterpretation of transcript data occurs at this stage. Sections 4.8.3.2 through 4.8.3.6 report domain-level convergence and divergence. Section 4.8.3.7 presents the structural typology. Section 4.8.3.8, titled 'Cross-Case Structural Pattern,' synthesizes cross-domain patterns.

### 4.8.3.1 Case Inventory and Expectation-Gap Classification

Table 32 presents the case inventory for all seven leadership interviews, drawing from finalized Case Abstraction Profiles (CAPs) produced through directed content analysis using the Leadership Risk Cognition Codebook (LRCC).

On the Expectation–Gap Tension Index, six cases fall in the High Expectation / High Gap category, with COG scores between 11 and 13 and GAP scores between 12 and 17, and one case — INT-02 — falls in the Low Expectation / High Gap category with COG 9/28 and GAP 18/28, the highest GAP score in the dataset.

**Table 32**
*Case Inventory Matrix*

| Interview | Role / Sector | Primary Risk Frame (ERF) | Primary Cognitive Expectation (COG) | Primary Observed Gap (GAP) | COG Score (/28) | GAP Score (/28) | Governance Formalization | Technocentric Framing | Expectation–Gap Quadrant |
|---|---|---|---|---|---|---|---|---|---|
| INT-01 | CISO iGaming | ERF-REG / ERF-REP | COG-HOL | TFR-PROB / GAP-BUS | 13 | 13 | High | High | High Expectation / High Gap |
| INT-02 | Director of Security Services Managed Services | ERF-SURV | COG-LXI / COG-IMP | TFR-PROB | 9 | 18 | Moderate | High | Low Expectation / High Gap |
| INT-03 | Senior Vice President, Cybersecurity Financial Services | ERF-FIN | COG-HOL | GAP-BUS | 13 | 16 | High | High | High Expectation / High Gap |



| Intervie w | Role / Sector | Primary Risk Frame (ERF) | Primary Cognitive Expectatio n (COG) | Primary Observe d Gap (GAP) | COG Scor e (/28) | GAP Scor e (/28) | Governance Formalizatio n | Technocentr ic Framing | Expectation– Gap Quadrant |
|---|---|---|---|---|---|---|---|---|---|
| INT-04 | Executive Director, Customer Success Cybersecurit y SaaS | ERF-FIN | COG-HOL | GAP-MYO | 12 | 12 | Moderate | Moderate | High Expectation / High Gap |
| INT-05 | Director / CISO Education & Public Service | ERF-FIN | COG-LXI / COG-IMP | TFR-PROB / GAP-LXI / GAP-ENTRY | 12 | 13 | High | High | High Expectation / High Gap |
| INT-06 | Chief Executive Officer Cybersecurit y SaaS | ERF-STRAT / ERF-SURV | COG-HOL / COG-STRAT | GAP-BUS | 12 | 17 | Moderate | High | High Expectation / High Gap |
| INT-07 | Director of Information Security (functionally CISO) Higher Education | ERF-SURV / ERF-FIN / ERF-REP | COG-LXI | GAP-BUS | 11 | 15 | Moderate | High | High Expectation / High Gap |

*Note.* The COG Score and GAP Score are aggregate salience totals derived from the Leadership Risk Cognition Codebook (LRCC), with a maximum of 28 per domain. Governance Formalization was assessed descriptively from the case narrative. Technocentric Framing reflects the combined salience of the TFR-PROB and TFR-SOLN codes. Because Likelihood × Impact reasoning appeared in all seven cases, the narrative details these findings rather than in a separate column. ERF, COG, and GAP codes reflect the highest-salience dominant codes per domain; secondary codes appear where two codes share equivalent salience.

Table 33 classifies the seven cases by typology quadrant and identifies two sub-patterns within the High Expectation/High Gap quadrant.

**Table 33**

*Expectation–Gap Typology Classification*

| Interview | COG Score (/28) | GAP Score (/28) | Typology Quadrant | Sub-Pattern | Profile Characteristic |
|---|---|---|---|---|---|
| INT-01 | 13 | 13 | High Expectation / High Gap | Symmetric | Expectation and gap salience are equal. The participant articulates the desired cognitive standard with the same elaboration and emphasis applied to diagnosing its absence. The governance architecture distributes **L × I** responsibility |



| Interview | COG Score (/28) | GAP Score (/28) | Typology Quadrant | Sub-Pattern | Profile Characteristic |
|---|---|---|---|---|---|
| | | | | | across a two-tier committee, moderating individual-level gap exposure. |
| INT-02 | 9 | 18 | Low Expectation / High Gap | Singular (sole outlier) | The only case in the dataset is outside the High Expectation quadrant. Accumulated practitioner-facing experience yields a highly elaborated gap diagnosis, with a comparatively compressed articulation of expectations. The nine-point asymmetry is the widest COG–GAP spread in the dataset. |
| INT-03 | 13 | 16 | High Expectation / High Gap | Gap-dominant | Gap observations exceed expectation articulations by three points. GAP-BUS reaches maximum salience (4) and functions as the primary explanatory construct for all observed practitioner shortfalls. Expectations remain high, but the institutionalized risk officer structure partially absorbs this pressure. |
| INT-04 | 12 | 12 | High Expectation / High Gap | Symmetric | Expectation and gap salience are equal. The participant defines desired professional cognition — holistic synthesis at executive-ready resolution — with equal clarity and precision to his gap diagnosis. Specialization-induced myopia (GAP-MYO) is the central organizing diagnostic construct. |
| INT-05 | 12 | 13 | High Expectation / High Gap | Near-symmetric (slight gap dominance) | A one-point gap advantage reflects methodological parity: the participant applies the same formulaic precision to both the expected standard and its observed absence. The expectation standard is the most formally specified in the dataset; the gap diagnosis mirrors it with equivalent exactness. |
| INT-06 | 12 | 17 | High Expectation / High Gap | Gap-dominant | Gap observations exceed expectation articulations by five points. GAP-BUS reaches maximum salience (4) and anchors the most theoretically elaborated gap diagnosis in the dataset. The participant's distinction between IT security and cybersecurity as professional identities constitutes the case's organizing argument. |
| INT-07 | 11 | 15 | High Expectation / High Gap | Gap-dominant | Gap observations exceed expectation articulations by four points. GAP-BUS reaches maximum salience (4) and co-occurs with high TFR-PROB and GAP-MYO. The gap-dominant configuration reflects a leader whose institutional experience — including a realized 2023 breach — centres on the consequences of professional shortfall rather than prescription of the idealized cognitive standard. |

*Note. COG Score and GAP Score are aggregate salience totals derived from the Leadership Risk Cognition Codebook (LRCC; maximum 28 per domain). This study categorizes High Expectation as COG ≥ 11 and High Gap as GAP ≥ 12. Sub-pattern classification based on the absolute difference between COG and GAP scores: Symmetric = |COG – GAP| ≤ 1; Near-symmetric = |COG – GAP| = 1–2; Gap-dominant = |COG – GAP| ≥ 3. No case falls in a low-gap quadrant.*

### 4.8.3.2 Enterprise Risk Framing (ERF Domain)

All seven leaders frame cybersecurity risk in enterprise terms. Financial impact (ERF-FIN)

appears in every case and functions as the universal referent for risk severity and the primary admission



criterion for escalation to senior leadership. The universality of this framing holds across sectors, organizational size, and governance maturity level.

Leaders in regulated sectors elaborate the financial frame through regulatory and audit exposure. INT-01, operating in the iGaming sector, records ERF-REG at salience 3 and describes a convergence of cybersecurity and compliance risk — virtually any security breach becomes a potential licensing exposure at the same time, elevating cyber risk to board-level visibility. INT-03, in multinational banking, identifies regulatory examination and audit scrutiny as the primary organizational attention-setters, with financial magnitude as the secondary escalation trigger. In both cases, regulatory framing amplifies rather than replaces financial framing.

INT-02, INT-04, and INT-06 each record ERF-SURV at salience 3, though the survival frame operates differently across cases: INT-02 organizes risk around service viability and client retention, INT-04 applies a binary executive filter — Does this risk jeopardize organizational survival? — and INT-06 frames survival as a strategic consequence of the risk–reward equation, the only case in which strategic framing (ERF-STRAT, Sal. 3) leads over all other ERF dimensions.  Regulatory framing is structurally absent from INT-04 (salience 0), consistent with a lightly regulated SaaS operating environment.

INT-05 presents the most distinctive ERF profile. Rather than organizing risk by category — financial, regulatory, or reputational — the participant applies an explicit likelihood × impact × controls formula in which risk type is analytically subordinate to methodological discipline. ERF codes register at salience 1 or below across all dimensions. This low aggregate ERF salience reflects a methodological rather than categorical orientation: the participant treats the formula as the risk definition, rendering categorical labelling secondary.

Reputational framing (ERF-REP) appears in INT-01 (Sal. 3), INT-02 (Sal. 2), and INT-07 (Sal. 2). INT-01 provides the most elaborated reputational evidence, describing reputational consequence as



potentially existential even when financial impact is technically limited. ERF-REP is absent from INT-03, INT-04, INT-05, and INT-06, indicating that several leaders organize risk primarily through financial, operational, or strategic lenses without foregrounding stakeholder perception as a distinct risk dimension. Across all seven cases, leaders position cybersecurity as one component of a multi-domain enterprise risk architecture rather than as its primary driver.

### 4.8.3.3 Risk Governance and Process (RGP Domain)

All seven leaders describe recognizable risk governance structures. Tiered escalation (RGP-ESC) appears in every case and carries the highest salience of any RGP code across the dataset. Leaders describe functionally analogous but structurally differentiated escalation pathways: a dual committee track (INT-01), domain-differentiated escalation triggers (INT-02), cross-functional risk committee integration (INT-03), triage-before-escalation filtering (INT-04), three distinct escalation routes (including board-initiated risk inquiry) in INT-05, direct escalation to the CEO and two senior technical leaders through a consistently applied but undocumented process (INT-06), and tiered severity escalation intensified following a realized breach (INT-07).

Likelihood × impact evaluation (RGP-LXI) appears in all seven cases, though the operationalizations differ substantially. INT-05 applies a personal formula: impact × likelihood, minus effective controls, adjusted for organizational risk tolerance. INT-01 describes a two-stage committee rating system with cross-functional validation. INT-03 describes a governance model in which dedicated risk officers assume quantitative **L × I** responsibility, providing financial and frequency data, while cybersecurity professionals contribute technical and qualitative input. INT-04 confirms collaborative financial impact evaluation through a structured triage process, without explicit probability notation. INT-02 acknowledges an **L × I** evaluation procedure but describes it as informal, undocumented, and insufficiently quantified. INT-06 acknowledges the absence of a formal likelihood methodology; instead,



the participant defaults to treating all risks as likely until they can prove otherwise. INT-07 describes **L × I** evaluation as aligned with the participant's personal risk formulation but not operationalized as a quantitatively codified institutional process. The data indicate that all seven organizations nominally apply **L × I** logic, but the depth, ownership, and codification of that application vary substantially across cases.

Cross-functional collaboration (RGP-COLL) appears across all seven cases at salience 1–3. Participation structures vary: COO, CTO, and Chief Compliance Officer involvement in committee validation (INT-01); domain-differentiated involvement by risk category (INT-02); risk officers supplying quantitative data that the cybersecurity team cannot independently generate (INT-03); CEO-level involvement at high-severity threshold in a compressed small-organization governance model (INT-04); coalitions of privacy, audit, and risk management groups in board-facing escalation (INT-05); organization-wide risk signal escalation embedded in team culture, with all members required to surface risk indicators regardless of function (INT-06); and cross-functional alignment with institutional safety and enterprise risk functions (INT-07).

Risk acceptance logic (RGP-ACC) appears explicitly in INT-01 (gold/silver/bronze tiering with documented accountability), INT-03 (four-option decision flow: mitigate, decommission, remediate, accept), INT-05 (board-owned risk tolerance as the organizational driver), and INT-06 (treat as likely until proven otherwise, as an operative cultural default). INT-07 records the risk acceptance logic at low salience—present but not formally codified. INT-02 and INT-04 treat acceptance implicitly, through leadership judgment rather than documented frameworks. Governance formalization ranges from high (INT-01, INT-03, INT-05) through moderate (INT-04, INT-07) to moderate-low (INT-02, INT-06).



**4.8.3.4 Expected Cognitive Work (COG Domain)**

All seven leaders articulate expectations regarding the cognitive work cybersecurity professionals should perform. Holistic, enterprise-wide reasoning (COG-HOL) appears in every case and reaches salience 3 or higher in five of the seven, with INT-04 recording the maximum salience of 4. Leaders expect professionals to synthesize technical findings with organizational context and produce an analysis that executive decision-makers can use without further translation. INT-05 records the lowest COG-HOL salience (1) in the dataset, anchoring the expectation in proportional **L × I** calibration rather than architectural scope; a difference in the standard applied, not in its height.

Likelihood estimation (COG-LXI) appears in all seven cases at salience 2 or 3. Leaders expect professionals to estimate the likelihood and defend that estimate with evidence. INT-05 frames this expectation most explicitly, stating that professionals must be able to account for the evidence on which they place their likelihood estimate, not merely state its outcome. INT-01 and INT-03 record COG-LXI at salience 2, where governance structures — committee validation and risk officer delegation, respectively — share or absorb the likelihood estimation function.

Business impact reasoning (COG-IMP) appears across all seven cases. INT-04, INT-05, and INT-06, all of which either lack dedicated risk officer structures or operate in environments where practitioners must influence executive decisions directly, record COG-IMP at salience 3. INT-03, where risk officers supply financial and frequency data, records COG-IMP at salience 1. The data indicate that this expectation is strongest when no structural support for impact quantification exists. INT-02 records COG-IMP at salience 2 and describes the expectation that professionals translate operational deficiencies into profit-and-loss terms. INT-07 records COG-IMP at salience 2, with the expectation scoped to senior staff who are required to produce business cases and justify risk trade-offs in organizational terms.



Tiered options reasoning (COG-OPT) appears at moderate salience in INT-01 (Sal. 2), INT-02 (Sal. 1), and INT-03 (Sal. 2). Leaders in these cases expect professionals to present alternative remediation options with associated risk trade-offs rather than single-path recommendations. COG-OPT registers at salience 0 in INT-04, INT-05, and INT-06, and at salience 1 in INT-07, where triage logic, the **L × I** formula, and seniority-stratified expectations, respectively, serve or limit the options-evaluation function.

Strategic alignment (COG-STRAT) appears at salience 3 in INT-06, at salience 2 in INT-01, INT-02, INT-03, INT-04, and INT-07, and at salience 1 in INT-05.

### 4.8.3.5 Observed Professional Gaps (GAP Domain)

Leaders report substantive professional deficiencies across all seven cases. Technical problem framing (TFR-PROB) registers the highest aggregate salience of any code in the dataset: 20 out of a maximum 28 across seven cases. It appears at salience 3 or higher in six of the seven cases — INT-01, INT-02, INT-03, INT-05, INT-06, and INT-07 — and at salience 2 in INT-04. Leaders describe a consistent pattern in which cybersecurity professionals define risk events as technical problems — proposing tool acquisition, patch deployment, or control gaps as the problem statement — rather than framing them as organizational risk events requiring likelihood estimation, impact analysis, and business-relevant communication.

Business-value blindness (GAP-BUS) appears in all seven cases, with an aggregate salience of 22/28. GAP-BUS reaches Sal. 4 in three cases (INT-03, INT-06, and INT-07). INT-03 records maximum salience (4) and the participant positions this as the primary explanatory construct for all observed professional shortfalls, describing the failure as both attitudinal and structural. INT-06 and INT-07 also record GAP-BUS at maximum salience (4). Across these three cases, participants elaborate on business-value blindness as the primary explanatory construct and the dataset's most severe professional gap. INT-01 and INT-02 record salience 3. INT-02 estimates that 90% or more of practitioners require



leadership to perform risk translation on their behalf and reports being unable to recall a professional submission adequately prepared in risk terms. INT-04 and INT-05 record GAP-BUS at salience 2, where the primary observed gap involves myopic specialization and L × I calibration failure, respectively.

Myopic, siloed thinking (GAP-MYO) appears in all seven cases at aggregate salience 18/28. INT-03, INT-04, INT-06, and INT-07 record salience 3. INT-04 describes specialization-induced myopia as an industry-wide structural pattern, using pre-formed language — "forest from the trees" — prepared before the interview, indicating a deeply held and broadly applied conviction. INT-03 frames this pattern as a failure to take a holistic risk management perspective, leading professionals to focus on narrow, minute-level technical concerns. INT-01, INT-02, and INT-05 record GAP-MYO at salience 2.

Absence of structured L × I reasoning (GAP-LXI) appears in six of seven cases. INT-02, INT-03, INT-05, and INT-06 each record a salience score of 3; INT-04 and INT-07 record a salience score of 2. INT-01 records salience 0 — the single exception in this code, consistent with the two-tier committee validation structure that distributes and compensates for individual L × I reasoning deficits. Entry-level professional immaturity (GAP-ENTRY) appears in four cases, with INT-05 recording the strongest evidence (Sal. 3), including observations from national sector-level professional forums indicating low professional community engagement with risk formalization. Scale miscalibration (GAP-SCAL) appears in four cases, with INT-02 providing the most elaborated evidence: the participant describes multiple concrete instances in which professionals assigned high-severity ratings based on the presence of a vulnerability or control gap, without contextualizing the finding within enterprise-scale likelihood and impact assessment.

### 4.8.3.6 Developmental Patterns (DEV Domain)

All seven leaders describe risk judgment as an acquired rather than innate professional capability. They identify developmental pathways through two primary mechanisms — formal training



and experiential learning — though they differ in which mechanism they emphasize and in how they assess its sufficiency.

Formal training (DEV-FORM) receives endorsement in all seven cases. INT-05 records the highest salience in the dataset (4) and prescribes professional certification in risk, privacy, and audit disciplines as the primary developmental remedy, noting that current formal training is "fairly weak." INT-01, INT-02, and INT-06 each record DEV-FORM at salience 3. INT-02 specifies a curriculum — business goal comprehension, financial literacy, risk framework literacy, and cost-benefit analysis — and identifies the appropriate delivery stage as the point of professional entry. INT-06 prescribes the most philosophically grounded curricular reform in the dataset, arguing that training must begin with the foundational concepts of 'cyber' and 'technology' as risk management constructs rather than technical identity markers. INT-03, INT-04, and INT-07 record DEV-FORM at salience 2 and assess formal training as capable of establishing conceptual foundations but insufficient for developing the contextual judgment the role requires.

Experiential learning (DEV-EXP) appears across all seven cases. INT-04 and INT-07 record DEV-EXP at salience 3. INT-04 describes the transition from technical to risk-based cognition as a mid-career developmental milestone achieved through organizational exposure rather than training participation. INT-07 describes this same transition as an organizational socialization process supported by cross-domain mentorship and business-facing exposure. INT-03 and INT-01 record salience 2; INT-03 endorses on-the-job learning as the primary mechanism for developing real risk judgment, while INT-01 treats experience as one channel within a multi-pathway developmental model. INT-06 records salience 2, framing mentored experiential learning as safe failure management rather than skill transfer. INT-02 and INT-05 record DEV-EXP at salience 1 and explicitly subordinate experience to formal training as a developmental mechanism.



Executive mandate (DEV-EXEC) appears in INT-01 (Sal. 2), INT-02 (Sal. 1), INT-03 (Sal. 1), INT-04 (Sal. 2), INT-06 (Sal. 2), and INT-07 (Sal. 1). INT-04 describes executive mandate as the primary organizational driver of risk-language adoption, indicating that professionals adopt risk-oriented language and reasoning because organizations demand it rather than through intrinsic professional motivation. INT-06 treats organizational culture as a developmental asset while simultaneously identifying the CISO scapegoat dynamic as a structural barrier to field-level progress—a dual rhetoric in which executive mandate both enables and constrains professional development. INT-07 records DEV-EXEC at salience 1, noting descriptively that institutional expectations shifted following the breach without framing that shift as a developmental theory. INT-05 records DEV-EXEC at salience 0 and does not invoke organizational culture or executive mandate as a developmental mechanism.

Cross-domain exposure (DEV-MULTI) appears at salience 2 in INT-01, INT-05, and INT-07, functioning as a complementary developmental pathway in each case. In INT-01, cross-domain exposure operates alongside formal training and experiential learning within a multi-pathway model. In INT-05, engagement with risk and privacy functions provides a compensatory, informal pathway, while formal credentialing remains the primary prescription. In INT-07, the participant identifies exposure across institutional risk, safety, and enterprise functions as a substantive corrective to siloed professional development. No leader identifies additional technical cybersecurity training as the corrective intervention for the professional gaps they describe.

### 4.8.3.7 Structural Typology

The Expectation–Gap Tension Index, calculated from CAP Section 4 for each case, classifies six interviews as High Expectation / High Gap (COG ≥ 11; GAP ≥ 12) and one interview — INT-02 (COG 9 / GAP 18) — as Low Expectation / High Gap. The High Expectation / High Gap cases include INT-01 (COG 13 / GAP 13), INT-03 (COG 13 / GAP 16), INT-04 (COG 12 / GAP 12), INT-05 (COG 12 / GAP 13), INT-06



(COG 12 / GAP 17), and INT-07 (COG 11 / GAP 15). INT-02 (COG 9 / GAP 18) forms the only Low Expectation / High Gap case, reflecting a leader whose gap diagnosis substantially exceeds her articulation of the expected cognitive standard and whose accumulated practitioner-facing experience produces high GAP elaboration relative to a comparatively compressed COG profile.

Within the High Expectation / High Gap quadrant, two sub-patterns appear across the six qualifying cases. INT-01 (COG 13 / GAP 13) and INT-04 (COG 12 / GAP 12) exhibit symmetric profiles; both participants articulate cognitive expectations and observed practitioner gaps with comparable elaboration and salience. INT-03 (COG 13 / GAP 16), INT-06 (COG 12 / GAP 17), and INT-07 (COG 11 / GAP 15) exhibit gap-dominant profiles, in which observed practitioner limitations substantially exceed the articulated cognitive standard, with gap observations carrying greater frequency, elaboration, and salience than expectation articulations. INT-05 (COG 12 / GAP 13) presents a near-symmetric configuration with slight gap dominance. No case falls in either low-gap quadrant.

### 4.8.3.8 Cross-Case Structural Pattern

The cross-case data indicate a consistent structural configuration across all seven leadership perspectives. Leaders in sectors ranging from iGaming and financial services to managed services, cybersecurity SaaS, and post-secondary education all articulate elevated cognitive expectations for cybersecurity professionals (COG scores: 9–13/28) and simultaneously report substantial and persistent failure to meet those expectations (GAP scores: 12–18/28). This co-occurrence holds across differences in organizational size, governance formalization, and sector regulation density.

To facilitate structured comparison across cases, Table 34 summarizes selected cross-case thematic features derived from the Leadership Risk Cognition Codebook (LRCC). The table consolidates the presence or absence of key risk-definition constructs, probabilistic framing elements, assumptions



regarding professional differentiation, and references to formalized training structures across the seven finalized Case Abstraction Profiles (CAPs).

**Table 34**

*Cross-Case Salience Summary for L × I Framing and Training*

| CAP | RGP-LXI | COG-LXI | COG-IMP | DEV-FORM | Explicit Personal L × I Formula |
|-----|---------|---------|---------|----------|--------------------------------|
| INT-01 | 3 | 2 | 2 | 3 | 0 |
| INT-02 | 2 | 2 | 2 | 3 | 0 |
| INT-03 | 2 | 2 | 1 | 2 | 0† |
| INT-04 | 2 | 2 | 3 | 2 | 0 |
| INT-05 | 3 | 3 | 3 | 4 | 1 |
| INT-06 | 2 | 3 | 3 | 3 | 0 |
| INT-07 | 2 | 3 | 2 | 2 | 1 |

Note. Values represent LRCC salience scores extracted from each Case Abstraction Profile (CAP). Cross-case synthesis includes seven finalized Case Abstraction Profiles (CAPs) from INT-01 through INT-07. "Explicit Personal **L × I** Formula" is coded as 1 only when the participant provides an explicit personal definition of risk as likelihood × impact (or an equivalent multiplicative formulation); 0 indicates no personal definition stated.

† INT-03 describes likelihood × impact as structurally delegated to risk officers rather than articulated as a personal operating definition.

In the ERF domain, financial impact functions as the universal referent for risk severity across all seven cases, with regulatory exposure adding salience in highly regulated sectors and methodological framing replacing categorical framing in INT-05. In the RGP domain, escalation structures are universally present, but operationalization of **L × I** evaluation differs substantially, from a formally delegated quantitative function (INT-03) to an informal, undocumented procedure (INT-02). In the COG domain, holistic integration (COG-HOL) appears in most cases. It reaches maximum salience in INT-04, while INT-05 and INT-07 place greater emphasis on explicit likelihood-by-impact reasoning.

In the GAP domain, technical problem framing (TFR-PROB) has the highest aggregate salience in the dataset (20/28) and appears at salience 3 or higher in six of the seven cases. Business-value blindness (GAP-BUS) follows at 22/28; INT-03, INT-06, and INT-07 each record maximum salience (4),



reinforcing business-value blindness as the most severe and cross-sectorally consistent professional gap in the dataset. Myopic reasoning (GAP-MYO) appears in all seven cases at an aggregate salience of 18/28. These three codes co-occur consistently across participant accounts. In the DEV domain, leaders divide between a formal training and credentialing emphasis (INT-01, INT-02, INT-05), an experiential learning emphasis (INT-03, INT-04, INT-07), and a bifocal model combining curricular reform with mentored experiential judgment development (INT-06), with no leader identifying additional technical cybersecurity training as the corrective intervention for the observed professional gaps.

The shared cross-case structure, therefore, remains straightforward: leadership expectations remain elevated, but professional discourse still falls short of those expectations in persistent, recognizable ways.

Across cases, leaders describe a field in which risk language and enterprise expectations are present, but the underlying cognitive work of likelihood-based reasoning, business translation, and comparative prioritization remains unevenly developed; descriptively, this cross-case pattern parallels the quantitative finding that professional competence depends not only on exposure to risk-related training, but also on whether risk becomes cognitively salient in practice. Chapter 5 examines the implications of this structural alignment across the quantitative and qualitative results.

### 4.9    Hypothesis Testing Summary

This Section summarizes the results of hypothesis testing based on the quantitative and qualitative analyses reported in this chapter. Consistent with the study's mixed-methods design, the analysis evaluated the hypotheses according to the methods specified in Chapter 3. To test $H_{11}$ and $H_{13}$, the NLP analysis reported in Section 4.2 measured the structural representation of risk management within training frameworks. The Structural Equation Modelling results reported in Sections 4.5 and 4.6



tested the hypothesized relationship ($H_{12}$) between training and professional competence. Unless otherwise specified, hypotheses evaluated using Structural Equation Modelling (SEM) are based on the single-factor specification of risk management competence (Model 1), as described in Section 4.5.7. The results of the qualitative leadership interview analysis in Section 4.8 tested the expectation–capability alignment hypothesis ($H_{14}$). The control group comparison reported in Section 4.7.1 tested Hypothesis $H_{15}$. Hypothesis dispositions are reported descriptively and without interpretation.

### 4.9.1  Hypothesis $H_{11}$: Structural Integration of Risk Management in Training Frameworks

**$H_{11}$:** *Cybersecurity training frameworks that integrate risk management principles exhibit a statistically significant representation of risk-related Knowledge, Skill, and Ability (KSA) statements (IV1) relative to other competency domains.*

**Hypothesis $H_{11}$:** The NLP-based analysis of Task, Knowledge, and Skill (TKS) statements tested Hypothesis $H_{11}$ within the NICE framework (see Section 4.2). Risk-related content accounted for 4.5% of high-confidence semantically matched statements and ranked 18th among 29 competency domains. Although risk-related statements were clearly identifiable when present, their overall activation frequency was substantially lower than that of foundational and operational competency areas.

These results indicate that the NICE framework contains identifiable risk management content; however, the hypothesis predicted that frameworks integrating risk management would exhibit statistically significant representation relative to other competency domains—not merely above-baseline presence. The evaluative standard for $H_{11}$ is therefore proportional significance, given risk management's normative role in enterprise security doctrine. Under this standard, 4.5% representation — modestly above the 3.4% equal-distribution baseline — is insufficient to constitute meaningful structural integration when cybersecurity governance (CGOV) claims 18.8%, security operations (SOPS)



claims 11.7%, and systems security architecture (SYSA) claims 8.7%. RISK ranks 18th of 29 domains. The data fails to support the hypothesis because the observed representation level lacks the structural weight required by an integrated risk management framework.

Disposition: **Not Supported**

### 4.9.2    Hypothesis $H_{12}$: Translation of Training into Professional Risk Management Competence

*$H_{12}$: If cybersecurity training integrates risk management competencies (IV1x), then professionals demonstrate higher risk-based behaviours (DV1–DV4), because training increases the conceptual salience and perceived relevance of risk management (MeV1), thereby strengthening real-world risk decision-making competence.*

**Hypothesis $H_{12}$**: The Structural Equation Model (Model 1), detailed in Section 4.5, was used to test Hypothesis $H_{12}$.

The direct path from IV1x to *RM_Competence* was statistically significant (β = 0.406, p < .001). The indirect effect through MeV1 was statistically significant (β = 0.223, p < .001). The total effect was statistically significant (β = 0.629, p < .001).

Disposition: **Supported**

### 4.9.3    Hypothesis $H_{13}$: Risk Management Underrepresentation in Cybersecurity Curricula

*($H_{13}$). If cybersecurity curricula inadequately address risk management, then the RISK competency (IV1) will be significantly underrepresented relative to other competency domains, indicating a curricular gap.*

**Hypothesis $H_{13}$:** The NLP-based classification and competency frequency analysis tested Hypothesis $H_{13}$ by examining TKS statements within the NICE framework. The structural convergence documented in Section 4.2.1 justifies treating the NICE findings as indicative of the broader family of



frameworks. Results of the competency distribution analysis indicated that the RISK competency accounted for a substantially smaller proportion of total statements than other domains. The NICE framework underrepresents the RISK competency relative to other domains. This underrepresentation was consistent with the broader cross-framework comparison logic described in Chapter 3. Structural Equation Modelling results do not directly test Hypothesis $H_{13}$.

Disposition: **Supported (based on comparative frequency and proportional analysis)**

### 4.9.4  Hypothesis $H_{14}$: Organizational Expectation–Capability Alignment in Risk Management

$H_{14}$: *Organizations implicitly expect cybersecurity professionals to apply structured likelihood–impact risk reasoning consistent with formal risk management doctrine.*

**Hypothesis $H_{14}$**:  The qualitative data reported in Section 4.8 serve as the basis for testing Hypothesis $H_{14}$.   Using control variables for seniority and leadership status, we invited sufficiently senior leaders to an additional interview and asked them questions. Section 4.6 contains supplementary quantitative results regarding seniority and prior learning. The study includes these for completeness, but prioritizes other data as the primary tests for this hypothesis.

The analysis examined whether leaders articulated expectations that cybersecurity professionals should perform structured risk reasoning aligned with likelihood–impact analysis, residual risk evaluation, and executive-level comparability.

Across seven leadership cases, organizational expectations consistently reflected structured risk-reasoning requirements. In multiple cases, leaders explicitly referenced likelihood × impact calculations, residual control effects, and risk tolerance thresholds (INT-01, INT-02, INT-05). In other cases, while the organizational structure distributed the computations of probabilistic loss (Likelihood x Impact) across departments, management still expected cybersecurity professionals to estimate likelihood/impact and



deliver analytically defensible assessments (INT-03). Even in the least formally structured governance environment in the dataset (INT-04), professionals were expected to synthesize risk in executive-ready, business-aligned terms.

Leadership expectations, therefore, are aligned with formal risk management doctrine, including likelihood calibration, impact quantification, benchmarking, and governance integration. However, these expectations did not consistently align with demonstrated professional competence, as evidenced by practitioner gaps observed across all seven interview cases and corroborated by the SEM findings.

Disposition: **Supported (Expectation Confirmed; Alignment Gap Identified)**

### 4.9.5 Hypothesis $H_{15}$: Benchmarking Cybersecurity Professionals Against Non-Cybersecurity Professionals

$H_{15}$: *Cybersecurity professionals will demonstrate significantly higher risk management competence than non-cybersecurity professionals.*

**Hypothesis $H_{15}$**: The control group comparison reported in Section 4.7.1 tested Hypothesis $H_{15}$. The analysis employed a three-stage sequence: measurement validity and cross-group portability (Stage 1), comparison of central tendency (Stage 2), and comparison of distributional dispersion (Stage 3).

Stage 1 confirmed that the eight-item foundational risk-reasoning scale (ASMT_1–3, PRIO_1–3, ACTN_1–2) demonstrated acceptable measurement properties in both populations. Single-group confirmatory factor analyses yielded a strong fit for the cybersecurity professional group (CFI = 0.998, RMSEA = 0.040, SRMR = 0.034) and acceptable fit for the control group (CFI = 0.945, RMSEA = 0.049, SRMR = 0.062). Cross-group metric invariance was confirmed (ΔCFI = −0.006, within the |ΔCFI| ≤ 0.010 threshold per (Cheung & Rensvold, 2002)), establishing the minimum condition for legitimate cross-group comparison. Internal consistency was strong across the combined dataset (Cronbach's α = 0.83).



Stage 2 compared the mean composite scores between the groups. Cybersecurity professionals exhibited a mean composite score of 3.82 (SD = 0.777, Median = 4.00); the control group exhibited a mean score of 3.72 (SD = 0.402, Median = 3.75). A Welch two-sample t-test indicated that this difference was not statistically significant (t = 1.271, p = .205). The associated effect size was trivial (Cohen's d = 0.16), falling below the conventional threshold for a small effect. Item-level testing across all eight indicators revealed no significant group difference on six of eight items, except for ACTN_2 (p = .014) and ASMT_2 (p = .053). The dominant pattern across items was one of cross-group equivalence.

Figure 15 displays the distribution of composite risk-reasoning scores using boxplots. While median values are similar across groups, the cybersecurity group exhibits substantially greater dispersion. Figure 15 presents kernel density estimates of the same scores, showing a broader spread in the cybersecurity group than in the control group for the eight-item foundational risk-reasoning composite (RM8). Dashed vertical lines indicate group means. The cybersecurity and control groups



exhibit nearly identical mean scores but markedly different variances, consistent with the variance-ratio test reported in Section 4.7.1.3 (F = 3.74, p < .001).

Stage 3 revealed a statistically significant and substantively large difference in distributional dispersion between groups, independent of the non-significant difference in central tendency. The cybersecurity professional group exhibited a standard deviation of 0.777, compared to 0.402 in the control group, yielding a variance ratio of 3.74 (p < .001). This heterogeneity of variance indicates that, although group means were comparable, the cybersecurity professional group exhibited markedly greater individual-level

**Figure 15**

*Foundational Risk-Reasoning Scores by Group*

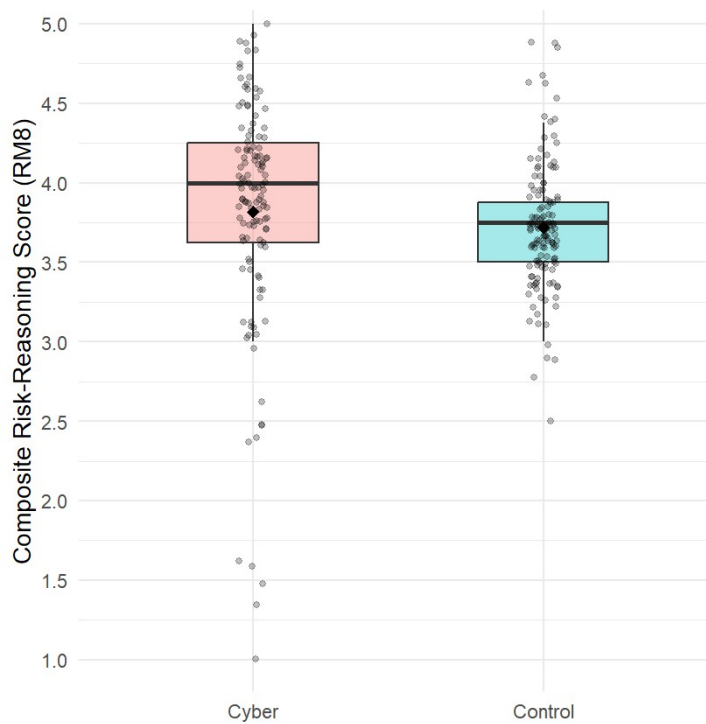

**Figure 15**

*Distribution of Foundational Risk-Reasoning*

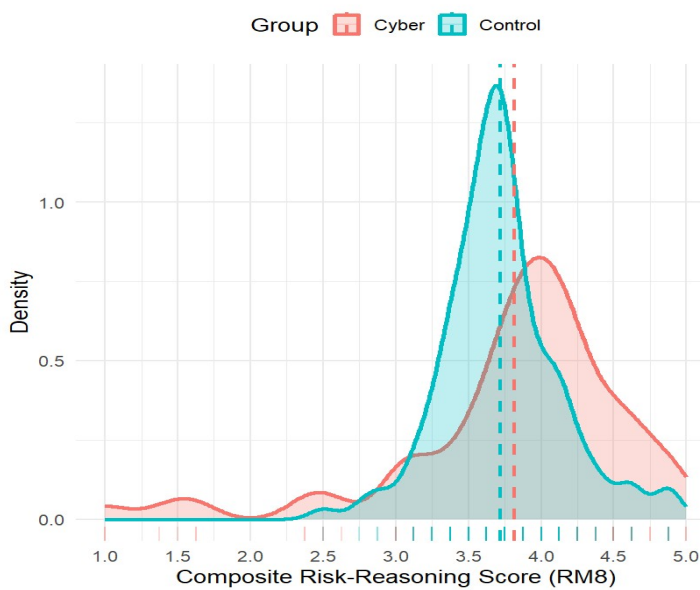



variability in foundational risk-reasoning responses.

The central hypothesis — that cybersecurity professionals would demonstrate significantly higher risk management competence than non-cybersecurity professionals — is not supported by the data. Neither the t-test nor the effect size estimate provides evidence of a statistically significant or practically meaningful group advantage. The finding that the control group demonstrates comparable mean-level foundational risk reasoning, as measured by a metrically valid and internally consistent scale, constitutes a substantive empirical result. The significant variance differential in Stage 3 constitutes a material contextual finding. Although $H_{15}$ remains unconfirmed, the greater dispersion among cybersecurity professionals demonstrates that their foundational risk reasoning is more heterogeneous—and foundational risk-reasoning responses vary substantially across individuals within that population—than that of the control group. Chapter 5 examines the theoretical implications of both the null mean result and the variance differential.

Disposition: **Not Supported**.

### 4.9.6 Summary of Supported and Unsupported Hypotheses

Table 35 summarizes the hypothesis dispositions. The analytical methods outlined in Sections 4.2, 4.5, 4.7, and 4.8 provided the basis for these evaluations. The study presents mediation results as statistical findings from the Structural Equation Model rather than defining them as standalone hypotheses.

**Table 35**

*Summary of Hypothesis Dispositions*

| Hypothesis | Description (Abbreviated) | Analytical Basis | Disposition |
|---|---|---|---|
| $H_{11}$ | Training frameworks that integrate risk management principles | NLP Content Analysis — NICE framework, TKS statement classification | **Not Supported** |
| | will exhibit a statistically significant representation of risk-related Knowledge, Skill, and Ability (KSA) statements (IV1) relative to other competency domains. | | |
| H₁₂ | Training → risk competence via conceptual salience (SEM, IV1x → MeV1 → *RM_Competence*) | SEM Model 1 — direct (β=.406), indirect (β=.223), total (β=.629); all p < .001 | **Supported** |
| H₁₃ | If cybersecurity curricula inadequately address risk management, then the RISK competency (IV1) will be significantly underrepresented relative to other competency domains, indicating a curricular gap. | NLP Classification + Competency Frequency Analysis, NICE Framework v2.0.0 | **Supported** |
| H₁₄ | Organizations expect structured **L × I** reasoning; expectation–capability gap. | Qualitative Leadership Interview Analysis (n=7) | **Supported (Expectation Confirmed; Alignment Gap Identified)** |
| H₁₅ | Cybersecurity professionals demonstrate higher RM competence than non-cyber professionals. | Cross-group CFA + Welch t-test + Variance ratio analysis (Section 4.7.1) | **Not Supported (null result theoretically informative; see Section 5.4)** |

The following chapter interprets these findings in relation to the study's theoretical framework, existing literature, and practical implications for cybersecurity training and leadership practice.

## 4.10   Chapter Summary

This chapter presented the study's empirical results using a mixed-methods design organized according to the sequence outlined in Chapter 3.

The chapter first reported findings from the NLP-based analysis of cybersecurity training frameworks, describing the frequency and distribution of risk management concepts within Task,



Knowledge, and Skill (TKS) statements. The analysis uncovers an empirical account of the structural gaps in how cybersecurity training artifacts represent risk management.

Survey data preparation and descriptive statistics followed, covering screening procedures, sample characteristics, and the observed indicators required for Structural Equation Modelling. These results establish the analytic sample and the measurement properties of the observed variables before model estimation.

Measurement model results were then reported, including standardized factor loadings, composite reliability estimates, and assessments of convergent and discriminant validity. The analysis evaluated risk management competence under both single-factor and second-order specifications. Chapter 5 examines the implications of this empirical compression, interpreting the observed 'factor collapse' as evidence of a holistic cognitive orientation among cybersecurity practitioners.

The chapter then presented results from the Structural Equation Model, including estimated direct, indirect, and total effects linking training exposure to risk-embedded content, perceived relevance of risk management, and risk management competence. This Section detailed the model fit indices, contrasted alternative measurement specifications, and outlined the impact of control variables.

Supplementary quantitative analyses were also reported, including descriptive group comparisons between leadership and non-leadership participants.

Finally, this chapter presents the qualitative findings from leadership interviews, highlighting the gap between executive expectations and practitioner reality. The thematic analysis identified key themes—including risk understanding, governance, prioritization, control selection, communication, and capability development—and summarized them descriptively to provide context for the quantitative results.



Taken together, this chapter reports the study's empirical findings across framework-level, respondent-level, and leadership discourse data sources. The following chapter integrates these results and examines their theoretical and practical implications for cybersecurity training, risk management competence, and leadership practice.



**Chapter 5 – Discussion, Implications, and Conclusions**

This chapter interprets the empirical findings presented in Chapter 4 in light of the theoretical

framework, research questions, and hypotheses that structured this inquiry. It draws out their

implications for training design, professional governance, and organizational risk management practice.

**5.1    Overview and Integrative Interpretation**

This study set out to resolve a practical paradox at the centre of contemporary cybersecurity

governance. Organizations increasingly expect cybersecurity professionals to perform enterprise risk

management: to estimate likelihood, assess strategic impact, compute expected loss using those joint

estimates, prioritize competing risks, and communicate in the risk-calculus language of enterprise risk

frameworks. Yet major cyber failures continue to occur despite substantial technical investment, and

the post-incident record consistently reveals not a shortage of technical controls but a failure of risk

judgment. This dissertation investigates the structural origin of that failure.

The answer this study returns is direct: the cybersecurity profession has organized itself around

its training, as a threat management discipline that borrows risk vocabulary, rather than as a formal risk

management discipline. This insight is not an observation about individual professionals or a criticism of

training effort. It is a structural claim about how training architecture constructed the profession, how

workforce frameworks encode professional competence, and what cognitive architecture those

frameworks install in the practitioners they shape

Four independent empirical analyses converge on this conclusion from different observational

angles. The NLP analysis of the NICE Framework v2.0.0 establishes it at the institutional level: the

framework that defines cybersecurity workforce competence does not encode risk-calculation cognition

at the operational level: the joint estimation of event likelihood and strategic organizational impact, and



the computation of expected loss from those two assessments. The SEM analysis establishes this at the cognitive level: practitioners internalize the framework's undifferentiated architecture, producing a single, compressed global factor, whereas formal risk management theory predicts four analytically distinct behavioural domains. The control-group comparison establishes that cybersecurity professionals demonstrate no measurable advantage over the general professional population in foundational risk reasoning. The leadership interviews establish it at the organizational level: senior leaders expect likelihood-impact reasoning from their teams, while the majority cannot themselves articulate the computational basis of that reasoning, because the same institutional architecture formed them.

The convergence of these four findings across independent methods is not incidental. It is evidence of a single causal mechanism operating across three institutional levels. This chapter names that mechanism the three-level institutional explanation and uses it as the organizing spine of the argument throughout.

### 5.1.1  *The Three-Level Institutional Explanation*

The explanation operates across three mutually reinforcing levels — institutional, cognitive, and organizational — each anchored in a distinct body of empirical evidence and connected to the next by a directional handoff supported by the data.

#### 5.1.1.1  LEVEL 1 — Institutional Architecture

*Primary evidence*: NLP analysis of NICE Framework v2.0.0 (2,111 TKS statements)

Cybersecurity workforce frameworks encode threat management within a deterministic cognitive architecture. The word "likelihood" appears zero times across 2,111 TKS statements. The word "probability" appears zero times. The Likelihood x Impact joint construct, the definitional core of enterprise risk as a computationally tractable concept, appears zero times. Critically, NICE frames



consequences (impacts) at the technical level—what the threat does to a system—rather than at the enterprise level that risk governance requires (data exfiltration, service disruption, system compromise), or at the enterprise-strategic level that formal risk management requires. Consequence recognition in NICE asks: what happens to the system? Enterprise risk management asks: What does this cost the organization, relative to its strategic objectives, financial exposure, and regulatory obligations? These are not the same question, and NICE provides no TKS-level requirement to translate. This architectural absence constitutes Level 1: the institutional architecture that determines what training cybersecurity professionals receive and, crucially, what frameworks never require them to reason about.

↓ *Level 1 produces the cognitive architecture that Level 2 practitioners internalize.* ↓

### 5.1.1.2 LEVEL 2 — Cognitive Internalization

*Primary evidence*: SEM Model 1 path model and DV collapse

Practitioners internalize exactly the architecture that Level 1 installs. The SEM training-salience-competence pathway demonstrates that training exposure elevates overall risk management competence (total standardized beta = 0.629). But the structure of this competency directly mirrors the framework's architecture: the four theoretically distinct behavioural domains of risk management collapse into a single, undifferentiated global factor. Practitioners develop a stronger global orientation toward risk management when they receive risk-integrated training; they do not develop the differentiated risk-calculation reasoning: likelihood estimation, strategic impact assessment, expected-loss computation, and comparative prioritization — across assessment, prioritization, communication, and action that enterprise practice requires, because the framework that shaped their training never differentiated or operationalized these operations.

↓ **Level 2 produces the competence profile that Level 3 organizations receive.** ↓



### 5.1.1.3  LEVEL 3 — Organizational Consequence

*Primary evidence*: Control group comparison (d = 0.16, p = .205) and leadership interviews (n = 7)

Organizations receive practitioners whose actual competence does not match what the training and credentialing system claims to produce. The control-group comparison establishes this at the population level: cybersecurity professionals are statistically indistinguishable from general professionals in foundational risk reasoning (Cohen's d = 0.16, p = .205). The leadership interviews establish it at the organizational expectation level: six of seven senior executives operate in the High Expectation / High Gap typology, expecting likelihood-impact reasoning from teams that the training system never systematically equipped them to perform. The assumption gap extends upward into leadership cognition itself: only 2 of 7 leaders articulate an explicit personal likelihood × impact formula, though all expect this reasoning from their teams. The same Level 1 architecture that moulded the teams now constrains the leaders who manage them.

### 5.1.2  The Convergence Argument

The theoretical power of this study derives not from any single finding but from the convergence of four independent methods on the same structural conclusion. The NLP analysis is not sample-dependent: the absence of likelihood and probability from 2,111 TKS statements is an architectural fact, not a statistical estimate. The SEM operates on survey data from 126 practitioners across Europe and North America (from different sectors). The control group comparison uses a different population, a different instrument administration, and a different statistical framework. The leadership interviews apply qualitative coding to seven case profiles. These four methods share no data, instruments, or analytical procedures. They share a conclusion.



That shared conclusion is not coincidental confirmation. It is evidence of a causal mechanism: Level 1 encodes threat management as risk management; Level 2 practitioners internalize that encoding faithfully; Level 3 organizations receive practitioners whose competence profile reflects the encoding, while expecting the profile that the vocabulary implies. The system is self-concealing. The same training architecture that creates the absence of risk-calculation cognition: the combined capacity to estimate likelihood probabilistically, assess impact at the strategic-enterprise level, and compute expected loss from those joint estimates, also provides the vocabulary that makes it invisible. Every actor operates consistently within the architecture, which is why the gap persists despite extensive professional investment and genuine organizational commitment to risk management.

Across all four evidence streams, risk management in cybersecurity functions as a linguistic construct rather than a consistently applied framework for risk-calculation, cognition, and decision-making: the vocabulary is embedded in training and professionally internalized; the calculus it implies is not.

### 5.1.3    LEVEL 1 — Institutional Architecture

This Section establishes what the institutional training architecture — meaning the authoritative, profession-wide frameworks that formally define cybersecurity workforce competence, principally the NICE Framework v2.0.0 — was built to produce. The NLP analysis of NICE v2.0.0 reveals that the operational vocabulary of cybersecurity workforce development does not include risk-calculation cognition (expected loss) reasoning. This result is the logical foundation of the three-level causal chain: the framework encodes threat management, and everything downstream follows.

### 5.2    The Training Architecture Problem (RQ1, RQ3; $H_{11}$, $H_{13}$)



Research Questions 1 and 3 (RQ1 & RQ3) address the structural representation of risk management in cybersecurity training frameworks and the identification of curricular gaps. $H_{11}$ predicted that training frameworks would exhibit statistically significant representation of risk-related KSA statements relative to other competency domains. $H_{13}$ predicted that the framework would significantly underrepresent the RISK competency relative to other domains, indicating a curricular gap.

### 5.2.1   *What the Evidence Demonstrates*

The NLP analysis of the NICE Framework v2.0.0 classified all 2,111 TKS statements across 29 competency domains using semantic embeddings based on semantic distance, with a high-confidence threshold of >=0.75. Risk management content produced 4.5% of high-confidence matches, ranking 18th among the 29 cybersecurity competency domains. This result directly addresses $H_{11}$: the framework does not significantly represent risk management content relative to other competency domains. It ranks in the lower third of all domains. $H_{11}$ is not supported. The evidence supports $H_{13}$: the NICE framework significantly underrepresents the RISK competency domain compared with the other domains, thereby confirming a critical curricular gap.

The finding of proportional underrepresentation, however, is not the primary result. The primary result is architectural. Across all 2,111 TKS statements, "likelihood" appears zero times. "Probability" appears zero times. The Likelihood x Impact = Risk compound construct, which, following Courtney (1977) and FIPS PUB 65 (Department of Commerce, 1979), is the definitional foundation of enterprise risk as a computationally tractable concept, appears zero times. No TKS statement requires likelihood estimation or probability modelling. No TKS statement requires probability risk-calculation cognition. No TKS statement requires formal risk prioritization. No TKS statement requires risk communication grounded in a likelihood-impact calculation. Impact language appears, but always as a



standalone consequence descriptor, never paired with a probability weight. The framework also contains no TKS-level requirement for risk-based prioritization across its full statement corpus, not merely within risk-labelled statements.

The 70.1% high-confidence match rate within identified RISK statements confirms that the classification instrument successfully identifies risk-related content when it exists. The absence of likelihood and probability is therefore not a methodological artifact: the framework does not contain these constructs, not merely the words for them.

### 5.2.2    What the Evidence Strongly Suggests

The zero-frequency results for likelihood and probability support a precise theoretical claim: NICE encodes consequence recognition and threat response within a deterministic cognitive architecture. What the framework teaches practitioners to ask is: "What bad things could happen, and what do we do about them?" The framework never requires practitioners to ask: 'How likely is each bad thing, how severe would it be in terms of harm to the organization's strategic objectives and financial position, and how does the product of those two estimates compare across competing risks to determine where to invest limited resources?' The first question produces threat-responsive practice. The second produces risk-reasoning practice. These are not versions of the same activity. They are systematically different cognitive operations producing systematically different decisions, particularly under resource constraints, where cybersecurity decisions can not mitigate everything.

The framework encodes three categories of activity related to risk but not constituting risk in the enterprise sense. First, consequence recognition: TKS statements reference impact, harm, damage, and effect, the outcome dimension of risk stripped of probability weighting. Second, threat and vulnerability enumeration: identifying what could cause harm and where exposure exists. This threat



and vulnerability enumeration is the precondition to risk reasoning, not risk reasoning itself. Third, control specification: implementing safeguards driven by threat identification or compliance requirements. None of these three activities requires a probability estimate. Practitioners can perform all three activities entirely within a deterministic, threat-enumeration cognitive framework. What they collectively omit is what this study terms *risk-calculation cognition*: the sequential capacity to estimate event likelihood, assess impact at the strategic-enterprise level, and compute their product as an expected-loss score that supports comparative prioritization, rational control selection, and evidence-based risk communication.

What does it mean that a practitioner can fully satisfy all three of these professional obligations without ever estimating a probability, without ever translating system-level impact into enterprise-strategic terms, and without ever computing a risk score that combines those two assessments?  First, certification training teaches cybersecurity practitioners to implement security controls (firewalls, patches, access restrictions, etc.) based on two triggers: either they identified a threat, or a compliance rule requires it. Neither trigger involves calculating expected loss (i.e., the probability of an event).

Second, regarding strategic impact: NICE's consequence recognition vocabulary is system-scoped. Practitioners trained within it learn to describe what a threat does to a system: the data lost, the service disrupted, the control bypassed. Training architecture does not train them to translate that technical consequence into the enterprise terms that risk governance requires: the financial exposure it creates, the regulatory liability it triggers, the strategic objective it threatens, or the board-level significance it carries. A practitioner can accurately characterize a data breach's technical footprint but cannot estimate its expected organizational cost. The impact term in the risk formula requires enterprise-strategic scoping; the training framework provides system-technical scoping.



Third, regarding risk calculation: A practitioner trained in this architecture never assembles those two assessments, however imprecisely formed, into a calculated risk score using a risk matrix. The transition from qualitative assessment (Likely × Major) to quantitative expected loss (Risk Score = 64, classification: High, treatment: Mitigate) is the operation that makes risk *comparable* across heterogeneous threats on a common scale. Without this computation, practitioners cannot meaningfully compare a frequently occurring low-impact event against a rare catastrophic one. They cannot rationally decide whether to invest in likelihood-reducing controls (preventive) or impact-reducing controls (detective/corrective), because they have no calculation to determine which dimension, if reduced, would move the risk score below appetite. And they cannot communicate *why* a risk is high.  Without an **L × I** basis, 'high' is an assertion, not a calculation.

A practitioner can go through their entire professional life operating within a mental model that works like this: "Some technical risk could happen → here is what it would look like → here is the control that stops it (by reducing its likelihood and impact)." That is a deterministic chain — if X then Y — with no risk calculation anywhere in it. It feels like risk management because it uses risk vocabulary, but it is actually closer to a checklist or decision tree than to a formal risk calculation.  The three-point framework, lacking true expected loss calculation, means that a practitioner trained within this deterministic architecture is not a poor risk manager. They are not a risk manager at all in the enterprise sense — they are a threat responder whose training has never required them to ask "how likely is this threat, how impactful would it be (to the enterprise) should it materialize (given its likelihood and impact), what is the risk, and is it worth spending limited resources on compared to the next risk on the list?" That comparative, probability-weighted question is what formal risk management requires and what this training architecture never installs.



The correct theoretical claim is therefore not that NICE inadequately represents risk management. It is that NICE is not a risk management framework. It is a threat management framework that uses risk vocabulary at the category level while encoding threat management operations at the TKS level. NICE does not produce poor risk prioritizers. It produces practitioners for whom risk prioritization, as a formal risk-calculation operation — requiring likelihood estimation, strategic-impact assessment, and the joint computation of expected loss — does not exist as a trained cognitive concept. This distinction between a framework with a gap and one with a different architectural purpose carries profoundly different implications for practice and policy. The leadership evidence at Level 3 reinforces this distinction. Leaders assume the framework has produced structured risk reasoning in their teams precisely because it contains risk vocabulary — the same vocabulary the NLP analysis reveals is present at the category level and absent at the operational level. The assumption gap is therefore not a leadership failure; it is the predictable cognitive consequence of a framework designed to supply the language of risk without the architecture of its calculation.

### 5.2.3    What the Evidence Implies for Practice

If the dominant cybersecurity workforce framework prioritizes threat management over risk-calculation (expected-loss) reasoning, the training programs built on it will reproduce the same cognitive architecture. Adding more risk-labelled content to a framework that remains architecturally deterministic will not yield risk-calculation cognition—the capacity to estimate likelihood, assess strategic impact, and compute expected loss. It will produce more risk vocabulary layered onto the same threat-management base. The industry must architecturally revise the framework at the TKS statement level: explicitly operationalize likelihood estimation; compute joint risk (likelihood x impact); prioritize comparative risk using expected loss or equivalent decision-theoretic criteria; and communicate risk-



calculation (expected-loss) conclusions that distinguish high-probability/low-impact from low-probability/high-impact scenarios. These are not additions to the current framework. They are foundational operations that the current framework lacks.

### 5.2.4    *$H_{11}$ and $H_{13}$: Hypothesis Disposition*

**$H_{11}$ — Not Supported:** Risk management content accounts for 4.5% of high-confidence TKS classifications, ranking 18th of 29 competency domains. The prediction that risk-related KSA statements would be significantly more represented than those in other domains is not supported. The not-supported disposition is strengthened, not weakened, by the amplifying finding: the underrepresentation is structural, reflecting the complete absence of probabilistic reasoning from the operational vocabulary. The framework does not adequately represent risk management. It does not represent probabilistic risk reasoning at all (a prerequisite for risk-calculation (expected-loss) reasoning).

**$H_{13}$ — Supported:** The NICE framework significantly underrepresents the RISK competency domain relative to all other domains, a finding consistent with $H_{13}$. The gap is structural rather than proportional: it lies not in the quantity of risk-related content but in the cognitive architecture that embeds it. $H_{11}$ Not Supported and $H_{13}$ Supported are convergent, not contradictory. While NICE includes risk management content, the framework does not represent it to a significant extent relative to other domains (rejecting $H_{11}$) and significantly underrepresents it relative to domain expectations (supporting $H_{13}$). Both dispositions point to the same architectural conclusion.

### 5.2.5    *Level 1 → Level 2 Handoff*

The institutional architecture established in this Section does not remain confined to the document level. Practitioners trained within this certification training system internalize its cognitive structure. The SEM findings in Section 5.3 demonstrate this internalization directly: the training-



salience-competence pathway shows that professional cognition absorbs the framework's signal, and the collapse of the four risk management dependent variables into a single undifferentiated factor shows that this internal signal mirrors the framework's undifferentiated architecture.

### 5.3 Risk Vocabulary Without Risk Cognition: The SEM Findings (RQ2; $H_{12}$)

The SEM findings reported in this section constitute the Level 2 evidence base, demonstrating how practitioners internalize the institutional architecture encoded in workforce frameworks as a specific, measurable cognitive compression pattern.

#### 5.3.1 LEVEL 2 — Cognitive Internalization

This Section demonstrates what Level 1 produces in practitioner cognition. The SEM establishes that the training-salience-competence pathway is real and significant: training shapes competence. But this resulting competence mirrors the framework's structure. Practitioners internalize an undifferentiated global orientation toward risk rather than the four analytically differentiated competencies that formal risk management requires.

Research Question 2 (RQ2) asks how well cybersecurity training programs equip professionals with the competencies required for risk management in real-world settings. $H_{12}$ predicted that training integration of risk management competencies would predict higher risk-based behaviours, mediated through the conceptual salience of risk. The SEM results support this hypothesis, but the structure of this increased competence and the anomaly within the dependent variables yield a richer, more troubling theoretical finding than the hypothesis framework alone would generate.

#### 5.3.2 What the Evidence Demonstrates



The evidence bearing on RQ2 spans three dimensions: global model fit, structural path estimates, and mediation structure. The following sections evaluate each dimension in turn.

### 5.3.2.1 Model Fit

Table 24 in Chapter 4 reports the global fit of Model 1 (n = 126). This fit supports strong model acceptance. Scaled chi-square (547) = 745.984 (unscaled: 709.114); CFI = 0.990; TLI = 0.992; RMSEA = 0.049, 90% CI [0.038, 0.059]; SRMR = 0.079. The CFI and TLI both exceed the >=0.95 threshold for an acceptable fit and approach the >=0.99 threshold for an excellent fit. The RMSEA falls below the 0.050 boundary for close fit, with the entire 90% confidence interval below 0.060. The SRMR falls within the 0.080 conventional cutoff. In WLSMV estimation with ordinal indicators, the chi-square statistic is not the primary evaluation criterion because its power scales with n, and models of this complexity can produce significant chi-square values under good fit. The joint multi-index pattern supports model acceptance without reservation.

### 5.3.2.2 The Training-Salience-Competence Pathway

Following the conventions of Chapter 4, standardized path coefficients (std.all) define the relationships in this model. The training-to-salience path (a) was beta = 0.453, p < .001 (unstandardized: 0.715, SE = 0.098, z = 7.290). Training exposure is a significant and robust predictor of conceptual salience. The standardized coefficient of 0.453 indicates a nearly half-standard-deviation increase in salience per standard deviation increase in training integration. The path establishes that professional cognition absorbs the training signal; subsequent paths address the mechanism by which this absorption produces differentiated competence.

The salience-to-competence path (b) was beta = 0.491, p < .001 (unstandardized: 0.484, SE = 0.110, z = 4.422). Conceptual salience significantly predicts overall risk management competence. This finding is the central mediation finding: it is not training exposure per se that generates competence,



but the degree to which risk management frameworks become cognitively active in professional practice. A practitioner for whom risk management remains a peripheral concept, regardless of the volume of training received, will not demonstrate the same competence as one for whom the same training has generated genuine cognitive salience.

The direct training-to-competence path (c-prime) was beta = 0.406, p < .001 (unstandardized: 0.632, SE = 0.129, z = 4.888). The significant direct effect indicates partial rather than full mediation. Section 5.6 develops the theoretical interpretation of this direct path in conjunction with the leadership interview findings. Briefly stated: the direct path represents a competence-transmission mechanism that operates independently of explicit cognitive salience — one that the leadership data identify as organizational risk discourse: the ambient circulation of risk expectations in professional environments where leaders consistently invoke risk language without being able to specify the cognitive operations it implies. Training shapes competence through two channels simultaneously: one through deliberate cognitive activation and the other through organizational absorption. The partial mediation structure does not qualify the main finding; it provides evidence of a second developmental pathway operating in parallel. The indirect (mediated) effect was beta = 0.223 (p < .001; unstandardized: 0.346, SE = 0.074, z = 4.670), accounting for 35.5% of the total effect (0.223/0.629). The total effect was beta = 0.629, p < .001 (unstandardized: 0.978, SE = 0.137, z = 7.136): a one-standard-deviation increase in training integration is associated with a 0.63 standard deviation increase in overall risk management competence, a large effect by conventional benchmarks. The sensitivity model (Model 1_sens) produced structurally consistent results under full covariate saturation (a = 0.547, b = 0.295, p = .004; c-prime = 0.434; indirect = 0.162, p = .001), confirming the mediation effect is not an artifact of unmeasured confounds. Notably, years of professional experience did not significantly predict conceptual salience after controlling for training exposure ($\beta$ = −0.078, p = .458), which rules out the most plausible alternative interpretation of



the mediation finding: that salience—and through it competence—is simply a product of career tenure rather than training design. How training frames risk shapes its internalization as a cognitively central construct, regardless of how long a professional has been in the field.

### 5.3.2.3 Epistemic Compression: The Dependent Variable Collapse

The most theoretically consequential finding in the SEM is not a path coefficient. It is the structure of the dependent variable. The hypothesized Model 2 specified *RM_Competence* as a second-order construct with four, first-order factors: risk assessment (DV1), risk prioritization (DV2), risk-informed action (DV3), and risk communication/risk justification (DV4), the four analytically distinct stages of the formal risk management cycle. The empirical data did not support this differentiated structure. The 4 first-order factors produced near-zero or negative residual variances (Heywood cases), indicating the constructs do not exhibit independent variance beyond their shared common factor. The four domains collapse into a single global latent factor.

This result demands a conventional psychometric interpretation, suggesting that either common method variance is too high or the theory incorrectly framed the construct as multidimensional. Two features of the data argue against it. First, the AVE for *RM_Competence* (0.492) approaches the conventional threshold, and composite reliability (CR = 0.972) confirms strong internal consistency; the measurement properties do not indicate a poorly specified instrument. Second, and more decisively, the NLP analysis provides an independent, method-distinct line of evidence pointing to the same conclusion: NICE encodes no differentiation among risk assessment, prioritization, communication, and action at the TKS level. A measurement artifact cannot explain why an independent analysis of a policy document shows the same structural absence as the survey constructs do. The convergence of these two independent evidence sources establishes that the collapse is not a measurement failure. It is epistemic



compression: the compression of a theoretically differentiated construct into an undifferentiated global self-perception, caused by a training architecture that never operationalized the differentiation.

The term "epistemic" is deliberate. Epistemic compression is a finding about how professionals know what they know about risk management, not merely about how much they know.  Epistemic compression is a condition in which training produces a global positive orientation toward risk activity without producing differentiated cognitive schemas for its component behavioural domains. Practitioners do not differentiate risk assessment from risk prioritization, nor either from risk communication or risk-informed action, not because these are not genuinely distinct activities, but because their professional formation never required them to distinguish them. Each of these domains requires a distinct cognitive operation: risk assessment requires the joint estimation of likelihood and strategic impact; risk prioritization requires the computation of expected loss and the comparison of scores across competing risks; risk communication requires the articulation of that calculation in terms accessible to non-technical stakeholders; and risk-informed action requires the selection of controls that rationally reduce either the likelihood or impact term — or both — to bring the risk score within appetite. The training architecture never differentiated these operations, so practitioners never developed separate cognitive schemas for them. The DV collapse in the SEM is the cognitive mirror of the TKS structure in NICE: the framework's lack of differentiation reproduces itself within professional cognition.

### 5.3.3    Prior Application as a Boundary Condition on Training Efficacy

The control variable results yield a finding that qualifies and deepens the main mediation interpretation. The MeV1 equation incorporates two prior learning controls: CV2_RM_LRN1, which measures completion of formal risk management training, and CV2_RM_LRN2, which measures regular



application of risk management learning in practice. CV2_RM_LRN1 was non-significant ($\beta$ = 0.127, p = .337). CV2_RM_LRN2 was significant and substantively large ($\beta$ = 0.432, p = .006), approaching the magnitude of the primary training exposure path (IV1x $\rightarrow$ MeV1; $\beta$ = 0.453).

This contrast is theoretically consequential. Completing formal training does not, on its own, predict cognitive salience after controlling for training exposure. Regularly applying risk management learning does, at a near-equivalent magnitude to training itself. The implication is that receipt of training does not unconditionally activate the training-to-salience pathway. It requires application as the activating condition. Training that is received but not applied in practice does not generate the cognitive salience that drives the mediated competence gain; it produces competence only through the direct pathway, which the partial mediation structure documents separately.

This finding is convergent with the structured practice effect reported in Section 4.7 (Cohen's d = 1.394 between structured and non-structured practitioners). Both results point to the same mechanism: deliberate application of risk reasoning in authentic professional contexts is what converts training exposure into the differentiated cognitive schemas that enterprise risk governance requires. The training architecture provides the vocabulary. An application is what installs the cognition. When a practitioner merely completes and files a training program (rather than embeds it in supervised practice contexts), they activate only the weaker direct path to cognitive salience, and the epistemic compression finding explains precisely why: without application, the cognitive differentiation between assessment, prioritization, action, and communication never emerges.

For credentialing purposes, this result reframes the meaning of a prerequisite condition. The relevant question is not whether a candidate has completed prior risk management training. It is whether they have applied it. CV2_RM_LRN1 answers the completion question; the data show it does not predict salience. CV2_RM_LRN2 answers the application question; the data show it nearly replicates



the effect of current training exposure. Credentials that assess knowledge without requiring

demonstrated application, on the evidence of this finding, certify exposure rather than activation.

### 5.3.4    *What the Evidence Strongly Suggests*

The parallel between the NLP finding and the SEM DV structure is the strongest piece of

evidence in the study. It is no coincidence that the training framework fails to differentiate risk

assessment, prioritization, communication, and action at the TKS level, and that practitioners trained

within that framework fail to differentiate these as distinct cognitive activities. The DV collapse is the

faithful internalization of the framework's own architecture. This DV collapse is the training-cognition

mirror: practitioners' cognitive structure reflects the operational structure of the framework that shaped

their formation.

A practitioner with a globally positive self-assessment of risk management competence who

cannot independently execute the stages of risk assessment (qualitative estimation of likelihood and

impact, and from that risk quantification), expected loss prioritization, and structured risk

communication will respond to threats in ways that feel consistent with risk management without

executing the analytic operations that distinguish risk management from threat management. They will

describe a threat as "high risk" based on impact severity rather than expected loss. They will

recommend controls based on threat recognition rather than comparative risk calculus. In aggregate,

professional practice defaults to heuristic, adversarial, and impact-focused reasoning — responding to

recognized threats with proportionate controls — rather than to the formal enterprise risk calculus of

probability-weighted loss that governance frameworks assume underlies those decisions. They will

communicate in risk vocabulary without the risk-calculus content that vocabulary implies — without a

likelihood estimate, a strategic-impact assessment, or a computed expected-loss figure that would



justify the 'high risk' characterization to a risk-literate audience. These three specific behaviours are precisely what the organizational failures motivating this dissertation exhibit.

The structured-practice finding illuminates the ceiling. Practitioners who engage in structured, deliberate risk management practice (operationalized as endorsing at least three of four structured-practice items at levels 4 or 5) demonstrate substantially higher competence than those who do not: Cohen's d = 1.394, 95% CI [1.003, 1.785] (n_structured = 68, n_non-structured = 58). This effect size (Cohen's d = 1.394) is the largest effect size in the study. It demonstrates that cognitive differentiation is achievable and that, as currently constituted, formal training does not reliably produce it. Training seeds cognitive salience (a = 0.453); deliberate practice in authentic contexts converts salience into differentiated behavioural capacity (d = 1.394).

### 5.3.5    *What the Evidence Implies for Practice*

The partial mediation finding confirms that investment in training yields measurable value. A one-standard-deviation increase in training integration yields a 0.63-standard-deviation advantage in competence. An apparent tension between the NLP and SEM findings warrants explicit resolution. The NLP measure (IV1) captures structural supply at the framework level — what the training architecture makes available across the practitioner population as a whole — while the SEM independent variable (IV1x) captures individual uptake: the degree to which a specific professional reports having encountered risk-integrated training in practice. The two operationalize the same theoretical construct at different levels; the 4.5% IV1 result explains why individual exposure is variable and why the salience pathway carries significant weight, while the $\beta = 0.629$ IV1x result confirms that where exposure does occur, competence responds accordingly. But the epistemic compression finding reveals the narrow concentration of that advantage: it elevates a global, undifferentiated competence score rather than the



differentiated domain-level capability that enterprise risk governance requires. Training investment produces more of what the framework encodes—a stronger threat-informed general risk orientation—rather than what enterprise practice requires: differentiated risk-calculation reasoning, the capacity to estimate likelihood, assess strategic impact, compute expected loss, and communicate those conclusions—across the four stages of the risk management cycle.

Curriculum designers must sequence risk-calculation cognition — event likelihood estimation, strategic-impact assessment, and their joint computation using a risk matrix — as a foundational cognitive skill before introducing risk management content, mirroring the way statistical reasoning precedes research methods in empirical disciplines. Professionals who lack the cognitive schema for expected-loss computation cannot meaningfully engage with risk management frameworks: they will absorb the vocabulary without the calculation. This phenomenon of language mimicry without underlying cognition is the condition this study documents at Level 2. Consequently, the professional 'vocabulary' of risk management functions as an aesthetic veneer rather than a functional decision-making toolset.

### 5.3.6    $H_{12}$: Hypothesis Disposition

**$H_{12}$ — Supported (partial mediation; epistemic compression as amplifying finding):** Training integration significantly predicts risk management competence both directly (beta = 0.406, p < .001) and indirectly through conceptual salience (beta = 0.223, p < .001). Total standardized effect = 0.629, a large effect by conventional benchmarks. 35.5% of the training's total effect operates through the cognitive activation mechanism.



**H$_{12}$ is supported:** The collapse of the DV structure into a single factor represents the theoretically meaningful finding of epistemic compression: a substantive result about the structure of professional risk cognition that confirms and amplifies the central claim of H$_{12}$.

### 5.3.7 Level 2 → Level 3 Handoff

If Level 1 determines what the training architecture installs, and Level 2 demonstrates that practitioners internalize exactly that architecture, Level 3 asks what the organizational consequences are of deploying practitioners whose competence profile reflects this internalization. The control group comparison and the leadership interview findings answer this question from two directions: one outward from the practitioner, and one inward toward organizational expectations. Both converge on the same structural gap.

## 5.4 The Competence Gap: Control Group Findings (RQ5, BQ1; H$_{15}$)

The control group findings translate the cognitive profile established at Level 2 into an organizational reality: the practitioners that cybersecurity training produces are not empirically distinguishable from non-specialists on foundational risk-reasoning measures—a result of theoretical rather than merely statistical significance.

### 5.4.1 LEVEL 3 — Organizational Consequence (practitioner side)

This Section demonstrates the organizational-level consequence of Levels 1 and 2, measured from the practitioner outward. If the workforce framework does not embed probabilistic reasoning (expected-loss risk cognition) (Level 1) and practitioners internalize it faithfully (Level 2), then cybersecurity professionals should demonstrate no advantage over general professionals in foundational risk reasoning. The control-group comparison directly tests this prediction and confirms it.



Research Question 5 (RQ5) asks how the foundational risk-reasoning competencies of cybersecurity-trained professionals compare to those of non-cybersecurity professionals. $H_{15}$ predicted that cybersecurity professionals would demonstrate significantly higher risk management competence. The control group finding is the most counterintuitive result in this study, and it demands the most precise interpretation.

### 5.4.2   What the Evidence Demonstrates

The control group comparison used a shared eight-item foundational scale assessing risk assessment (ASMT_1-3), risk prioritization (PRIO_1-3), and risk-informed action (ACTN_1-2): the basic cognitive operations of likelihood estimation, comparative prioritization, and proportionate response that are definitional to risk management across all formal enterprise frameworks. The cybersecurity sample (n = 126) produced M = 3.82, SD = 0.78. The control group of general professionals (n = 133) produced a mean of M = 3.72 and a standard deviation of SD = 0.40. Welch t-test: t(185.16) = 1.271, p = .205, Cohen's d = 0.16, 95% CI [-0.054, 0.251]. The mean difference is statistically non-significant. The effect size is trivial. $H_{15}$ is not supported.

The null mean result conceals a variance pattern that is itself a substantive finding. The cybersecurity SD of 0.78 is nearly twice that of the control group's SD of 0.40, yielding a variance ratio of approximately 3.74 (p < .001). The training system does not consistently elevate foundational risk reasoning. It produces a heterogeneous population: some practitioners develop genuine risk reasoning capacity; the majority remain at or near the general professional baseline.

The distribution of competence tiers within the primary sample confirms this. Using *RM_Competence* factor scores extracted from Model 1 and standardized relative to the sample mean: 10.3% fall at or below -1.00 SD (Low Differentiation); 36.5% fall between -0.99 and 0.00 SD (Below



Average); 41.3% fall between 0.00 and +0.99 SD (Above Average); and 11.9% fall at or above +1.00 SD

(High Differentiation). Approximately 1 in 8 cybersecurity professionals reaches the high differentiation

threshold. The training system produces a small elite, a large undifferentiated middle, and a vulnerable

tail performing below the general professional baseline.

### 5.4.3    *What the Evidence Strongly Suggests*

The null result is not an absence of evidence. A Cohen's d of 0.16 falls well below any practical

significance threshold, and the study meets simulation-based SEM adequacy criteria for the obtained

sample size (Wolf et al., 2013). The study's high statistical power precludes attributing the null finding to

an insufficient sample size. It reflects a genuine absence of a specialization advantage, not a failure to

detect an existing effect. It is a specific and theoretically important piece of evidence: the logical

outcome, at the population level, of the institutional architecture identified at Level 1. If the training

framework does not encode risk-calculation cognition (joint likelihood estimation, strategic impact

assessment, and expected-loss computation) at the operational level, practitioners trained within it

should not demonstrate superior probabilistic reasoning relative to individuals who received no such

training. That is precisely what the data show. The failure of $H_{15}$ is not a finding against the study's

central claim. It is its most direct empirical confirmation.

The variance dispersion pattern strengthens this interpretation. The cybersecurity population

includes a minority who demonstrate genuine risk-reasoning capacity, confirming that the competence

ceiling is achievable. The minority status of this group confirms that, as currently constituted, formal

training does not reliably produce the competency they exhibit. The mechanism is consistent with the

SEM: practitioners who encounter genuine likelihood-impact reasoning demands in their organizations

and engage in structured practice develop differentiated competence. Those who receive the same



training in environments where risk vocabulary is performative rather than operational do not. The training system produces variance rather than a consistent elevation because the Level 1 framework never required probabilistic (expected-loss) reasoning to distinguish high from low performers.

The BQ1 answer follows directly: approximately 11.9% of cybersecurity professionals demonstrate high risk-management competence, differentiated from the non-structured-practice subgroup by Cohen's d = 1.394. Mediation analysis of the cross-group comparison further illuminates why even this minority advantage exists: conceptual salience accounts for the competence difference between cybersecurity and non-cybersecurity professionals ($\beta$ = .413, p = .049). The advantage is not structural. It does not reflect a different cognitive architecture for risk reasoning — but rather a contextual one: cybersecurity professionals inhabit environments where the institutional framework consistently foregrounds risk vocabulary, thereby elevating salience and, consequently, competence scores. The profession produces risk-competent individuals, but not through its formal training architecture. It produces them despite that architecture, through experience, organizational context, and deliberate practice that compensates for what the framework fails to install.

### 5.4.4   What the Evidence Implies for Practice

Organizations that assume cybersecurity credentials certify competence in risk reasoning are operating on an empirically unsupported premise. Credentialing programs, including CISSP, CISM, and CRISC, present themselves as risk-informed. A certification program that tests knowledge of risk frameworks without testing the capacity to estimate likelihood and impact to calculate risk, compute expected loss, prioritize competing risks, and communicate risk-quantified conclusions (specifying expected loss and its **L × I** basis) to non-technical audiences certifies familiarity with risk vocabulary. It is



not certifying risk management competence. The null specialization effect is the empirical statement of this distinction at the population level.

The practical consequence is that organizations systematically staff risk governance roles with professionals whose foundational risk-reasoning competence is not demonstrably superior to that of the general professional population, while governing those roles as if the credential confirmed that competence. The governance risk this creates is not technical in nature. It is cognitive.

### 5.4.5    $H_{15}$ and BQ1: Hypothesis Disposition

**$H_{15}$ — Not Supported (null result theoretically informative; see Section 5.4.2):**

Cybersecurity professionals do not demonstrate significantly higher foundational risk-reasoning competence than non-cybersecurity professionals (d = 0.16, p = .205, 95% CI [-0.054, 0.251]). The formal hypothesis is not supported. The null result is not a finding of absence; it is the empirically most important finding in the study, confirming, through independent benchmarking evidence, that the Level 1 architectural absence reproduces itself as a Level 3 population-level competence equivalence.

**BQ1 — Answered:** 11.9% of the primary sample falls at or above +1 SD on the *RM_Competence* latent factor (High Differentiation). The data reveal a d = 1.394 gap between this group and the non-structured-practice subgroup, highlighting the ceiling-breaking power of deliberate practice. The training system does not reliably produce high-differentiation practitioners. It produces approximately 1 in 8.

One might argue that the null result reflects measurement limitations. The scale is too generic to capture the specific risk reasoning that cybersecurity training develops. The foundational scale measures the basic cognitive operations of risk reasoning that are definitional to enterprise risk management across all formal frameworks: estimating likelihood, weighting impact, calculating risk,



prioritizing competing risks, and selecting proportionate responses. These operations are not domain-specific. They constitute the pre-theoretical core of risk as an analytical framework—a foundation shared by enterprise risk management, actuarial science, and every domain that employs formal risk reasoning. If cybersecurity professionals have genuinely internalized risk management principles and not merely risk vocabulary, they should demonstrate superiority in precisely these operations. The NLP analysis explains why they do not: the training frameworks that shaped their professional formation never required them to internalize probabilistic reasoning (and thus expected-loss cognition). The null result is the logical outcome of the Level 1 architectural absence. The instrument's sensitivity is confirmed by the cybersecurity population's greater variance (SD = 0.78 vs. 0.40), indicating that it detects competence differences when they exist. The problem is not that the scale cannot detect cybersecurity-specific risk reasoning. The problem is that such reasoning is not present at the population level.

### 5.5    Leadership, Governance, and the Assumption Gap (RQ4; $H_{14}$)

Where the preceding section measured the Level 3 consequence from the practitioner outward, this section measures it from the leadership expectation inward — revealing an assumption gap that operates independently of practitioner competence and compounds its effect.

#### 5.5.1    *LEVEL 3 — Organizational Consequence (expectation side)*

This Section demonstrates the organizational-level consequence of Levels 1 and 2, measured from the leadership expectation inward. If practitioners do not possess the risk-reasoning competence that organizations require (as established from the practitioner side in Section 5.4), the leadership interviews reveal why organizations do not recognize this: the same Level 1 architecture that shapes



practitioners also shapes leaders. The assumption gap is the system operating exactly as its institutional architecture predicts.

Research Question 4 (RQ4) asks whether organizational mandates implicitly require cybersecurity professionals to apply structured likelihood-impact risk reasoning and whether this demand mirrors their actual skill sets. $H_{14}$ predicted that organizations assign responsibilities consistent with formal strategic risk management frameworks while assuming professionals demonstrate differentiated risk management competence.

### 5.5.2    What the Evidence Demonstrates

Across seven leadership interviews spanning CISOs, a CEO, a Senior Vice President, an Executive Director, and a Director of Security Services, six cases fell in the High Expectation / High Gap typological quadrant. The remaining case (INT-02) was coded as Low Expectation/High Gap, confirming that even when organizational expectations are formally modest, the gap between expectations and practitioner capability remains substantial.

Leaders describe the cognitive work they expect from their teams with striking consistency. In all seven cases, they demand risk-calculation cognition (**L x I** as expected loss) reasoning (COG-LXI) at moderate to high salience. Leaders INT-04, INT-05, and INT-06 expect high salience for business-impact articulation (COG-IMP). Holistic or architectural reasoning (COG-HOL) is the dominant cognitive expectation in INT-01, INT-03, INT-04, and INT-06. Leaders do not describe expecting compliance checklists or vulnerability counts. They describe expecting professionals to translate technical conditions into enterprise risk language: to say, "this threat carries a 40% probability of materializing in the next 12 months and would generate direct financial exposure of $2.3 million and regulatory penalty risk of a further $800,000" rather than "this vulnerability is rated High on CVSS."



The cross-case salience summary provides the starkest and most theoretically consequential evidence in the leadership dataset. COG-LXI — the expectation that professionals estimate event likelihood and compute a risk score from its product with strategic impact — appears in all seven cases at salience 2 or 3. No leader is indifferent to risk-calculation cognition. The universality of this expectation is itself a significant finding: across every organizational sector, every seniority level, and every governance architecture in this sample, leaders expect their teams to produce outputs from $L \times I$ reasoning (risk calculations) without being able to articulate the calculus that produces them.

What they cannot do, in five of seven cases, is name that expectation. Only INT-05 and INT-07 articulate an explicit personal likelihood × impact formula. INT-03 describes $L \times I$ as structurally delegated to risk officers rather than personally operationalized. The remaining four cases score zero on the explicit personal $L \times I$ articulation code. This distinction — between expecting the outputs of risk calculation cognition and being able to name the cognitive process that produces those outputs — is the operational definition of the assumption gap. Leaders hold expectations without the vocabulary to specify them, test for them, or recognize their absence.

Survey data from the primary sample reinforce this pattern at the practitioner level: only 35% of cybersecurity professional respondents selected an explicit likelihood–impact definition when asked to characterize how they understand their work in practice, while 31% described cybersecurity as preventing compromise and 20% as detecting attacks. Most practitioners, like most leaders, do not frame their own professional activity in $L \times I$ terms — which is precisely why the assumption that such reasoning underlies cybersecurity practice persists unchallenged within organizations.

### 5.5.3    What the Evidence Strongly Suggests



The defining pattern in the interview data is not the expectation itself but its unexamined character. Leaders assume that the credentials and training their teams hold have developed the cognitive capacity for the full risk-calculation sequence: estimating event likelihood, translating technical consequences into strategic-organizational impact terms, and computing the product of those two assessments as a defensible expected-loss figure. They do not test this assumption. These leaders overlook a structural failure operating at three levels simultaneously: the frameworks used to train their teams do not require likelihood estimation, strategic-impact assessment, or risk-matrix computation that combines those two assessments into a comparable, actionable score. Leaders assume all three operations are present. The training architecture installs none of them.

The theoretical significance of this finding lies not in the existence of the expectation but in the absence of the vocabulary to name it. A leader who can state "I expect my team to estimate event likelihood, translate technical impact into a strategic cost figure, and derive a comparable risk score from their product" can, at minimum, detect the absence of that operation. They can ask for it during the hiring process. They can specify it in performance standards. They can recognize when a practitioner has substituted a CVSS severity rating for an expected-loss calculation. A leader who expects that cognitive output but cannot name its component operations has no diagnostic instrument. When practitioners fail to deliver risk-calculation cognition, the nameless expectation produces a nameless failure—one that leaders consistently misattribute. The interview data reflect this misattribution pattern directly: leaders describe practitioner deficiencies as communication problems (professionals cannot translate technical findings into executive language), experience problems (professionals lack the seniority to think at the organizational level), or maturity problems (professionals have not yet developed risk judgment). None of these attributions identifies the structural source of the failure: that the training system never required practitioners to perform the three-part cognitive sequence —



likelihood estimation, strategic-impact assessment, and expected-loss computation — that the role demands. Because leaders cannot name what is missing, they cannot locate where it went missing. The assumption gap is self-concealing not merely because vocabulary circulates without cognition, but also because the absence of naming capacity prevents it from being diagnosed at the institutional level where it originated.

This finding is not a criticism of individual leaders. It is evidence of the same structural condition identified by the NLP analysis at Level 1. The same institutional and training architecture that shaped the teams also formed the leaders who manage them. They absorbed risk vocabulary without risk-calculation cognition: they use the language of likelihood estimation without having estimated likelihoods as a formal professional operation; they speak of impact in enterprise terms without having systematically translated system-level consequences into strategic costs; and they invoke risk as a prioritization criterion without having applied a risk matrix to derive a score that renders competing risks comparable on a common scale. They described expecting "risk-based decision-making" from their teams, yet were unable to specify what the full computational operation of risk calculation entails — because they were never explicitly required to perform any of its three component steps.

This entire cycle of systemic reinforcement is the self-concealing character of the three-level mechanism. The concealment operates simultaneously at all three levels of the missing skill. When leaders use the vocabulary of likelihood estimation without performing it, the vocabulary of strategic impact without translating technical consequences to enterprise terms, and the vocabulary of risk calculation without applying a risk matrix, they signal to their teams that these operations are occurring. Professionals who learn to reproduce the same vocabulary while practising threat-enumeration routines receive positive organizational feedback: they are seen as risk-literate because they speak the language fluently, not because they are performing the operations the language implies. The feedback loop



reinforces the conflation at all three levels. A leader who says "we need to assess the risk of this threat" without computing the **L × I** score has, unintentionally, modelled the very absence of the installed training architecture. The industry selects its next generation of leaders from a population that has internalized this conflation across all three operations as professional expertise. The structural misclassification becomes invisible because every actor in the system operates consistently within the architecture that produced it, using the complete vocabulary of risk calculation while performing none of its definitive steps. The three-level mechanism is therefore also a three-level vocabulary failure: the framework never names risk calculation as a required cognitive operation; practitioners never acquire the term as a professional standard; leaders never develop the language to demand it specifically, and without that language, no actor at any level can locate the gap that all three levels are simultaneously producing.

This mechanism has a direct, measurable impact on practitioner competence. The SEM cross-group analysis established that conceptual salience accounts for the modest competence advantage cybersecurity professionals hold over the general professional population ($\beta = .413$, $p = .049$). The interview data describe the precise mechanism that produces this salience advantage: organizational environments in which leaders consistently invoke risk vocabulary, frame cybersecurity as an enterprise risk function, and signal that risk judgment is central to the role. Leadership discourse is not merely an organizational narrative; it is the primary organizational driver of the cognitive salience that the SEM identifies as the active ingredient in competence formation. The implication is that leadership development in risk-calculation cognition is not a downstream consequence of fixing training architecture; it is a parallel and necessary lever, because leaders who cannot themselves perform the full sequence: estimating likelihood, assessing strategic impact, and computing expected loss — cannot sustain the organizational discourse that activates it in their teams.



### 5.5.4    *What the Evidence Implies for Practice*

Persistent cyber failures, breaches that occur despite extensive technical investment, and incidents that cascade into enterprise crises because risk was not quantified before the event are the observable outputs of this self-reinforcing system. They are not failures of individual competence. They are the predictable consequence of an institutional architecture that produces threat-responsive professionals, equips them with risk vocabulary, deploys them in roles that require risk-reasoning capacity, and supplies leadership with an assumption that the training system has already resolved the gap.

Addressing this gap at the organizational level requires distinguishing threat management responsibilities from risk management responsibilities in job descriptions and performance management, explicit competency assessment at the point of hire for roles carrying risk governance responsibilities and replacing the credential-as-competence assumption with direct evidence.

### 5.5.5    *$H_{14}$: Hypothesis Disposition*

$H_{14}$ — **Supported (Expectation Confirmed; Alignment Gap Identified):** All seven organizations assign cybersecurity professionals responsibilities requiring differentiated risk-calculation cognition (likelihood-impact) reasoning: risk escalation, board-level risk communication, strategic risk prioritization, and governance participation. All seven leaders describe expectation structures consistent with formal strategic risk management frameworks. The interview data coding identified a high-salience gap between organizational expectations and professional capability in six of seven cases on the Expectation–Gap Tension Index (INT-02, classified as Low Expectation/High Gap). Critically, COG-LXI — the expectation of risk-calculation cognition — appears in all seven cases: the assumption gap is universal even where its overall intensity is lower.



**H$_{14}$ is supported**: The amplifying finding, that many leaders cannot articulate the computational basis of the reasoning they expect, confirming the assumption gap extends upward into leadership cognition itself, strengthens rather than qualifies this support. It is not surprising that leadership expects professionals to possess risk-calculation cognition already.  The profession uses risk-language.

### 5.6    Theoretical Contributions

The three-level institutional explanation, Institutional Architecture -> Cognitive Internalization -> Organizational Consequence, is this study's primary theoretical contribution. The unit of contribution is the causal chain, not any single finding in isolation. Multi-method convergence on a single causal mechanism provides a level of theoretical confidence that no single-method study can achieve. The four theoretical contributions named below all derive from and are subsidiary to this overarching structural claim.

### 5.6.1    *Institutional Misclassification as a Structural Condition*

The first theoretical contribution is a reframing of how the cybersecurity profession is institutionally classified. The mainstream position held by professional standards bodies, training frameworks, organizational job descriptions, and academic literature treats cybersecurity as a field that already belongs to risk management, with the only problem being remediable training gaps. This framing is problematic because it treats the structural condition as remediable.

This study's evidence supports a different claim: the institutional architecture organizes cybersecurity as a threat management discipline that has merely borrowed risk vocabulary. The borrowing reflects a historical trajectory in which institutions layered risk language onto a practice community that had already defined itself by threat identification and control implementation. The risk vocabulary is genuine and functional within the profession's self-understanding. The risk cognition



implied by the vocabulary is absent from the operational architecture of the dominant training framework.

The distinction between a discipline with gaps and one suffering structural misclassification dictates fundamentally different paths for practice and policy. Treating it as a gap problem leads to curriculum reform: adding risk content to frameworks that remain architecturally deterministic. Treating it as a structural misclassification leads to institutional redesign: reconstituting the cognitive architecture of professional formation around risk-calculation cognition — the joint estimation of likelihood and strategic impact, and the derivation of expected loss from those assessments — as a foundational competency, which requires acknowledging that current frameworks are not risk-management frameworks and building new ones on different foundations.

The novelty of this contribution lies not in the observation that cybersecurity has a risk management problem, frequently noted in practitioner literature, but in the precision and empirical grounding of the structural mechanism: NLP analysis at the TKS statement level demonstrates architectural absence; SEM demonstrates cognitive reproduction; cross-group benchmarking demonstrates population-level competence consequence; and leadership interviews demonstrate organizational assumption gap. No prior empirical study has traced this causal chain across four independent methods.

### 5.6.2    *Epistemic Compression as a Competency-Based Learning Theory Extension*

The second theoretical contribution extends Competency-Based Learning Theory (CBLT). CBLT predicts that declarative knowledge, when applied in authentic contexts with feedback, transfers into procedural competence organized into differentiated domains corresponding to the distinct cognitive operations required by each domain. The SEM DV collapse is anomalous on a standard CBLT account.



Epistemic compression, as this study defines it, extends CBLT by explaining why differentiation fails to emerge. It describes the condition in which the cognitive framework that training produces does not reflect the actual logic of the risk management domain its design seeks to cultivate. Training produces a global positive orientation toward risk management activity without producing the differentiated cognitive schemas that would allow distinct behavioural domains to emerge. The compression is not a learning failure in the conventional sense. It is the faithful internalization of a training architecture that never differentiated the domains in the first place.

This extension is generalizable. Any professional domain in which the authoritative training architecture fails to represent the cognitive structure of professional practice should exhibit the same compression dynamic. The measurement of epistemic compression via SEM, using the collapse of theoretically differentiated second-order constructs as an indicator, offers a generalizable diagnostic methodology. The most directly comparable case is medical training, where research has demonstrated that medical education can develop pattern-recognition capacity without equipping students with the probabilistic reasoning required for uncertainty management in novel clinical presentations (Blumenthal-Barby & Krieger, 2015; Gaissmaier & Gigerenzer, 2008). The cybersecurity finding closely parallels this structure.

Webb et al. (2014) document three endemic deficiencies in information security risk management practice (perfunctory risk identification, estimation without situational evidence, and intermittent assessment), and attribute these to inadequate organizational-level situation awareness. Baiden (2024) demonstrates that organizations with robust risk management integration achieve superior cybersecurity outcomes, with risk management serving as the dominant mediating factor in the effectiveness of cybersecurity strategy. Neither study identifies the cognitive mechanism that produces Webb's deficiencies or accounts for Baiden's differential outcomes at the practitioner level. This study's



finding of epistemic compression — the collapse of four theoretically differentiated risk-competency domains into a single undifferentiated global orientation — provides that mechanism. Practitioners who cannot cognitively differentiate risk assessment from prioritization, communication, and risk-informed action will perform each of Webb's three deficiencies as a predictable output: they will identify risks perfunctorily because they lack the schema to decompose threat states into probability-weighted exposure estimates; they will estimate without systematic evidence because their training never installed the cognitive operation of structured likelihood estimation; and they will treat assessment as an event rather than a process because continuous monitoring requires the differentiated competencies that compression forecloses. Baiden's differential outcomes reflect the organizational-level consequence of whether that cognitive architecture is present or absent. This study provides the individual-level explanation for why the gap Baiden identifies is not self-correcting.

### 5.6.3    Nameless Expectation as a Self-Concealment Mechanism

The third theoretical contribution is the concept of nameless expectation as a structural self-concealment mechanism, derived from the leadership interview data and convergently supported by the SEM partial mediation structure. It is not smaller in theoretical scope than the preceding contributions; it closes the explanatory loop that the other three open. The assumption gap documented in this study is not simply that leaders expect more than practitioners deliver. It is that leaders hold cognitive expectations they cannot specify, which produces a diagnostic incapacity that makes the gap structurally invisible.

Where a leader can name the cognitive operation they require — "I need a likelihood estimate, a strategic impact figure, and their product expressed as an expected loss" — the organizational system can test for it, hire for it, and develop it. Where the expectation is present but unnamed — "I need risk-



based decision-making" — the system lacks an instrument to detect its absence. Failure is reattributed to proximate and visible causes (communication, experience, maturity) rather than to the distal structural cause (training architecture).

This mechanism receives independent quantitative support from the SEM partial mediation structure. The direct training-to-competence path ($\beta = 0.406$) accounts for 64.5% of the total training effect—the portion that does not operate through explicit cognitive salience. If competence were transmitted exclusively through deliberate cognitive activation, the salience pathway would fully mediate the training effect, and the direct path would be non-significant. The significant direct path indicates that a second transmission mechanism is present: practitioners absorb competence from organizational environments in which risk language circulates at the leadership level, independent of whether they or their leaders can articulate what cognitive operations that language implies. The SEM and the interview data are therefore not parallel findings from different methods — they are the same finding measured from two directions. The interview data identify the mechanism; the SEM quantifies its effect. Nameless expectation is not only a qualitative organizational phenomenon. It is a measurable developmental pathway.

This mechanism is generalizable beyond cybersecurity to any professional domain in which the dominant training architecture adopts a field's vocabulary without embedding its cognitive operations. The vocabulary creates the expectation. Because these cognitive operations are absent, the system fails to meet the expectation. The naming incapacity (itself a product of the same training architecture) prevents a diagnosis of the gap. The result is a self-reinforcing system in which the structural failure is invisible to every actor whose position would otherwise enable them to correct it.

### 5.6.4    Training-Cognition Mismatch as a Diagnostic Construct



The fourth theoretical contribution is the interpretation of the null specialization effect as evidence of training-cognition mismatch: the domain that training claims to produce does not correspond to the cognitive operations that define competence in that domain. Cybersecurity training develops threat identification, control specification, and vulnerability management capacity. It claims, through its risk vocabulary, to also develop risk reasoning capacity. The null specialization effect demonstrates that cognitive outcomes at the population level do not sustain this claim.

Training-cognition mismatch describes the condition in which a training architecture is internally valid yet fails to develop the competencies it explicitly encodes. Still, those competencies do not correspond to the cognitive operations required by professional practice. The construct provides a diagnostic framework for evaluating whether domain-specific credentialing produces domain-specific cognition, operationalizable via the cross-group benchmarking methodology demonstrated in this study. This operationalizability is itself a methodological contribution that warrants explicit naming. The sequential research design employed in this study — NLP analysis of authoritative workforce frameworks to identify structural encoding, SEM of practitioner populations to detect cognitive internalization, and cross-group benchmarking against a non-specialist baseline to assess population-level competence consequence — constitutes a generalizable diagnostic methodology for identifying training-cognition mismatch in any professional domain. This three-stage approach extends to any field in which a dominant workforce architecture adopts technical or conceptual vocabulary faster than it embeds the analytic reasoning that vocabulary implies. The methodology bridges education research, professional competency science, and applied domain expertise in a way that single-method studies cannot, and its replication across other technical professions — data science, artificial intelligence engineering, software architecture — represents a tractable and consequential research programme.

## 5.7    Practical Implications



The three-level institutional explanation translates into specific practical implications for three constituencies, each mapped to the institutional level where intervention must occur.

### 5.7.1    For Framework Developers and Standards Bodies (Level 1)

NICE v2.0.0, ECSF, and SPARTA require architectural revision at the TKS statement level. Not the addition of risk-related vocabulary or category headings: risk vocabulary is already present. The specific requirement is explicit operationalization of risk calculation (risk-calculation cognition) using probabilistic reasoning and impact estimation relative to an organization's strategic goals as a required cognitive competency. TKS statements must require the estimation of event likelihood using probability-based or frequency-based methods; the estimation of impact relative to strategic goals; the joint computation of risk as a function of likelihood and strategic (or business) impact; comparative prioritization of competing risks using expected loss or equivalent decision-theoretic criteria; and communication of risk conclusions that distinguish high-probability/low-impact from low-probability/high-impact scenarios. Without these operational requirements at the TKS level, any framework revision will reproduce the structural condition this study identifies: risk vocabulary layered onto a deterministic cognitive architecture. Framework developers must embed risk calculation and probabilistic reasoning as a required cognitive operation across multiple work roles, rather than confining them to specialist risk analyst positions.

### 5.7.2    For Professional Education Providers (Level 2)

Certification programs, including CISSP, CISM, CRISC, and equivalent credentials, must be evaluated against the specific criterion of whether their assessment mechanisms require candidates to demonstrate expected-loss cognition, including probabilistic (and strategic-impact) reasoning. The presence of risk vocabulary in credential frameworks does not establish that the assessment mechanism



tests risk (exposure) cognition. A program that tests knowledge of risk frameworks without testing the capacity to estimate likelihood (qualitatively), estimate strategic impact (qualitatively), compute expected loss quantitatively (using a risk table), prioritize competing risks, and communicate risk-exposure justifications and conclusions to non-technical audiences is certifying familiarity with risk vocabulary, not risk management competence.

Education providers must sequence risk calculation (including probabilistic) reasoning as a foundational cognitive skill before introducing risk management content. Professionals who lack the cognitive schema for risk probability x strategic impact estimation cannot meaningfully engage with risk management frameworks: they will absorb the vocabulary without the cognition. This sequencing imperative is supported empirically: in the structural model, prior formal application of risk management (CV2_RM_LRN2) produced a standardized effect on perceived conceptual salience ($\beta$ = 0.432) that approached the magnitude of the primary training exposure path ($\beta$ = 0.453), indicating that pre-existing risk-reasoning exposure functions as a near-equivalent antecedent to formal training itself. This result treats prior cognitive preparation not as a background covariate but as a substantive prerequisite condition for training efficacy. Certification assessments should include quantitative risk-estimation scenarios, prioritization exercises based on likelihood-impact trade-offs, and cost-benefit analyses of competing mitigation strategies.

### 5.7.3    For Organizations Deploying Cybersecurity Professionals (Level 3)

Organizations should explicitly audit the alignment between the risk-reasoning competencies they expect of cybersecurity professionals and the cognitive training they have actually received. The assumption that credentialing produces risk reasoning capacity is empirically unsupported. For each role carrying risk governance responsibility, organizations should identify the specific probabilistic reasoning



operations required and gather evidence through structured competency assessment, not credential review, that the incumbent can perform them.

The High Expectation / High Gap pattern in six of seven cases indicates that the most significant governance risk in cybersecurity organizations may not be technical. It may be cognitive. The organizational response should include structured competency assessment at the point of hire for senior roles, explicit investment in training in likelihood-impact reasoning as a standalone cognitive capability, and an explicit distinction, at the job description and performance management levels, between threat management responsibilities and risk management responsibilities. The profession has treated these as synonymous. The evidence in this study demonstrates the organizational cost of that conflation.

### 5.8    Limitations and Responses

Any empirical study operates within methodological constraints that bound the scope of its claims; the following subsections identify the primary limitations of this study and explain the design and analytical decisions taken to mitigate their effects.

#### 5.8.1    Sample Size and Representativeness

The primary sample (n = 126) and the control group (n = 133) are modest relative to the scale of the theoretical claims (though they satisfy simulation-based SEM adequacy criteria for models with comparable structure and loading magnitudes (Wolf et al., 2013)). Variance-covariance matrix non-positive-definiteness warnings in both models indicate that the operation is near the boundary of parameter identification under WLSMV estimation. The NLP analysis is not sample-dependent: the absence of likelihood and probability from 2,111 TKS statements is an architectural fact. The interview findings are interpretive. The control-group comparison yields a null effect with a near-zero effect size, which is informative in a way a near-significant result would not be. The statistical comparison between



cybersecurity professionals and general professionals produced a result so small it is essentially zero (Cohen's d = 0.16), which, by conventional benchmarks, is trivial. The groups are, for practical purposes, identical in their foundational risk reasoning. This near-zero result is more useful as evidence than a result of, say, d = 0.20 or 0.22 with p = .06 — "almost significant." A near-significant result would leave open the interpretation that the effect is real. Still, the sample was too small to detect it, inviting the response "we just needed more participants." A near-zero effect size closes that door — it says the difference is not hiding just beyond the significance threshold, it is genuinely absent. The convergence of four independent evidence streams mitigates the inferential risk associated with any single component.

### 5.8.2   Common Method Variance

The same respondents completed all survey-based SEM components within a single instrument. The WLSMV estimator with ordinal indicator treatment corrects for distributional assumptions that amplify common method bias. The control variable structure introduces variation not shared with the primary construct items. Common method variance cannot explain why NICE contains zero likelihood statements, why leadership interviews reveal a universal assumption gap, or why the control group comparison yields a null result. These findings are method-independent and collectively corroborate the SEM story.

### 5.8.3   Cross-Sectional Design

The SEM cannot establish causal directionality. The training-to-salience-to-competence sequence is theoretically motivated and consistent with the data, but the data cannot rule out alternative orderings. The theoretical motivation is strong: established CBLT research demonstrates that deliberate training precedes competence transfer. The sensitivity model under full covariate saturation



does not substantially alter the path structure. A longitudinal design would strengthen causal attribution and is the highest-priority recommendation for future research.

### 5.8.4    *Generalizability of the NLP Analysis*

This study analyzed only the NICE Framework v2.0.0. NICE is the most authoritative and widely adopted cybersecurity workforce development framework globally, and its structural properties are the strongest single available indicator of the cognitive architecture the profession installs in its practitioners. Whether ECSF, SPARTA, or other frameworks exhibit the same structural pattern has not been tested empirically. This limitation is real. Extending the NLP analysis to ECSF and SPARTA is proposed as the highest-priority future research direction.

### 5.9    Future Research

The findings of this study open several productive lines of inquiry that lie beyond its current scope; four directions are identified below as priorities for extending and stress-testing the theoretical and empirical contributions made here.

### 5.9.1    *Multi-Framework NLP Comparison*

This study recommends extending the NLP analysis to ECSF and SPARTA to test whether the structural absence of risk calculation and likelihood reasoning at the TKS level is an artifact of NICE's particular design history or a generalizable feature of cybersecurity workforce development architecture globally. If ECSF, developed with explicit reference to ISO 31000 and COSO, encodes risk calculation and probabilistic reasoning at the operational level, whereas NICE does not, the comparative analysis provides causal leverage on which specific framework design decisions produce versus suppress risk cognition, and offers a concrete model for architectural revision.



### 5.9.2    Longitudinal Competence Development Study

A longitudinal study tracking cohorts through formal training programs, measuring risk-reasoning competence at intake, post-training, and at 12- and 24-month follow-up, would directly test the causal sequence inferred from cross-sectional data. The study should include a structured-practice intervention arm testing whether deliberate probability-estimation exercises and risk calculation introduced after initial training can close the competence gap. The primary outcome should include both the global *RM_Competence* factor and a task-based assessment of likelihood estimation, risk calculation, and risk prioritization, to test whether structured practice generates the cognitive differentiation that training alone does not.

### 5.9.3    Organizational Competence Audit Methodology

The High Expectation / High Gap typology identifies a practical need for organizational-level tools to diagnose the gap between expected and demonstrated risk-reasoning capacity in cybersecurity teams. Future research should develop and validate a structured competence audit instrument that draws on the RGP, COG, and GAP coding framework developed for this study, allowing organizations to assess whether their cybersecurity professionals can perform the specific cognitive operations required by their governance frameworks.

### 5.9.4    The Prioritization Deficit

Without risk calculation (the capacity to estimate event likelihood, assess strategic and organizational impact, and derive an expected-loss score from their product), practitioners cannot perform formal risk prioritization or make rational control-selection decisions. Prioritization requires a common metric: the **L × I** score. Without it, practitioners must rank threats based on their impressions



of severity, a process that systematically reproduces the biases this study documents. The consequence, systematic decision bias, overweighting high-consequence low-probability events and underweighting moderate-consequence high-probability events, is theoretically predictable and practically significant. Future research should test whether this bias is empirically demonstrable in practitioner populations using vignette-based experimental designs, and whether it is attenuated in populations with explicit training in probabilistic reasoning, controlling for domain expertise.

### 5.9.5    Leadership Risk Cognition as a Competence Multiplier

The finding that only two of seven senior cybersecurity leaders articulate an explicit personal likelihood × impact formula — despite all seven managing teams whose work they describe in precisely those terms — raises a research question that extends beyond practitioner development. If the same Level 1 institutional architecture formed leaders and practitioners alike, and if leadership discourse is the primary organizational driver of practitioner salience (as the $\beta = .413$ cross-group mediation result implies), then leadership risk cognition that includes risk calculation and a probabilistic reasoning capacity may function as a competence multiplier across entire teams. Future research should test whether leaders who demonstrate explicit likelihood-impact reasoning produce higher *RM_Competence* scores in their direct reports, controlling for training exposure and organizational sector. A leadership intervention study (introducing structured **L × I** risk calculus reasoning development at the executive and director level and measuring downstream practitioner competence at 12-month follow-up) would directly test whether the assumption gap can be closed from the top of the organizational hierarchy rather than solely through curriculum reform at the framework level.

### 5.10   Conclusion



This study began with a practical puzzle and arrived at a structural answer. The question was why cybersecurity professionals, despite extensive training and credentialing, continue to exhibit risk management failures that drive recurring organizational crises. The answer is not that professionals lack motivation, effort, or technical skill. The answer is that the profession was never institutionally organized to produce the risk management cognition it claims to produce.

The three-level institutional explanation provides the causal account. Level 1: The NICE Framework encodes consequence recognition (at the technical system level) and threat response within a deterministic architecture that lacks the three-part competency risk calculation, which requires likelihood estimation, strategic-impact assessment, and the joint computation of these into an expected-loss score. Level 2: Practitioners trained within this architecture internalize its undifferentiated cognitive structure, producing a single global competence factor rather than the differentiated risk reasoning formal enterprise frameworks require. Level 3: organizations receive practitioners whose competence profile reflects this internalization, while leaders, themselves shaped by the same Level 1 architecture, assume the training system has already produced the risk-calculation capacity — likelihood estimation, strategic-impact assessment, and expected-loss computation — that the vocabulary implies.

The convergence of NLP structural absence, SEM epistemic compression, null specialization effect, and leadership assumption gap does not permit a partial interpretation. It supports a single structural conclusion: The cybersecurity profession has taken institutional form as a threat management discipline that uses risk vocabulary. This organizational structure is encoded in frameworks, reproduced in cognition, invisible to leadership, and consequential for organizations.

The appropriate response is not curriculum reform at the margins. It is the explicit institutional acknowledgement that cybersecurity, as currently constituted, is not a risk management profession, and the deliberate redesign of professional formation to make it one. This dissertation provides empirical



evidence that such a redesign is necessary, a theoretical framework for understanding why the current architecture produces the outcomes it does, and specific research and practice directions for beginning the work.

### 5.10.1 Summary: Hypothesis Disposition Table

**Table 36**

*Hypothesis Disposition*

| Hyp. | RQ | Disposition | Key Empirical Basis |
|------|-----|-------------|---------------------|
| $H_{11}$ | RQ1 | Not Supported | Risk content = 4.5% of TKS, ranked 18th of 29 domains. 0 occurrences of "likelihood"; 0 of "probability"; 0 **L × I** constructs across 2,111 TKS statements. Strategic-impact assessment at the enterprise level: absent. Risk-matrix computation (**L × I** score): absent.  Not significantly represented relative to other domains. |
| $H_{13}$ | RQ3 | Supported | The NICE framework significantly underrepresented the RISK competency relative to all other domains. The gap is structural (probabilistic reasoning is architecturally absent at the TKS level), not proportional. |
| $H_{12}$ | RQ2 | Supported — Partial Mediation | β: a=0.453, b=0.491, c'=0.406, indirect=0.223, total=0.629 (all p < .001). 35.5% of the total effect mediated through salience. DV collapse = epistemic compression (not measurement failure). |
| $H_{14}$ | RQ4 | Supported — Expectation Confirmed; Gap Identified | 6/7 cases: High Expectation / High Gap. COG-LXI expected at moderate–high in all 7 cases. Only 2/7 leaders articulate an explicit personal **L × I** formula. Assumption gap extends into leadership cognition. |
| $H_{15}$ | RQ5 | Not Supported - (null result theoretically informative; see Section 5.4) | t(185.16)=1.271, p=.205, d=0.16. Cybersecurity professionals are statistically indistinguishable from the general population. Null result is the logical Level 3 consequence of Level 1 architectural absence. |
| BQ1 | BQ1 | Answered | 11.9% at ≥+1 SD (High Differentiation). d=1.394 [1.003, 1.785] structured vs. non-structured practice. Formal training does not reliably produce high-differentiation practitioners. |



# References


Adams, M., & Makramalla, M. (2015). Cybersecurity skills training: An attacker-centric gamified approach. *Technology Innovation Management Review*. https://timreview.ca

AlDaajeh, S., Saleous, H., Alrabaee, S., Barka, E., Breitinger, F., & Raymond Choo, K.-K. (2022). The role of national cybersecurity strategies on the improvement of cybersecurity education. *Computers & Security*, *119*, 102754. https://doi.org/10.1016/j.cose.2022.102754

Aldasoro, I., Gambacorta, L., Giudici, P., & Leach, T. (2022). The drivers of cyber risk. *Journal of Financial Stability*, *60*, 100989. https://doi.org/10.1016/j.jfs.2022.100989

Alina, S. (2012). Risk management: An integrated approach to risk management and assessment. *Annals of Faculty of Economics, 2*, *1*, 776–781.

Almeida, F. (2025). Comparative analysis of EU-based cybersecurity skills frameworks. *Computers & Security*, *151*, 104329. https://doi.org/10.1016/j.cose.2025.104329

American Psychological Association. (2016). *APA: Psychologists should obtain informed consent from research participants*. Https://Www.Apa.Org. https://www.apa.org/news/press/releases/2014/06/informed-consent

American Psychological Association. (2017, June 1). *Ethical principles of psychologists and code of conduct*. Https://Www.Apa.Org. https://www.apa.org/ethics/code

American Psychological Association. (2020). *Publication Manual of the American Psychological Association (7th Ed.)* (7th Edition). American Psychological Association. https://apastyle.apa.org/products/publication-manual-7th-edition

Anderson, R., Barton, C., Boehme, R., Clayton, R., Ganan, C., Grasso, T., Levi, M., Moore, T., Savage, S., & Vasek, M. (2019, January). *Measuring the changing cost of cybercrime*.





Anderson, S., & Williams, T. (2018). Cybersecurity and medical devices. *Computer Standards & Interfaces*, *56*(C), 134–143. https://doi.org/10.1016/j.csi.2017.10.001

Antonucci, D. (2017). *The cyber risk handbook: Creating and measuring effective cybersecurity capabilities* (1st edn). Wiley. https://doi.org/10.1002/9781119309741

Armin, J., Thompson, B., Ariu, D., Giacinto, G., Roli, F., & Kijewski, P. (2015, August). 2020 Cybercrime economic costs: No measure no solution. *2015 10th International Conference on Availability, Reliability and Security Proceedings*. 2015 10th International Conference on Availability, Reliability and Security (ARES). https://doi.org/10.1109/ARES.2015.56

Arrow, K. J., & Lind, R. C. (1970). Uncertainty and the evaluation of public investment decisions. *The American Economic Review*, *60*(3), 364–378. JSTOR.

Assante, M. J., & Tobey, D. H. (2011). Enhancing the cybersecurity workforce. *IT Professional*, *13*(1), 12–15. https://doi.org/10.1109/MITP.2011.6

Babbie, E. R. (2013). *The basics of social research* (4th ed). Thomson/Wadsworth.

Baiden, L. A. (2024). *Exploring the mediating role of risk management in the effective implementation of comprehensive cybersecurity strategies* [Doctoral Dissertation]. National University.

Baron, R. M., & Kenny, D. A. (n.d.). *The Moderator-Mediator Variable Distinction in Social Psychological Research: Conceptual, Strategic, and Statistical Considerations*.

Beasley, M. S., Branson, B. C., Braumann, E. C., & Pagach, D. P. (2023). Understanding the Ecosystem of Enterprise Risk Governance. *The Accounting Review*, *98*(5), 99–128. https://doi.org/10.2308/TAR-2020-0488

Benson, G. S., Akyildiz, I. F., & Appelbe, W. F. (1990). A formal protection model of security in centralized, parallel, and distributed systems. *ACM Transactions on Computer Systems*, *8*(3), 183–213. https://doi.org/10.1145/99926.99928





Bicak, A., Xiang, M., Liu, & Murphy, D. (2014). *Cybersecurity curriculum development: Introducing specialties in a graduate program*. https://doi.org/10.13140/2.1.2466.1440

Blumenthal-Barby, J. S., & Krieger, H. (2015). Cognitive Biases and Heuristics in Medical Decision Making: A Critical Review Using a Systematic Search Strategy. *Medical Decision Making*, *35*(4), 539–557. https://doi.org/10.1177/0272989X14547740

Braumann, E. C. (2018). Analyzing the role of risk awareness in enterprise risk management. *Journal of Management Accounting Research*, *30*(2), 241–268. https://doi.org/10.2308/jmar-52084

Braumann, E. C., Grabner, I., & Posch, A. (2020). Tone from the top in risk management: A complementarity perspective on how control systems influence risk awareness. *Accounting, Organizations and Society*, *84*, 101128. https://doi.org/10.1016/j.aos.2020.101128

Cabaj, K., Domingos, D., Kotulski, Z., & Respício, A. (2018). Cybersecurity Education: Evolution of the Discipline and Analysis of Master Programs. *Computers & Security*, *75*, 24–35. https://doi.org/10.1016/j.cose.2018.01.015

Cains, M. G., Flora, L., Taber, D., King, Z., & Henshel, D. S. (2022). Defining Cyber Security and Cyber Security Risk within a Multidisciplinary Context using Expert Elicitation. *Risk Analysis*, *42*(8), 1643–1669. https://doi.org/10.1111/risa.13687

Campbell, S. (2005). Determining Overall Risk. *Journal of Risk Research*, *8*(7–8), Article 7–8. https://doi.org/10.1080/13669870500118329

Caulkins, B., Marlowe, T., & Reardon, A. (2018). Cybersecurity Skills to Address Today's Threats. *Proceedings of the AHFE 2018 International Conference on Human Factors in Cybersecurity, Advances in Intelligent Systems and Computing*, *782*, 187–192. http://link.springer.com/10.1007/978-3-319-94782-2





Cerullo, M. J. (1985). General Controls in Computer Systems. *Computers & Security*, *4*(1), 33–45.

    https://doi.org/10.1016/0167-4048(85)90007-0

Chandrinos, T.-A. (2023). *Analysis of Frameworks/Methods for Information Security Risk Management*

    [Thesis, University of Piraeus].

    https://dione.lib.unipi.gr/xmlui/bitstream/handle/unipi/15481/Chandrinos_mte2131.pdf?seque

    nce=1&isAllowed=y

Cheung, G. W., & Rensvold, R. B. (2002). Evaluating Goodness-of-Fit Indexes for Testing Measurement

    Invariance. *Structural Equation Modeling: A Multidisciplinary Journal*, *9*(2), 233–255.

    https://doi.org/10.1207/S15328007SEM0902_5

Choo, K.-K. R. (2011). The Cyber Threat Landscape: Challenges and Future Research Directions.

    *Computers & Security*, *30*(8), 719–731. https://doi.org/10.1016/j.cose.2011.08.004

Chowdhury, N., & Gkioulos, V. (2021). Key Competencies for Critical Infrastructure Cyber-security: A

    Systematic Literature Review. *Information & Computer Security*, *29*(5), 697–723.

    https://doi.org/10.1108/ICS-07-2020-0121

Chowdhury, N., Katsikas, S., & Gkioulos, V. (2022). Modeling Effective Cybersecurity Training

    Frameworks: A Delphi Method-Based Study. *Computers & Security*, *113*, 102551.

    https://doi.org/10.1016/j.cose.2021.102551

Cohen, J. (1988). Set Correlation and Contingency Tables. *Applied Psychological Measurement*, *12*(4),

    425–434. https://doi.org/10.1177/014662168801200410

COSO. (2004, September). *Enterprise Risk Management – Integrated Framework*. Committee of

    Sponsoring Organizations of the Treadway Commission.

    https://doi.org/doi://10.1002/9781119180012.ch3





Courtney, R. H. (1977). Security Risk Assessment in Electronic Data Processing Systems. *Proceedings of the June 13-16, 1977, National Computer Conference on - AFIPS '77*, 97. https://doi.org/10.1145/1499402.1499424

Courtney, R. H. (1982). A Systematic Approach to Data Security. *Computers & Security*, *1*(2), 99–112. https://doi.org/10.1016/0167-4048(82)90003-7

Craigen, D., Diakun-Thibault, N., & Purse, R. (2014). Defining Cybersecurity. *Technology Innovation Management Review*, *4*(10), 13–21. https://doi.org/10.22215/timreview/835

Creswell, J. W. (2014). *Research Design: Qualitative, Quantitative, and Mixed Methods Approaches* (3. ed., [Nachdr.]). SAGE Publishing.

Creswell, J. W., & Clark, V. L. P. (2018). *Designing and conducting mixed methods research* (Third Edition). SAGE.

Crumpler, W., & Lewis, J. A. (2019). *The Cybersecurity Workforce Gap* (pp. 1–10). Center for Strategic and International Studies.

Dawson, J., & Thomson, R. (2018). The Future Cybersecurity Workforce: Going Beyond Technical Skills for Successful Cyber Performance. *Frontiers in Psychology*, *9*, 744. https://doi.org/10.3389/fpsyg.2018.00744

Deloitte. (2019). The Changing Faces of Cybersecurity—Closing the Cyber Risk Gap. *Toronto Financial Services Alliance*.

Deloitte. (2022). The Skills-based Organization: A New Operating Model for Work and the Workforce. *Deloitte Insights*. https://www2.deloitte.com/content/dam/insights/articles/us175310_consulting-the-skills-based-org-report/DI_The-skills-based-organization-report.pdf





Department of Commerce. (1974). *Guidelines for Automatic Data Processing Physical Security and Risk Management* (NBS FIPS 31; FIPS PUB 31, p. 92). National Institute of Standards and Technology (U.S.). https://doi.org/10.6028/NBS.FIPS.31

Department of Commerce. (1979). *Guideline for Automatic Data Processing Risk Analysis* (NBS FIPS 65; FIPS PUB 65, p. 32). National Institute of Standards and Technology (U.S.). https://doi.org/10.6028/NBS.FIPS.65

Đorđević, I., Scharf, E., Raptis, D., & Gran, B. A. (2002). *Suitability of Risk Analysis Methods for Security Assessment of Large-Scale Distributed Computer Systems* (Vol. 2002/026). Institutt for energiteknikk.

Dreyer, P., Jones, T., Klima, K., Oberholtzer, J., Strong, A., Welburn, J., & Winkelman, Z. (2018). *Estimating the Global Cost of Cyber Risk: Methodology and Examples*. RAND Corporation. https://doi.org/10.7249/RR2299

Easterly, J. (2023, May 7). *The Attack on Colonial Pipeline: What We've Learned & What We've Done Over the Past Two Years | CISA*. National Coordinator for Critical Infrastructure Security and Resilience. https://www.cisa.gov/news-events/news/attack-colonial-pipeline-what-weve-learned-what-weve-done-over-past-two-years

Eisenberg, M. B. (2008). Information Literacy: Essential Skills for the Information Age. *DESIDOC Journal of Library & Information Technology*, *28*(2), Article 2. https://doi.org/10.14429/djlit.28.2.166

European Commission. (2019). *A Proposal for a European Cybersecurity Taxonomy*. Publications Office, Joint Research Centre. https://data.europa.eu/doi/10.2760/106002

European Union Agency for Cybersecurity. (2022). *ECSF, European Cybersecurity Skills Framework*. Publications Office. https://data.europa.eu/doi/10.2824/859537




Falco, G. J., Eling, M., Jablanski, D., Weber, M., Miller, V., Gordon, L. A., Wang, S. S., Schmit, J., Thomas,

    R., Elvedi, M., Maillart, T., Donavan, E., Dejung, S., Durand, E., Nutter, F., Scheffer, U., Arazi, G.,

    Ohana, G., & Lin, H. (2019). Cyber Risk Research Impeded by Disciplinary Barriers. *Science*,

    *366*(6469), Article 6469. https://doi.org/10.1126/science.aaz4795

Falco, G. J., & Rosenbach, E. (2022). *Confronting Cyber Risk: An Embedded Endurance Strategy for*

    *Cybersecurity* (1st edn). Oxford University Press.

    https://doi.org/10.1093/oso/9780197526545.001.0001

Fandi, A. H., & Mhawi, R. A. (2021). Evaluation of Cyber Security Management in light of the Technology

    Acceptance Model. *International Journal of Research in Social Sciences and Humanities*, *11*(3).

    https://doi.org/10.37648/ijrssh.v11i03.029

Ferreira, M. P., Santos, J. C., De Almeida, M. I. R., & Reis, N. R. (2014). Mergers & acquisitions research: A

    bibliometric study of top strategy and international business journals, 1980–2010. *Journal of*

    *Business Research*, *67*(12), Article 12. https://doi.org/10.1016/j.jbusres.2014.03.015

Florackis, C., Louca, C., Michaely, R., & Weber, M. (2023). Cybersecurity Risk. *The Review of Financial*

    *Studies*, *36*(1), 351–407. https://doi.org/10.1093/rfs/hhac024

Fornell, C., & Larcker, D. F. (1981). Evaluating Structural Equation Models with Unobservable Variables

    and Measurement Error. *Journal of Marketing Research*, *XVIII*, 39–50.

Fruhlinger, J. (2020, February 12). *Equifax Data Breach FAQ: What Happened, Who Was Affected, What*

    *was the Impact?* CSO Online. https://www.csoonline.com/article/567833/equifax-data-breach-

    faq-what-happened-who-was-affected-what-was-the-impact.html

Gaissmaier, W., & Gigerenzer, G. (2008). Statistical illiteracy undermines informed shared decision

    making. *Zeitschrift Für Evidenz, Fortbildung Und Qualität Im Gesundheitswesen*, *102*(7), 411–

    413. https://doi.org/10.1016/j.zefq.2008.08.013




Galvez, S. M., & Guzman, I. R. (2008). Social Cognitive Theory: Information Security Awareness and

Practice. *AMCIS 2008 Proceedings, Americas Conference on Information Systems (AMCIS)*.

https://aisel.aisnet.org/

Gates, A. Q., Salamah, S., & Longpre, L. (2014). *Roadmap for Graduating Students with Expertise in the

Analysis and Development of Secure Cyber-Systems*. *Computer Science (University of Texas at El

Paso)*.

Gates, S., Nicolas, J.-L., & Walker, P. L. (2012). Enterprise Risk Management: A Process for Enhanced

Management and Improved Performance. *Management Accounting Quarterly*, *13*(3), pp 28-38.

Gee, J. P. (2014). *An introduction to discourse analysis: Theory and method* (Fourth edition). Routledge.

George, B. J. (1985). Contemporary Legislation Governing Computer Crimes. *Criminal Law Bulletin*, *21*(5),

389–412.

Giorgetto, S. A. (2021). Risk and Crisis Management. An Overview. *Economia Aziendale Online -, Vol 12*,

1-12 Pages. https://doi.org/10.13132/2038-5498/12.1.1-12

Glau, K., Scherer, M., & Zagst, R. (Eds). (2015). *Innovations in Quantitative Risk Management* (Vol. 99).

Springer International Publishing. https://doi.org/10.1007/978-3-319-09114-3

Goldberg, J. (1987). Some Principles and Techniques for Designing Safe Systems. *ACM SIGSOFT Software

Engineering Notes*, *12*(3), 17–19. https://doi.org/10.1145/29934.29936

Goode, J. (2016, October). Towards A Comparison of Training Methodologies on Employee's

Cybersecurity Countermeasures Awareness and Skills in Traditional vs. Socio-Technical

Programs. *KSU Conference on Cybersecurity Training, Research and Practice*. KSU Conference on

Cybersecurity Training.





Goode, J. (2018). *Comparing Training Methodologies on Employee's Cybersecurity Countermeasures Awareness and Skills in Traditional vs. Socio-Technical Programs* [Doctoral Dissertation]. Nova Southeastern University.

Gordon, L. A., Loeb, M. P., & Tseng, C.-Y. (2009). Enterprise risk management and firm performance: A contingency perspective. *Journal of Accounting and Public Policy*, *28*(4), 301–327. https://doi.org/10.1016/j.jaccpubpol.2009.06.006

Gossy, G. (2008). *Stakeholder Rationale for Risk Management: Implications for Corporate Finance Decisions*. Westdeutscher Verlag GmbH.

Guetzkow, J., Lamont, M., & Mallard, G. (2004). What Is Originality in the Humanities and the Social Sciences? *American Sociological Review*, *69*, 190–212.

Gutterman, A. (2023). Stakeholder Theory. *SSRN Electronic Journal*. https://doi.org/10.2139/ssrn.4387595

Haag, S., Siponen, M., & Liu, F. (2021). Protection Motivation Theory in Information Systems Security Research: A Review of the Past and a Road Map for the Future. *ACM SIGMIS Database: The DATABASE for Advances in Information Systems*, *52*(2), 25–67. https://doi.org/10.1145/3462766.3462770

Hagen, J., & Sutter, B. (2019, June 1). *The #1 Reason Small Businesses Fail—And How to Avoid It | SCORE* [Business Information]. SCORE. https://www.score.org/resource/blog-post/1-reason-small-businesses-fail-and-how-avoid-it

Hair, J. F., Page, M., & Brunsveld, N. (2019). *Essentials of Business Research Methods* (4th edn). Routledge. https://doi.org/10.4324/9780429203374

Hair, J. F., Ringle, C. M., & Sarstedt, M. (2011). PLS-SEM: Indeed a Silver Bullet. *Journal of Marketing Theory and Practice*, *19*(2), 139–152. https://doi.org/10.2753/MTP1069-6679190202





Halliday, S., Badenhorst, K., & von Solms, R. (1996). A Business Approach to Effective Information Technology Risk Analysis and Management. *Information Management & Computer Security*, *4*(1), 19–31. https://doi.org/https://doi.org/10.1108/09685229610114178

Harman, H. H. (1976). *Modern Factor Analysis* (3rd edn). University of Chicago Press.

Henderson, D. J., Wayne, S. J., Shore, L. M., Bommer, W. H., & Tetrick, L. E. (2008). Leader-member Exchange, Differentiation, and Psychological Contract Fulfillment: A Multilevel examination. *Journal of Applied Psychology*, *93*(6), Article 6. https://doi.org/10.1037/a0012678

Hess, W. (2021). *Enterprise Risk Management as a Measurement of Cybersecurity Effectiveness: A Correlational Study* [Doctoral Dissertation]. Capella University.

Hsieh, H.-F., & Shannon, S. E. (2005). Three Approaches to Qualitative Content Analysis. *Qualitative Health Research*, *15*(9), 1277–1288. https://doi.org/10.1177/1049732305276687

Hubbard, D. W. (2009). *The Failure of Risk Management: Why It's Broken and How to Fix It* (1st ed). John Wiley & Sons, Incorporated.

Hubbard, D. W. (2014). *How to Measure Anything: Finding the Value of Intangibles in Business* (Third edition). John Wiley & Sons, Inc.

Hubbard, D. W., & Seiersen, R. (2017). *How to Measure Anything in Cybersecurity Risk* (First Edition). John Wiley & Sons, Inc.

Hubbert, S. (2012). *Essential Mathematics for Market Risk Management* (2nd ed). Wiley.

IBM. (2024a). *Cost of a Data Breach Report 2024* (No. 2024; Cost of a Data Breach Report, p. 46). IBM Corporation. https://www.ibm.com/reports/data-breach

IBM. (2024b). *IBM's Cybersecurity Education Initiatives*. IBM. https://www.ibm.com/new/announcements/addressing-cybersecurity-skills-gap-higher-education





International Organization for Standardization. (2018). *Risk Management Guidelines (ISO 31000:2018)*

    (ISO 31000:2018). International Organization for Standardization.

    https://www.iso.org/standard/65694.html

International Organization for Standardization. (2022). *ISO/IEC 27001:2022 Information security,*

    *cybersecurity and privacy protection—Information security management systems—*

    *Requirements*. International Organization for Standardization.

    https://www.iso.org/standard/82875.html

ISC[2]. (2019). *Future of Work: Forecasting Emerging Technologies' Impact on Work in the Next Era of*

    *Human-Machine Partnerships*. Institute for the Future / Dell Technologies.

    https://legacy.iftf.org/realizing2030-futureofwork/

ISC[2]. (2024). *Global Cybersecurity Workforce Prepares for an AI-Driven World* ((ISC)[2] Cybersecurity

    Workforce Study, p. 50). Institute for the Future / Dell Technologies.

Jankensgård, H. (2019). A Theory of Enterprise Risk Management. *Corporate Governance: The*

    *International Journal of Business in Society*, *19*(3), 565–579. https://doi.org/10.1108/CG-02-

    2018-0092

Johnson III, R. (2018, August 14). 60 Percent of Small Companies Close Within 6 Months of Being

    Hacked. *Cybercrime Magazine*. https://cybersecurityventures.com/60-percent-of-small-

    companies-close-within-6-months-of-being-hacked/

Joint Task Force Interagency Working Group. (2020). *Security and Privacy Controls for Information*

    *Systems and Organizations* (Revision 5, NIST SP 800-53 (Rev 5)). National Institute of Standards

    and Technology. https://doi.org/10.6028/NIST.SP.800-53r5

Jones, J. (2006). An Introduction to Factor   Analysis of Information Risk  (FAIR). *Risk Management*

    *Insight LLC*, 1–76.





Jurafsky, D., & Martin, J. H. (2021). *Speech and language processing: An introduction to natural language processing, computational linguistics, and speech recognition* (3. ed. [Nachdr.]). Prentice Hall.

Kang, Y., Cai, Z., Tan, C.-W., Huang, Q., & Liu, H. (2020). Natural Language Processing (NLP) in Management Research: A Literature Review. *Journal of Management Analytics*, *7*(2), 139–172. https://doi.org/10.1080/23270012.2020.1756939

Khansa, L., & Liginlal, D. (2007). The Influence of Regulations on Innovation in Information Security. *13th Americas Conference on Information Systems, AMCIS 2007*, 180.

Klumpes, P. (2023). Coordination of cybersecurity risk management in the U.K. insurance sector. *The Geneva Papers on Risk and Insurance - Issues and Practice*, *48*(2), 332–371. https://doi.org/10.1057/s41288-023-00287-9

Knight, F. H. (1921). *Risk, Uncertainty, and Profit* (1957 Reprint). Hart, Schaffner & Marx; Houghton Mifflin Company.

Krippendorff, K. (2004). *Content Analysis: An Introduction to Its Methodology (second edition)*. Sage Publications.

Landwehr, C. E. (1981). Formal Models for Computer Security. *ACM Computing Surveys*, *13*(3), 247–278. https://doi.org/10.1145/356850.356852

Langhan, M. L., Goldman, M. P., & Tiyyagura, G. (2022). Can Behavior-Based Interviews Reduce Bias in Fellowship Applicant Assessment? *Academic Pediatrics*, *22*(3), 478–485. https://doi.org/10.1016/j.acap.2021.12.017

Li, Y., Hills, T., & Hertwig, R. (2020). A Brief History of Risk. *Cognition*, *203*, 104344. https://doi.org/10.1016/j.cognition.2020.104344

Libicki, M. C., Ablon, L., & Webb, T. (2015). *The Defender's Dilemma: Charting a Course Toward Cybersecurity*. RAND Corporation.





Lowrance, W. W. (1976). *Of Acceptable Risk: Science and the Determination of Safety*.

Maleh, Y., Sahid, A., & Belaissaoui, M. (2021). A Maturity Framework for Cybersecurity Governance in
Organizations. *EDPACS*, *63*(6), 1–22. https://doi.org/10.1080/07366981.2020.1815354

Manadhata, P. K., & Wing, J. M. (2011). An Attack Surface Metric. *IEEE Transactions on Software
Engineering*, *37*(3), 371–386. https://doi.org/10.1109/TSE.2010.60

Marotta, A., & Madnick, S. E. (2020). Analyzing the Interplay Between Regulatory Compliance and
Cybersecurity. *SSRN Electronic Journal*. https://doi.org/10.2139/ssrn.3542563

Mavrogiorgos, K., Mavrogiorgou, A., Kiourtis, A., Zafeiropoulos, N., Kleftakis, S., & Kyriazis, D. (2022a).
Automated Rule-Based Data Cleaning Using NLP. *2022 32nd Conference of Open Innovations
Association (FRUCT)*, 162–168. https://doi.org/10.23919/FRUCT56874.2022.9953810

Mavrogiorgos, K., Mavrogiorgou, A., Kiourtis, A., Zafeiropoulos, N., Kleftakis, S., & Kyriazis, D. (2022b).
Automated Rule-Based Data Cleaning Using NLP. *2022 32nd Conference of Open Innovations
Association (FRUCT)*, 162–168. https://doi.org/10.23919/FRUCT56874.2022.9953810

McNeil, A. J., Frey, R., & Embrechts, P. (2015). *Quantitative Risk Management: Concepts, Techniques and
Tools* (Revised edition). Princeton University Press.

Mikes, A., & Kaplan, R. S. (2013). Managing Risks: Towards a Contingency Theory of Enterprise Risk
Management. *SSRN Electronic Journal*. https://doi.org/10.2139/ssrn.2311293

Miles, M. B., Huberman, A. M., & Saldaña, J. (2014). *Qualitative data analysis: A methods sourcebook*
(Third edition). SAGE Publications, Inc.

Mitchell, R. (2015). *Web scraping with Python: Collecting data from the modern web* (First edition).
O'Reilly.

Moore, E., Likarish, D., Bastian, B., & Brooks, M. (2020). An Institutional Risk Reduction Model for
Teaching Cybersecurity. In L. Drevin, S. Von Solms, & M. Theocharidou (Eds), *Information*





*Security Education. Information Security in Action* (Vol. 579, pp. 18–31). Springer International Publishing. https://doi.org/10.1007/978-3-030-59291-2_2

Morgan, D. L. (2014). Pragmatism as a Paradigm for Social Research. *Qualitative Inquiry*, *20*(8), 1045–1053. https://doi.org/10.1177/1077800413513733

National Commission for the Protection of Human Subjects of Biomedical and Behavioral Research. (1979). *The Belmont report: Ethical principles and guidelines for the protection of human subjects of research*. U.S. Department of Health and Human Services. https://www.hhs.gov/ohrp/regulations-and-policy/belmont-report/read-the-belmont-report/index.html

Ndlela, M. N. (2018). A Stakeholder Approach to Risk Management. In M. N. Ndlela, *Crisis Communication* (pp. 53–75). Springer International Publishing. https://doi.org/10.1007/978-3-319-97256-5_4

Nedaei, B. H. N., Rasid, S. Z. A., Sofian, S., Basiruddin, R., & Kalkhouran, A. A. N. (2015). A Contingency-Based Framework for Managing Enterprise Risk. *Global Business and Organizational Excellence*, *34*(3), 54–66. https://doi.org/10.1002/joe.21604

Neumann, P. G. (1986). On Hierarchical Design of Computer Systems for Critical Applications. *IEEE Transactions on Software Engineering*, *SE-12*(9), 905–920. https://doi.org/10.1109/TSE.1986.6313046

Neumann, P. G. (1989). The Computer-Related Risk of the Year: Misplaced Trust in Computer Systems. *Proceedings of the Fourth Annual Conference on Computer Assurance, 'Systems Integrity, Software Safety and Process Security*, 9–13. https://doi.org/10.1109/CMPASS.1989.76030

NICCS. (2024, October). *National Initiative for Cybersecurity Careers and Studies (NICCS): Catalog of over 14,000 Cybersecurity Training Courses & Career Pathway Tools*. CISA.GOV. https://niccs.cisa.gov/





Nieles, M., Dempsey, K., & Pillitteri, V. Y. (2017). *An Introduction to Information Security* (NIST SP 800-12r1; NIST SP 800-12r1, p. NIST SP 800-12r1). National Institute of Standards and Technology. https://doi.org/10.6028/NIST.SP.800-12r1

Nikolić, B., & Ružić-Dimitrijević, L. (2009). Risk Assessment of Information Technology Systems. *Issues in Informing Science and Information Technology, Issues in Informing Science and Information Technology*, *6*, 21.

NIST. (2011). *NIST Managing Information Security Risk: Organization, Mission, and Information System View (NIST SP 800-39)* (NIST SP 800-39; NIST SP 800-39, p. 88). National Institute of Standards and Technology (U.S.). https://nvlpubs.nist.gov/nistpubs/Legacy/SP/nistspecialpublication800-39.pdf

NIST. (2012). *Guide for Conducting Risk Assessments* (Joint Task Force Transformation Initiative NIST SP 800-30r1; 0 edn, NIST SP 800-30r1, p. 95). National Institute of Standards and Technology. https://doi.org/10.6028/NIST.SP.800-30r1

NIST. (2018a). *Framework for Improving Critical Infrastructure Cybersecurity, Version 1.1* (NIST CSWP 04162018; NIST CSWP 04162018, p. 55). National Institute of Standards and Technology. https://doi.org/10.6028/NIST.CSWP.04162018

NIST. (2018b). *Risk Management Framework for Information Systems and Organizations: A System Life Cycle Approach for Security and Privacy* (NIST SP 800-37r2; NIST SP 800-37r2 Joint Task Force Transformation Initiative, p. 183). National Institute of Standards and Technology. https://doi.org/10.6028/NIST.SP.800-37r2

NIST. (2023). *NICE Framework Competency Areas: Introduction and Proposed List* (NIST NICE Framework Competency Areas; NIST NICE Framework Competency Areas, p. 8). National Institute of Standards and Technology.





https://www.nist.gov/system/files/documents/2023/06/14/NICEFramework_CompetencyAreas_List.pdf

NIST. (2024). *Framework for Improving Critical Infrastructure Cybersecurity, Version 2.0* (NIST CSWP 29; NIST CSWP 29, p. 32). National Institute of Standards and Technology. https://doi.org/10.6028/NIST.CSWP.29

NIST Glossary. (2025). *Glossary: Computer Security Resource Center* (Computer Security Resource Center). National Institute of Standards and Technology. https://csrc.nist.gov/glossary

Oltramari, A., & Kott, A. (2018). Towards a Reconceptualisation of Cyber Risk: An Empirical and Ontological Study. *Journal of Information Warfare*, *17*(1), Article 1.

Orceyre, M. J., Courtney, R. H., & Bolotsky, G. R. (1978). *Considerations in the Selection of Security Measures for Automatic Data Processing Systems* (NBS SP 500-33; 0 edn, Department of Commerce, p. 40). National Bureau of Standards. https://doi.org/10.6028/NBS.SP.500-33

Ormazabal, G., Barth, M. E., Klausner, M., Rajan, M. V., Reichelstein, S., Reiss, P. C., Taylor, D. J., Arif, S., Malenko, A., & McCall, A. L. (2010). *The Role of the Board in Corporate Risk Oversight*. https://api.semanticscholar.org/CorpusID:15479227

Pandey, S., Pandey, S. K., & Miller, L. (2017a). Measuring Innovativeness of Public Organizations: Using Natural Language Processing Techniques in Computer-Aided Textual Analysis. *International Public Management Journal*, *20*(1), 78–107. https://doi.org/10.1080/10967494.2016.1143424

Pandey, S., Pandey, S. K., & Miller, L. (2017b). Measuring Innovativeness of Public Organizations: Using Natural Language Processing Techniques in Computer-Aided Textual Analysis. *International Public Management Journal*, *20*(1), 78–107. https://doi.org/10.1080/10967494.2016.1143424




Parekh, G., DeLatte, D., Herman, G. L., Oliva, L., Phatak, D., Scheponik, T., & Sharman, A. T. (2018).

    Identifying Core Concepts of Cybersecurity: Results of Two Delphi Processes. *IEEE Transactions*

    *on Education*, *61*(1), Article 1. https://doi.org/10.1109/TE.2017.2715174

Parker, D. B. (1981). *Computer Security Management*. Prentice-Hall.

Patton, M. Q. (2015). *Qualitative Research & Evaluation Methods: Integrating Theory and Practice*

    (Fourth edition). SAGE.

Peirce, C. S. (1878). How to Make Our Ideas Clear. *Popular Science Monthly*, *12*, 286–302.

Perera, A. A. S. (2019). Enterprise Risk Management – International Standards and Frameworks.

    *International Journal of Scientific and Research Publications (IJSRP)*, *9*(7), 7.

    https://doi.org/10.29322/IJSRP.9.07.2019.p9130

Petersen, R., Santos, D., Smith, M. C., Wetzel, K. A., & Witte, G. (2020). *NIST Workforce Framework for*

    *Cybersecurity (NICE Framework)* (NIST NICE). National Institute of Standards and Technology.

    https://doi.org/10.6028/NIST.SP.800-181r1

Phillips, E., & Pugh, D. (2015). *How to Get a PhD: A Handbook for Students and their Supervisors* (6th

    edn). McGraw-Hill Education. https://books.google.ca/books?id=xMkvEAAAQBAJ

Piesarskas, E., Hajný, J., Levillain, O., Grigaliunas, S., Versinskiene, E., Bruze, E., & Zylius, R. (2019).

    *SPARTA Cybersecurity Skills Framework* (Strategic Programs for Advanced Research and

    Technology in Europe (SPARTA), p. 83). ENISA.

    https://www.enisa.europa.eu/publications/european-cybersecurity-skills-framework-ecsf

Podsakoff, P. M., MacKenzie, S. B., Lee, J.-Y., & Podsakoff, N. P. (2003). Common Method Biases in

    Behavioral Research: A Critical Review of the Literature and Recommended Remedies. *Journal of*

    *Applied Psychology*, *88*(5), 879–903. https://doi.org/10.1037/0021-9010.88.5.879




Quon, T. K., Zeghal, D., & Maingot, M. (2012). Enterprise Risk Management and Firm Performance. *Procedia - Social and Behavioral Sciences*, *62*, 263–267. https://doi.org/10.1016/j.sbspro.2012.09.042

Ralston, P. A. S., Graham, J. H., & Hieb, J. L. (2007). Cyber Security Risk Assessment for SCADA and DCS networks. *ISA Transactions*, *46*(4), 583–594. https://doi.org/10.1016/j.isatra.2007.04.003

Rashid, A., Danezis, G., Chivers, H., Lupu, E., Martin, A., Lewis, M., & Peersman, C. (2018). Scoping the Cyber Security Body of Knowledge. *IEEE Security & Privacy*, *16*(3), Article 3. https://doi.org/10.1109/MSP.2018.2701150

Ricci, S., Hajny, J., Piesarskas, E., Parker, S., & Janout, V. (2020). Challenges in Cyber Security Education. *International Journal of Information Security and Cybercrime*, *9*(2), 7–11. https://doi.org/10.19107/IJISC.2020.02.01

Richardson, H. A., Simmering, M. J., & Sturman, M. C. (2009). A Tale of Three Perspectives: Examining Post Hoc Statistical Techniques for Detection and Correction of Common Method Variance. *Organizational Research Methods*, *12*(4), 762–800. https://doi.org/10.1177/1094428109332834

Ross, R., & Pillitteri, V. (2024). *Protecting Controlled Unclassified Information in Nonfederal Systems and Organizations* (NIST SP 800-171r3; NIST SP 800-171r3, p. 120). National Institute of Standards and Technology (U.S.). https://doi.org/10.6028/NIST.SP.800-171r3

Saggar, R., & Singh, B. (2019). Drivers of Corporate Risk Disclosure in Indian Non-financial Companies: A Longitudinal Approach. *Management and Labour Studies*, *44*(3), 303–325. https://doi.org/10.1177/0258042X19851918

Saldaña, J. (2016). *The Coding Manual for Qualitative Researchers*. SAGE. https://books.google.ca/books?id=bCf7zgEACAAJ





SANS / GIAC. (2025). *Attract, Hire, and Retain Mid-Level Cybersecurity Roles* (p. 20). SANS / GIAC.

https://www.sans.org/mlp/2024-attract-hire-retain-midlevel-cybersecurity-roles/

Sarkar, D., & Sarmah, A. K. (2024). Analyzing the Nexus Between Entrepreneurship and Business

Mathematics—A Comprehensive Study on Strategic Decision-making, Financial Modeling, and

Risk Assessment in Small and Medium Enterprises (SMEs). *International Journal of Research and

Review*, *11*(2), 40–51. https://doi.org/10.52403/ijrr.20240206

Scala, N. M., Reilly, A. C., Goethals, P. L., & Cukier, M. (2019). Risk and the Five Hard Problems of

Cybersecurity. *Risk Analysis*, *39*(10), Article 10. https://doi.org/10.1111/risa.13309

Schall, P. D. (2019). *Examining the Relationship between formal RMF training and perceptions of RMF

effectiveness, sustainability and commitment in RMF practitioners* [Doctoral Dissertation,

University of the Cumberlands]. https://rmf.org/wp-content/uploads/2018/12/RMF-

Dissertation.pdf

Schall, P. D., & Oni, O. (2019). Examining the relationship between formal RMF training and perceptions

of RMF effectiveness, sustainability and commitment in RMF practitioners. *Cyber Security: A

Peer-Reviewed Journal*, *3*(1), 25. https://doi.org/10.69554/ZYSZ3480

Schatz, D., Bashroush, R., & Wall, J. (2017). Towards a More Representative Definition of Cyber Security.

*The Journal of Digital Forensics, Security and Law*. https://doi.org/10.15394/jdfsl.2017.1476

Schinagl, S., & Shahim, A. (2020). What do we Know about Information Security Governance?: From the

Basement to the Boardroom: Towards Digital Security Governance. *Information & Computer

Security*, *28*(2), 261–292. https://doi.org/10.1108/ICS-02-2019-0033

Shefrin, H. (2016). *Behavioral Risk Management*. Palgrave Macmillan US.

https://doi.org/10.1057/9781137445629





Shreeve, B., Hallett, J., Edwards, M., Anthonysamy, P., Frey, S., & Rashid, A. (2021). "So if Mr Blue Head here clicks the link…" Risk Thinking in Cyber Security Decision Making. *ACM Transactions on Privacy and Security*, *24*(1), 1–29. https://doi.org/10.1145/3419101

Skomra, W. (2017). Risk Management as Part of Crisis Management Tasks. *Foundations of Management*, *9*(1), 245–256. https://doi.org/10.1515/fman-2017-0019

Sommestad, T., Karlzén, H., & Hallberg, J. (2015). A Meta-Analysis of Studies on Protection Motivation Theory and Information Security Behaviour: *International Journal of Information Security and Privacy*, *9*(1), 26–46. https://doi.org/10.4018/IJISP.2015010102

Stine, K., Quinn, S., Witte, G., & Gardner, R. K. (2020). *Integrating Cybersecurity and Enterprise Risk Management (ERM)*. National Institute of Standards and Technology. https://doi.org/10.6028/NIST.IR.8286

Sumner, P., Day, J., & Mahoney, M. (2024, October 13). *Cybersecurity: An Evolving Governance Challenge*. Harvard Law School Forum on Corporate Governance. https://corpgov.law.harvard.edu/2020/03/15/cybersecurity-an-evolving-governance-challenge/

Sun, N., Zhang, J., Rimba, P., Gao, S., Zhang, L. Y., & Xiang, Y. (2019). Data-Driven Cybersecurity Incident Prediction: A Survey. *IEEE Communications Surveys & Tutorials*, *21*(2), 1744–1772. https://doi.org/10.1109/COMST.2018.2885561

Sunde, S. J. (2017). Assurance and Cyber Risk Management. In *The Cyber Risk Handbook: Creating and Measuring Effective Cybersecurity Capabilities* (1st edn, pp. 271–280). Wiley. https://doi.org/10.1002/9781119309741.ch18

Švábenský, V., Vykopal, J., & Čeleda, P. (2020). What Are Cybersecurity Education Papers About? A Systematic Literature Review of SIGCSE and ITiCSE Conferences. *Proceedings of the 51st ACM*





*Technical Symposium on Computer Science Education*, 2–8.

https://doi.org/10.1145/3328778.3366816

Swanson, M., Hash, J., & Bowen, P. (2006). *Guide for Developing Security Plans for Federal Information Systems* (NIST SP 800-18r1; 0 edn, p. NIST SP 800-18r1). National Institute of Standards and Technology. https://doi.org/10.6028/NIST.SP.800-18r1

Szterenfeld, A. (2022). Cybersecurity Solutions for a Riskier World. *ThoughtLab*, 85.

Taylor, B. J. (2005). Risk Management Paradigms in Health and Social Services for Professional Decision Making on the Long-Term Care of Older People. *British Journal of Social Work*, *36*(8), 1411–1429. https://doi.org/10.1093/bjsw/bch406

Teddlie, C., & Tashakkori, A. (2009). *Foundations of Mixed Methods Research: Integrating Quantitative and Qualitative Approaches in the Social and Behavioral Sciences*. SAGE.

Timmermans, S., & Tavory, I. (2012). Theory Construction in Qualitative Research: From Grounded Theory to Abductive Analysis. *Sociological Theory*, *30*(3), 167–186. https://doi.org/10.1177/0735275112457914

Uddin, Md. H., Ali, Md. H., & Hassan, M. K. (2020). Cybersecurity hazards and financial system vulnerability: A synthesis of literature. *Risk Management*, *22*(4), 239–309. https://doi.org/10.1057/s41283-020-00063-2

Vance, J. (2023, November 30). Cash Flow Management for Small Businesses: Expert Tips. *Preferred CFO*. https://preferredcfo.com/cash-flow-reason-small-businesses-fail/

Varga, S., Brynielsson, J., & Franke, U. (2021). Cyber-threat perception and risk management in the Swedish financial sector. *Computers & Security*, *105*, 102239. https://doi.org/10.1016/j.cose.2021.102239





Wang, S. S. (2017). Optimal Level and Allocation of Cybersecurity Spending: Model and Formula. *SSRN Electronic Journal*. https://doi.org/10.2139/ssrn.3010029

Wang, S. S. (2019). Integrated framework for information security investment and cyber insurance. *Pacific-Basin Finance Journal*, *57*, 101173. https://doi.org/10.1016/j.pacfin.2019.101173

Wangen, G. (2016). An Initial Insight into Information Security Risk Assessment Practices. *2016 Federated Conference on Computer Science and Information Systems (FedCSIS)*, 999–1008.

Webb, J., Ahmad, A., Maynard, S. B., & Shanks, G. (2014). A situation awareness model for information security risk management. *Computers & Security*, *44*, 1–15. https://doi.org/10.1016/j.cose.2014.04.005

WEF. (2025, February 3). *World Economic Forum's Initiative on Bridging the Cyber Skills Gap:* World Economic Forum. https://initiatives.weforum.org/bridging-the-cyber-skills-gap/home

Wei, X. (2018). Research on Monitoring and Control for Supply Chain Finance Risk. *Proceedings of the 2017 5th International Education, Economics, Social Science, Arts, Sports and Management Engineering Conference (IEESASM 2017)*. Proceedings of the 2017 5th International Education, Economics, Social Science, Arts, Sports and Management Engineering Conference (IEESASM 2017). https://doi.org/10.2991/ieesasm-17.2018.118

White, J. M. (2014). *Security Risk Assessment: Managing Physical and Operational Security*. Butterworth-Heinemann is an imprint of Elsevier.

Wilkinson, I. C. (2020). *Cybersecurity Using Risk Management Strategies of U.S. Government Health Organizations* [Doctoral Dissertation]. Walden University.

Williams, L. J., Hartman, N., & Cavazotte, F. (2010). Method Variance and Marker Variables: A Review and Comprehensive CFA Marker Technique. *Organizational Research Methods*, *13*(3), 477–514. https://doi.org/10.1177/1094428110366036





Wolf, E. J., Harrington, K. M., Clark, S. L., & Miller, M. W. (2013). Sample Size Requirements for Structural Equation Models: An Evaluation of Power, Bias, and Solution Propriety. *Educational and Psychological Measurement*, *73*(6), 913–934. https://doi.org/10.1177/0013164413495237

Wood, H. M., & Kimbleton, S. R. (1979). Access control mechanisms for a network operating system. *1979 International Workshop on Managing Requirements Knowledge (MARK)*, 821–830. https://doi.org/10.1109/MARK.1979.8817205

Woodward, B., Imboden, T., & Martin, N. L. (2013). An Undergraduate Information Security Program: More than a Curriculum. *An Undergraduate Information Security Program: More than a Curriculum*, *24*(1), 9.

Zhao, X., Lynch, J. G., & Chen, Q. (2010). Reconsidering Baron and Kenny: Myths and Truths about Mediation Analysis. *Journal of Consumer Research*, *37*(2), 197–206. https://doi.org/10.1086/651257

Zwilling, M. (2022). Trends and Challenges Regarding Cyber Risk Mitigation by CISOs—A Systematic Literature and Experts' Opinion Review Based on Text Analytics. *Sustainability*, *14*(3), 1311. https://doi.org/10.3390/su14031311




**Appendices**

**Appendix A**

**Leadership Interview Guide**

**Leadership Perspectives on Risk Reasoning and Cybersecurity**

<u>Purpose</u>

This semi-structured interview investigates senior leaders' risk mental models, their integration of cybersecurity into enterprise risk architecture, and the specific analytical rigour they expect from the cybersecurity workforce. The protocol elicits naturalistic leadership reasoning by avoiding prompts that trigger prescribed, 'textbook' answers. Questions function as prompts to guide conversation; selective follow-up probes allow for deeper pursuit of clarification when a participant's response warrants further detail.

<u>Protocol Design</u>

The interview is semi-structured. The interview protocol presents primary questions to all participants in a set sequence, minimizing order-effect bias. Optional probes are used at the interviewer's discretion when a primary response is brief, ambiguous, or when deeper elaboration would add analytical value. I omitted probes when the participant's primary response was already substantive. I fixed the order of primary questions; supplementary follow-up questions may vary across cases based on the direction of the conversation.

<u>Participant Eligibility</u>

Participants hold, or have held, a senior leadership role in which they directly supervised cybersecurity professionals or managed cybersecurity teams. These roles include Chief Information Security Officer, Chief Executive Officer, Vice President, Director, and Executive Director. Participants need not be technical cybersecurity practitioners; the study specifically seeks a leadership perspective rather than practitioner self-assessment.

<u>Ethical and Consent Note</u>

Participation is voluntary. Participants are informed before the interview that: (a) the session will be recorded and transcribed; (b) their responses will be anonymized in all reported findings; (c) they may decline to answer any question or withdraw at any time without consequence. Identifiable information — including organization name, role, and jurisdiction — is retained only in the confidential project file and does not appear in the dissertation.

**Preliminary Context (Recorded for Classification — Not Reported)**

The following items establish participant and organizational context for analytical classification. While responses inform the case profile, the reporting process anonymizes all findings before publication.

| Industry sector | |
|---|---|
| **Organizational size (approximate headcount)** | |
| **Geographic location (country/region)** | |
| **Leadership role while supervising cybersecurity professionals** | |

## Section 1 — How Leaders Conceptualize Risk

**Q1.** When you think about risk in your organization, what does that concept generally mean to you?

**Q2.** How does risk typically enter senior-level discussions or decisions in your organization?

**Q3.** Are there certain kinds of risks that tend to receive more attention or urgency than others — and if so, what drives that?

> **Optional probe:** Are there times when different risks must be compared against each other, and does that require a different kind of analysis than evaluating one issue in isolation?

## Section 2 — What Happens When a Risk Is Identified

**Q4.** When a significant risk is identified in your organization, walk me through what typically happens next—from the moment it is identified to the point at which a decision is made.

> **Optional probe:** Is there usually a point where someone formally estimates how likely the risk is to materialize, or how severe the consequences could be?

**Q5.** Who tends to be involved at each of those stages — and does that change depending on whether the risk is financial, operational, regulatory, or cybersecurity-related?

> **Optional probe:** If a similar level of exposure arose from a regulatory issue versus a cybersecurity issue, would the process look similar or different?

## Section 3 — Cybersecurity's Role in Risk Decision-Making

**Q6.** Where does cybersecurity tend to fit within your organization's overall approach to risk and decision-making?

**Q7.** When cybersecurity professionals bring a risk issue or recommendation to senior leadership, what kind of analysis or judgment do you expect them to have already completed before the conversation?

> **Follow-up:** In your experience, does the way cybersecurity professionals present issues to leadership typically match what you believe that role ideally requires, or is there sometimes a gap?

**Q8.** Can you describe a time when a cybersecurity recommendation felt genuinely well-prepared — where you came away confident the professional had really thought it through? What made it feel that way?



> **Additional probe:** Did it help you understand how likely the issue was to materialize, how serious it could be, or how it compared to other risks you were managing at the time?

**Q9.** And the opposite — can you recall a time when a cybersecurity recommendation felt incomplete or underdeveloped? Without naming anyone, what was missing?

## Section 4 — Professional Development, Expectations, and Gaps

**Q10.** When cybersecurity professionals underestimate or overlook something in their decision-making, in your experience, what does that tend to be?

**Q11.** In your view, is managing organizational risk something central to the cybersecurity role, or more of something that emerges from their technical responsibilities?

**Q12.** How do cybersecurity professionals in your organization tend to develop sound judgment over time — through formal training, experience on the job, mentorship, or something else?

> **Optional probe:** In your experience, do cybersecurity professionals tend to appreciate or actively value risk management as essential to their work?

**Q13.** If you could change one thing about how cybersecurity professionals are trained or developed — specifically in their ability to support risk-based decision-making — what would it be?

## Closing

Is there anything else about how cybersecurity and organizational risk intersect at the leadership level that you believe is important to understand?

**Interviewer notes:**



# Appendix B

## Survey Item — SEM Variable Mapping

This appendix provides a complete mapping of survey items to the latent variables and observed covariates specified in the Structural Equation Model. All item text is reproduced verbatim from the Qualtrics survey instrument (Cybersecurity Professional Practice). This Appendix presents the response scales and Qualtrics block identifiers exactly as they appeared in the instrument.

The measurement model comprises IV1x (4 items), MeV1 (6 items), and four first-order dependent-variable constructs, DV1–DV4 (5 items each), which load onto the second-order latent factor *RM_Competence*. The structural equation model includes CV1 and CV2 as observed covariates, accounting for variance unrelated to the primary predictors. Two distinct instrument blocks provide the DV items, capturing both the technical and managerial dimensions of the NICE framework: QID17 uses a 5-point Agree/Disagree scale; QID18 uses a 5-point Never–Always frequency scale. Three QID17 items and two QID18 items measure each DV construct, providing a multi-indicator assessment of the participant's mental model.

| Item Code | Block | Survey Item (verbatim) | Response Scale | SEM Variable |
|---|---|---|---|---|
| **IV1x** Training Exposure — Self-Reported Exposure to Risk-Embedded Training Content (4 items) | | | | |
| **IV1x_1** | QID23 | My cybersecurity training covered how to compare alternative safeguards based on their expected effects on business outcomes. | Agree / Disagree (5-point) | **IV1x** |
| **IV1x_2** | QID23 | My training provided examples of weighing how often an issue might occur against how disruptive it could be. | Agree / Disagree (5-point) | **IV1x** |
| **IV1x_3** | QID23 | My training explained how to rank work items based on their expected effect on business operations or outcomes. | Agree / Disagree (5-point) | **IV1x** |
| **IV1x_4** | QID23 | My training linked technical cybersecurity activities to organization-level objectives. | Agree / Disagree (5-point) | **IV1x** |
| **MeV1** Conceptual Salience — Perceived Relevance of Risk Management to Cybersecurity Work (6 items) | | | | |
| **MeV1_1** | QID24 | Topics from my cybersecurity training are useful in everyday decisions in my role. | Agree / Disagree (5-point) | **MeV1** |
| **MeV1_2** | QID24 | In my work, topics from cybersecurity training feel directly | Agree / Disagree (5-point) | **MeV1** |



| Item Code | Block | Survey Item (verbatim) | Response Scale | SEM Variable |
|:---:|:---:|---|---|:---:|
| | | applicable rather than abstract concepts. | | |
| **MeV1_3** | QID24 | Using topics from cybersecurity training helps my team or work group make better decisions. | Agree / Disagree (5-point) | **MeV1** |
| **MeV1_4** | QID24 | Cybersecurity training content connects to our organization's objectives. | Agree / Disagree (5-point) | **MeV1** |
| **MeV1_5** | QID24 | The training I've received helps me anticipate the downstream effects of different options. | Agree / Disagree (5-point) | **MeV1** |
| **MeV1_6** | QID24 | Applying what I learned helps reduce time or resource demands in practice. | Agree / Disagree (5-point) | **MeV1** |
| **DV1** Risk-Based Assessment — Likelihood × Impact Reasoning in Professional Practice (5 items) | | | | |
| **DV1_1** | QID17 | When deciding whether to act, changes in how often an issue could occur OR how disruptive it might be as an incident influence my decisions. | Agree / Disagree (5-point) | **DV1** |
| **DV1_2** | QID17 | When new information changes the likelihood of a problem, it affects how I assess its priority. | Agree / Disagree (5-point) | **DV1** |
| **DV1_3** | QID17 | Changes in operating conditions shape analysts' perceptions of the severity of potential consequences. | Agree / Disagree (5-point) | **DV1** |
| **DV1_4** | QID18 | Before deciding, I gather enough information to judge both how often something might occur and how disruptive it could be. | Never–Always (5-point) | **DV1** |
| **DV1_5** | QID18 | When information is limited, I make reasonable assumptions to estimate how often something might happen and how disruptive it could be. | Never–Always (5-point) | **DV1** |
| **DV2** Risk-Based Prioritization — Ordering Actions by Expected Risk Severity (5 items) | | | | |
| **DV2_1** | QID17 | When something looks urgent but has a lower business impact, I | Agree / Disagree (5-point) | **DV2** |



| Item Code | Block | Survey Item (verbatim) | Response Scale | SEM Variable |
|---|---|---|---|---|
| | | prioritize based on impact rather than (ticket) urgency alone. | | |
| **DV2_2** | QID17 | When expected operational effects change, I revisit earlier prioritization decisions. | Agree / Disagree (5-point) | **DV2** |
| **DV2_3** | QID17 | Decision makers use differences in expected outcomes to justify prioritizing one item over another. | Agree / Disagree (5-point) | **DV2** |
| **DV2_4** | QID18 | I deprioritize items with lower expected effect, even if they attract a lot of attention within the organization. | Never–Always (5-point) | **DV2** |
| **DV2_5** | QID18 | When two tasks compete, I choose the one that causes the least potential disruption. | Never–Always (5-point) | **DV2** |
| **DV3 Control** Selection — Justified by Likelihood or Impact Reduction (5 items) | | | | |
| **DV3_1** | QID17 | Differences in how safeguards affect likelihood OR consequences influence which controls I recommend. | Agree / Disagree (5-point) | **DV3** |
| **DV3_2** | QID17 | Control recommendations differ depending on whether they reduce likelihood, reduce impact, or both. | Agree / Disagree (5-point) | **DV3** |
| **DV3_3** | QID18 | If a safeguard doesn't meaningfully change outcomes, I look for alternative approaches. | Never–Always (5-point) | **DV3** |
| **DV3_4** | QID18 | I consider whether an action mainly reduces how often an issue occurs or how severe it is when it does occur. | Never–Always (5-point) | **DV3** |
| **DV3_5** | QID18 | I document my choice of safeguards, so others understand their expected effect. | Never–Always (5-point) | **DV3** |
| **DV4 Risk** Communication — Explaining Trade-offs and Residual Risk to Stakeholders (5 items) | | | | |
| **DV4_1** | QID17 | After safeguards are selected, I consider what outcomes remain possible if they do not work as expected. | Agree / Disagree (5-point) | **DV4** |



| Item Code | Block | Survey Item (verbatim) | Response Scale | SEM Variable |
|---|---|---|---|---|
| **DV4_2** | QID17 | When discussing security decisions with non-technical colleagues, I explain trade-offs in terms of consequences and outcomes. | Agree / Disagree (5-point) | **DV4** |
| **DV4_3** | QID18 | I summarize key choices in terms of expected outcomes and remaining exposure. | Never–Always (5-point) | **DV4** |
| **DV4_4** | QID18 | I pay attention to conditions that could cause us to revisit a decision later. | Never–Always (5-point) | **DV4** |
| **DV4_5** | QID18 | I check back after implementation to see whether the outcomes matched expectations. | Never–Always (5-point) | **DV4** |
| **CV1** Years of Experience — Observed Covariates (2 items) | | | | |
| **CV1_YEARS** | QID14 | How many total years have you worked primarily in cybersecurity? | Ordinal: 0–1 / 2–4 / 5–9 / 10–14 / 15–19 / 20+ | **CV1 (observed)** |
| **CV1_SENIORITY** | QID15 | Which best describes your seniority level? | Ordinal: Individual Contributor / Team Lead / Manager / Director / Executive | **CV1 (observed)** |
| **CV2** Prior Risk Management Learning — Observed Covariates (3 items) | | | | |
| **CV2_RM_LRN1** | QID22 | I have completed formal instruction focused on decision-making under uncertainty. | Agree / Disagree (5-point) | **CV2 (observed)** |
| **CV2_RM_LRN2** | QID22 | I regularly apply concepts from structured decision-making or governance frameworks. | Agree / Disagree (5-point) | **CV2 (observed)** |
| **CV2_RM_LRN_CHK1** | QID19 | Which of the following sources have contributed to your understanding of risk management? (Select all that apply): No structured training / On-the-job mentoring / Self-directed learning / Standards-based training (ISO 31000, NIST RMF, FAIR, ERM) / | Categorical multi-select; recoded binary (0 = no structured training, 1 = any | **CV2 (observed)** |



| Item Code | Block | Survey Item (verbatim) | Response Scale | SEM Variable |
|-----------|-------|------------------------|----------------|--------------|
|           |       | Industry certifications / Formal coursework | structured source) |              |

**Note.** Item codes correspond to Qualtrics data export tags as specified in the QSF instrument file. The QID17 Agree/Disagree scale uses the following anchors, providing a standardized metric for the primary salience units: 1 = Strongly Disagree, 3 = Neutral, 5 = Strongly Agree. The QID18 Never–Always scale was anchored as follows: 1 = Never, 3 = Sometimes, 5 = Always. The model treats CV1 items as ordinal observed covariates and utilizes a binary version of CV2_RM_LRN_CHK1 to account for structured training history.

The model treats CV1_YEARS and CV1_SENIORITY as ordinal observed covariates, accounting for the variance associated with participant tenure and seniority. The model recoded CV2_RM_LRN_CHK1 as a binary covariate (0 = no structured training endorsed; 1 = any structured training endorsed) before model estimation.



**Appendix C**

Leadership Interviews - Leadership Risk Cognition Codebook

*This study used this codebook to analyze all leadership interview transcripts.*

## Preamble

This codebook was applied deductively to all leadership interview transcripts. The study's theoretical framework provides five domains to categorize the 28 codes: Competency-Based Learning Theory and enterprise risk management principles. Codes map directly onto the constructs examined in the quantitative model, enabling construct alignment between the qualitative and quantitative phases of the mixed-methods analysis.

The study employed a deductive approach, developing codes before data collection and applying them consistently across all cases (Hsieh & Shannon, 2005). The research design excluded emergent codes during analysis, relying solely on the pre-defined LRCC codebook. The analysis began by dividing each transcript into 18–30 meaningful idea units to ensure coding consistency. Salience scores were assigned using the behavioural anchors in Section 2 of this appendix. The Section 3 formula and threshold table provided the basis for the Expectation–Gap Tension Index scores reported in Section 4.8.

## Section 1 — Code Definitions (28 Codes, Five Domains)

The following tables present the authoritative definition for each code in the Leadership Risk Cognition Codebook. Definitions specify the code's conceptual scope and the application condition — what must be present in a transcript unit for the code to apply. Each definition provides the level of precision necessary to support consistent inter-rater application.

### Domain A — Enterprise Risk Framing (ERF)

| Domain A — Enterprise Risk Framing (ERF) | |
|---|---|
| **Code** | **Definition** |
| **ERF-SURV** | The interviewee framed Risk as a direct threat to business survival, operational continuity, or organizational viability. Applies when a leader explicitly positions a risk in terms of whether the organization can continue to function, serve clients, or remain solvent. |
| **ERF-FIN** | Risk framed in financial or revenue terms. Includes explicit reference to monetary loss, profit-and-loss impact, cost of remediation, ransom payment, or revenue disruption as the primary organizational consequence of a risk event. |
| **ERF-REG** | The interviewee framed Risk as regulatory, compliance, or audit exposure. Includes licensing consequences, regulatory penalty, examination scrutiny, statutory obligation, or breach of compliance frameworks as the primary organizational consequence. |
| **ERF-REP** | The interviewee framed Risk as reputational or public trust damage. Includes brand erosion, media coverage, client perception, or loss of stakeholder confidence as the primary or secondary organizational consequence of a risk event. |
| **ERF-STRAT** | The interviewee framed Risk as interference with strategic business objectives. Includes disruption to product delivery, competitive positioning, organizational mission, or strategic planning as the framing lens for risk consequence. |

## Domain B — Risk Governance and Process (RGP)

| Domain B — Risk Governance and Process (RGP) | |
|---|---|
| **Code** | **Definition** |
| **RGP-LXI** | Explicit likelihood × impact evaluation in governance practice. Applies when a leader describes a process — formal or informal — in which both the probability of occurrence and the organizational consequence of a risk event are assessed together by the professional as the basis for risk rating or prioritization. |
| **RGP-COLL** | Cross-functional collaboration in risk evaluation or decision-making. Applies when a leader describes deliberate involvement of personnel from multiple organizational functions — finance, compliance, operations, executive leadership — in assessing, rating, or acting on identified risks. |
| **RGP-ESC** | Formal or semi-formal escalation pathways and risk committee structures. Applies when a leader describes how risks move from operational identification through tiered review to executive or board-level decision-making, including the conditions and thresholds that trigger escalation. |
| **RGP-COMP** | Cross-domain risk comparison or portfolio-level risk reasoning. Applies when a leader describes comparing risks across different organizational categories — cybersecurity versus financial, operational versus strategic — or reasoning about relative priority across a risk registry or portfolio. |
| **RGP-ACC** | Risk acceptance and residual risk logic. Applies when a leader describes a process by which the organization formally or informally acknowledges that a risk remains after mitigation, accepts the residual exposure, and assigns accountability for that decision. |

## Domain C — Expected Cognitive Work (COG)

| Domain C — Expected Cognitive Work (COG) | |
|---|---|
| **Code** | **Definition** |
| **COG-LXI** | Expectation that cybersecurity professionals estimate likelihood as part of their risk analysis contribution. Applies when a leader articulates a normative standard that practitioners should assess or defend the probability of a risk event occurring, not merely identify that a vulnerability exists. |
| **COG-IMP** | Expectation that cybersecurity professionals understand and communicate business impact in organizational terms. Applies when a leader articulates that practitioners should translate technical findings into financial, operational, regulatory, or strategic consequences relevant to executive decision-makers. |
| **COG-HOL** | Expectation of a holistic, enterprise-wide, or architecturally informed risk perspective. Applies when a leader articulates that practitioners should consider the full organizational context of a risk — across |
| | functions, domains, and consequence dimensions — rather than within the technical boundary of the incident or vulnerability. |
| **COG-COMP** | Expectation of comparative risk reasoning. Applies when a leader articulates that practitioners should evaluate risks relative to each other, to the organizational risk registry, to industry benchmarks, or to competing priorities — rather than assessing each risk in isolation. |
| **COG-OPT** | Expectation that practitioners present tiered remediation options with associated risk trade-offs. Applies when a leader articulates a standard in which professionals provide decision-makers with alternative response pathways — at different cost, risk-reduction, or residual-risk levels — rather than a single recommendation. |
| **COG-RES** | Expectation that practitioners evaluate residual risk after controls are applied. Applies when a leader articulates that professionals should assess the organizational exposure that remains following proposed or implemented mitigations and whether that residual exposure is acceptable. |
| **COG-STRAT** | Expectation that practitioners situate risk analysis within broader strategic business objectives. Applies when a leader articulates that professionals should connect their risk outputs to organizational strategy, revenue objectives, regulatory environment, or competitive positioning. |

## Domain D — Observed Professional Gaps (GAP)

| Domain D — Observed Professional Gaps (GAP) | |
|---|---|
| **Code** | **Definition** |
| **GAP-TFR (parent)** | Technocentric Risk Framing — structural parent code. Applied at the case level only where neither TFR-PROB nor TFR-SOLN individually captures the full pattern. In practice, TFR-PROB and TFR-SOLN are the applied codes; GAP-TFR is the organizing construct. |
| **TFR-PROB** | GAP subcode — Technical-only problem framing. Applies when a leader describes practitioners defining risk events in exclusively technical terms: framing a vulnerability, incident, or control gap as a technical problem to be solved rather than as an organizational risk event requiring likelihood and impact assessment and business-relevant communication. |
| **TFR-SOLN** | GAP subcode — Tool-centric remediation logic. Applies when a leader describes practitioners proposing tool acquisition, technology deployment, or the addition of a control layer as the default remediation response, without conducting root-cause analysis or examining process-level or organizational contributors. |
| **GAP-MYO** | Siloed or myopic thinking. Applies when a leader describes practitioners applying domain-specific cognitive frameworks without enterprise-wide contextualization — perceiving individual technical vulnerabilities or operational gaps rather than the organizational risk event and its cross-functional consequences. |
| **GAP-BUS** | Business-value blindness. Applies when a leader describes practitioners failing to understand, access, or reason within the financial and strategic logic that governs executive decision-making, producing outputs that are technically accurate but organizationally irrelevant or unintelligible to senior leadership. |
| **GAP-SCAL** | Lack of scale awareness. Applies when a leader describes practitioners assigning risk severity based on the presence of a vulnerability or control gap, without calibrating that assignment to enterprise-scale likelihood and impact — resulting in miscalibrated severity ratings that do not reflect organizational exposure. |



| GAP-LXI | Absence of structured likelihood × impact reasoning. Applies when a leader describes practitioners failing to apply formal or semi-formal **L × I** analysis to their risk assessments, even in governance contexts where such analysis is structurally expected or required. |
|---|---|
| GAP-ENTRY | A risk mindset is not natural at the entry level. Applies when a leader describes the absence of risk management orientation as a foundational condition of early-career cybersecurity professionals — attributing it to professional socialization, educational preparation, or the absence of formal risk formation at the point of entry. |

*Note. GAP-TFR (Technocentric Risk Framing) is the structural parent code for TFR-PROB and TFR-SOLN. It appears only as an organizing label. TFR-PROB and TFR-SOLN are the applied codes in all five cases; GAP-TFR does not contribute to the Tension Index GAP score. The seven codes that contribute to the GAP score are: TFR-PROB, TFR-SOLN, GAP-MYO, GAP-BUS, GAP-SCAL, GAP-LXI, and GAP-ENTRY.*

### Domain E — Development of Risk Judgment (DEV)

| Domain E — Development of Risk Judgment (DEV) | |
|---|---|
| **Code** | **Definition** |
| **DEV-EXP** | Experiential learning is the primary driver of risk judgment development. Applies when a leader describes on-the-job exposure — incident participation, cross-functional engagement, accumulated decision-making experience — as the mechanism through which cybersecurity professionals develop contextual risk reasoning. |
| **DEV-FORM** | Need for formal business or risk training. Applies when a leader articulates that structured educational intervention — courses, certifications, credential programs in risk, finance, regulatory compliance, or business management — is necessary to develop risk-competent cybersecurity professionals. |
| **DEV-MULTI** | Cross-domain exposure as a development mechanism. Applies when a leader describes deliberate exposure to functions outside cybersecurity—risk management, privacy, audit, finance, compliance, and executive leadership—as the pathway through which practitioners develop an enterprise-level risk perspective. |
| **DEV-EXEC** | Executive mandate or organizational culture as the shaping force for risk mindset. Applies when a leader describes organizational expectations, escalation standards, leadership demands, or cultural norms — rather than training or experience — as the mechanism that drives cybersecurity professionals to adopt risk-oriented language and reasoning. |

## Section 2 — Salience Scoring Scale

Salience scores reflect the weight and emphasis the participant assigns to a theme in the transcript—not the analyst's judgment of its theoretical importance. Each code receives one salience score per case, drawn from the behavioural anchors below. Salience is assigned after the full code frequency table is complete, using frequency, elaboration, and the prompted/unprompted character of the evidence together.



**Table 37**

*Salience Scoring Behavioural Anchors*

| Score | Label | Behavioural Anchor |
|-------|-------|--------------------|
| **0** | None | The code concept is absent from the transcript in any form. The theme is neither mentioned by the interviewee nor implied or described by the participant. |
| **1** | Low | The concept appears once, in direct response to an interviewer's question, with no elaboration beyond the minimum required to answer. The participant does not return to this theme independently at any later point in the transcript. |
| **2** | Moderate | The concept appears once or twice. The participant responds to a question and offers some elaboration — an example, a qualification, a consequence — beyond the minimum the question required. OR the concept appears in a follow-up answer that meaningfully extends an earlier prompted response. Evidence is consistent and clear, but remains primarily limited to prompted questions. |
| **3** | High | The concept appears in two or more distinct idea units across at least two different question contexts. At least one instance is unprompted — the participant raises the concept independently, extends substantially beyond the scope of the question, or returns to the theme in a different Section of the interview. The participant may use emphatic language ('always,' 'critical,' 'fundamentally,' 'never'). |
| **4** | Central / Recurring | The concept is a dominant thread across the interview. It appears in three or more idea units, is raised unprompted at multiple points, and the participant treats it as a core organizing principle of their argument. Evidence is frequently unprompted, emotionally weighted, analytically specific, or accompanied by detailed illustrative examples that substantially exceed what the question required. Salience 4 is rare—apply conservatively. |

> **Prompted vs. Unprompted Gating Rule (Required Before Assigning Salience 3 or 4)**
>
> Before assigning salience 3 or 4, the analyst must confirm that at least one instance of the code meets the unprompted criterion: the participant raised the concept without direct interviewer prompting or substantially exceeded what the question required. If all instances are proportionate responses to direct questions and the participant does not return to the theme independently, the coding rubric caps the salience score at 2. The Interview Questions Document (Document 3 in the analytical protocol ingestion order) serves as the reference standard for determining prompted/unprompted status.

# Section 3 — Expectation–Gap Tension Index

The Tension Index produces a COG score and a GAP score for each case by summing the salience values of the seven COG codes and the seven GAP codes, respectively. These scores classify each case into one of four typological quadrants. CAP Section 4 records both scores and classification scores for each case.



**Table 38**

*Tension Index Formula*

| Index | Formula | Codes Included (only) |
|---|---|---|
| **Total COG Salience Score** | Sum of salience values across all 7 COG codes (Maximum: 28) | COG-LXI + COG-IMP + COG-HOL + COG-COMP + COG-OPT + COG-RES + COG-STRAT |
| **Total GAP Salience Score** | Sum of salience values across all 7 GAP codes (Maximum: 28) | TFR-PROB + TFR-SOLN + GAP-MYO + GAP-BUS + GAP-SCAL + GAP-LXI + GAP-ENTRY |

**Table 39**

*Classification Thresholds*

| COG Score | GAP Score | Classification |
|---|---|---|
| **≥ 10 / 28** | **≥ 10 / 28** | **HIGH EXPECTATION / HIGH GAP** |
| ≥ 10 / 28 | < 10 / 28 | High Expectation / Low Gap |
| < 10 / 28 | ≥ 10 / 28 | Low Expectation / High Gap |
| < 10 / 28 | < 10 / 28 | Low Expectation / Low Gap |

**Counting Rule**

Only the 14 codes listed above (7 COG + 7 GAP) contribute to the Tension Index; the calculation excludes ERF, RGP, and DEV codes from both COG and GAP scores. The analysis excludes the GAP-TFR parent code to ensure more granular results at the sub-code level; only TFR-PROB and TFR-SOLN contribute to the GAP score. CAP Section 4 requires both raw scores and the resulting classification label for every entry.

# Section 4 — Numeric Salience Vector (Fixed Code Order)

CAP Section 5 for every case record one salience value (0–4) for each of the 28 codes in the fixed order below. The fixed sequence is applied identically across all cases to enable cross-case comparison of the complete salience profile. The final report includes zero-frequency codes as 0 rather than omitting them from the analysis. Colour coding in the table below indicates domain membership.



**Table 40**

*Fixed Code Sequence for Numeric Salience Vector*

| Domain A — ERF | Domain B — RGP | Domain C — COG | Domain D — GAP | Domain E — DEV |
|---|---|---|---|---|

| # | Code | Dom. | | # | Code | Dom. |
|---|---|---|---|---|---|---|
| 1 | ERF-SURV | A | | 15 | COG-OPT | C |
| 2 | ERF-FIN | A | | 16 | COG-RES | C |
| 3 | ERF-REG | A | | 17 | COG-STRAT | C |
| 4 | ERF-REP | A | | 18 | TFR-PROB | D |
| 5 | ERF-STRAT | A | | 19 | TFR-SOLN | D |
| 6 | RGP-LXI | B | | 20 | GAP-MYO | D |
| 7 | RGP-COLL | B | | 21 | GAP-BUS | D |
| 8 | RGP-ESC | B | | 22 | GAP-SCAL | D |
| 9 | RGP-COMP | B | | 23 | GAP-LXI | D |
| 10 | RGP-ACC | B | | 24 | GAP-ENTRY | D |
| 11 | COG-LXI | C | | 25 | DEV-EXP | E |
| 12 | COG-IMP | C | | 26 | DEV-FORM | E |
| 13 | COG-HOL | C | | 27 | DEV-MULTI | E |
| 14 | COG-COMP | C | | 28 | DEV-EXEC | E |

*Note. This order is fixed and identical across all seven cases. The sequence is not varied. CAP Section 5 covers all 28 codes in this order with no omissions. Salience 0 is a valid and required entry for codes absent from a transcript.*





Appendix D

Leadership Interview Coding Summary

**INT-01 · Chief Information Security Officer · iGaming / Online Gambling**

## Participant Profile

| Field | Value |
| --- | --- |
| Interview ID | INT-01 |
| Anonymized Role | Chief Information Security Officer |
| Industry | iGaming / Online Gambling |
| Organization Size | ~290 employees |
| Region | Malta / Europe |
| Codebook Applied | LRCC |

## Domain Salience Summary

| Domain | Codes Present | Total Salience | Max Possible | % of Max |
| --- | --- | --- | --- | --- |
| A — ERF | 3 of 5 | 8 | 20 | 40% |
| B — RGP | 5 of 5 | 12 | 20 | 60% |
| C — COG | 7 of 7 | 13 | 28 | 46% |
| D — GAP | 5 of 7 | 13 | 28 | 46% |
| E — DEV | 4 of 4 | 9 | 16 | 56% |
| **TOTAL** | **24 of 28** | **55** | **112** | **49%** |

**Expectation–Gap Tension Index:** COG Total = 13 | GAP Total = 13 | COG–GAP Spread = 0



**Typological Classification:** High Expectation Case; High Gap Evidence; Symmetric Spread — neither expectation inflation nor gap minimization is present. Both domains are evidence-grounded and of equal salience weight.

---

### Dominant Themes

Five themes, derived from Step 5 thematic synthesis and Step 8 Case Abstraction Profile.

- **Mature governance architecture, uneven professional quality:** INT-01's organization operated a formally institutionalized, two-tier risk governance structure featuring a dedicated IT security risk committee and a broader enterprise risk committee, with cross-functional composition that included the COO, CTO, and Compliance Officer. The architecture embedded a two-stage likelihood × impact rating process, a Jira-tracked risk registry, and structured escalation pathways to the board level. Despite this structural maturity, the participant identified persistent and recurrent deficiencies in professional risk cognition that permeated the governance process — a condition framed not as a governance failure but as a workforce capability gap.

- **Technocentric framing as the primary and generalized professional gap:** The dominant gap pattern identified by INT-01 was the consistent tendency of cybersecurity professionals to conceptualize risks as technical problems requiring tool-based remediation, while failing to conduct root cause analysis or to recognize process-level and operational dimensions — such as 24/7 SOC coverage — as potentially more consequential risk factors. A specific, internally sourced organizational case provided grounding for this pattern involving SIEM and SOAR investment recommendations that omitted the coverage dimension and emerged across multiple teams as a characteristic behavioural norm of tool-culture security thinking.

- **Business-impact blindness as a cognitive scope failure:** A conceptually distinct but related gap concerned with professionals' failure to understand the full organizational consequences of security events, particularly disproportionate reputational exposure. INT-01 framed this not as a knowledge deficit but as a structural limitation in cognitive scope — distinguishing between professionals who "appreciate" and those who genuinely "understand" that a publicly disclosed minor breach could be existentially consequential to the firm. This framing elevated GAP-BUS from an attitudinal observation to a diagnosis of bounded risk reasoning.

- **Risk management as emergent from, rather than definitional to, the cybersecurity role:** When directly asked whether risk management was central to the cybersecurity professional's role, INT-01 characterized it as something that necessarily accompanies the role rather than as its defining purpose. This position



introduced a theoretically significant moderating tension. While the participant held strong professional cognitive expectations (COG-HOL, COG-LXI) and delivered a clear developmental prescription, he simultaneously described an active executive practice of enforcing role-scope boundaries and deflecting responsibility away from compliance and audit functions. This posture structurally constrained the organizational environment in which professionals might develop broader risk reasoning.

- **Business-level formal training as the highest-priority developmental intervention:** INT-01's developmental prescription was unambiguous and delivered without hedging: formal training in iGaming regulatory and compliance literacy represented the single most important change to professional development. This prescription emerged from a coherent logical chain — the participant characterized experiential learning within a single organization as too slow and event-contingent; multi-channel external exposure was valued but insufficient; the participant identified formal, business-level training as the necessary corrective for the gap between professionals' technical orientation and the enterprise risk demands of a heavily regulated operating environment.

### Analytical Paragraph for the INT-01 Interview

INT-01, a Chief Information Security Officer at a mid-size iGaming firm operating within a heavily regulated European licensing environment, led a risk governance architecture of demonstrable institutional maturity, characterized by a two-tier committee structure with fixed cross-functional membership, a formal two-stage likelihood × impact rating protocol, and a tiered remediation recommendation framework that explicitly allocated residual risk accountability to decision-makers. The governance domain (B — RGP) was the most densely coded in the transcript (salience 12; 60% of maximum), with collaborative risk process (RGP-COLL, salience 4) functioning as the organizational default across all risk categories rather than as a contextual practice. Notwithstanding this structural strength, INT-01 identified persistent professional deficiencies operating beneath and within the governance architecture. The dominant gap pattern was technocentric problem framing: professionals consistently defined cybersecurity risks as technical problems requiring tool-based solutions, failed to conduct root cause analysis, and neglected operational dimensions — such as continuous SOC coverage — as potentially primary risk factors. This pattern, documented through a specific organizational case and explicitly generalized across teams, achieved a gap salience of 13 out of 28, symmetric with the cognitive expectation score. A secondary and analytically related gap — business-impact blindness — was framed by the participant not as attitudinal inattention but as a structural failure of cognitive scope, evidenced by professionals' inability to recognize the disproportionate reputational consequences of even minor security events. The participant's primary cognitive expectation of professionals was holistic, enterprise-encompassing risk framing before escalation, a



requirement articulated across all three phases of the interview and operationally defined as pre-escalation consultation with compliance, finance, and operational functions. A theoretically significant tension emerged in the developmental domain: INT-01 endorsed a multi-channel development model and identified business-level formal training as the single highest-priority intervention, while simultaneously maintaining an executive boundary management practice that confined professionals' operational scope to technical risk functions — a structural constraint that partially limits the organizational conditions under which broader risk judgment might develop.



**INT-02 · Director of Security · IT Infrastructure & Managed Services**

### Participant Profile

| Field | Value |
|---|---|
| Interview ID | INT-02 |
| Anonymized Role | Director of Security Services |
| Industry | IT Infrastructure, Cybersecurity & Managed Services |
| Organization Size | ~150 employees |
| Region | Canada (Toronto, Ontario) |
| Codebook Applied | LRCC |

### Domain Salience Summary

| Domain | Codes Present | Total Salience | Max Possible | % of Max |
|---|---|---|---|---|
| A — ERF | 5 of 5 | 10 | 20 | 50% |
| B — RGP | 5 of 5 | 7 | 20 | 35% |
| C — COG | 5 of 7 | 9 | 28 | 32% |
| D — GAP | 7 of 7 | 18 | 28 | 64% |
| E — DEV | 3 of 4 | 5 | 16 | 31% |
| **TOTAL** | **25 of 28** | **49** | **112** | **44%** |

**Expectation–Gap Tension Index:** COG Total = 9 | GAP Total = 18 | COG–GAP Spread = −19

**Typological Classification:** Low Expectation Case; High Gap Evidence; Large Negative Spread — the participant's documented experience of practitioner failure (GAP = 18/28, 64% of maximum) substantially outweighs her articulated professional expectations (COG = 9/28, 32% of maximum), consistent with a leader who has normalised the gap as a structural feature of the profession rather than an individual performance failure.



## Dominant Themes

Five themes derived from Step 5 thematic synthesis and Step 8 Case Abstraction Profile.

- **Compound and normalized professional gap:** INT-02 documented a multi-dimensional failure of risk reasoning that she had absorbed into her daily practice as a routine compensatory function: technocentric problem framing (TFR-PROB, Salience 4), absence of structured likelihood × impact reasoning, scale unawareness, and business value blindness consistently co-occurred across practitioner submissions. The participant did not treat the gap as episodic or individually attributable; the participant characterized it as pervasive, quantified it at '90% or more' of practitioners, and extended it to the most junior professional tier (Tier 1 SOC analysts), framing it as a foundational socialization failure built into the profession rather than a remedial performance issue.

- **Inability to recall a single adequate practitioner risk submission:** When invited by the interview protocol to offer a positive example of well-prepared risk reasoning from a practitioner, INT-02 offered none, stating explicitly that she did not remember a time when a submission was well prepared and she was satisfied with it, unless it originated from senior management downward rather than from practitioners upward. This reversal of the expected information flow is analytically distinctive and directly relevant to the dissertation's argument about the training integration deficiency, as it establishes the gap not as an occasional phenomenon but as empirically unbroken within the participant's professional experience.

- **TFR-PROB as the transcript's organizing diagnostic frame:** Technocentric problem framing achieved the highest salience of any single code in the interview (Salience 4, the only code at that level), appearing across five distinct idea units spanning four separate interview questions. The participant's recurring formulation — that professionals see a technical problem to solve rather than an organizational risk to evaluate — functioned as the master explanation for why business translation fails, why scale awareness is absent, and why the professionals bypass likelihood × impact reasoning. GAP-BUS (Salience 3), GAP-SCAL (Salience 3), and GAP-LXI (Salience 3) each operated as downstream consequences of the same foundational framing error.

- **Formal training as the primary and specifically prescribed developmental remedy:** INT-02 articulated the clearest and most prescriptive developmental position of the transcript: The participant described formal training as 'the main thing,' an evaluatively comparative formulation that explicitly ranked it above both experiential learning (characterised as unreliable and event-contingent) and coaching (critiqued as quality-dependent on the mentor's own risk perspective). The prescribed curriculum was specific and structured across three stages: business goal comprehension, risk reasoning principles at each



organizational level, and goal mapping through a risk management lens. This training design aligns directly with the cognitive expectations articulated elsewhere in the interview.

- **Governance nominally present but operationally informal:** The organization's risk governance process contained identifiable structural elements — a tiered escalation pathway, domain-differentiated director-level involvement, and a named likelihood × impact evaluation stage — but the participant acknowledged each element as informal and undocumented. The participant stated directly that the **L × I** process failed to provide a clear path for the immediate quantification of risk cost, yielding a governance profile characterized by nominal presence without operational formalization. This internal tension between architecture and execution distinguished the case from both mature governance models and fully absent ones.

### Analytical Paragraph for the INT-02 Interview

INT-02, a Director of Security Services at a mid-sized North American managed services provider, framed enterprise risk through a business-survival logic that anchored her risk prioritisation consistently across the interview: financial exposure, reputational damage, and service continuity each functioned as escalation thresholds, with the company's survival treated as the ultimate risk referent (ERF-SURV, Salience 3, the domain's most emphatic code). The organization's governance architecture incorporated a tiered escalation structure, domain-differentiated stakeholder involvement, and a named likelihood × impact evaluation stage, yet the participant immediately qualified that evaluation as informal and non-quantitative — a governance profile of nominal presence without operational rigour (Domain B, salience 7/20, 35%) that distinguished this case from organizations with a fully embedded risk process. Domain C (Expected Cognitive Work) was pragmatically selective (salience 9/28, 32%): the participant named risk calculation, business impact articulation, and presentation of response options as the three core professional requirements, while COG-COMP and COG-RES received zero salience, indicating that her expectations were instrumentally sufficient rather than analytically comprehensive. The gap domain was the transcript's most analytically saturated (salience 18/28, 64%), organized around a single dominant finding: TFR-PROB achieved the only Salience 4 code in the interview, grounded in an inability to recall a single practitioner submission she considered risk-adequate without her intervention, a quantified assertion that 90% or more of practitioners lacked the required level of risk thinking, and an unprompted extension of this gap to entry-level Tier 1 SOC analysts — establishing the deficit as foundational rather than senior-level. GAP-BUS, GAP-SCAL, and GAP-LXI each reached Salience 3 as functionally related downstream failures of the same technocentric framing orientation. Developmentally, the participant prescribed formal training — delivered at the point of



professional entry — in business goals, financial reasoning, and risk management mapping as the primary corrective, explicitly discounting experiential and coaching-based pathways as unreliable substitutes.



**INT-03 · Senior Vice President, Cybersecurity · Financial Services**

## Participant Profile

| Field | Value |
|---|---|
| Interview ID | INT-03 |
| Anonymized Role | Senior Vice President, Cybersecurity |
| Industry | Financial Services (Banking) |
| Organization Size | ~80,000 employees (multinational) |
| Region | New York-based; global operations |
| Codebook Applied | LRCC |

## Domain Salience Summary

| Domain | Codes Present | Total Salience | Max Possible | % of Max |
|---|---|---|---|---|
| A — ERF | 4 of 5 | 9 | 20 | 45% |
| B — RGP | 5 of 5 | 9 | 20 | 45% |
| C — COG | 7 of 7 | 13 | 28 | 46% |
| D — GAP | 6 of 7 | 16 | 28 | 57% |
| E — DEV | 3 of 4 | 5 | 16 | 31% |
| **TOTAL** | **25 of 28** | **52** | **112** | **46%** |

**Expectation–Gap Tension Index:** COG Total = 13 | GAP Total = 16 | COG–GAP Spread = −3

**Typological Classification:** High Expectation Case; High Gap Evidence; Moderate Negative Spread — gap observations exceed and are more emphatic than expectation articulations, consistent with a participant whose interview style is primarily diagnostic rather than prescriptive. The primary driver of the asymmetry is GAP-BUS (Salience 4), the sole code in the interview to reach maximum salience, with COG-HOL (Salience 3) as the primary expectation counterpart.



### Dominant Themes

Five themes derived from Step 5 thematic synthesis and Step 8 Case Abstraction Profile.

- **Financial and regulatory logic as the governing risk frame:** INT-03 organized their enterprise risk framing around a single, consistently applied logic: financial magnitude determines which risks reach senior-level attention and which recommendations succeed or fail. ERF-FIN achieved Salience 3 — the highest in Domain A — anchored by the recurring formulation that 'money always wins the game,' and illustrated how the institution removed a security control because its trading-latency cost exceeded its protective value. Regulatory and audit exposure (ERF-REG, Salience 2) provided the organizational attention driver for priority-setting, with an anticipatory internal governance posture (finding issues before regulators do), establishing the mature organizational context in which he diagnosed professional gaps. ERF-REP was absent, a notable omission for a financial services institution that likely reflects this participant's internal operational orientation rather than institutional indifference.

- **Business value and monetary reasoning gap as the primary and organizing professional deficit:** GAP-BUS achieved Salience 4 — the only code in the interview to do so — appearing across four distinct idea units in four separate question zones with consistent emphatic language across all of them. The participant framed the deficit as simultaneously attitudinal and structural: cybersecurity professionals do not engage with the financial logic that governs executive risk decisions, and the role's information architecture structurally denies them access to the asset valuations and frequency data required for quantitative likelihood × impact reasoning. This dual framing distinguished the gap from a simple training deficiency and carried direct remediation implications, as resolving the structural access problem requires organizational redesign rather than training intervention alone. The HFT case, volunteered spontaneously, provided the interview's most concrete institutional illustration of the gap's operational consequences.

- **The COG-HOL / GAP-MYO asymmetry as the interview's most consistent analytical pattern:** The most analytically coherent pair of findings in the transcript was the symmetrical opposition between COG-HOL (Salience 3) as the primary professional expectation and GAP-MYO (Salience 3, Frequency 5) as the most frequently observed failure mode. The participant articulated holistic, enterprise-wide, architecturally informed reasoning as the foundational standard for professional competence — naming it as the answer to what he would change most if he could change one thing — while simultaneously documenting that professionals consistently demonstrate the precise opposite: narrow, minute-level, scope-limited thinking that treats individual technical decisions as isolated personal burdens rather than as components of a collaborative enterprise risk governance process. The architectural training sub-dimension of COG-HOL was introduced spontaneously, without any architectural framing in the interviewer's question, elevating its analytical significance as a self-generated professional requirement.



- **Governance maturity as a context that does not automatically produce risk-capable practitioners:** The organizational governance environment described by INT-03 was the most institutionally sophisticated of the interview series: dedicated risk officers embedded in each business unit with direct access to frequency data and financial valuations, formal risk committees with systematic escalation rules, a four-option risk treatment taxonomy, and consistent process governance across risk types. The participant activated all five RGP codes (Domain B, Salience 9/20, 45%) and formally institutionalized the LxI function. The analytically significant implication is that the analysis diagnosed professional gaps within — not against — this mature environment, establishing that well-governed organizations do not automatically produce risk-capable cybersecurity practitioners and that structural governance maturity may even compound the gap by delegating quantitative risk reasoning to specialist risk officers, thereby normalizing cybersecurity professionals' non-engagement with financial risk quantification.

- **The spontaneous CISSP closing insight as profession-wide convergent evidence:** The analytically most distinctive unit in the transcript was INT-03-25: a closing observation volunteered after the interview had formally concluded, with no question prompt whatsoever, in which the participant cited CISSP examination failure rates as evidence that technology-focused reasoning bias operates not only within his organization but across the profession at the credentialing level. This observation (INT-03-25) was the only entirely unprompted unit in the interview. Its content positioned a widely recognized professional certification as diagnostic evidence that the professional credentialing process reproduced and validated at the profession-wide level the same gap diagnosed through organizational observation, elevating TFR-PROB from an organizational anecdote to a cross-institutional, credential-anchored pattern finding. The participant simultaneously endorsed experiential learning as the primary mechanism for developing contextual risk judgment and critiqued formal cybersecurity training as insufficient for that purpose.

### Analytical Paragraph for the INT-03 Interview

INT-03, a Senior Vice President of Cybersecurity at a large multinational financial services institution with global operations, articulated a risk governance model of considerable institutional sophistication — featuring dedicated risk officers embedded in each business unit, formal risk committees with rule-governed escalation thresholds, a four-option risk treatment framework, and a structurally anticipatory regulatory posture — while simultaneously delivering the most analytically dense and internally coherent gap diagnosis in the interview series. The enterprise risk framing (Domain A, Salience 9/20) organized subsequent professional diagnoses around a single governing logic: financial magnitude determines senior-level attention and executive decision outcomes, as captured in the participant's recurring observation that money always wins the game. Against that referential



standard, GAP-BUS emerged as the sole Salience 4 code in the interview and the interview's primary explanatory construct, diagnosed as both an attitudinal failure — professionals do not engage with the monetary logic of executive risk decisions — and a structural access problem, in that the information architecture of the cybersecurity role denies practitioners the asset valuations and frequency data that full likelihood × impact reasoning requires. The primary cognitive expectation articulated against this gap was COG-HOL (Salience 3): holistic, enterprise-wide, architecturally informed risk reasoning, explicitly named as the single most important professional competency change the participant would implement, and described as routinely violated by a professional community that defaults to narrow, minute-level, technology-focused problem framing (GAP-MYO, Frequency 5, Salience 3). The governance domain was notable for a structural delegation that may compound the gap: the quantitative LxI function was formally assigned to risk officers rather than cybersecurity teams, normalizing the latter's non-engagement with financial risk quantification and making the GAP-LXI deficit partly an organizational design feature rather than a purely individual training failure. Developmentally, the participant endorsed experiential, on-the-job learning as the primary acquisition mechanism for contextual risk judgment, critiqued formal cybersecurity training as generically insufficient for developing that judgment, and — in the interview's sole entirely unprompted unit — cited CISSP examination failure rates as convergent profession-wide evidence that the technology-focused reasoning bias diagnosed in organizational practice is reproduced and institutionally validated at the credentialing level.



## INT-04 · Executive Director of Customer Success · Cybersecurity SaaS

### Participant Profile

| Field | Value |
|---|---|
| Interview ID | INT-04 |
| Anonymized Role | Executive Director of Customer Success |
| Industry | Cybersecurity SaaS |
| Organization Size | ~47 employees (small enterprise) |
| Region | United States (Chicago, IL; global operations) |
| Codebook Applied | LRCC |

### Domain Salience Summary

| Domain | Codes Present | Total Salience | Max Possible | % of Max |
|---|---|---|---|---|
| A — ERF | 3 of 5 | 7 | 20 | 35% |
| B — RGP | 4 of 5 | 8 | 20 | 40% |
| C — COG | 5 of 7 | 12 | 28 | 43% |
| D — GAP | 6 of 7 | 12 | 28 | 43% |
| E — DEV | 4 of 4 | 8 | 16 | 50% |
| **TOTAL** | **22 of 28** | **47** | **112** | **42%** |

**Expectation–Gap Tension Index:** COG Total = 12  |  GAP Total = 12  |  COG–GAP Spread = 0

**Typological Classification:** High Expectation Case; High Gap Evidence; Perfectly Symmetric Spread — COG and GAP scores are identical (12/28 each, both clearing the ≥10 threshold), indicating a participant who holds mature and consistently articulated professional expectations while documenting equally consistent and well-evidenced failure to meet them. Both expectations and gaps are calibrated by career stage rather than applied as a uniform professional standard, producing symmetry through calibration rather than through indifference.



### Dominant Themes

Five themes derived from Step 5 thematic synthesis and Step 8 Case Abstraction Profile.

- **Financial and survival primacy as the organizing enterprise risk frame:** INT-04 framed enterprise risk almost exclusively through financial and survival lenses, with ERF-FIN achieving Salience 3 as the most recurrent single ERF code — appearing across four distinct units and three separate interview questions — and the Q3 assertion that financial impact is 'almost always' prioritised above all other risk categories, including cybersecurity, constituting the clearest leadership risk prioritisation statement in the transcript. ERF-SURV reinforced this frame through an operative risk taxonomy that distinguished survival risk from all other risk types as the primary decision-relevant category for a small enterprise. ERF-REG and ERF-REP were both absent—a structurally distinctive pattern for a cybersecurity SaaS provider with data-handling and customer obligations—indicating that compliance and reputational exposure were absorbed into the financial frame rather than treated as independent risk categories.

- **Holistic integration as the interview's dominant and organizing cognitive standard:** COG-HOL achieved Salience 4 — the highest single-code score in this analysis and the only code to reach that level — appearing across five distinct idea units in three separate interview Sections through multiple independent mechanisms: the perceived-vs-real risk filter, the triage-from-multiple-perspectives standard, the 3 am executive-consumable incident report standard, and the developmental threshold conditioned on professional seniority. The participant employed a medical analogy — the CTI specialist as a surgeon diagnosing everything as a surgical problem, regardless of whether surgery is the appropriate intervention — to illustrate why specialization-induced framing constitutes a failure of professional competence rather than a failure of effort. COG-IMP (Salience 3) operated as the second anchor of the expectation profile, with the participant explicitly stating 'I preach' impact evaluation before escalation, signalling a deeply held professional conviction rather than a context-specific observation.

- **Specialization-induced myopia as a pre-formed, industry-wide structural gap:** GAP-MYO (Salience 3) was the transcript's dominant gap code, and its analytical distinctiveness derived not merely from salience but from provenance: the participant explicitly noted that this observation was prepared before the interview, having written it in advance as the answer to the gap question, a level of pre-formation that elevated analytical confidence substantially above interview-prompted reflection. The characterization was unambiguous in its industry-wide scope — 'very, very common across the entire industry' — and the participant specifically identified the mechanism as familiarity-induced domain capture: professionals trained in a particular technical subdomain (CTI, vulnerability management, incident response) apply that domain's diagnostic template to every risk situation regardless of its organizational relevance. TFR-PROB (Salience 2) represented this pattern at the product level, and GAP-BUS and GAP-LXI (both Salience 2)



operated as its downstream consequences: domain-constrained framing produced output that did not translate into real organizational risk and did not reflect structured likelihood × impact reasoning.

- **Developmentally stratified gap expectations and the empirical absence of formal risk programmes:** A distinguishing feature of this case was the participant's explicit calibration of risk management expectations by career stage: he did not require entry-level professionals to meet holistic risk cognition standards, with the expectation reserved for mid-career and senior practitioners. This stratification structurally normalized the gap at the entry level, framing it as an acceptable developmental state rather than a universal professional failure. The developmental profile (Domain E, Salience 8/16, 50% — the strongest of any domain) was anchored by DEV-EXP (Salience 3) and grounded empirically in an extended FBI airport narrative that demonstrated how analytical discipline in incident reporting directly reduced the likelihood that a risk would materialize. The participant simultaneously stated that he had never observed a formal risk programme implemented at any enterprise, positioning organic career experience as the de facto and universal development pathway—a career-length empirical observation directly relevant to the dissertation's thesis about training integration deficiencies.

- **Risk as organizational survival language and executive mandate as the adoption driver:** The participant delivered an entirely unsolicited closing statement — with no question prompt — framing risk as the language cybersecurity professionals must adopt to survive in non-cybersecurity-focused conversations, a formulation that condensed the interview's central argument and positioned risk cognition as an organizational survival competency rather than a supplementary analytical skill. DEV-EXEC (Salience 2) provided the mechanism: the participant attributed professional adoption of risk management language not to intrinsic valuation but to executive mandate, particularly C-level pressure in large enterprises to translate cybersecurity work into agreed-upon risk frameworks. COG-STRAT (Salience 2), reinforced by the same unprompted closing statement, established strategic alignment — rendering technical findings in terms legible to non-cybersecurity organizational actors — as both a cognitive expectation and an externally enforced professional survival condition.

---

### Analytical Paragraph for the INT-04 Interview

INT-04, an Executive Director of Customer Success at a small cybersecurity SaaS enterprise, articulated a risk governance model that reflected the compressed, collaborative, and financially driven realities of the small-company context: tiered but informal escalation cadences, CEO-inclusive risk review rhythms, and a SWAT-response mechanism for high-impact events (Domain B, Salience 8/20, 40%), operationally coherent without being formally elaborate. The participant hierarchically organized the enterprise risk framing around financial and



survival primacy. (ERF-FIN achieved Salience 3 across four units, with the participant explicitly asserting that the organization almost always prioritizes financial impact over cybersecurity) while ERF-REG and ERF-REP were both absent, establishing that compliance and reputational exposure were absorbed into the financial survival frame rather than treated as independent risk categories. Against this referential standard, COG-HOL emerged as the sole Salience 4 code in the analysis and the interview's organizing cognitive expectation: the organization required mid-career and senior professionals to synthesise technical, organisational, and risk domains into executive-ready analysis available at any hour, a standard the participant articulated through five independent mechanisms including a medical analogy, a triage norm, and an explicit developmental threshold. The dominant observed gap, GAP-MYO (Salience 3), operated as the systematic violation of that standard: the participant pre-formed and explicitly documented the observation that domain specialists generate diagnostics calibrated to their trained technical framework, regardless of organizational relevance — a pattern characterized as industry-wide in scope and structurally reproduced across the profession. Development of risk judgment was attributed primarily to organic career experience — the participant reported having never observed a formal risk programme at any enterprise — with executive mandate rather than intrinsic motivation identified as the proximate driver of risk language adoption, and an unsolicited closing formulation framing risk cognition as an organizational survival language whose adoption professionals must achieve to remain relevant in non-cybersecurity conversations.



**INT-05 · CISO / Sr. Director · Post-Secondary Education & Public Service**

### Participant Profile

| Field | Value |
|---|---|
| **Interview ID** | INT-05 |
| **Anonymized Role** | CISO / Sr. Director |
| **Industry** | Post-Secondary Education and Public Service |
| **Organization Size** | ~$100M/year operating budget; ~500–600 staff |
| **Region** | Canada — Vancouver and Kamloops, British Columbia |
| **Codebook Applied** | LRCC |

### Domain Salience Summary

| Domain | Codes Present | Total Salience | Max Possible | % of Max |
|---|---|---|---|---|
| A — ERF | 2 of 5 | 3 | 20 | 15% |
| B — RGP | 5 of 5 | 11 | 20 | 55% |
| C — COG | 6 of 7 | 13 | 28 | 46% |
| D — GAP | 5 of 7 | 13 | 28 | 46% |
| E — DEV | 3 of 4 | 7 | 16 | 44% |
| **TOTAL** | **21 of 28** | **47** | **112** | **42%** |

**Expectation–Gap Tension Index:** COG Total = 13 | GAP Total = 13 | COG–GAP Spread = 0

**Typological Classification:** High Expectation Case; High Gap Evidence; Perfectly Symmetric Spread — COG and GAP scores are identical (13/28 each, both well above the ≥10 threshold). The symmetry reflects a participant who holds demanding, quantitative, and operationally specific analytical expectations and documents equally robust, multi-evidenced evidence of their systemic non-fulfilment. Domain B (Risk Governance & Process) is the strongest domain (55% of maximum), reflecting a participant whose primary orientation is methodological and governance-architectural.



## Dominant Themes

Five themes derived from Step 5 thematic synthesis and Step 8 Case Abstraction Profile.

- **Method-anchored, type-agnostic enterprise risk framing:** INT-05 organized his enterprise risk framing around a single analytical formula — likelihood × impact minus effective controls, bounded by organizational risk tolerance — that he presented as personally internalized and applicable uniformly across all risk categories, irrespective of type. This methodological orientation explains the sparse ERF profile (Domain A, Salience 3/20, 15%): rather than building a taxonomy of named risk categories, the participant treated risk type as analytically agnostic, with the formula doing the classifying work that categorical framing does in other accounts. ERF-SURV, ERF-REP, and ERF-STRAT were absent — the absence of reputational framing being particularly notable given the post-secondary sector context, where breach-related reputational exposure is a recognized governance concern — while ERF-FIN appeared descriptively as an observation about what drives executive urgency rather than as the participant's own primary risk lens.

- **Multi-tiered governance architecture as the most densely populated domain:** Domain B (Risk Governance & Process) was the highest-salience domain in the transcript (Salience 11/20, 55%), with all five RGP codes active and RGP-ESC achieving Salience 3 across six distinct units — the highest frequency of any single code in the interview. INT-05 articulated a tripartite escalation typology: audit-to-board, IS-department-to-executive, and board-initiated top-down pathways, with the board explicitly positioned as the legitimate locus of risk ownership and accountability. RGP-LXI (Salience 3) was co-dominant with RGP-ESC, reflecting the participant's normative investment in **L × I** as a governance prerequisite: likelihood × impact calibration was framed not as a professional aspiration but as the necessary precondition for any intelligent executive discussion. Peer benchmarking through tools such as BitSight (RGP-COMP, Salience 2) extended the governance expectation into the comparative domain, requiring that IS risk be rendered board-comparable and positioned within the full organizational risk registry.

- **Quantitative and evidence-anchored cognitive expectations as co-dominant professional standards:** COG-LXI (Salience 3) and COG-IMP (Salience 3) emerged as co-dominant expectation codes, tightly coupled across three units each and framed as prerequisites rather than aspirational standards. The organization required professionals to defend their likelihood placements empirically and to quantify impact in financial, resource, and operational duration terms. The participant's own phrasing was 'How can you support your belief that the likelihood is wherever you've placed it?' COG-COMP (Salience 2) extended these demands into the comparative and benchmarking domain, and COG-RES (Salience 2) added the residual dimension: the organization expected professionals to articulate how applied controls had reduced the likelihood to an acceptable level, not merely identify a threat. COG-OPT was absent,



distinguishing this participant from those who additionally require tiered trade-off option presentation as a governance contribution.

- **Concrete, multi-evidenced gap documentation with a Socratic demonstration at its apex:** The gap domain (Domain D, Salience 13/28, 46%) was structured around three codes, each achieving Salience 3: TFR-PROB, GAP-LXI, and GAP-ENTRY. TFR-PROB appeared across four distinct units spanning three response contexts, with the most vivid illustration being the malware count example — technical metrics deployed in an executive briefing where they carried no organizational decision-making value — and the 'loosey-goosey' characterization of IS risk analysis as the substitution of technical rationalizations for structured assessment. GAP-LXI reached its highest analytical development in the termination audit unit, where INT-05 conducted a real-time Socratic interrogation of a professional's unsubstantiated 'high risk' designation, modelling the correct likelihood calibration that the professional had failed to perform — the only instance in this dataset where a gap is not merely named but analytically demonstrated through a reconstructed professional dialogue. GAP-ENTRY (Salience 3, Frequency 4) was grounded across four units with unusual breadth of evidence, culminating in sector-level community evidence from national post-secondary professional forums (CUCCIO, Canhite) that established the gap as a collective disposition rather than an individual or organizational phenomenon.

- **Emphatic formal certification prescription and the complete absence of executive-mandate development logic:** DEV-FORM achieved Salience 4 — the highest single-code salience score in the transcript — spanning four units that moved from diagnosis (formal training assessed as 'fairly weak'), through sector-level structural corroboration (low professional interest in risk formalization at national forums), to an unambiguous prescriptive imperative: 'get certified in risk', explicitly extended to privacy and audit disciplines as the complete professional development remedy. DEV-MULTI (Salience 2) was endorsed both as an informal compensatory mechanism — cross-domain exposure through direct engagement with risk and privacy functions — and as the explicit content of the formal prescription, establishing cross-domain credentialing as the structural objective rather than a supplement. The participant acknowledged DEV-EXP (Salience 1) as the current dominant pathway but explicitly positioned it as an insufficient substitute; the truncated sentence 'it's better if you get formal education, but —' communicated a preference hierarchy without ambiguity. DEV-EXEC was absent (a structurally significant omission given the participant's own senior position), consistent with a theory of professional development grounded in individual credentialing rather than organizational culture or executive mandate.



### Analytical Paragraph for the INT-05 Interview

INT-05, a Director and CISO with career experience spanning post-secondary education and public service sectors in British Columbia, Canada, anchored his professional framework in an explicit likelihood × impact minus effective controls formula that he applied as a type-agnostic standard across all organizational risk categories — producing the sparsest enterprise risk framing profile in the dataset (Domain A, Salience 3/20, 15%) and reflecting a methodological rather than categorical orientation in which the formula does the classifying work that named risk categories perform in other accounts. The governance domain was the most densely populated in the transcript (Domain B, Salience 11/20, 55%), featuring a tripartite escalation typology, the board as the explicit locus of risk ownership, mandatory peer benchmarking to achieve board-comparability, and RGP-LXI at Salience 3 framing likelihood × impact calibration as the precondition for any intelligent executive discussion. Against that governance standard, the participant articulated co-dominant quantitative cognitive expectations — COG-LXI and COG-IMP each at Salience 3, framed as prerequisites rather than aspirations — and documented their systematic non-fulfilment across a gap profile (Domain D, Salience 13/28) organized around three codes at Salience 3: TFR-PROB (technical metrics deployed without organizational decision-making value, across four units), GAP-LXI (structured **L × I** reasoning absent, most richly evidenced through a Socratic real-time demonstration of the correct calibration process that the professional had failed to perform), and GAP-ENTRY (analyst-level analytical incapacity evidenced at the individual, organizational, and sector-community levels through national post-secondary professional forums). An unambiguous normative hierarchy characterized the developmental domain: DEV-FORM achieved Salience 4 — the highest code salience in the interview — grounded in a diagnosis of current formal training as 'fairly weak,' sector-level structural corroboration of low professional interest in risk formalization, and a prescriptive imperative to obtain professional certification across risk, privacy, and audit disciplines; DEV-EXEC was absent, consistent with a theory of risk judgment development grounded in individual credentialing rather than organizational culture or executive mandate, and analytically notable given the participant's direct observation that many professionals do not spend time on risk management.



**INT-06 · Chief Executive Officer (CEO) · Cybersecurity SaaS**

## Participant Profile

| Field | Value |
|---|---|
| Interview ID | INT-06 |
| Anonymized Role | Chief Executive Officer (CEO) |
| Industry | ICT — Cybersecurity Awareness, Training & Culture (SaaS) |
| Organization Size | ~46 FTEs; ~1,500 client organisations; ~1 million end users |
| Region | Canada (Fredericton, New Brunswick) |
| Codebook Applied | LRCC |

## Domain Salience Summary

| Domain | Codes Present | Total Salience | Max Possible | % of Max |
|---|---|---|---|---|
| A — ERF | 4 of 5 | 8 | 20 | 40% |
| B — RGP | 5 of 5 | 9 | 20 | 45% |
| C — COG | 4 of 7 | 12 | 28 | 43% |
| D — GAP | 6 of 7 | 17 | 28 | 61% |
| E — DEV | 3 of 4 | 7 | 16 | 44% |
| **TOTAL** | **22 of 28** | **53** | **112** | **47%** |

**Expectation–Gap Tension Index:** COG Total = 12 | GAP Total = 17 | COG–GAP Spread = −5

**Typological Classification:** High Expectation Case; High Gap Evidence; Moderate Negative Spread — both COG (12/28) and GAP (17/28) clear the ≥10 threshold, placing this case firmly in the High Expectation / High Gap quadrant. The negative spread of five points reflects a participant whose gap observations are more emphatic, more extensively elaborated, and more thoroughly evidenced than his articulation of cognitive expectations, consistent with a diagnostic rather than prescriptive interview posture. Domain D achieves 61% of maximum possible salience, the highest single-domain percentage in the series.



## Dominant Themes

Five themes derived from Step 5 thematic synthesis and Step 8 Case Abstraction Profile.

- **Strategic, entrepreneurial risk framing organized around trade-off and context:** INT-06 defined risk from the opening unit as the cost dimension of a risk-reward equation — 'always contextual, always relational to what's the benefit' — a formulation that recurred across the full interview as the unifying interpretive lens through which the participant evaluated professional performance. ERF-STRAT and ERF-SURV each reached Salience 3, co-anchoring a risk frame that was entrepreneurial and adaptive rather than compliance-driven or reputational: Professionals must assess risk relative to the business value at stake, recognizing that cognitive miscalibration—not procedural non-compliance—is the dominant organizational failure mode. ERF-REP was absent despite the participant leading a SaaS platform serving regulated industries. At the same time, ERF-REG and ERF-FIN each appeared only once — the latter solely in a municipality case study rather than in relation to the participant's own organization.

- **Four active cognitive expectations at equal and high salience, organized around strategic outcome-orientation:** The COG domain exhibited a distinctive uniformity among its active codes (Salience 12/28, 43%), with COG-LXI, COG-IMP, COG-HOL, and COG-STRAT all reaching Salience 3, while COG-COMP, COG-OPT, and COG-RES were absent, producing an expectation profile that was strategic and qualitative rather than procedural or comparative. COG-LXI was grounded in Protection Motivation Theory and the diagnosis that likelihood calibration is the primary failure point in professional risk reasoning. COG-HOL was the most frequently coded expectation (four units), operationalized through both a negative exemplar (the adversarial imagination deficit and the failure of systemic thinking) and a positive exemplar in the Kerry Frey SIGNET framework. COG-STRAT achieved its most complete articulation in the 'department of know-how' reframing, which repositioned cybersecurity as a mission-enabler operating through outcome-driven logic: if we do X, we reduce Y, which enables Z.

- **Business blindness is the interview's dominant, recurring, and most emphatically evidenced gap:** GAP-BUS achieved Salience 4 — the only code in this case to do so, and the structuring analytical claim across the entire transcript. INT-06 argued that the cybersecurity field produces IT security professionals rather than cybersecurity professionals, with 'cyber' defined etymologically as encompassing people, technology, and control: the consequent absence of business management grounding, liberal arts awareness, and empathy for leadership decision contexts produces practitioners unable to translate technical findings into executive-actionable intelligence. This argument spanned four distinct units across Q10, Q12, and Q13. It drew on both concrete negative evidence — the $2M firewall recommendation that



ended a professional's career at a major institution by ignoring the business case — and cross-industry empirical data on IT project failure rates and AI adoption outcomes. The formulation 'People are hard' condensed the diagnostic claim in analytically precise terms and appeared as an emphatic, self-generated summary.

- **Likelihood miscalibration diagnosed through cognitive science theory and evidenced as a structural organizational failure:** GAP-LXI (Salience 3) was developed in this transcript at a theoretical depth not matched elsewhere in the interview series. INT-06 did not merely assert that professionals fail to estimate likelihood accurately — he developed an explanatory theory grounded in optimism bias and novelty bias as systematic cognitive distortions, identified likelihood calibration as 'where we die' in organizational risk reasoning, and proposed a practical governance workaround: treat all risks as likely until they can be proven not to be. TFR-PROB (Salience 3) provided the structural explanation, with the IT security professional identity creating the cognitive conditions for technical-only framing; GAP-MYO (Salience 3) extended the diagnosis in two directions — the misanthropic user-blaming syndrome that treats human factors as problems to be automated away, and the failure of adversarial imagination that leaves professionals unable to conceive of attack vectors outside their trained frame.

- **Bifocal developmental prescription and the CISO scapegoat dynamic as structural developmental barriers:** The developmental prescription was bifocal and philosophically grounded. DEV-FORM (Salience 3) reached its most elaborated form in the closing unit, where INT-06 prescribed teaching the etymological meaning of 'cyber' (people, technology, control) and 'technology' (techne + logos: the art and skill of building or using something, combined with careful consideration of what is gained or lost) as the foundational curriculum for risk management in cybersecurity — an unusual prescription grounded in disciplinary philosophy rather than competency frameworks or certification requirements. DEV-EXP (Salience 2) positioned mentored experiential learning as safe failure management: mentors create the conditions in which practitioners can make and learn from mistakes without career-limiting consequences, with the AI threat to mentorship traditions introduced unprompted as a structural risk to professional judgment development. DEV-EXEC (Salience 2) appeared in dual form — organizational culture as a positive enabler and, more emphatically, the CISO scapegoat dynamic as a structural barrier: short tenure, disproportionate compensation, and blame-absorber positioning collectively constitute a governance design that structurally disincentivizes the development of strategic risk management orientation.



### Analytical Paragraph for the INT-06 Interview

INT-06, a Chief Executive Officer leading a cybersecurity awareness and training SaaS platform serving approximately 1,500 client organizations (~1 million people), framed enterprise risk throughout the interview as the cost dimension of a strategic trade-off equation — always contextual, always relational to organizational objectives — producing an ERF profile (Salience 8/20, 40%) anchored by ERF-STRAT and ERF-SURV at Salience 3 each and absent of reputational framing, consistent with an entrepreneurial risk calculus in which strategic consequence rather than compliance or public trust drives risk prioritization. The governance architecture (Domain B, Salience 9/20, 45%) reflected the informality of a small organization: all five RGP codes were active, but none exceeded Salience 2, and the participant explicitly acknowledged the absence of formal likelihood methodology, describing an informal precautionary heuristic — treat all threats as likely until proven otherwise — as the operative substitute. Against that referential frame, the participant articulated four active cognitive expectations at uniform Salience 3 — COG-LXI, COG-IMP, COG-HOL, and COG-STRAT — each internally coherent and organized around the strategic outcome-orientation captured in the 'department of know-how' reframing and the outcome-driven logic of if-we-do-X-reduce-Y-enable-Z. GAP-BUS (Salience 4) served as the sole organizing explanatory claim, structuring the gap domain (Domain D, Salience 17/28, 61%—the highest single-domain percentage in the case series): the field trains IT security professionals rather than cybersecurity professionals, with 'cyber' defined etymologically as encompassing people, technology, and control, and the consequent absence of business management grounding, liberal arts awareness, and empathy for leadership decision context was evidenced across four units through a concrete career-limiting negative exemplar and cross-industry project failure data. GAP-LXI (Salience 3) extended the diagnosis into theorized cognitive territory — optimism bias and novelty bias as systematic distortions of likelihood calibration — while GAP-MYO (Salience 3) documented the misanthropic user-blaming syndrome and adversarial imagination deficit as downstream manifestations of the same technical-only professional identity. The developmental prescription was bifocal: DEV-FORM (Salience 3) proposed formal curriculum reform grounded in the philosophical etymology of 'cyber' and 'technology,' while DEV-EXP (Salience 2) endorsed mentored experiential learning as safe failure management, with the structural CISO scapegoat dynamic — short tenure, high compensation, blame-absorber positioning — identified as the organizational barrier that systematically prevents the development of genuine strategic risk management orientation.



**INT-07 · CISO / Director · Higher Education**

## Participant Profile

| Field | Value |
|---|---|
| Interview ID | INT-07 |
| Anonymized Role | CISO / Director of Information Security Services (functionally CISO) |
| Industry | Higher Education / Post-Secondary |
| Organization Size | ~9,000 FTE |
| Region | Canada — Ontario |
| Codebook Applied | LRCC |

## Domain Salience Summary

| Domain | Codes Present | Total Salience | Max Possible | % of Max |
|---|---|---|---|---|
| A — ERF | 3 of 5 | 6 | 20 | 30% |
| B — RGP | 5 of 5 | 12 | 20 | 60% |
| C — COG | 5 of 7 | 11 | 28 | 39% |
| D — GAP | 6 of 7 | 15 | 28 | 54% |
| E — DEV | 3 of 4 | 8 | 16 | 50% |
| **TOTAL** | **22 of 28** | **52** | **112** | **46%** |

**Expectation–Gap Tension Index:** COG Total = 11 | GAP Total = 15 | COG–GAP Spread = −4

**Typological Classification:** High Expectation Case; High Gap Evidence; Moderate Negative Spread — both COG (11/28) and GAP (15/28) clear the ≥10 threshold, placing this case in the High Expectation / High Gap quadrant. The negative spread of four points reflects a participant whose gap observations are more extensively evidenced and theorized than his articulation of cognitive expectations, consistent with a diagnostic and culturally explanatory interview posture. Domain B (Risk Governance & Process) is the strongest domain (60% of maximum), reflecting institutional seniority and direct ownership of post-breach governance redesign.



## Dominant Themes

Five themes derived from Step 5 thematic synthesis and Step 8 Case Abstraction Profile.

- **The participant theorizes business value blindness as a cultural and structural deficit:** GAP-BUS achieved Salience 4 — the only code in the transcript to do so — appearing across five distinct idea units and functioning as the central organizing argument of the gap analysis. INT-07 distinguished this case by not merely observing the gap but theorising its causes through a tri-partite explanatory framework: the autodidactic professional formation tradition of cybersecurity (professionals who trained themselves and developed strong self-reliance norms), an educational heritage grounded in technical rather than business disciplines, and a formal risk training landscape calibrated to financial-sector contexts that does not translate into IT/cybersecurity operational realities. This causal account — cultural, developmental, and curricular simultaneously — constitutes the most fully elaborated gap explanation in the study and distinguishes this case from those in which participants observed business blindness without structural explanation.

- **Technocentric professional output and the 'three pages on technology' formulation:** Two consecutive units substantiated TFR-PROB (Salience 3) through a formulation of unusual diagnostic precision: cybersecurity staff presentations typically devote three pages to technology and half a paragraph to risk analysis, a characterization that captures both the directional imbalance and the magnitude of the gap in a single, vivid phrase. Unit 17 extended this critique spontaneously beyond the question prompt, establishing TFR-PROB as a held conviction rather than a prompted acknowledgment. GAP-MYO (Salience 2) appeared in the specific context of professionals explicitly avoiding available business analysis resources — not because those resources were absent from the institution, but because the professional culture of self-reliance actively discouraged help-seeking from adjacent functions.

- **Reactive governance maturation: breach-triggered executive recalibration:** Domain B (Risk Governance & Process) was the most densely populated domain in the transcript (Salience 12/20, 60%), with all five RGP codes active and RGP-COMP (Salience 3) and RGP-ACC (Salience 3) as the co-dominant governance themes. The most analytically significant governance narrative was the 2023 breach and its institutional consequences: before the breach, executive risk cognition framed cyber threats as comparatively minor relative to physical continuity risks such as weather events; the CBC news story that followed the breach catalyzed a rapid reframing, elevating cyber risk to institutional parity with physical threats. This reactive rather than anticipatory governance maturation — driven by realized event and media exposure rather than prospective risk analysis — raised analytically important questions about whether formal training could achieve comparable cognitive change under non-crisis conditions.



- **Seniority-stratified cognitive expectations that formalize the entry-level gap:** The COG profile (Domain C, Salience 11/28, 39%) was pragmatic and seniority-calibrated rather than architecturally comprehensive. The primary cognitive expectation, COG-LXI (Salience 3), was explicitly conditioned on professional seniority: the organization required a formal likelihood × impact analysis for senior staff but not for junior or entry-level practitioners. This stratification, while pragmatically defensible in an operational context, formally normalized the gap at the entry level rather than addressing it as a developmental target. COG-IMP and COG-OPT (Salience 2) established complementary expectations: the participant required professionals to articulate business impact in time and cost terms and to identify specific trade-offs to enable new investments. COG-HOL and COG-COMP were absent, indicating that holistic architectural thinking and cross-type comparative risk reasoning were not among this participant's stated professional expectations.

- **Dual developmental endorsement and a pointed critique of finance-centric training:** Domain E (Development of Risk Judgment) was the second-strongest domain in proportional terms (Salience 8/16, 50%), with DEV-EXP (Salience 3) and DEV-FORM (Salience 3) co-equal as the primary developmental mechanisms. A candid autobiographical observation anchored DEV-EXP — the participant's experiential background — and he acknowledged this without defensiveness, as the historical norm for the profession. DEV-FORM operated on two distinct registers: optimism about the emerging institutionalization of formal pathways, including SANS certifications and cybersecurity master's programmes, and pointed critique of existing risk training as too finance-sector-oriented and insufficiently adapted to IT/cybersecurity operational realities. The most prescriptive developmental recommendation — training that meets IT professionals where they are, delivered spontaneously beyond the Q17 prompt — prioritized curricular contextualization over training volume. DEV-EXEC was absent, consistent with the INT-05 case and a professional-credentialing rather than culture-change theory of risk judgment development.

### Analytical Paragraph for the INT-07 Interview

INT-07, a Director of Information Security Services functioning in a CISO capacity within a large Canadian higher education institution, operated within a governance architecture that was the most densely populated domain in the transcript (Domain B, Salience 12/20, 60%): all five RGP codes were active, RGP-COMP and RGP-ACC each reached Salience 3, and the governance narrative was organized around a defining institutional event — a 2023 data breach and its subsequent CBC news coverage — that catalysed the elevation of cyber risk to institutional parity with physical threats, demonstrating that executive risk cognition at this institution matured reactively through realized events rather than prospectively through training or anticipatory analysis. The



enterprise risk framing (Domain A, Salience 6/20, 30%) was pluralistic but undominated, drawing on financial, reputational, and survival/continuity frames at equal salience. At the same time, ERF-REG was absent—a notable omission for a Canadian institution with material obligations under FIPPA, PHIPA, and research data regulation. Against that institutional backdrop, the gap domain (Domain D, Salience 15/28, 54%) was organized around GAP-BUS at Salience 4 as the sole organizing diagnostic claim, theorised not as individual deficiency but as the product of a tri-partite structural explanation: autodidactic professional formation, technical rather than business educational heritage, and a formal risk training landscape calibrated to financial-sector contexts that does not translate into cybersecurity operational realities. The participant substantiated TFR-PROB (Salience 3) by noting that staff presentations typically devote 3 pages to technology and only 1/2 a paragraph to risk analysis. Cognitive expectations (Domain C, Salience 11/28, 39%) were pragmatic and seniority-stratified: COG-LXI was the primary expectation, but it was explicitly conditioned on professional seniority, formally normalizing the gap at the entry level rather than addressing it as a developmental target. Developmentally, DEV-EXP and DEV-FORM co-anchored the prescription at Salience 3 each — the latter most forcefully in the spontaneous critique that existing risk training was too finance-sector-oriented and needed to meet IT professionals where they are, a call for curricular contextualization rather than increased training volume, with DEV-EXEC absent and DEV-MULTI endorsed through cross-functional mentorship from adjacent organizational units.



**Appendix E**

NLP Classification Pipeline — Technical Specification

This appendix provides the technical specification of the Natural Language Processing (NLP) pipeline used to classify 2,111 Task, Knowledge, and Skill (TKS) statements from the NIST NICE framework against 29 competency categories. It supplements Section 3.4 of the methodology chapter and provides the operational detail necessary for reproducibility and auditability of the classification results reported in Chapter 4.

<u>**Section 1 — Pipeline Overview**</u>

The pipeline processed all 2,111 TKS statements in the NIST NICE Framework v2.0.0 across four sequential stages, consuming one API call per statement.

| Step | Stage | Description | Output |
|------|-------|-------------|--------|
| 1 | **Taxonomy Construction** | The research design operationalized each NICE competency category as a structured taxonomy entry comprising: a plain-language label; a conceptual definition; a discriminative keyword list; inclusion criteria; exclusion criteria with redirection rules; and priority disambiguation notes. The complete taxonomy was serialized to JSON and injected verbatim into every classification prompt. | JSON taxonomy object (29 entries) |
| 2 | **LLM-assisted NLP classification Classification** | Each TKS statement was submitted to gpt-5-mini-2025-08-07 (OpenAI) via the structured output API (responses.parse), with a Pydantic schema that constrained the model to produce exactly 29 scored outputs per statement. Each output comprised three fields: relevance_score (0.0–1.0), confidence_score (0.0–1.0), and a natural-language justification. The prompt included explicit relevance anchors and disambiguation rules. The pipeline processed fifteen statements concurrently, identifying overlapping nodes in the COG and GAP domains; the script persisted results to a pickle file after each statement to ensure crash recovery and full resumability. | 29 × {relevance, confidence, justification} per statement |
| 3 | **Post-Processing & Cutoff Application** | A 0.45 relevance cutoff was applied uniformly across all 29 categories. The filtering logic set all scores below 0.45 to zero, along with their associated labels and justifications. Scores at or above 0.75 constitute the high-confidence tier used for all primary analysis and reported findings. The pipeline assigned the UNKN code only when no category score met the 0.45 cutoff. Sensitivity testing across the 0.70–0.80 range confirmed that rank ordering of competency categories was stable. | Tiered classification per statement; UNKN where applicable |



| Step | Stage | Description | Output |
|------|-------|-------------|--------|
| 4 | Measurement & Ranking | The pipeline computed the total number of TKS statements classified at each tier for each competency category. A high-confidence (≥ 0.75) statement count determined the category rank, ensuring that only the most reliable data drove the final hierarchy. The RISK category was the primary focus: its rank among the 29 categories, its absolute count, and the presence or absence of probabilistic reasoning terms ("likelihood", "probability", "impact") within classified statements were the primary empirical outputs. | Ranked frequency table; RISK category analysis |

### Section 2 — Classification Output Schema

Each TKS statement received 29 structured output objects (one per competency category), each containing three typed fields enforced by Pydantic schema validation via OpenAI's 'responses.parse' API.

| Field | Type | Description | Role in Analysis |
|-------|------|-------------|------------------|
| relevance_score | Float [0.0–1.0] | The degree to which the TKS statement addresses the competency category. Anchors: 0.0 = irrelevant; 0.25 = tangential; 0.50 = partial; 0.75 = strong; 1.00 = central. | Primary basis for tier assignment and category ranking. |
| confidence_score | Float [0.0–1.0] | Model's certainty in its relevance assessment, independent of the relevance score. Distinguishes "weak but confident" from "weak and uncertain" classifications. | Used in sensitivity analysis; not part of primary tier logic. |
| justification | String | The language model provides a natural-language explanation to justify each assigned relevance score. Records the specific language in the TKS statement that triggered or suppressed classification. | Enables post-hoc auditability of any individual classification decision. |

### Section 3 — Scoring Tier Structure

The pipeline applies a four-tier relevance structure. The classifier assigns discrete labels during classification; post-processing then applies the 0.45 cutoff, eliminating all weak-match scores from reported findings. All primary analysis and reported findings use the high-confidence tier exclusively.

| Tier | Relevance Threshold | Label | Analytical Treatment |
|------|---------------------|-------|----------------------|
| Excluded | < 0.25 | 0.0 (irrelevant) | Score set to 0. The coding logic did not assign the statement to any category because it failed to meet the 0.45 salience threshold. |



| Tier | Relevance Threshold | Label | Analytical Treatment |
|------|---------------------|-------|----------------------|
| **Weak Match** | 0.25 – 0.44 | 0.5 (weak) | Assigned a discrete label of 0.5 by the classifier. Zeroed out by the post-processing cutoff (0.45) and excluded from all reported findings. |
| **Moderate Match** | 0.45 – 0.74 | 0.5 (moderate) | Survives the post-processing cutoff. Retained for supplementary completeness analysis only. Excluded from all primary distribution calculations and reported findings. |
| **High-Confidence** | ≥ 0.75 | 1.0 (strong match) | Used for all primary analysis, category frequency counts, percentage-of-corpus calculations, and RISK category ranking reported in Chapter 4. |
| **Unclassified** | No score ≥ 0.45 across all 29 categories | UNKN | The statement is coded UNKN and excluded from competency frequency analysis. |

### Section 4 — Taxonomy Entry Structure

Each of the 29 competency category entries in the JSON taxonomy comprised six fields. The full taxonomy was injected verbatim into every classification prompt to ensure definitional consistency across all 2,111 TKS statements.

| JSON Field | Description | Function |
|------------|-------------|----------|
| **id/name** | Short code and plain-language label (e.g., "RISK" / "Cyber Risk Management"). | Identifies the category in the structured output schema and in all exported data files. |
| **description** | Conceptual scope of the competency in 2–4 sentences. | Provides the classifier with the intended boundaries of the category. |
| **keywords** | Discriminative keyword list specific to this category. | Guides relevance scoring toward language distinctive to the competency, reducing cross-category spillover. |
| **include_if** | Conditions that must be present for a TKS statement to receive a strong relevance score. | Operationalizes "what counts" as belonging to the category. |
| **exclude_if** | Conditions under which apparent relevance is overridden, with redirection to the more appropriate category. | Prevents multi-domain statements from generating spurious high-confidence assignments. |
| **priority_notes** | Disambiguation rule for statements that could plausibly belong to multiple categories. | Implements the prompt instruction to prefer specific domains over broad ones (e.g., prefer CLOU or DATA over CYBR). |



**Section 5 — Configuration Reference**

Two source files (src/classify.py and src/post_processing.py) define all tunable parameters for the analysis. This appendix records the exact (language) model version to support reproducibility.

| Parameter | Value | Purpose |
|---|---|---|
| **MODEL** | gpt-5-mini-2025-08-07 | The classification pipeline uses an exact versioned model for all calls, ensuring consistent results across the dataset. Version-pinned to ensure reproducibility. |
| **MAX_WORKERS** | 15 | Maximum concurrent classification threads. Peak concurrent API calls: 15. |
| **STRONG_THRESHOLD** | 0.75 | Relevance score boundary for high-confidence classification (label = 1.0). Used in all primary analyses. |
| **WEAK_THRESHOLD** | 0.25 | Relevance score boundary for weak-match classification (label = 0.5). The post-processing cutoff subsequently eliminates statements in this tier. |
| **RELEVANCE_CUTOFF** | 0.45 | Post-processing threshold applied uniformly across all categories. Scores below this value are zeroed out and excluded from all reported findings. |

**Classification Auditability**

Because the pipeline persisted a natural-language justification for every classification decision, researchers can verify any TKS statement's assignment post-hoc by inspecting the recorded justification text. This field documents the specific language in the statement that triggered or suppressed the relevance score, providing a direct audit trail from raw TKS text to category assignment to reported competency frequency.

The post-processing step produces a timestamped export directory containing CSV files and a manifest.json file; this manifest records the exact configuration parameters used to generate the reported findings. The manifest permanently associates the reported findings with the exact pipeline configuration that produced them.

The multi-label design — in which a single TKS statement may receive high-confidence scores in multiple categories simultaneously — reflects genuine multi-domain content rather than classification conflict and eliminates the need for forced single-category tie-breaking rules.